\documentclass[12pt,epsf,amssymb]{article}

\usepackage{graphicx}
\usepackage{amsfonts}
\usepackage{amsmath}
\usepackage[usenames]{color}
\usepackage{amssymb}
\usepackage{amsthm}
\usepackage[mathscr]{eucal} 
\usepackage{url}

\usepackage[all]{xy}

\def\N{\mathbb{N}}
\def\Z{\mathbb{Z}}
\def\Q{\mathbb{Q}}
\def\R{\mathbb{R}}
\def\C{\mathbb{C}}

\begin{document}

\baselineskip 0.6cm
\newcommand{\nequiv}{\mbox{\ooalign{\hfil/\hfil\crcr$\equiv$}}}
\newcommand{\nsupset}{\mbox{\ooalign{\hfil/\hfil\crcr$\supset$}}}
\newcommand{\nni}{\mbox{\ooalign{\hfil/\hfil\crcr$\ni$}}}
\newcommand{\nin}{\mbox{\ooalign{\hfil/\hfil\crcr$\in$}}}

\newcommand{\vev}[1]{ \left\langle {#1} \right\rangle }
\newcommand{\bra}[1]{ \langle {#1} | }
\newcommand{\ket}[1]{ | {#1} \rangle }
\newcommand{\Dsl}{\mbox{\ooalign{\hfil/\hfil\crcr$D$}}}
\newcommand{\Slash}[1]{{\ooalign{\hfil/\hfil\crcr$#1$}}}
\newcommand{\EV}{ {\rm eV} }
\newcommand{\KEV}{ {\rm keV} }
\newcommand{\MEV}{ {\rm MeV} }
\newcommand{\GEV}{ {\rm GeV} }
\newcommand{\TEV}{ {\rm TeV} }

\def\diag{\mathop{\rm diag}\nolimits}
\def\tr{\mathop{\rm tr}}

\def\Spin{\mathop{\rm Spin}}
\def\SO{\mathop{\rm SO}}
\def\O{\mathop{\rm O}}
\def\SU{\mathop{\rm SU}}
\def\U{\mathop{\rm U}}
\def\Sp{\mathop{\rm Sp}}
\def\SL{\mathop{\rm SL}}

\def\change#1#2{{\color{blue}#1}{\color{red} [#2]}\color{black}\hbox{}}

\theoremstyle{definition}
\newtheorem{thm}{Theorem}[subsection]
\newtheorem{defn}[thm]{Definition}
\newtheorem{exmpl}[thm]{Example}
\newtheorem{props}[thm]{Proposition}
\newtheorem{lemma}[thm]{Lemma}
\newtheorem{rmk}[thm]{Remark}
\newtheorem{notn}[thm]{Notation}
\newtheorem{qstn}[thm]{Question}
\newtheorem{excs}[thm]{Exercise}
\newtheorem{cor}[thm]{Corollary}
\newtheorem{anythng}[thm]{}

 \begin{titlepage}
  
% \begin{flushright}
% Madrid-xxx \\
% IPMU15-0097
% \end{flushright}
 
  \vskip 1cm
 \begin{center}
  
  {\large \bf Notes on Gepner Construction}
  
  \vskip 1.2cm
 
  Kanade Nishikawa\footnote{K.N. has graduated from the U. Tokyo
  and is working for a private company.} and Taizan Watari
 
 \vskip 0.4cm
%  
%  {\it $^1$Madrid
% \\[2mm]
% 
%    $^3$
 {\it    Kavli Institute for the Physics and Mathematics of the Universe, 
    University of Tokyo, Kashiwa-no-ha 5-1-5, 277-8583, Japan }
%   }

 \vskip 1.5cm
   
\abstract{In this note, we describe the Gepner construction purely as orbifold (rather than a $\beta$-prescription or orbifold-like process).
This is done by distinguishing orbifolding by a chiral symmetry from that
by a left-right diagonal symmetry, and also by exploiting
an extra degree of freedom available in orbifolds of Type 0 SCFTs. We also
describe by using the Arf theory how the extra freedom in Type 0 SCFT
orbifolds in general (not just in the Gepner construction) should be implemented
in higher genus amplitudes. This note also determines when
a combinatorial data set of a Gepner construction has a mirror Gepner
construction, and also provides two thinking frameworks to determine
the SCFT vertex operators of the Gepner model SCFTs with phase ambiguity
and branch cuts under control. 
} 
 \end{center}
 \end{titlepage}

% \begin{flushright}
% Version \today
% \end{flushright}

\tableofcontents
 
%%%%%%%%%%%%%%%%%%%%%%%%%%%%%%%%%%%%%%%%%%%%%%%%%%%%%
\section{Introduction}
 \label{sec:Intro}
%%%%%%%%%%%%%%%%%%%%%%%%%%%%%%%%%%%%%%%%%%%%%%%%%%%%%

The Gepner construction is a procedure to specify a modular invariant
SCFT for each choice of combinatorial data of certain
kinds \cite{Gepner:1987vz};
the SCFTs constructed that way are called the Gepner models. 
It is often regarded that the construction is well-understood and
well-documented, at least by the standard of the string theory community. 
The present authors were not fully satisfied by the expositions that they
looked at, however. One of the purposes of this article is to provide
a brushed-up presentation of the construction. 

There are three aspects in a large fraction of existing literatures
on the Gepner construction that the present authors intend to improve.
One is the fact that the construction is referred to in the literatures
as the $\beta$-prescription, the method of simple currents, or an orbifold-like
procedure, but not as orbifold (e.g.,
\cite[p.769]{Gepner:1987qi}, %% Gepner spacetime supersymmetry
\cite[p.209]{Recknagel:1997sb}, %% Reck Schom
\cite[\S5.6]{Blumenhagen:2009zz}, \cite[p.270]{Eguchi:2015book}).
We observe that this confusion follows
from mixing up orbifolding by chiral symmetry and left-right diagonal
symmetry (the 1st paragraph of \S \ref{ssec:from0B-toII}). In this article,
we present the Gepner construction as orbifold, by partly following
the observation in Ref. \cite[App. A]{Fuchs:1991vu},
by separating the orbifolding by diagonal symmetries and chiral symmetries,
and by adding a little more conceptual clarification on combinatorial
data for the construction ({\it Conceptual clarification} in
pp. \pageref{pg:cc-start}--\pageref{pg:cc-end} and
{\it Notes} in p. \pageref{pg:notes}). 
The other is the lack of a systematic construction algorithm applicable
to all the choices of the combinatorial data that are regarded as descent
in one article or another; one algorithm ($\beta$-prescription) covers
certain choices of the combinatorial data, while another algorithm
(orbifold-like) is for other choices. This issue is also resolved in the
presentation in this article. That is partly due to separation of the
chiral and diagonal orbifoldings; the other vital element turns out
to be the extra freedom (besides the discrete torsion)
available in orbifolds of Type 0 SCFTs \cite{Intriligator:1990ua}
(see an {\it observation} in
pp. \pageref{pg:IV-st}--\pageref{pg:IV-fin}, {\it Reinterpretation}
in
pp. \pageref{pg:reintrp-AfrPhase-k=0-st}--\pageref{pg:reintrp-AfrPhase-k=0-fin}
and {\it Notes} in p. \pageref{pg:notes} in \S \ref{ssec:from0B-toII}).
We also describe how this extra freedom should be implemented
systematically in higher genus amplitudes of a Type 0 SCFT orbifold
by using the Arf theory as language. 

The construction in this article does not deal exclusively with the SCFTs
that can be used for Type II string compactification with spacetime
supersymmetry, so a broader class of SCFTs is covered. 
This not only reflects a change in the scientific motivations since late
1980's. By studying the condition for spacetime supersymmetry only at
a later stage in construction, one can separate the orbifolding
by a left-right diagonal symmetry from that by a chiral symmetry.
This contributes to the conceptual
clarification referred to already, and also to distinguishing
which conditions on a set of combinatorial data for Gepner construction
are for theoretical consistency or for spacetime supersymmetry.
Section \ref{ssec:Gepn-Ncrit-0SCFT} explains 
how a Type 0 SCFT is given for any choice of combinatorial data\footnote{
Although some Gepner models are easily interpreted \cite{Witten:1993yc}
as special limits of non-linear-sigma-model SCFTs, not all the Gepner
model SCFTs may have such an interpretation. We will stick to purely
algebraic treatment of the 2D SCFTs in this article. 
} % 
set 
$\{ (k_i, R_i)\}_{i=1,\cdots,r}$, $\Gamma \subset {\cal G}$, $\epsilon$
and $p_0$ (see p. \pageref{pg:Gepner-cmb-daga-list}); let ${\cal S}_0$ be
the set of all possible choices of such a combinatorial data set. 
For the Type 0 SCFT to be converted to a Type II SCFT with the left-mover
and right-mover GSO projections given by the U(1) symmetry of the left-moving
and right-moving $N=2$ superconformal algebras, the combinatorial data
has to be in the subset ${\cal S}_{\rm II} \subset {\cal S}_0$
characterized by the extra conditions (\ref{eq:cond-in-type0-2bb},
\ref{eq:def-fcn-y0}, \ref{eq:cond-for-specFlow23-inGepner-A},
\ref{eq:cond-Gepner-survive-specBy1-2}), as we will see in
section \ref{ssec:from0B-toII}.

The second purpose of this article concerns mirror symmetry. 
When a class of combinatorial data (or geometric data) provides a
construction of SCFTs, one may ask whether their mirror SCFTs
are also constructed by the same class of combinatorial data or not.
If they are, what is the map from the set of combinatorial
(or geometric) data to itself?
We will see in section \ref{sec:mirror} that the mirror of
the Gepner model for a choice in ${\cal S}_0$ is {\it not}
necessarily the Gepner model of any choice in ${\cal S}_0$. 
It is only for a choice in ${\cal S}_{(\ref{eq:rltn-y0-y0vee})} \subset {\cal S}_0$,
where the subset ${\cal S}_{(\ref{eq:rltn-y0-y0vee})}$ is characterized by
the condition (\ref{eq:rltn-y0-y0vee}),  
that the corresponding Gepner model has a mirror Gepner construction
for some choice in ${\cal S}_0$. The subset ${\cal S}_{\rm II}$ is contained in
${\cal S}_{(\ref{eq:rltn-y0-y0vee})}$, and the mirror Gepner data of
a choice in ${\cal S}_{\rm II}$ [resp. ${\cal S}_{(\ref{eq:rltn-y0-y0vee})}$]
is found within ${\cal S}_{\rm II}$ [resp. ${\cal S}_{(\ref{eq:rltn-y0-y0vee})}$].
A systematic algorithm to determine a mirror set of Gepner data in
${\cal S}_{(\ref{eq:rltn-y0-y0vee})}$ is also found
(\ref{def:Union-of-Gamma-y-NS},
\ref{eq:mrr-discTrs-a}--\ref{eq:mrr-discTrs-d}) in section \ref{sec:mirror}.

The third purpose of this article is about the SCFT vertex operator
of the Gepner model SCFTs. It has been said that the SCFT vertex operator
should be determined by bosonization, spectral flow, or simple current
method by using the SCFT vertex operator of the minimal tensor model,
and yes indeed, they are the right directions to go. Very few papers
go beyond those slogans, however, to write down a precise enough thinking
framework to the extent that we can deal with the complex phases and branch
cuts of the SCFT vertex operators with confidence.\footnote{
The present authors hope only to maintain the level of rigor desirable
in the string theory community (not in the VOA community) in this article. 
The presentation in section \ref{ssec:CFTop-by-bos-lattice} will surely
not meet the quality in the VOA community so that it may be regarded as
a blueprint for reformulation by professional mathematicians. 
} %
Sections \ref{ssec:CFTop-by-sc-ext} and \ref{ssec:CFTop-by-bos-lattice}
intend to offer such thinking frameworks, with the simple-current method
and U(1)--parafermion decomposition, respectively.  Both are applicable
to the Type 0 SCFTs with the data in ${\cal S}_0$, not just those in
${\cal S}_{\rm II}$. The latter (in section \ref{ssec:CFTop-by-bos-lattice})
is complete, however, only for minimal tensor models with $\{ (k_i, R_i)\}$
where all $k_i$'s are odd; a loose end remains in the cases where the
classical {\it simple current fixed-point resolution} problem is relevant. 
Motivation for this study is described in a little more
detail in the first three pages of section \ref{sec:CFTop}. Section 
\ref{ssec:CFTop-by-sc-ext} has its summary (with reference to
key equation and conditions) in its last paragraph. 
A summary of section \ref{ssec:CFTop-by-bos-lattice} is found
in its opening one paragraph and the last 1.5 pages.

{\bf For busy readers:} Sections \ref{sec:Gepn-Data},
\ref{sec:mirror} and \ref{sec:CFTop} are for the three purposes/questions
described above. Details in one section are often irrelevant in another.
Sections \ref{ssec:CFTop-by-sc-ext} and \ref{ssec:CFTop-by-bos-lattice}
are also independent from one another.
Those who are familiar with the Gepner construction might jump directly
to the formulae or discussion on subtleties in
sections \ref{sec:MTMc-Type0}--\ref{sec:Gepn-Data} that have already been
referred to in this Introduction.

Those who are not interested in Gepner construction might still read
the discussion in {\it observation} in
pp. \pageref{pg:IV-st}--\pageref{pg:IV-fin} and
{\it Reinterpretation} in
pp. \pageref{pg:reintrp-AfrPhase-k=0-st}--\pageref{pg:reintrp-AfrPhase-k=0-fin}
as a general statement on orbifold theory of 2D SCFT.
Although section \ref{ssec:CFTop-by-sc-ext} is written for Gepner
construction, it will not be difficult to read it for a general abelian
orbifold of 2D (non-holomorphic) SCFT.

%%%%%%%%%%%%%%%%%%%%%%%%%%%%%%%%%%%%%%%%%%%%%%%%%%%%%
\section{Minimal Tensor Model Critical Type 0B SCFTs}
\label{sec:MTMc-Type0}
%%%%%%%%%%%%%%%%%%%%%%%%%%%%%%%%%%%%%%%%%%%%%%%%%%%%%

Section \ref{ssec:N=2mm-SCFT} only serves the purpose of
quoting well-known facts and setting notations in this article. 
We begin in section \ref{ssec:mtm-crit-SCFT} with reviewing the
description of the $N=2$ minimal tensor model {\it critical} SCFTs
in \cite[App. A]{Fuchs:1991vu}. The presentation in
Ref. \cite[app. A]{Fuchs:1991vu} is based on discrete torsion
introduced in \cite{Vafa:1986wx}
(cf \cite{Sharpe:1999pv}, \cite{Sharpe:1999xw}, \cite{Sharpe:2000ki}), 
so that we have a theoretical foundation for consistency even for higher-genus
amplitudes \cite{Vafa:1986wx}. We also observe that combinatorial data
for the Gepner construction have not been given appropriate mathematical
characterization, and that that was why the literature has avoided
referring to the Gepner construction as an orbifold; so we propose
in section \ref{ssec:mtm-crit-SCFT} a characterization for
the combinatorial data that we think is appropriate.

%%%%%%%%%%%%%%%%%%%%%%%%%%%%%%%%%%%%%%%%%%%%%%%%%%%%
\subsection{A Brief Summary on Type 0B Minimal Model SCFTs}
\label{ssec:N=2mm-SCFT}
%%%%%%%%%%%%%%%%%%%%%%%%%%%%%%%%%%%%%%%%%%%%%%%%

{\bf Vertex operator superalgebra:}
There is a series of vertex operator superalgebras (SVOAs) labeled
by $k=1,2,\cdots$, called the $N=2$ {\it minimal model SVOA of level-}$k$.
In the data $(V, Y, {\bf 1}, \omega)$ of any one of the SVOAs in this series,
the operator algebra contains an $N=2$ superconformal algebra (SCA),
and all the states in $V$ are generated on top of the vacuum state
${\bf 1} \in V$ by the oscillator modes of the current operators of the
$N=2$ SCA. The central charge of the Virasoro algebra is
\begin{align}
  c = \frac{3k}{k+2}.
  \label{eq:min-m-VirC}
\end{align}
We will also use the notation $\bar{k}:=k+2$.
The $\Z/2\Z$-grading-even part of the SVOA is known to be the
coset VOA 
\begin{align}
  \hat{\mathfrak{su}}(2)_k \otimes \hat{\mathfrak{so}}(2)_1 / \U(1)_{k+2};
  \label{eq:coset-VOA}
\end{align}
here, $\hat{\mathfrak{g}}_k$ stands for the Affine $\mathfrak{g}$ Lie
algebra of level-$k$. The $\hat{\mathfrak{so}}(2)_1$ VOA is the same as
the $\U(1)_2$ VOA, and the $\U(1)_N$ VOA is the same as the
lattice-$\langle +2N \rangle$ VOA for all $N \in \N_{>0}$.

The NS-type (i.e., untwisted) irreducible representations
and the Ramond-type (i.e., SVOA-$\Z/2\Z$-twisted) irreducible
representations of the $N=2$ minimal model SVOA of level-$k$ are labeled by
\begin{align}
  {\cal F}_{N}^{k.{\rm fnd}} & \; := \left\{ 
  (\ell, m) \; | \; \ell \in \{ 0, 1,\cdots, k\}, \quad
  m \in \{ -\ell, -\ell+2,\cdots, \ell-2,\ell\} \right\},  \; 
  \label{eq:N=2mm-list-repr} \\
  {\cal F}_R^{k.{\rm fnd}} & \; := \left\{ (\ell,m) \; | \;
        \ell \in \{0,1,\cdots, k\}, \quad
        m \in \{-\ell+1, -\ell+3,\cdots, \ell-1, \ell+1 \} \right\}.
   \nonumber      
\end{align}
The representation space $M_{\ell,m}^{NS/R}$ is decomposed into
\begin{align}
  M_{\ell,m} & \; := M^{NS}_{\ell,m} = {\cal L}^\ell_{m,0} \oplus {\cal L}^\ell_{m,2},
  \qquad (\ell,m) \in {\cal F}_{N}^{k.{\rm fnd}}
  \label{eq:N=2mm-SVOArep-VOArep-rltn} \\
  M_{\ell,m} & \; := M^{R}_{\ell,m} = {\cal L}^\ell_{m,1} \oplus {\cal L}^\ell_{m,3},
  \qquad (\ell,m) \in {\cal F}_R^{k.{\rm fnd}}
  \nonumber 
\end{align}
where ${\cal L}^\ell_{m,s}$ is the irreducible representation space of the
coset VOA (\ref{eq:coset-VOA}) that fits into the following decomposition:
\begin{align}
  M_\ell^{\hat{\mathfrak{su}}(2)_k} \otimes W_{-s}^{\U(1)_2} \cong
  \oplus_{m \in \Z/2\bar{k}\Z}^{(\ell + m + s \in 2\Z)}
  \left( {\cal L}^\ell_{m,s} \otimes W_{-m}^{\U(1)_{\bar{k}}} \right), \qquad 
  \ell \in \{ 0,1,\cdots, k\}, \quad s \in \Z/4\Z;
  \label{eq:dcmp-N=2-minimalM-SVOA-repr-2-bosCst-reprs} 
\end{align}
$M_\ell^{\hat{\mathfrak{su}}(2)_k}$ is the spin-$(\ell/2)$ irreducible
representation of the Affine
Lie algebra $\hat{\mathfrak{su}}(2)_k$ (with the range $\ell \in \{ 0,1,
\cdots, k\}$), and $W_s^{\U(1)_2}$ and $W_m^{\U(1)_{\bar{k}}}$
(with $s \in \Z/4\Z$ and $m \in \Z/2\bar{k}\Z$) are the irreducible
representations of the $\U(1)_2$ and $\U(1)_{\bar{k}}$ VOAs, respectively.
For $m \in \Z/2\bar{k}\Z$ not in the range of (\ref{eq:N=2mm-list-repr})
even modulo $+2\bar{k}\Z$, we should read the representation
${\cal L}^\ell_{m,s}$ of the VOA (\ref{eq:coset-VOA}) in
the decomposition (\ref{eq:dcmp-N=2-minimalM-SVOA-repr-2-bosCst-reprs}) as  
\begin{align}
 {\cal L}^\ell_{m,s} \cong {\cal L}^{k-\ell}_{\bar{k}+m,s+2}. 
\label{eq:field-id}
\end{align}

The ground state of the irreducible representation $M_{\ell,m}$
is in ${\cal L}^\ell_{m,0} \subset M_{\ell,m}$ for $(\ell,m) \in
{\cal F}_{N}^{k.{\rm fnd}}$, and in ${\cal L}^{\ell}_{\ell+1,1} \subset
M_{\ell,\ell+1}$ for $(\ell,m) = (\ell,\ell+1) \in {\cal F}_R^{k.{\rm fnd}}$. 
For all other $(\ell,m)$ in ${\cal F}_{R}^{k.{\rm fnd}}$, one state
in ${\cal L}^\ell_{m,1}$ and another in ${\cal L}^\ell_{m,3}$ forms a degenerate
pair of ground states of $M_{\ell,m}$. Their conformal weights ($L_0$-eigenvalue)
and the U(1) charge ($J_0$-eigenvalue) are
\begin{align}
  h^\ell_{m,s} = \frac{\ell(\ell+2)-m^2}{4\bar{k}} + \frac{s^2}{8}, \qquad
  q^\ell_{m,s} = \frac{m}{\bar{k}} - \frac{s}{2},
\label{eq:h&q-N=2mm-each-VOA-repr}
\end{align}
where we should use $\ell,m$ in (\ref{eq:N=2mm-list-repr}) and
$s$ in the range $\{-1,0,1,2\}$ (than in $\Z/4\Z$). All the states
in ${\cal L}^\ell_{m,s}$ have conformal weights and the U(1) charges
that are $(h^\ell_{m,s}+\Z, q^\ell_{m,s} + 2\Z)$. More information
is available from the character formula in \cite{Matsuo:1986cj}. 

The isomorphism (\ref{eq:field-id}) motivates to introduce
a representation $M_{\ell,m}$ of the $N=2$ minimal model SVOA
for the labels $(\ell,m)$ in a double cover, for example, 
% [memo: alternative $\ell-2\bar{k}+2 \leq m \leq \ell$ and
% $\ell -2\bar{k}+3 \leq m \leq \ell+1$]
%
\begin{align}
{\cal F}_{\text{NS}}^{k.{\rm dbl}}& \; := \{(\ell,m)|\ell,m\in\Z,0\leq \ell \leq k, -(k+2)<m\leq k+2, \;\; \ell +m \in 2\Z\}\\
{\cal F}_{\text{R}}^{k.{\rm dbl}}& \; := \{(\ell, m)|\ell,m\in\Z,0\leq \ell \leq k, -(k+2)<m\leq k+2, \;\;  \ell+m \in 1+2\Z\},
\end{align}
or equivalently, 
\begin{align}
 {\cal F}_{NS}^{k.{\rm dbl}'} & := 
 \left\{ (\ell,m) \; | \; \ell \in \{ 0,1,\cdots, k\}, \; m \in \Z/2(k+2)\Z, \; \ell + m \in 2\Z \right\}, \\
 {\cal F}_{R}^{k.{\rm dbl}'} & := 
 \left\{ (\ell,m) \; | \; \ell \in \{ 0,1,\cdots, k\}, \; m \in \Z/2(k+2)\Z, \; \ell + m \in 1+ 2\Z \right\}. 
\end{align}
The representation $M_{\ell,m}$ for $(\ell,m)$ in ${\cal F}_{NS/R}^{k.{\rm dbl}}$
but not in ${\cal F}_{NS/R}^{k.{\rm fnd}}$ are set equal to $M_{k-\ell,m\pm (k+2)}$
where either $(k-\ell, m + (k+2))$ or $(k-\ell, m-(k+2))$ must be found 
in ${\cal F}_{NS/R}^{k.{\rm fnd}}$. With this convention understood, the coset-VOA 
decomposition (\ref{eq:dcmp-N=2-minimalM-SVOA-repr-2-bosCst-reprs}) holds true
for the whole $\ell,m$ in ${\cal F}_{NS/R}^{k.{\rm dbl}}$. 

The characters
\begin{align}
  \chi^\ell_{m,s}(\tau) := \left\{ \begin{array}{ll}
    {\rm Tr}_{{\cal L}^\ell_{m,s}} \left[
      q^{L_0-\frac{c}{24}} \right], & \ell+m+s \in 2\Z, \\
   0 & \ell+m+s \in 1 +2\Z, \end{array} \right.
    \qquad \quad
  \chi^{\ell}_{m,s}(\tau) = \chi^{k-\ell}_{m+ \bar{k},s+2}(\tau), 
\end{align}
where $q := e^{2\pi i \tau}$, transform under ${\rm SL}_2\Z$ 
linear fractional transformations on $\tau$ as
\begin{align}
  \chi^\ell_{m,s}(\tau+1) & \; = e^{2\pi i \left( h_{m,s}^\ell - \frac{c}{24} \right)}
  \chi^\ell_{m,s}(\tau), \\
  \chi^\ell_{m,s}(-1/\tau) & \; = S^{(\mathfrak{su}(2),k)}_{\ell,\ell'}
  S^{(\U(1),2)}_{s,s'} (S^{(\U(1),k+2)}_{m,m'})^{cc} \; \chi^{\ell'}_{m',s'}(\tau),
\end{align}
where the superscript
${}^{cc}$ stands for complex conjugation, and
% $h^s_{\ell,m}$ for the mod-$+\Z$ value of the conformal weights
% in (\ref{eq:h&q-N=2mm-each-VOA-repr}).
% The
the matrices $S^{(\mathfrak{su}(2),k)}$, $S^{(\U(1),2)}$ and $S^{(\U(1),k+2)}$ are 
\begin{align}
  S^{(\mathfrak{su}(2),k)}_{\ell,\ell'} & \; := \sqrt{\frac{2}{k+2}} \sin
  \left( \pi \frac{(\ell+1)(\ell'+1)}{k+2} \right),  \qquad
    \ell, \; \ell' \in \{ 0,1,\cdots, k\}, \\
  S^{(\U(1),N)}_{x,x'} & \; := \frac{1}{\sqrt{2N}}
    \left( e^{-2\pi i \frac{xx'}{2N}} \right),  \qquad x,x' \in \Z/(2N\Z).
\end{align}
The sum is over the range $s' \in \Z/4\Z$ and $m' \in \Z/(2\bar{k}\Z)$,
and $\ell' \in \{0,1,\cdots, k\}$. 

The characters $\chi^\ell_{m,s}(\tau)$ are for irreducible representations
of the VOA (\ref{eq:coset-VOA}); we will also use the characters of
the irreducible representations of the $N=2$ minimal model SVOA later. 
\begin{align}
  {\rm ch}_{\ell,m} := {\rm Tr}_{M_{\ell,m}}[q^{L_0 - c/24}],
  \qquad {\rm ch}_{\ell,m}^{-} := \left\{ \begin{array}{ll}
    \ell-m: \quad {\rm even}: & \chi^\ell_{m,0}-\chi^\ell_{m,2}, \\
    \ell-m: \quad {\rm odd}: & \chi^\ell_{m,1}-\chi^\ell_{m,3}. 
    \end{array} \right. 
\end{align}
Note that ${\rm ch}_{k-\ell,m+\bar{k}}^- = - {\rm ch}_{\ell,m}^-$. 

\vspace{3mm}

{\bf Type 0 Superconformal Field Theory:}
Fix some positive integer $k$, and one may think of listing up all
the modular invariant and unitary Type 0 SCFTs\footnote{
\label{fn:SCFT-min}
We mean by an {\it SCFT} a set of data including
$({\cal H}_{\rm tot}^{\rm pre}, Y_{\rm tot}, {\bf 1}\tilde{\bf 1}, \omega,
\tilde{\omega})$, where ${\cal H}_{\rm tot}^{\rm pre}$ is the fermionic
(or pre-gauged) Hilbert space,
$Y_{\rm tot}: {\cal H}_{\rm tot}^{\rm pre} \rightarrow
{\rm End}({\cal H}_{\rm tot}^{\rm pre})
[[z,z^{-1},\bar{z},\bar{z}^{-1}, (z\bar{z})^\R]]$ the state-operator
correspondence, ${\bf 1}\tilde{\bf 1} \in {\cal H}_{\rm tot}^{\rm pre}$
the vacuum state, and $\omega \in {\cal H}_{\rm tot}^{\rm pre}$
[resp. $\tilde{\omega} \in {\cal H}_{\rm tot}^{\rm pre}$] the state s.t.
$Y_{\rm tot}(\omega)$ [resp. $Y_{\rm tot}(\tilde{\omega})$]
is the left-moving [resp. right-moving] energy-momentum tensor current. 
The left-moving [resp. right moving] superchiral algebra ${\cal A}$
[resp. $\widetilde{\cal A}$], which consists of states in
${\cal H}_{\rm tot}^{\rm pre}$ where $\tilde{L}_0 = 0$ [resp. $L_0 = 0$], 
should be an $N\geq 1$ SVOA, and ${\cal H}_{\rm tot}^{\rm pre}$ is the direct
sum of untwisted representations of ${\cal A} \otimes \widetilde{\cal A}$.
Its Type 0 (or post-gauged) SCFT, whose Hilbert space is denoted by
${\cal H}_{\rm tot}^{\rm post}$
is the one obtained by gauging the diagonal fermion parity $(-1)^{F_L-F_R}$
from the fermionic theory with ${\cal H}_{\rm tot}^{\rm pre}$; it is not that
a chiral fermion parity is gauged, so there is no NS--R or R--NS sector in
the Type 0 SCFT ${\cal H}_{\rm tot}^{\rm post}$. 
The $\Z/2\Z$-even part of the SVOA ${\cal A}$ [resp. $\widetilde{\cal A}$]
is ${\cal A} \cap {\cal H}_{\rm tot}^{\rm post}$ [resp. 
  $\widetilde{\cal A} \cap {\cal H}_{\rm tot}^{\rm post}$].
One may also prepare a vector space ${\cal H}_{\rm tot}^{\rm dbl}$ so that
it contains both ${\cal H}_{\rm tot}^{\rm pre}$ and ${\cal H}_{\rm tot}^{\rm post}$
and the representation of the superchiral algebras
${\cal A}$ and $\widetilde{\cal A}$ is manifest. 

We say that an SCFT is {\it modular invariant} when the partition function
of ${\cal H}_{\rm tot}^{\rm post}$ is modular invariant. 
} %
whose left-moving
and right-moving superchiral algebras ${\cal A}$ and $\widetilde{\cal A}$
both contain the $N=2$ level-$k$
minimal model SVOA. The study of the list of such SCFTs has been
pioneered, for example, by \cite{Gepner:1987vz} and  \cite{Qiu:1987ux},
and completed in \cite{Gannon:1996hp}. 

There is a series of such Type 0 SCFTs, called the $A_n$-series;
one Type 0 SCFT in this series is found for any $k=1,2,\cdots$,
with the relation $\bar{k}=n+1$.
There is another series of such Type 0 SCFTs, called the $D_n$-series;
one Type 0 SCFT in this series is found for any even $k \geq 4$. 
The relation is $\bar{k}=2n-2$. Besides the two series of Type 0 SCFTs,
there are Type 0 SCFTs named 
$E_6$ (where $\bar{k}=12$),
$E_7$ (where $\bar{k}=18$) and
$E_8$ (where $\bar{k}=30$), respectively. 
All the unitary and modular invariant Type 0 SCFTs listed
in \cite{Qiu:1987ux} are obtained as orbifolds of those
$A_n$, $D_n$ or $E_r$ Type 0 SCFTs. 

The Type 0 SCFT in the $A_{k+1}$ series has the following Hilbert space
${\cal H}_{\rm tot}^{\rm post}$,
\begin{align}
  ({\cal H}_{tot}^{\rm post})^{\oplus 2} & \; =
  \left( \oplus_{\ell=0}^k
  \oplus_{m \in \Z/2\bar{k}\Z}^{(\ell-m \;{\rm even})} \oplus_{s=0,2}^{\in \Z/4\Z}
  {\cal L}_{m,s}^\ell \otimes \widetilde{\cal L}_{m,s}^\ell \right) \oplus
  \left( \oplus_{\ell=0}^k
  \oplus_{m \in \Z/2\bar{k}\Z}^{(\ell - m \;{\rm odd})}\oplus_{s=1,3}^{\in \Z/4\Z}
        {\cal L}_{m,s}^\ell \otimes \widetilde{\cal L}_{m,s}^\ell \right) ,
\end{align}
whose associated other Hilbert spaces are 
\begin{align}
  ({\cal H}_{tot}^{\rm pre})^{\oplus 2} & \; = \left( \oplus_{\ell=0}^k
  \oplus_{m \in \Z/2\bar{k}\Z}^{(\ell - m \;{\rm even})}
  M_{\ell,m} \otimes \widetilde{M}_{\ell,m} \right), \\
  ({\cal H}_{tot}^{\rm dbl})^{\oplus 2} & \; = ({\cal H}_{tot}^{\rm pre})^{\oplus 2}
  \oplus
  \left( \oplus_{\ell=0}^k
  \oplus_{m \in \Z/2\bar{k}\Z}^{(\ell - m \;{\rm odd})}
  M_{\ell,m} \otimes \widetilde{M}_{\ell,m} \right) .   \nonumber 
\end{align}
Here, the representation
spaces $\widetilde{M}_{\ell,m}$ and $\widetilde{\cal L}_{m,s}^\ell$ are those
of the right-mover (i.e. anti-holomorphic) $N=2$ level-$k$ minimal model SVOA,
and the right-mover coset VOA (\ref{eq:coset-VOA}), respectively. 

In any of the Type 0 SCFTs in the $D_n$ series and the $E_r$ series,
the representation space is of the form
\begin{align}
  ({\cal H}_{tot}^{\rm post})^{\oplus 2} & \; = \left( \oplus_{\ell,\tilde{\ell}=0}^k
  \oplus_{m \in \Z/2\bar{k}\Z} \oplus_{s \in \Z/4\Z}^{(\ell - m-s \;{\rm even})}
  ({\cal L}^\ell_{m,s} \otimes \widetilde{\cal L}^{\tilde{\ell}}_{m,s}
  )^{\oplus L_{\ell,\tilde{\ell}}^{R}} \right) , 
\end{align}
where the integer-valued matrix $L^{R}_{\ell,\tilde{\ell}}$ is
\begin{align}
  R = D_n, \quad n \equiv 0 \; {\rm mod~}2: & \quad 
  L^{R}_{\ell,\tilde{\ell}} = \delta_{\ell,0}^{{\rm mod}2}
   \left( \delta_{\ell,\tilde{\ell}} + \delta_{\tilde{\ell},2n-4-\ell}\right), \\
   R = D_n, \quad n \equiv 1 \; {\rm mod~}2:  & \quad
   L^{R}_{\ell,\tilde{\ell}} =
   \delta_{\ell,0}^{{\rm mod}2} \delta_{\ell,\tilde{\ell}}
   + \delta_{\ell,1}^{{\rm mod}2} \delta_{\tilde{\ell},2n-4-\ell},    
\end{align}
and
\begin{align}
  L^{E_6}_{\ell,\tilde{\ell}} & \; =
  (\delta_{\ell,0}+\delta_{\ell,6})
  (\delta_{\tilde{\ell},0}+\delta_{\tilde{\ell},6}) +
  (\delta_{\ell,3}+\delta_{\ell,7})
  (\delta_{\tilde{\ell},3}+\delta_{\tilde{\ell},7}) +
  (\delta_{\ell,4}+\delta_{\ell,10})
  (\delta_{\tilde{\ell},4}+\delta_{\tilde{\ell},10}), \\
  L^{E_7}_{\ell,\tilde{\ell}} & \; = 
  (\delta_{\ell,0}+\delta_{\ell,16})
  (\delta_{\tilde{\ell},0}+\delta_{\tilde{\ell},16}) +
  (\delta_{\ell,4}+\delta_{\ell,12})
  (\delta_{\tilde{\ell},4}+\delta_{\tilde{\ell},12}) +
  (\delta_{\ell,6}+\delta_{\ell,10})
  (\delta_{\tilde{\ell},6}+\delta_{\tilde{\ell},10}) \nonumber \\
  & \qquad + \delta_{\ell,8}\delta_{\tilde{\ell},8}
  + \delta_{\ell,8}(\delta_{\tilde{\ell},2}+\delta_{\tilde{\ell},14})
  + (\delta_{\ell,2}+\delta_{\ell,14}) \delta_{\tilde{\ell},8}, \\ 
  L^{E_8}_{\ell,\tilde{\ell}} & \; =
  (\delta_{\ell,0}+\delta_{\ell,10}+\delta_{\ell,18}+\delta_{\ell,28})
  (\delta_{\tilde{\ell},0}+\delta_{\tilde{\ell},10}+\delta_{\tilde{\ell},18}
  +\delta_{\tilde{\ell},28}) \nonumber \\
 & \qquad  +   (\delta_{\ell,6}+\delta_{\ell,12}+\delta_{\ell,16}+\delta_{\ell,22})
  (\delta_{\tilde{\ell},6}+\delta_{\tilde{\ell},12}+\delta_{\tilde{\ell},16}
  +\delta_{\tilde{\ell},22}). 
\end{align}
Note that $L^{R}_{\ell,\tilde{\ell}}=0$ for any pair
with an odd $\ell - \tilde{\ell}$. Those Type 0 SCFTs,
where the left-moving and right-moving superchiral algebras
contain the $N=2$ minimal model SVOA of level-$k$,
are labeled by the data $(k, R)$; $R$ is one of
$A_n$, $D_n$ and $E_r$. 

%%%%%%%%%%%%%%%%%%%%%%%%%%%%%%%%%%%%%%%%%%%%%%%
\subsection{Minimal Tensor Model Critical Type 0B SCFTs}
\label{ssec:mtm-crit-SCFT}
%%%%%%%%%%%%%%%%%%%%%%%%%%%%%%%%%%%%%%%%%%%%%%%%%

By combining multiple modular invariant Type 0 $N=2$ minimal model SCFTs
with the flat-space-target non-linear sigma model (NLSM) SCFT, it is
possible to construct a modular invariant Type 0 SCFT with the
critical\footnote{
Such a Type 0 SCFT becomes a string vacuum by combining it with
the $\R^{1,1}$-target NLSM SCFT and the $bc\beta\gamma$ ghost system. 
} %
central charge $(c,\tilde{c}) = (12,12)$.
We will review the construction of the critical Type 0 SCFT by
almost following \cite[app. A]{Fuchs:1991vu} and adding conceptual
clarification when necessary; we will also see that the construction is
interpreted as an orbifold.  We will call those Type 0 SCFTs as $N=2$
{\it minimal tensor model critical Type 0 SCFTs} in this article; the word
{\it critical} is used here to imply that the flat-space-target NLSM SCFT
is included to fill the necessary Virasoro central charges $(12,12)$ in
the non-light-cone sector of superstring theory. 

{\bf Flat-space-target NLSM SCFT:} For the purpose stated above, think
of a set of data $\{(k_i, R_i)\}_{i=1,\cdots, r}$ such that 
\begin{align}
  \hat{c} \in \{1,2,3,4\},   \label{eq:cond-cpx-dim-Z}
\end{align}
where 
\begin{align}
  \hat{c} := \sum_{i=1}^r \frac{c_i}{3} = \sum_{i=1}^r \frac{k_i}{k_i+2}. 
\end{align}
The remaining central charges $12-3\hat{c}$ is provided by
the $\R^{8-2\hat{c}}$-target NLSM SCFT.
The characters of the $\R^{8-2\hat{c}}$-target NLSM SVOA are of the form of 
\begin{align}
  \frac{q^{-\frac{p^2}{2}}}{\eta^{8-2\hat{c}}(\tau)} \; \chi_{s_0}(\tau),
   \qquad p \in \R^{8-2\hat{c}}, \quad s_0 \in G_{D_{4-\hat{c}}},  
\end{align}
where $\chi_{s_0}(\tau)$ is due to the excited states of
free fermion oscillators, and the other factors due to those
of free boson oscillators. Here, 
\begin{align}
 G_{D_n} & \; := \left\{ \begin{array}{ll}
     n \; {\rm even}: & \Z/2\Z \times \Z/2\Z = \{ 0, {\rm sp},  {\rm sp}', {\rm vct} \}, \qquad {\rm sp}+{\rm sp}' = {\rm vct}, \\
     n \; {\rm odd}: & \Z/4\Z = \{ 0, \; {\rm sp}, \; {\rm vct}, \; {\rm sp}' \}, \qquad {\rm vct} = 2 {\rm sp}, \; {\rm sp}' = 3 {\rm sp},
 \end{array} \right.
 \label{eq:disc-grp-Dn}
\end{align}
is the discriminant group and 
\begin{align}     
 \left\{ \begin{array}{ll}
   n \; {\rm even}: &  
   ({\rm sp}, {\rm sp}) = ({\rm sp}', {\rm sp}') = \frac{n}{4} + \Z, \quad
   ({\rm sp}, {\rm sp}') = \frac{n-2}{4}+ \Z, \\
   n \; {\rm odd}: & 
   ({\rm sp}, {\rm sp}) = \frac{n}{4} +\Z
  \end{array} \right.
\end{align}
the $\Q/\Z$-valued discriminant bilinear form, respectively, of the 
$D_n$ lattice.\footnote{
Although the range of the value of $(4-\hat{c})$ of interest
is not where the lattice $D_{4-\hat{c}}$ is of intrinsically $D$-type than
of $A$-type, there is nothing wrong in extrapolating the discriminant group
and the discriminant form above to the range of $(4-\hat{c})$ of our interest.
} %
Note that 
\begin{align}
  ({\rm vct},s) = \left\{ \begin{array}{ll}
    0+\Z, & s = 0, {\rm vct} \in G_{D_n}, \\
    \frac{1}{2}+\Z, & s = {\rm sp}, \; {\rm sp}' \in G_{D_n}
  \end{array} \right.
  \label{eq:Dn-disc-form-wVct}
\end{align}
regardless of whether $n$ is even or odd; we will use this fact later. 

{\bf The partition function} of the critical Type 0 $N=2$ minimal tensor
model SCFT is set as follows:
\begin{align}
  Z^{\rm crt}_{m.t.m.0B} :=
   \frac{q^{\frac{p_L^2}{2}} \bar{q}^{\frac{p_R^2}{2}}}{|\eta|^{2(8-2\hat{c})}}
  & \;
  \sum_{(b_1,\cdots, b_r) }^{ \in (\Z/2\Z)^r}   
  \sum_{\tilde{s}_0 \in}^{G_{D_{4-\hat{c}}}}
  \chi_{\tilde{s}_0+b_{\rm tot} {\rm vct}}\bar{\chi}_{\tilde{s}_0}
   \prod_{i=1}^r \sum_{\ell_i, \tilde{\ell}_i} L^{R_i}_{\ell_i, \tilde{\ell}_i}
  \sum_{\tilde{m}_i \in }^{\Z/2\bar{k}_i\Z} \sum_{\tilde{s}_i \in }^{\Z/4\Z}
  \chi_{\tilde{m}_i, \; \tilde{s}_i+2 b_i}^{\ell_i}
  \bar{\chi}^{\tilde{\ell}_i}_{\tilde{m}_i, \; \tilde{s}_i}
    \nonumber   \\
  & \;  \frac{1}{2^{2r}} \sum_{(c_1,\cdots, c_r)}^{\in (\Z/2\Z)^r}
  \mathbb{E}\left[ - \sum_{j=1}^r c_j \left( ({\rm vct} , \tilde{s}_0)
      + \frac{\tilde{s}_j}{2} \right)
    \right];  \label{eq:partFcn-m.t.m.+Mkwsk-type0-1} 
\end{align}
the character $\bar{\chi}$ of a representation of a right-mover VOA
uses $\bar{q} = e^{-2 \pi i \bar{\tau}}$ instead of $q$. 
\begin{align}
 \mathbb{E}\left[ x \right] := e^{2\pi i x }, 
\end{align}
$b_{\rm tot} := \sum_{i=1}^r b_i$, and 
$L_{\ell_i,\tilde{\ell}_i}^{R_i}$ is the integer-valued matrix of the
Type 0 level-$k_i$ $N=2$ minimal model SCFT reviewed
in section \ref{ssec:N=2mm-SCFT}. 
The sum over $c_1,\cdots, c_r \in \Z/2\Z$ and the factor $2^{-r}$
makes sure that all the states in this theory have
$s_i \equiv 2({\rm vct},s_0)_{D_{4-\hat{c}}}$ mod 2
in the right-moving representations; all the $(r+1)$ tensored
representations of the right-moving VOAs are therefore either in the NS-type
representations altogether, or in the R-type representations altogether. 
Moreover,
\begin{align}
  s_i \equiv \tilde{s}_i \equiv 2({\rm vct}, \tilde{s}_0)_{D_{4-\hat{c}}}
  \equiv 2 ({\rm vct}, s_0)_{D_{4-\hat{c}}} \qquad {\rm mod~}2
  \label{eq:cond-in-type0}
\end{align}
with $s_i = \tilde{s}_i + 2b_i$ and $s_0 = \tilde{s}_0 + b_{\rm tot} {\rm vct}$ 
for all the states in this theory. 
For a given set of $\{\tilde{\ell}_i\}$, $\{\ell_i\}$ and $\{\tilde{m}_i\}$,
there are $2^{r+1}$ choices of $(\tilde{s}_0;\tilde{s}_{j=1,\cdots, r})$ that
contribute to the partition function, while there are $2^r$ more sums
over the choices of $b_1, \cdots, b_r \in \Z/2\Z$
(and hence over the choices of $(s_0;s_{i=1,\cdots, r})$).

The Hilbert space $({\cal H}_{\rm tot}^{\rm post})^{\rm crt}_{m.t.m.}$ is
determined by the
prescription (\ref{eq:partFcn-m.t.m.+Mkwsk-type0-1}); the vector space
$({\cal H}_{\rm tot}^{\rm pre})^{\rm crt}_{m.t.m.}$ is otbained by dropping the
R-type representations, and relaxing the relation
$s_0=\tilde{s}_0 + b_{\rm tot}{\rm vct}$ to
$s_0 \in \tilde{s}_0 + \Z {\rm vct} \subset G_{D_{4-\hat{c}}}$;
those $2^{r+1} \times 2^{r+1}$ $(s_0;s_i;\tilde{s}_0;\tilde{s}_i)$ still
satisfy (\ref{eq:cond-in-type0})
because of (\ref{eq:Dn-disc-form-wVct}). The left-moving
superchiral algebra of this $({\cal H}_{\rm tot}^{\rm pre})^{\rm crt}_{m.t.m.}$
is the product of those of the $(r+1)$ SCFTs; the same is true with
the right-moving superchiral algebra. This is enough to conclude that
this construction gives rise to a Type 0 SCFT.

{\bf Conceptual clarification:}
\label{pg:cc-start}
Let us pay attention to formal aspects of the variables (such as $c_j$'s,
$b_i$'s, $\tilde{m}_i$'s and $\tilde{s}_i$'s) that are being summed.
First, the sets of data
\begin{align}
  \mu_R := (\tilde{s}_0; \tilde{m}_i; \tilde{s}_i), \qquad
  \mu_L := (s_0;m_i; s_i)
\end{align}
are regarded as elements of\footnote{
\label{fn:improvements-in-extra-Gepner-review}
Clear conceptual distinction between $G^{\rm crt}$ and
${\cal G}^{\rm crt}_{\rm bos}$ is one of improvements that
are introduced in this article. 
In many reviews on Gepner constructions
(e.g., \cite{Fuchs:1991vu}, \cite[\S4.1]{Recknagel:1997sb}),
$2\beta'_{(i)}$ in this article is introduced as an object in $G^{\rm crt}$
denoted by $\beta_i$. Many reviews also introduce an object
$(1;\vec{1}_{r\;{\rm elements}}; \vec{1}_{r\; {\rm elements}})$
denoted by $\beta_0$ without specifying whether it is in
$G^{\rm crt}$ or ${\cal G}^{\rm crt}_{\rm bos}$, and use it
for computation of worldsheet fermion numbers
(we also do in
(\ref{eq:def-beta0-Gepnr-fermionNmbr-test}, \ref{eq:cond-in-type0-2a},
\ref{eq:def-Fr}))
%
% (as we also do in (\ref{eq:cond-in-type0-2a0}) and 
%
and for keeping track of spectral flow of half-integral units 
(we do in (\ref{eq:turn-into-spectFlow}) for flow of integral units)
on one hand, and they also use it for a possible generator of the orbifold
as in (\ref{eq:def-beta*-Gepnr-typical-genrtr}), on the other. 
%
% p. \pageref{pg:Gepnr-generator-beta0}), and also
% 
In fact, it is not possible to guarantee that the $\Q/\Z$-valued
inner products (\ref{eq:inn-prod-on-Glatt}) and
(\ref{eq:inn-prod-on-maxOrbGrp-crtBos}) are well-defined without
making distinction between $G^{\rm crt}$ and ${\cal G}^{\rm crt}_{\rm bos}$. 
Moreover, our choice of a generator $\beta'_0 \in {\cal G}^{\rm crt}_{\rm bos}$
in (\ref{eq:def-betaPrm-0}) is not the same as $\beta'_*$
in (\ref{eq:def-beta*-Gepnr-typical-genrtr}) when $\hat{c}$ is even;
with the convention that ${\cal G}^{\rm crt}_{\rm bos}$ is embedded into
$G^{\rm crt}$ through (\ref{eq:def-embd-Gepn-orbGrp-2-lattDiscGrp}), 
the latter has no relation with spectral flows of odd integral units. 
} % 
\begin{align}
     G^{\rm crt} & \; := G_{D_{4-\hat{c}}} \times \prod_{i=1}^r \left( G_{\U(1)_{\bar{k}_i}} \times G_{D_1} \right)
       = G_{D_{4-\hat{c}}} \times \prod_{i=1}^r \left( \Z/2\bar{k}_i\Z  \times \Z/4\Z \right).   
\end{align}
The $\Q/\Z$-valued combination
% \footnote{
% %
% The Kazama--Suzuki SVOAs \cite{Kazama:1988qp}
% are for a Lie algebra $\mathfrak{g}$, $\mathfrak{h} \subset \mathfrak{g}$
% and $\bar{k} \in \N_{> h(\mathfrak{g})}$; the $N=2$ minimal model SVOA of
% level-$k$ is regarded as a special case, with
% $\mathfrak{g}=\mathfrak{su}(2)$, $\mathfrak{h} = \U(1)$ and
% $\bar{k} = k+h(\mathfrak{g})$.
% There, $s_i$ is in the discriminant group $G_{D_{\dim_\C (G/H)_i}}$,
% and $(s'_i,s_i)_{D_1}$ is generalized to $(s'_i, s_i)_{D_{\dim_\C(G/H)_i}}$. 
% $m_i$ is in the weight lattice of $\mathfrak{h}$, and
% $\bar{k}_i^{-1} \cdot m'_im_i/2$ is replaced by
% $(kH-h(\mathfrak{h}))^{-1} \cdot (m'_i,m_i)_{\mathfrak{h}}$. 
% } %
%
\begin{align}
  \mu' \bullet \mu := (s'_0, s_0)_{D_{4-\hat{c}}} - \sum_{i=1}^r \frac{m'_i m_i}{2\bar{k}_i} + \sum_{i=1}^r (s'_i, s_i)_{D_1}
  \label{eq:inn-prod-on-Glatt}
\end{align}
for $\mu, \mu' \in G^{\rm crt}$ is the discriminant bilinear form on
$G^{\rm crt}$, with the flip in sign on the factors $\U(1)_{k_i+2}$.

Secondly, the elements 
$(b_1,\cdots, b_r)$ and $(c_1,\cdots, c_r)$ are in the group $(\Z/2\Z)^r$, which is further regarded 
as the subgroup $\Gamma_0^{\rm crt}$ of 
\begin{align}
      {\cal G}_{\rm bos}^{\rm crt} & \; := \Z/2\Z \times \prod_{i=1}^r \left( \Z/\bar{k}_i \Z \times \Z/2\Z \right);  
\end{align}
$(c_1,\cdots, c_r) \in (\Z/2\Z)^r$ is regarded as 
$x = \sum_i c_i \beta'_{(i)} \in \Gamma^{\rm crt}_0 \subset {\cal G}^{\rm crt}_{\rm bos}$, where 
\begin{align}
  \beta'_{(i)} := (1; \; \vec{0}_{r\;{\rm elements}}; \;\vec{0},1_i,\vec{0} \; ) \in {\cal G}^{\rm crt}_{\rm bos} , \qquad
  i = 1,\cdots, r. 
\end{align}
For any element $x = (x_0; x_i; x'_i) \in {\cal G}^{\rm crt}_{\rm bos}$,
let $2x \in G^{\rm crt}_{\rm bos}$ denote the element of the form
\begin{align}
  2x := (x_0 {\rm vct}; 2x_i; 2x'_i);
  \label{eq:def-embd-Gepn-orbGrp-2-lattDiscGrp}
\end{align}
note that ${\rm vct} \in G_{D_{4-\hat{c}}}$ is never of the form of
${\rm vct}=2s$ for $s \in G_{D_{4-\hat{c}}}$ when $\hat{c}$ is even,
but $x \in {\cal G}^{\rm crt}_{\rm bos}$ here, not in $G^{\rm crt}$. 
Then it follows that 
\begin{align}
  (2x) \bullet \mu = (x_0 {\rm vct}, s_0)_{D_{4-\hat{c}}} - \sum_{i=1}^r \frac{2x_i m_i}{2\bar{k}_i} 
  + \sum_{i=1}^r \frac{2x'_i s_i}{4} \in \Q/\Z
  \label{eq:pairing-maxOrbGrp-charge-crtBos}
\end{align}
for any $x \in {\cal G}^{\rm crt}_{\rm bos}$ and $\mu \in G^{\rm crt}$.
We also define
\begin{align}
  2x\bullet y := \frac{x_0 y_0}{2} - \sum_{i=1}^r \frac{2x_i y_i}{2\bar{k}_i}
  + \sum_{i=1}^r \frac{2x'_i y'_i}{4} \in \Q/\Z
  \label{eq:inn-prod-on-maxOrbGrp-crtBos}
\end{align}
for $x=(x_0;x_i;x'_i)$ and $y=(y_0;y_i;y'_i)$ in ${\cal G}^{\rm crt}_{\rm bos}$.
All of (\ref{eq:inn-prod-on-Glatt}, \ref{eq:pairing-maxOrbGrp-charge-crtBos},
\ref{eq:inn-prod-on-maxOrbGrp-crtBos}) have well-defined
$\Q/\Z$-value, while $x \bullet \mu$ and $x \bullet y$ do not
for $\mu \in G^{\rm crt}$ and $x,y \in {\cal G}^{\rm crt}_{\rm bos}$.
Distinction among (\ref{eq:inn-prod-on-Glatt},
\ref{eq:pairing-maxOrbGrp-charge-crtBos},
\ref{eq:inn-prod-on-maxOrbGrp-crtBos}) has not been made clearly
in the literature. 
\label{pg:cc-end}

{\bf Orbifold interpretation:}
It is possible to rewrite the partition function $Z^{\rm crt}_{m.t.m.0B}$
in (\ref{eq:partFcn-m.t.m.+Mkwsk-type0-1}) by using the notations
introduced above so that we give an orbifold interpretation to the theory
in question \cite[app. A]{Fuchs:1991vu}; we then have a foundation for
theoretical consistency of this theory
beyond the genus-1 amplitudes in string theory.
The rewritten form of the partition function is this: 
\begin{align}
  Z^{\rm crt}_{m.t.m.0B} =
   \frac{q^{\frac{p_L^2}{2}} \bar{q}^{\frac{p_R^2}{2}}}{|\eta|^{2(8-2\hat{c})}}
  & \;
  \sum_{y}^{ \in \Gamma^{\rm crt}_0}  \sum_{\mu \in}^{G^{\rm crt}}
   \sum_{\lambda, \tilde{\lambda}} L_{\lambda, \tilde{\lambda}}
  \chi_{\mu + 2y}^{\lambda} \bar{\chi}^{\tilde{\lambda}}_{\mu}
  \frac{1}{2^{2r}} \sum_{x \in}^{\Gamma^{\rm crt}_0}
  \mathbb{E}\left[ - 2x \bullet (\mu + y) \right]
  \epsilon(x, y),
  \label{eq:partFcn-m.t.m.+Mkwsk-type0-2}
\end{align}
where $\lambda = (\ell_1,\ell_2, \cdots, \ell_r)$,
$\tilde{\lambda} = (\tilde{\ell}_1,\cdots, \tilde{\ell}_r)$, and
$L_{\lambda,\tilde{\lambda}} := \prod_{i=1}^r L_{\ell_i, \tilde{\ell}_i}^{R_i}$. 
\begin{align}
  \chi^\lambda_{\mu_L} := \chi_{s_0} \prod_{i=1}^r \chi^{\ell_i}_{m_i, s_i}, \qquad
  \bar{\chi}^{\tilde{\lambda}}_{\mu_R} := \bar{\chi}_{\tilde{s}_0} \prod_{i=1}^r
    \bar{\chi}^{\tilde{\lambda}_i}_{\tilde{m}_i,\tilde{s}_i}.
\end{align}
For $x = \sum_{j=1}^r c_j \beta'_{(j)}$ and $y = \sum_{i=1}^r b_i \beta'_{(i)}$
in $\Gamma^{\rm crt}_0 \subset {\cal G}^{\rm crt}_{\rm bos}$, the discrete
torsion is set as \cite[p.256 (below (A.8))]{Fuchs:1991vu} 
\begin{align}
  \epsilon(x,y) = \mathbb{E} \left[ \sum_{i,j} c_j b_i
    \langle \beta'_{(j)}, \beta'_{(i)} \rangle \right], \qquad
  \langle \beta'_{(j)}, \; \beta'_{(i)} \rangle :=
      \frac{1}{2}(1-\delta_{ij}) + \Z.
  \label{eq:disc-tors-4-m.t.m.-by-FKS}
\end{align}
Before we talk about the orbifold interpretation of the rewritten form
of the partition function, let us confirm that the rewritten form is
the same as the previous expression.
\begin{align}
  - 2x \bullet \mu -2x\bullet y & \; =_{+\Z} - c_{\rm tot}({\rm vct}, s_0)_{D_{4-\hat{c}}}
  - \frac{c_{\rm tot}b_{\rm tot}}{2} - \sum_{j=1}^r c_j \frac{s_j+b_j}{2}, \\
  & \;  =_{+\Z}
  - \sum_{j=1}^r c_j \left( ({\rm vct}, s_0)_{D_{4-\hat{c}}} + \frac{s_j}{2}
   \right) - \sum_{j=1}^r\sum_{i=1}^r c_jb_i\frac{1+ \delta_{ij}}{2} \nonumber
\end{align}
indeed. 

The theory with the partition function $Z^{\rm crt}_{m.t.m.0B}$ in 
(\ref{eq:partFcn-m.t.m.+Mkwsk-type0-2}) is regarded as an orbifold
by the group $\Gamma^{\rm crt}_0 \subset {\cal G}^{\rm crt}_{\rm bos}$
with the discrete torsion $\epsilon$ in (\ref{eq:disc-tors-4-m.t.m.-by-FKS})
of the original theory whose partition function is
\begin{align}
  Z^{\rm crt}_{\rm org} := \frac{q^{\frac{p_L^2}{2}} \bar{q}^{\frac{p_R^2}{2}}}
  {|\eta|^{2(8-2\hat{c})}}
  & \; \frac{1}{2^r} \sum_{\mu \in}^{G^{\rm crt}}
   \sum_{\lambda, \tilde{\lambda}} L_{\lambda, \tilde{\lambda}}
   \chi_{\mu}^{\lambda} \bar{\chi}^{\tilde{\lambda}}_{\mu}.
   \label{eq:partFcn-m.t.m.+Mkwsk-pureDiag}
\end{align}
The original theory is the tensor product of the $\R^{8-2\hat{c}}$-target NLSM
Type 0 SCFT and the $r$ modular invariant Type 0 SCFTs, so the Virasoro
central charges add up to $(12-3\hat{c})+3\hat{c}=12$. 
The Hilbert space $({\cal H}_{\rm tot})^{\rm crt}_{\rm org}$ is constructed by
simply taking the tensor product of $({\cal H}_{\rm tot}^{\rm post})$
of those $(r+1)$ Type 0 SCFTs, so the partition function is the product
of those $(r+1)$ modular invariant SCFTs, and is modular invariant.  
The choice of $\epsilon: \Gamma^{\rm crt}_0 \times \Gamma^{\rm crt}_0
\rightarrow S^1$ in (\ref{eq:disc-tors-4-m.t.m.-by-FKS})
satisfies the conditions \cite{Font:1989gq}
\begin{align}
 \epsilon(x,y_1+y_2) & \; = \epsilon(x,y_1)\epsilon(x,y_2),  \label{eq:cond-disc-tors-FIQS-a} \\
 \epsilon(y,x) & \; = \epsilon(x,y)^{-1},  \label{eq:cond-disc-tors-FIQS-b} \\
 \epsilon(x,x) & \; = 1  \label{eq:cond-disc-tors-FIQS-c}
\end{align}
for any $x,y, y_1, y_2 \in \Gamma$ on a discrete torsion;  
the conditions (\ref{eq:cond-disc-tors-FIQS-a}--\ref{eq:cond-disc-tors-FIQS-c})
are equivalent to the ones derived in the original Ref. \cite{Vafa:1986wx}. 
Therefore, theoretical consistency of the orbifold theory
with (\ref{eq:partFcn-m.t.m.+Mkwsk-type0-2}) is guaranteed
at all higher genera by the theoretical consistency of the original
theory with (\ref{eq:partFcn-m.t.m.+Mkwsk-pureDiag}). 

The original CFT with (\ref{eq:partFcn-m.t.m.+Mkwsk-pureDiag}) 
before the orbifold by $\Gamma_0^{\rm crt}$ is the tensor product
of $(r+1)$ Type 0 SCFTs, but it is no longer easy to see the trace
of the $N=2$ SCAs of the $(r+1)$ theories. The left-mover
chiral algebra of the original CFT
with (\ref{eq:partFcn-m.t.m.+Mkwsk-pureDiag}) is
$\otimes_{i=0}^r {\cal A}_i^+$, where ${\cal A}_i^+$ is the
$\Z/2\Z$-even part of the left-mover superchiral algebra ${\cal A}_i$
of the $i$-th SCFT. Any double extension of $\otimes_{i=0}^r({\cal A}_i^+)$
does not contain the diagonal superconformal current
$\sum_{i=0}^r G_i$ of the $(r+1)$ SCFTs. The orbifold by
$\Gamma_0^{\rm crt} \cong (\Z/2\Z)^r$ made it possible. 

It is colloquially said (cf footnote \ref{fn:SCFT-min}), in general, that a Type 0 SCFT with ${\cal H}_{\rm tot}^{\rm post}$ is where a diagonal fermion parity is gauged from a fermionic SCFT
with ${\cal H}_{\rm tot}^{\rm pre}$; to be more precise, both the left-moving
superchiral algebra and the right-moving superchiral algebra have $\Z/2\Z$
automorphism symmetry associated with their SVOA $\Z/2\Z$ grading,
and the diagonal $\Z/2\Z =: (\Z/2\Z)_{\rm SVOA.frm.prty}$ of this
$\Z/2\Z \times \Z/2\Z$ is gauged;
we call the generator of this $\Z/2\Z$ the {\it diagonal SVOA fermion
  parity} 
(a fermion number $F_L\in \Z$ (a $\Z$-grading) does not have
to be defined). The CFT with $Z_{\rm org}^{\rm crt}$
in (\ref{eq:partFcn-m.t.m.+Mkwsk-pureDiag})
before the orbifold is therefore regarded as a CFT with a gauging by
$(\Z/2\Z)^{r+1}_{\rm SVOA.frm.prty}$, each of which is generated by the diagonal
SVOA fermion parity of one of the $(r+1)$ tensored SCFTs.
So, this CFT before the orbifold has the quantum symmetry
(\cite{Vafa:1989ih}, \cite{Bhardwaj:2017xup})   
$(\Z/2\Z)^{r+1}$ (the dual group of the gauged diagonal SVOA fermion parity
$(\Z/2\Z)^{r+1}_{\rm SVOA.frm.prty}$, sensitive to the
twistedness of the representations). 
The orbifold group $\Gamma_0^{\rm crt} \cong (\Z/2\Z)^r$ has been chosen
as the traceless subgroup of the quantum symmetry group.
So,  the CFT with $Z_{m.t.m.0B}^{\rm crt}$
in (\ref{eq:partFcn-m.t.m.+Mkwsk-type0-2}) after the orbifold
has just gauged the diagonal SVOA fermion parity in the universal
part of $(\Z/2\Z)^{r+1}_{\rm SVOA.frm.prty}$. 
This reasoning explains intuitively why the SCFT
with the partition function $Z^{\rm crt}_{m.t.m.0B}$
in (\ref{eq:partFcn-m.t.m.+Mkwsk-type0-1},
\ref{eq:partFcn-m.t.m.+Mkwsk-type0-2}) is an orbifold by
$\Gamma_0^{\rm crt} \cong (\Z/2\Z)^r$ of the CFT with $Z_{\rm org}^{\rm crt}$
in (\ref{eq:partFcn-m.t.m.+Mkwsk-pureDiag}).

%%%%%%%%%%%%%%%%%%%%%%%%%%%%%%%%%%%%%%%%%%%%%%%%%%%%%
\section{Gepner Model SCFTs}
\label{sec:Gepn-Data}
%%%%%%%%%%%%%%%%%%%%%%%%%%%%%%%%%%%%%%%%%%%%%%%%%%%%%%

%%%%%%%%%%%%%%%%%%%%%%%%%%%%%%%%%%
\subsection{Type 0B Gepner Model Critical SCFTs}
\label{ssec:Gepn-crit-0SCFT}
%%%%%%%%%%%%%%%%%%%%%%%%%%%%%%%%%%

Let us think of taking an orbifold of the tensored $(r+1)$ SCFTs
(i.e., the original CFT with $Z_{\rm org}^{\rm crt}$
in (\ref{eq:partFcn-m.t.m.+Mkwsk-pureDiag})) by a group larger than
$\Gamma_0^{\rm crt} \subset {\cal G}^{\rm crt}_{\rm bos}$ to construct a
Type 0 SCFT with the critical Virasoro central charge
$(c,\tilde{c}) = (12,12)$. 
We choose the orbifold group to be of the form 
\begin{align}
  \Gamma^{\rm crt} := \left\{ b_i \beta'_{(i)} + n_I \gamma_I \in {\cal G}^{\rm crt}_{\rm bos} \; | \; b_i \in \Z/2\Z, \; n_I \in \Z \right\}
    \subset {\cal G}^{\rm crt}_{\rm bos}, 
\end{align}
where $\gamma_I \in {\cal G}^{\rm crt}_{\rm bos}$ are additional generators,
and we assume that the set of elements of $\Gamma^{\rm crt}$ that are trivial
in the middle $r$ elements (in $\prod_{i=1}^r (\Z/\bar{k}_i\Z)$) agrees
with $\Gamma^{\rm crt}_0$. 
The discrete torsion $\epsilon: \Gamma^{\rm crt} \times
\Gamma^{\rm crt} \ni (x,y) \longmapsto \epsilon(x,y) \in S^1$ has to be
\cite{Vafa:1986wx} subject to the conditions
(\ref{eq:cond-disc-tors-FIQS-a}--\ref{eq:cond-disc-tors-FIQS-c}) for
consistency of the theory obtained in the orbifold construction
by $(\Gamma^{\rm crt}, \epsilon)$;
it is parametrized by $\langle \beta'_{(i)}, \gamma_J \rangle \in \Q/\Z$
and $\langle \gamma_I, \gamma_J \rangle \in \Q/\Z$ as 
\begin{align}
  \epsilon(x,y) & \; = \exp \left[ c_j b_i \langle \beta'_{(j)},\beta'_{(i)}\rangle
    + c_j \langle \beta'_{(j)}, \gamma_I \rangle m_I
    + n_J \langle \gamma_J, \beta'_{(i)} \rangle b_i
    + n_J \langle \gamma_J, \gamma_I \rangle m_I \right],  \\
  &  \qquad x = c_j \beta'_{(j)} + n_J \gamma_J, \quad
  y = b_i \beta'_{(i)} + m_I \gamma_I,   \nonumber
\end{align}
where $\langle -, -\rangle: \Gamma^{\rm crt} \times \Gamma^{\rm crt}
\rightarrow \Q/\Z$ is an anti-symmetric bilinear form (the
conditions (\ref{eq:cond-disc-tors-FIQS-a},
\ref{eq:cond-disc-tors-FIQS-b})) s.t. $\langle x,x \rangle = 0_{+\Z} \in \Q/\Z$
(the condition (\ref{eq:cond-disc-tors-FIQS-c})). 

{\bf Constraints on $\epsilon$:} 
The orbifold theory by $(\Gamma^{\rm crt}, \epsilon)$ is not necessarily
regarded as a Type 0 SCFT whose superchiral algebras
$\prod_{i=0}^r {\cal A}_i$ and $\prod_{i=0}^r\widetilde{\cal A}_i$
contain
the product of the superchiral algebras of the $(r+1)$ tensored SCFTs. 
The condition that they do when $\Gamma^{\rm crt}$ is larger than
$\Gamma^{\rm crt}_0$, is translated to (\ref{eq:cond-in-type0-1b})
besides (\ref{eq:disc-tors-4-m.t.m.-by-FKS}) as we will see below.  

The partition function of this orbifold CFT is (cf \cite{Gepner:1987vz})
\begin{align}
 Z^{\rm crt}_{\rm Gepn.0B} & = 
   \frac{q^{\frac{p_L^2}{2}} \bar{q}^{\frac{p_R^2}{2}}}{|\eta|^{2(8-2\hat{c})}}
    \sum_{y}^{ \in \Gamma^{\rm crt}}  \sum_{\mu_R \in}^{G^{\rm crt}}
   \sum_{\lambda, \tilde{\lambda}} \frac{L_{\lambda, \tilde{\lambda}}}{2^r}
  \chi_{\mu_R + 2y}^{\lambda} \bar{\chi}^{\tilde{\lambda}}_{\mu_R}
  \frac{1}{|\Gamma^{\rm crt}|} \sum_{x \in}^{\Gamma^{\rm crt}}
  \mathbb{E}\left[ - 2x \bullet (\mu_R + y) \right]
  \epsilon(x, y).
  \label{eq:partFcn-Gepner+Mkwsk-type0}
\end{align}
For $y \in \Gamma^{\rm crt}$ and for a given $\lambda$ and $\tilde{\lambda}$,
therefore, the states that survive in this CFT
(i.e., in $({\cal H}_{\rm tot}^{\rm post})_{\rm Gepn.}^{\rm crt}$)
are those with
\begin{align}
&
  -2\beta'_{(j)} \bullet (\mu_R + b_i \beta'_{(i)} + m_I \gamma_I )
  + \langle \beta'_{(j)} , b_i \beta'_{(i)} + m_I \gamma_I \rangle \in \Z,
  \qquad {}^\forall j=1,\cdots, r, \label{eq:cond-Gepner-proj-crt-type0-a} \\
&
  -2\gamma_J \bullet (\mu_R + b_i \beta'_{(i)} + m_I \gamma_I)
  + \langle \gamma_J, b_i \beta'_{(i)} + m_I \gamma_I \rangle \in \Z,
  \qquad {}^\forall J.
  \label{eq:cond-Gepner-proj-crt-type0-b}
\end{align}
For the discussion below, note that
\begin{align}
 \mu_L = \mu_R + 2y = \mu_R + b_i 2\beta'_{(i)} + m_I 2\gamma_I. 
\end{align}

We impose \cite[app. A]{Fuchs:1991vu} the following condition (ia, ib) on
the choice of $(\Gamma^{\rm crt}, \epsilon)$: 
when a theory contains the states in the representation with
$\lambda, \tilde{\lambda}$ and $(\mu_L, \mu_R)$ in the spectrum, then 
\begin{itemize}
\item [(ia)] the spectrum also {\it contain} all the $2^{r+1}$ representations
  of $\otimes_{i=0}^r ({\cal A}_i^+) \otimes_{i=0}^r (\widetilde{\cal A}_i^+)$
  with the same $\lambda$ and $\tilde{\lambda}$, but with $(\mu'_L, \mu_R')$
  that are different from $(\mu_L,\mu_R)$ by
  ${}^\forall \delta \tilde{s}_i \in 2\Z/4\Z$ ($i=1,\cdots, r$) and
  ${}^\forall \delta \tilde{s}_0 \in G_{D_{4-\hat{c}}}$ s.t.
  $(\delta \tilde{s}_0, {\rm vct}) = 0_{+\Z} \in \Q/\Z$, while
  keeping $\mu'_L-\mu'_R = \mu_L - \mu_R$ (i.e., fixed $b_i$ and $m_I$),  
\item [(ib)] the spectrum also {\it contain} all the $2^r$ representations
  with the same $\lambda, \tilde{\lambda}$ and $\mu_R$, but with $\mu'_L$
  that are different from $\mu_L$ by $b_i 2\beta'_{(i)}$,
  ${}^\forall b_i \in \Z/2\Z$.  
\end{itemize}
The condition (ib) makes sure that $[\otimes_{i=0}^r \widetilde{\cal A}_i ]^+$
acts on $({\cal H}_{\rm tot}^{\rm post})^{\rm crt}_{\rm Gepn.}$.
%
% here, $\widetilde{\cal A}_i^+$ [resp. ${\cal A}_i^+$] is the even part of
% the $\Z/2\Z$-grading of the right-moving superchiral algebra
% $\widetilde{\cal A}$ [resp. left-moving one ${\cal A}$] of the
% $i$-th ($i=0,1,\cdots, r$) tensored SCFT. 
% 
The condition (ia) guarantees that the algebra $\otimes_{i=0}^r
[{\cal A}_i \times \widetilde{\cal A}_i]^+$ acts on
$({\cal H}_{\rm tot}^{\rm post})^{\rm crt}_{\rm Gepn.}$. 

The conditions (ia, ib) are enough to make sure that the resulting CFT is a
Type 0 SCFT with the diagonal $N=2$ SCAs in both of the superchiral
algebras ${\cal A}$ and $\widetilde{\cal A}$.
The vector space $({\cal H}_{\rm tot}^{\rm pre})^{\rm crt}_{\rm Gepn.}$
should be set by dropping the Ramond--Ramond representations, and extending
from $({\cal H}_{\rm tot}^{\rm post})^{\rm crt}_{\rm Gepn.}$ for
$Z^{\rm crt}_{{\rm Gepn}.0B}$ by including the states with $\lambda,
\tilde{\lambda}$, $\mu_L+({\rm vct};\vec{0};\vec{0})$, $\mu_R$ when
$({\cal H}_{\rm tot}^{\rm post})^{\rm crt}_{\rm Gepn.}$ contains the states with
$\lambda, \tilde{\lambda}$, $\mu_L$ and $\mu_R$. The algebra
$\otimes_{i=0}^r({\cal A}_i \times \widetilde{\cal A}_i)$ acts on
$({\cal H}_{\rm tot}^{\rm pre})^{\rm crt}_{\rm Gepn.}$ then. 

The condition (ia) is satisfied automatically for any $\gamma_I$ chosen
from ${\cal G}^{\rm crt}_{\rm bos}$ in fact. Indeed, pick one set of
$\mu_R$, $b_i$ and $n_I$ (i.e., one set of $(\mu_L, \mu_R)$) that satisfies
both (\ref{eq:cond-Gepner-proj-crt-type0-a})
and (\ref{eq:cond-Gepner-proj-crt-type0-b}). Then $2^{r+1}$ choices of
$\mu'_R$ described in the condition (ia) also satisfy
both (\ref{eq:cond-Gepner-proj-crt-type0-a})
and (\ref{eq:cond-Gepner-proj-crt-type0-b}).
The condition (ib) is equivalent to the condition that
$b_i \in \Z/2\Z$ should drop out from 
(\ref{eq:cond-Gepner-proj-crt-type0-a}) and
(\ref{eq:cond-Gepner-proj-crt-type0-b}); this determines 
the parameters of the discrete torsion as follows:
\begin{align}
  \langle \beta'_{(j)}, \beta'_{(i)} \rangle  % _{{\rm if~}j \neq i}
    = 2\beta'_{(j)} \bullet \beta'_{(i)},
  \qquad
  \langle \gamma_J, \beta'_{(i)}\rangle = 2 \gamma_J \bullet \beta'_{(i)} \in \Q/\Z.
  \label{eq:cond-in-type0-1b}
\end{align}
This generalizes\footnote{
Ref. \cite[p.256]{Fuchs:1991vu} spelled out the idea (ia, ib), and
also implemented it for a pair of generators within
$\Gamma^{\rm crt}_0 \subset \Gamma^{\rm crt}$ to have the constraint
(\ref{eq:disc-tors-4-m.t.m.-by-FKS}). Ref. \cite{Fuchs:1991vu} also says
in its last paragraph (p.256) that ``care is to be taken...'' and
``we have to ensure ...'' even when $\Gamma^{\rm crt}$ is larger than
$\Gamma^{\rm crt}_0$. We just did that here, without restricting our attention
to the cases with spacetime supersymmetry. 
} %
the choices/solutions (\ref{eq:disc-tors-4-m.t.m.-by-FKS})
for the orbifold group $\Gamma_0^{\rm crt}$ to a more general
orbifold group $\Gamma^{\rm crt}$. To summarize, a part of the
parameters of the discrete torsion is determined completely
by (\ref{eq:disc-tors-4-m.t.m.-by-FKS}, \ref{eq:cond-in-type0-1b}),
whereas the parameters $\langle \gamma_I, \gamma_J \rangle$ remain free
(except the bilinear, anti-symmetric and vanishing diagonal properties
(\ref{eq:cond-disc-tors-FIQS-a}--\ref{eq:cond-disc-tors-FIQS-c})).
No constraint on the choice of the group $\Gamma^{\rm crt}$ follows from the
conditions (ia, ib).

We call the orbifold theory subject to the constraint
(\ref{eq:disc-tors-4-m.t.m.-by-FKS}, \ref{eq:cond-in-type0-1b})
a {\it Gepner model critical Type 0B SCFT}. The word {\it critical}
is included to imply that the critical Virasoro central charge has been
filled by the flat-space-target NLSM SCFT.

%%%%%%%%%%%%%%%%%%%%%%%%%%%%%%%%%%%%%%%%%%%%%
\subsection{Type 0 Gepner Model Non-critical SCFTs}
\label{ssec:Gepn-Ncrit-0SCFT}
%%%%%%%%%%%%%%%%%%%%%%%%%%%%%%%%%%%%%%%%%%%%%

We have seen toward the end of section \ref{ssec:mtm-crit-SCFT} that
a Type 0 minimal tensor model critical SCFT may be regarded
as a $(\Z/2\Z)^r$ orbifold of the tensor product of $(r+1)$ Type 0 SCFTs;
the $(\Z/2\Z)^r$ orbifold group is chosen within $(\Z/2\Z)^{r+1}_{\rm qnt.}$,
each one of which is the quantum symmetry group of the gauging of the
diagonal SVOA fermion parity in one of the $(r+1)$ SCFTs.
It may thus be possible to think of an intermediate step in this
orbifold (gauging) by $(\Z/2\Z)^r \subset (\Z/2\Z)^{r+1}_{\rm qnt.}$:
in the first step, take an orbifold by $(\Z/2\Z)^{r-1}$ acting only on
the $r$ tensored internal SCFTs (so the $\R^{8-2\hat{c}}$-target
NLSM Type 0 SCFT gets untouched); do the remaining $\Z/2\Z$ orbifold
in the second step. The first step yields an internal Type 0 SCFT with
the central charge $(c,\tilde{c}) = (3\hat{c},3\hat{c})$ tensored with
the $\R^{8-2\hat{c}}$-target Type 0 SCFT. This observation suggests that
a Type 0 SCFT may be constructed from $r$ Type 0 SCFTs without
imposing a constraint that the sum of their central charges is
the critical value, or an integer (as assumed in (\ref{eq:cond-cpx-dim-Z})). 
This observation is not just for the case with the orbifold group
$\Gamma_{0}^{\rm crt}$ but with arbitrary $\Gamma^{\rm crt}$ introduced at
the beginning of section \ref{ssec:Gepn-crit-0SCFT}. 
We elaborate more on this possibility in this section \ref{ssec:Gepn-Ncrit-0SCFT}. 

{\bf Preparation:}
To start off, note that there is arbitrariness in a given orbifold
construction with $(\Gamma^{\rm crt},\epsilon)$ how we choose extra generators
$\gamma \in \Gamma^{\rm crt} \subset {\cal G}^{\rm crt}_{\rm bos}$ besides
the generators of $\Gamma^{\rm crt}_0$. By exploiting the fact
that $\Gamma^{\rm crt}_0$ is always contained in $\Gamma^{\rm crt}$, 
we may choose all the extra generators to be in the form of 
$\gamma = (p_0(\vec{y});\vec{y};\vec{0})$ (i.e., $y'_i=0\in \Z/2\Z$ for
$i=1,\cdots, r$) without a loss of generality, where  
\begin{align}
  \vec{y} \in {\cal G} := \prod_{i=1}^r \Z/\bar{k}_i\Z.
  \label{eq:def-calG-4int}
\end{align}
Moreover, there is a structure $\Gamma^{\rm crt} \cong \Gamma^{\rm crt}_0 \times \Gamma$, where $\Gamma$ is the image of $\Gamma^{\rm crt}$ under the projection
\begin{align}
  {\cal G}^{\rm crt}_{\rm bos} & \longrightarrow {\cal G} =
  \prod_{i=1}^r \Z/\bar{k}_i\Z, \\
  \Gamma^{\rm crt} & \longrightarrow \Gamma. \nonumber 
\end{align}
Here, we use the assumption at the beginning of
section \ref{ssec:Gepn-crit-0SCFT} that the kernel of the projection
$\Gamma^{\rm crt} \rightarrow {\cal G}$ is $\Gamma^{\rm crt}_0$. The assumption
also implies that the projection (once through $\Gamma^{\rm crt}$) to the
space-time part $\Z/2\Z$ of ${\cal G}^{\rm crt}_{\rm bos}$,  
\begin{align}
  p_0: \Gamma \ni \vec{y} \longmapsto p_0(\vec{y}) \in \Z/2\Z, 
\end{align}
is a homomorphism. We will use the notation $\gamma(\vec{y}) :=
(p_0(\vec{y});\vec{y};\vec{0}) \in (\Gamma^{\rm crt}_0 \times \Gamma) \subset
{\cal G}^{\rm crt}_{\rm bos}$. 

{\bf Organize the spectrum into SVOA representations:} 
Let us organize the spectrum of the critical Type 0 SCFT captured in
$Z^{\rm crt}_{{\rm Gepn}. 0B}$ in (\ref{eq:partFcn-Gepner+Mkwsk-type0}) in a way
the action of the left-moving and right-moving superchiral algebras
is manifest. 
\begin{align}
 Z^{\rm crt}_{{\rm Gepn}. 0B} & \; = 
   \frac{q^{\frac{p_L^2}{2}} \bar{q}^{\frac{p_R^2}{2}}}{|\eta|^{2(8-2\hat{c})}}
    \sum_{\vec{y}}^{ \in \Gamma}  \sum_{\vec{\tilde{m}} \in}^{ \prod_{i=1}^r (\Z/2\bar{k}_i\Z)}
   \sum_{\lambda, \tilde{\lambda}} \frac{L_{\lambda, \tilde{\lambda}}}{2^{r+1}}
   \left( 
        {\rm ch}_{\vec{\tilde{m}}+ 2\vec{y}}^{\lambda}
        \overline{{\rm ch}}^{\tilde{\lambda}}_{\vec{\tilde{m}}}
        + (-1)^{p_0(\vec{y})}{\rm ch}_{\vec{\tilde{m}}+ 2\vec{y}}^{\lambda,-}
        \overline{{\rm ch}}^{\tilde{\lambda},-}_{\vec{\tilde{m}}}
   \right) \label{eq:partFcn-Gepner+Mkwsk-type0-SVOArepr}  \\
 &  \qquad \frac{1}{|\Gamma|} \sum_{\vec{x} \in}^{\Gamma}
  \mathbb{E}\left[ - 2\vec{x} \bullet (\vec{\tilde{m}} + \vec{y}) \right]
  \epsilon'(\vec{x}, \vec{y}). \nonumber
\end{align}
Here, $\vec{x}$, $\vec{y}$, $\vec{\tilde{m}}$ have only $r$ non-zero components,
and only the middle term of (\ref{eq:pairing-maxOrbGrp-charge-crtBos},
\ref{eq:inn-prod-on-maxOrbGrp-crtBos}) contributes in
$2\vec{x} \bullet \vec{\tilde{m}}$ and $2\vec{x} \bullet \vec{y}$ above. 
We use the characters of the product of the $(r+1)$ {\it super}chiral
algebras ($\otimes_i {\cal A}_i$)  
\begin{align}
  {\rm ch}_{\vec{m}}^{\lambda} & \; := \left\{ \begin{array}{ll}
    \ell_i + m_i \; {\rm even~for~} {}^\forall i, &
    (\chi_0 + \chi_{{\rm vct}}) \prod_{i=1}^r {\rm ch}_{\ell_i,m_i}, \\
    \ell_i + m_i \; {\rm odd~for~} {}^\forall i, &
    (\chi_{\rm sp} + \chi_{{\rm sp}'}) \prod_{i=1}^r {\rm ch}_{\ell_i,m_i} \\
    {\rm otherwise}, & 0,
  \end{array} \right.    \\
  {\rm ch}_{\vec{m}}^{\lambda,-} & \; := \left\{ \begin{array}{ll}
    \ell_i + m_i \; {\rm even~for~} {}^\forall i, &
    (\chi_0 - \chi_{{\rm vct}}) \prod_{i=1}^r
        {\rm ch}^{-}_{\ell_i,m_i}, \\
    \ell_i + m_i \; {\rm odd~for~} {}^\forall i, &
    (\chi_{\rm sp} - \chi_{{\rm sp}'}) \prod_{i=1}^r
        {\rm ch}^{-}_{\ell_i,m_i}, \\
      {\rm otherwise}, &    0,
  \end{array} \right.     
\end{align}
instead of those of the product of the $(r+1)$ chiral algebras
($\otimes_i ({\cal A}_i^+)$).
The characters $\overline{\rm ch}^{\tilde{\lambda}}$ and
$\overline{\rm ch}^{\tilde{\lambda},-}$ are those with the argument
$q=e^{2\pi i\tau}$ replaced by $\bar{q}=e^{-2\pi i \bar{\tau}}$. 

Although it is intuitively clear that the partition function
$Z^{\rm crt}_{{\rm Gepn}.0B}$ in (\ref{eq:partFcn-Gepner+Mkwsk-type0})
can be reorganized as in (\ref{eq:partFcn-Gepner+Mkwsk-type0-SVOArepr})
with a manifest structure of the action of the superchiral algebras,
technical steps to see this are as follows. The phase factor in
the orbifold projection in (\ref{eq:partFcn-Gepner+Mkwsk-type0})
is decomposed as follows:
\begin{align}
 \mathbb{E}\left[ -2x \bullet (\mu_R + y) \right] & \; \epsilon(x,y) =  
 \mathbb{E}\left[-2c_j \beta'_{(j)} \bullet \mu_R \right]
 \mathbb{E}\left[-2n_J \gamma(\vec{x}_J) \bullet \mu_R \right] \nonumber \\
& \mathbb{E}\left[c_j(-2\beta'_{(j)} \bullet \beta'_{(i)} + \langle \beta'_{(j)}, \beta'_{(i)}\rangle ) b_i \right] 
 \mathbb{E}\left[n_J(-2\gamma(\vec{x}_J) \bullet \beta'_{(i)} + \langle \gamma(\vec{x}_J), \beta'_{(i)}\rangle ) b_i \right] \nonumber \\
& \mathbb{E}\left[c_j(-2\beta'_{(j)}\bullet \gamma(\vec{y}_I) + \langle \beta'_{(j)}, \gamma(\vec{y}_I)\rangle )m_I\right] \nonumber \\
 & \mathbb{E}\left[n_J(-2\gamma(\vec{x}_J) \bullet \gamma(\vec{y}_I) + \langle \gamma(\vec{x}_I), \gamma(\vec{y}_I)\rangle) m_I \right].
 \label{eq:orbifold-phase-Type0-crit-bos}
\end{align}
%
% Here, and in (\ref{eq:partFcn-Gepner+Mkwsk-type0-SVOArepr})
% $2\vec{x}\bullet \vec{m}$ and $2\vec{x}\bullet \vec{y}$
% for $\vec{x}$, $\vec{y} \in {\cal G}$ and
% $\vec{m} \in \prod_{i=1}^r (\Z/2\bar{k}_i\Z)$ are
% the same as $2x\bullet \mu$ and $2x\bullet y$
% for $x,y\in {\cal G}^{\rm crt}_{\rm bos}$ and $\mu \in G^{\rm crt}$
% except that the sum is only over the middle $r$ elements. 
% 
The condition (\ref{eq:cond-in-type0-1b}) implies that the phase factors
in the 2nd and 3rd lines\footnote{
The phases in the 3rd line are equal to 
$-2\beta'_{(j)} \bullet \gamma(\vec{y}) - \langle \gamma(\vec{y}), \beta'_{(j)} \rangle =_{(\ref{eq:cond-in-type0-1b})} -4 \beta'_{(j)} \bullet \gamma(\vec{y})
= -p_0(\vec{y}) \in \Z$. 
} %
are trivial.
So, the average over $\delta x = c_j \beta'_{(j)}
\in \Gamma^{\rm crt}_0$ results in projection into the choices of
$(\tilde{s}_0;\vec{\tilde{s}})$ that are all in the NS-sector or in the
R-sector, leaving only the averaging over $\gamma(\vec{x})$
(with $\vec{x} \in \Gamma$) as 
in (\ref{eq:partFcn-Gepner+Mkwsk-type0-SVOArepr}).
Now, we are left with two phase factors (the 2nd factor in the 1st line
and that of the 4th line); both phases are common to all the
$2^{r+1}\times 2^r$ choices of the NS--NS sector
$(\tilde{s}_0;\vec{\tilde{s}})$ and $(s_0; \vec{s})$ and
$2^{r+1}\times 2^r$ choices in the R--R sector, for a given
$\lambda, \tilde{\lambda}$, $\vec{\tilde{m}}$,
$\vec{y}, \vec{x} \in \Gamma$ but with varying $(b_i)$. 
This explains why the partition function is given in the combination
of $({\rm ch}^\lambda_{\vec{\tilde{m}}+2\vec{y}}
\overline{\rm ch}^{\tilde{\lambda}}_{\vec{\tilde{m}}} + (-1)^{p_0(\vec{y})}
         {\rm ch}^{\lambda,-}_{\vec{\tilde{m}}+2\vec{y}}
         \overline{\rm ch}^{\tilde{\lambda},-}_{\vec{\tilde{m}}})/2$. 

The factor
$\epsilon': \Gamma \times \Gamma \ni (\vec{x}, \vec{y}) \longmapsto
\epsilon'(\vec{x},\vec{y})\in  S^1$
in (\ref{eq:partFcn-Gepner+Mkwsk-type0-SVOArepr}) should be determined
as follows: 
\begin{align}
  \mathbb{E}\left[ -2\gamma(\vec{x}) \bullet (\mu_R + \gamma(\vec{y}))
    \right] \epsilon(\gamma(\vec{x}), \gamma(\vec{y}))
  =: \mathbb{E}\left[ -2 \vec{x}\bullet (\vec{\tilde{m}} + \vec{y}) \right]
  \epsilon'(\vec{x},\vec{y}). 
\end{align}
Now, the rest is just a computation using the 2nd and 6th factors
in (\ref{eq:orbifold-phase-Type0-crit-bos}): 
\begin{align}
  \epsilon'(\vec{x}, \vec{y}) = \mathbb{E}\left[
    \langle \gamma(\vec{x}), \gamma(\vec{y}) \rangle
    - \frac{p_0(\vec{x}) p_0(\vec{y})}{2}
    - \frac{p_0(\vec{x}) 2({\rm vct}, \bar{s}_0(\tilde{\lambda}, \vec{\tilde{m}}))}
           {2} \right], 
\end{align}
where the first two phases are from the 6th factor
in (\ref{eq:orbifold-phase-Type0-crit-bos}) and the last one is from
the 2nd factor in (\ref{eq:orbifold-phase-Type0-crit-bos}). 
Although $\tilde{s}_0(\tilde{\lambda},\vec{\tilde{m}}) \in G_{D_{4-\hat{c}}}$ can be
determined from $\tilde{\lambda}$ and $\vec{\tilde{m}}$ only mod $+ {\rm vct}$,
what really matters in the expression above is whether $\tilde{s}_0$ is
NS-type (in $\{ 0, {\rm vct}\}$) or R-type (in $\{{\rm sp}, {\rm sp}'\}$). 
So, the phase $\epsilon'(\vec{x},\vec{y})$ is well-defined
for $\vec{x}, \vec{y} \in \Gamma$ and $(\tilde{\lambda},\vec{\tilde{m}})$.

{\bf The Arf theory, and an observation:}
\label{pg:IV-st}   The {\it Arf theory} is a
quantum field theory on 1+1-dimensions whose Hilbert space is
${\cal H}_{\rm tot} \cong \C$, without any excited states or without
a non-trivial Hamiltonian. Its partition function is set as
\begin{align}
 Z_{\rm Arf}[s] = (-1)^{{\rm Arf}(s)} = \mathbb{E}\left[ {\rm Arf}(s)/2 \right],
\end{align}
where $s$ ranges over $2^{2g}$ different spin structures on a
Riemann surface $\Sigma$ of genus $g$; 
the {\it Arf invariant} ${\rm Arf}: s\longmapsto {\rm Arf}(s) \in \Z/2\Z$
is given by
\begin{align}
  {\rm Arf}(s) := \#\left\{ \begin{array}{l}
    0{\rm~modes~of~a~Majorana~Weyl~
      spinor} \\ {\rm ~on~}\Sigma
    {\rm ~with~the~spin~structure~}s \end{array} \right\}
  \quad {\rm mod~}2 \in \Z/2\Z. 
\end{align}
The Arf theory without a renormalization group flow is regarded
as a CFT, with the Virasoro central charges $(c,\tilde{c}) = (0,0)$,
and $\omega = \tilde{\omega} = 0 \in {\cal H}_{\rm tot}$. 
For more information on the Arf invariant, and the Arf theory, and their
use in quantum field theory in general, see \cite{Karch:2019lnn},
\cite{Ji:2019ugf}, the papers that the two papers refer to, and those
that refer to them. 

As we see below, the two factors $\epsilon'(\vec{x},\vec{y})$ and
$(-1)^{p_0(\vec{y})}\epsilon'(\vec{x},\vec{y})$
in the partition function (\ref{eq:partFcn-Gepner+Mkwsk-type0-SVOArepr}) 
can be rewritten in terms of the original $\epsilon$ in \cite{Vafa:1986wx}
and the Arf theory (cf \cite[footnote 28]{Cheng:2024noz}).
Because the Arf theory is defined also for Riemann
surfaces of a higher genus, and because the discrete torsion
$\epsilon$ for higher genus amplitudes is related to
the phases $\epsilon(\vec{x}, \vec{y})$ in the genus-1 amplitude
through the factorization limit of the Riemann surfaces
(see \cite{Vafa:1986wx}), this translation allows us to formulate
the discrete torsion phases in an orbifold of SCFTs (than CFTs)
for higher genus Riemann surfaces. This is, therefore,
to update\footnote{
Ref. \cite[\S\S3--4]{Intriligator:1990ua} observes that a discrete torsion
in an orbifold of an SCFT can be acompanied by an extra information
$K: \Gamma \rightarrow \Z/2\Z$ compared with orbifolds of bosonic CFTs; 
this agrees with the homomorphism $p_0: \Gamma \rightarrow \Z/2\Z$ here
($K_g$ for $g \in \Gamma$ corresponds to $p_0(y)$ for $y \in \Gamma$ here). 
What is denoted by $\tilde{\epsilon}$
in \cite[\S\S3--4]{Intriligator:1990ua}
corresponds to $\epsilon^{00}$ here. The phases $\epsilon^{00}$ and
$\epsilon^{01}$ in (\ref{eq:def-4discTrs-a}, \ref{eq:def-4discTrs-b})
here have been captured in \cite[(3.20), (3.21)]{Intriligator:1990ua}. 
See also \cite[p.7]{Kreuzer:1995yi}.
% 
% We (the authors) could not find other work in this direction other than
% Ref. \cite{Intriligator:1990ua}. That may only be due to our ignorance;
% we welcome comments.
} %
 Ref. \cite{Intriligator:1990ua}. 

In the genus-1 amplitude (\ref{eq:partFcn-Gepner+Mkwsk-type0-SVOArepr}), 
the four complex phases, $\epsilon'$ for the NS-NS sector,
$\epsilon'$ for the R-R sector, 
$\epsilon' (-1)^{p_0(\vec{y})}$ for the NS-NS sector, and 
$\epsilon' (-1)^{p_0(\vec{y})}$ for the R-R sector are
for the four=$2^2$ distinct choices of the spin structure $s$;
the corresponding spin structures are denoted by
$s=00$, $01$, $10$ and $11$, respectively.\footnote{
The group $H^1(\Sigma_g; \Z/2\Z) \cong (\Z/2\Z)^{\oplus 2g}$
acts on the set of spin structures on $\Sigma_g$ faithfully and
transitively, but one of the spin structures needs to be used as a reference
to assign one of $(\Z/2\Z)^{2g}$ to each spin structure.
} %
The four phases are 
\begin{align}
  \epsilon^{00}(\vec{x},\vec{y}) & \; :=
  \epsilon(\gamma(\vec{x}), \gamma(\vec{y}))
  \mathbb{E}\left[ - \frac{p_0(\vec{x})p_0(\vec{y})}{2} \right]
  \Leftarrow   \epsilon'(\vec{x},\vec{y})|_{({\rm vct},\tilde{s}_0)=0+\Z} ,
  \label{eq:def-4discTrs-a} \\
  \epsilon^{01}(\vec{x},\vec{y}) & \; :=
  \epsilon(\gamma(\vec{x}), \gamma(\vec{y}))
  \mathbb{E}\left[ - \frac{p_0(\vec{x})p_0(\vec{y})}{2}
    - \frac{p_0(\vec{x})}{2} \right]
  \Leftarrow   \epsilon'(\vec{x},\vec{y})|_{({\rm vct},\tilde{s}_0)=1/2 + \Z} ,
    \label{eq:def-4discTrs-b} \\
  \epsilon^{10}(\vec{x},\vec{y}) & \; :=
  \epsilon^{00}(\vec{x},\vec{y})(-1)^{p_0(\vec{y})} =
  \epsilon(\gamma(\vec{x}),\gamma(\vec{y})) \mathbb{E}\left[
    - \frac{p_0(\vec{x})p_0(\vec{y})}{2}-\frac{p_0(\vec{y})}{2} \right],
    \label{eq:def-4discTrs-c} \\
  \epsilon^{11}(\vec{x},\vec{y}) & \; := 
\epsilon^{01}(\vec{x},\vec{y})(-1)^{p_0(\vec{y})} = 
  \epsilon(\gamma(\vec{x}),\gamma(\vec{y})) \mathbb{E}\left[
    - \frac{p_0(\vec{x})p_0(\vec{y})}{2}-\frac{p_0(\vec{y})}{2}
    -\frac{p_0(\vec{x})}{2} \right].  
    \label{eq:def-4discTrs-d}
\end{align}
It is not hard to see that they $\epsilon^{\hat{\delta}_2\hat{\delta}_1}
=: \epsilon^s$ (where $s = \hat{\delta}_2\hat{\delta}_1
\in \Z/2\Z \times \Z/2\Z$ in the convention just above
(\ref{eq:def-4discTrs-a})) are
\begin{align}
  \epsilon^{\hat{\delta}_2\hat{\delta}_1}(\vec{x},\vec{y})
  = \epsilon(\gamma(\vec{x}), \;
  \gamma(\vec{y}) ) (-1)^{{\rm Arf}[\hat{\delta}_2+p_0(\vec{x}),\hat{\delta}_1+p_0(\vec{y})]}(-1)^{{\rm Arf}[\hat{\delta}_2, \hat{\delta}_1]},
  \label{eq:def-4discTrs-byArf}
\end{align}
because ${\rm Arf}[\hat{\delta}_2,\hat{\delta}_1]
= \hat{\delta}_2 \hat{\delta}_1 \in \Z/2\Z$ when $g=1$ (in the convention
just above (\ref{eq:def-4discTrs-a})). 

One can verify that all of the following relations on
$\{ \epsilon^{\hat{\delta}_2\hat{\delta}_1} = \epsilon^s \}$
by using the properties of $\epsilon$ and $p_0$:
\begin{align}
  \frac{\epsilon^{**}(\vec{x}_1+\vec{x}_2,\vec{y})}{\epsilon^{**}(0,\vec{y})}
  & \; =
  \frac{\epsilon^{**}(\vec{x}_1,\vec{y})}{\epsilon^{**}(0,\vec{y})}
    \frac{\epsilon^{**}(\vec{x}_2,\vec{y})}{\epsilon^{**}(0,\vec{y})} \qquad
         {\rm for~each~of~} ** \in \{ 00, 01, 10, 11\},
         \label{eq:cond-disc-tors-S-a} \\
    \epsilon^{00}(0,\vec{y}) & \; = \epsilon^{01}(0,\vec{y}) =1, \qquad 
\epsilon^{10}(0,\vec{y}) , \quad \epsilon^{11}(0,\vec{y}) \in \{ \pm 1\}, 
    \label{eq:cond-disc-tors-S-b}\\
    \frac{\epsilon^{00}(\vec{x},\vec{y})}{\epsilon^{00}(0,\vec{y})} & \; =
    \frac{\epsilon^{10}(\vec{x},\vec{y})}{\epsilon^{10}(0,\vec{y})}, \qquad
    \frac{\epsilon^{01}(\vec{x},\vec{y})}{\epsilon^{01}(0,\vec{y})}  =
    \frac{\epsilon^{11}(\vec{x},\vec{y})}{\epsilon^{11}(0,\vec{y})},
    \label{eq:cond-disc-tors-S-g}
\end{align}
and 
\begin{align}
  \epsilon^{00}(\vec{x}+\vec{y},\vec{y}) = \epsilon^{10}(\vec{x},\vec{y}), & \quad
  \epsilon^{10}(\vec{x}+\vec{y},\vec{y}) = \epsilon^{00}(\vec{x},\vec{y}),    \label{eq:cond-disc-tors-S-c}\\
  \epsilon^{01}(\vec{x}+\vec{y},\vec{y}) = \epsilon^{01}(\vec{x},\vec{y}), & \quad
  \epsilon^{11}(\vec{x}+\vec{y},\vec{y}) = \epsilon^{11}(\vec{x},\vec{y}),    \label{eq:cond-disc-tors-S-d}\\
  \epsilon^{00}(-\vec{y},\vec{x}) = \epsilon^{00}(\vec{x},\vec{y}), & \quad
  \epsilon^{11}(-\vec{y},\vec{x}) = \epsilon^{11}(\vec{x},\vec{y}),    \label{eq:cond-disc-tors-S-e}\\
  \epsilon^{10}(-\vec{y},\vec{x}) = \epsilon^{01}(\vec{x},\vec{y}), & \quad
  \epsilon^{01}(-\vec{y},\vec{x}) = \epsilon^{10}(\vec{x},\vec{y}).    \label{eq:cond-disc-tors-S-f}
\end{align}
Conversely, it is also possible to see that the conditions
(\ref{eq:cond-disc-tors-S-a}--\ref{eq:cond-disc-tors-S-g})
and (\ref{eq:cond-disc-tors-S-b}--\ref{eq:cond-disc-tors-S-f}) 
almost determine the four maps $\{ \epsilon^{00}, \epsilon^{01},
\epsilon^{10}, \epsilon^{11}\}$ to be of the form given in
(\ref{eq:def-4discTrs-a}--\ref{eq:def-4discTrs-d}) for some
$\epsilon$ subject to (\ref{eq:cond-disc-tors-FIQS-a}--\ref{eq:cond-disc-tors-FIQS-c}) and a homomorphism $p_0: \Gamma \rightarrow \Z/2\Z$. The only
freedom is to modify $\epsilon^{11}$ by an overall $(\pm 1)$; 
this is equivalent to having the last factor
$(-1)^{{\rm Arf}[\hat{\delta}_2,\hat{\delta}_1]}$ in (\ref{eq:def-4discTrs-byArf}) 
or not, and this changes Type 0B SCFT to a Type 0A SCFT. This means
that there is very little room to generalize the phase
(\ref{eq:def-4discTrs-byArf}) beyond possibly dropping
the factor $(-1)^{{\rm Arf}[s]}$. 

We only say that the phase factors $\{ \epsilon^s \}$ in the partition
funciton / path integral are {\it governed} by the Arf invariant
and the original discrete torsion $\epsilon$ here. We will
take one step further at the end of this section \ref{ssec:Gepn-Ncrit-0SCFT},
however.

Here is a quick lesson from the argument above, which is
not just about the Gepner construction but about orbifold construction
among Type 0 SCFTs in general.
The discrete torsion phase in an orbifold
by a group $\Gamma$ is specified by $\epsilon: \Gamma \times \Gamma \rightarrow S^1$ subject to the conditions (\ref{eq:cond-disc-tors-FIQS-a}--\ref{eq:cond-disc-tors-FIQS-c}), the same as in \cite{Vafa:1986wx}, and by
a homomorphism $p_0: \Gamma \rightarrow \Z/2\Z$. Besides the
phase $\epsilon({\bf x},{\bf y})$ for
$({\bf x}, {\bf y}) \in\prod^{2g} \Gamma$, we have a non-trivial
phase $(-1)^{{\rm Arf}[s+p_0({\bf x},{\bf y})]}$ or
$(-1)^{{\rm Arf}[s+p_0({\bf x},{\bf y})]}(-1)^{{\rm Arf}[s]}$ for
the spin structure $s \in H^1(\Sigma, \Z/2\Z) = (\Z/2\Z)^{\oplus 2g}$;
\label{pg:IV-fin}
here, $p_0:\Gamma \rightarrow \Z/2\Z$ has been naturally extended to
$p_0: \prod^{2g} \Gamma \rightarrow (\Z/2\Z)^{\oplus 2g}$.  

\vspace{5mm}

{\bf Non-critical Gepner Type 0 SCFTs:} As we have already explained
the idea at the beginning of this section \ref{ssec:Gepn-Ncrit-0SCFT}, 
one may construct a Type 0 SCFT from $r$ $N=2$ minimal model SCFTs
without the $\R^{8-2\hat{c}}$-target SCFT, 
instead of constructing a Type 0 SCFT with the critical Virasoro
central charge $(c,\tilde{c})=(12,12)$, .
Here is a set of combinatorial data: \label{pg:Gepner-cmb-daga-list}
\begin{itemize}
\item 
$\{ (k_i, R_i) \}_{i=1,\cdots, r}$:
each $(k_i, R_i)$ specifies an unitary and
modular invariant Type 0 SCFT reviewed at the end of
section \ref{ssec:N=2mm-SCFT}. 
\item an orbifold group $\Gamma \subset {\cal G}$, where
  ${\cal G}$ is in (\ref{eq:def-calG-4int}), 
\item a discrete torsion $\epsilon:\Gamma \times \Gamma \rightarrow S^1$
  satisfying (\ref{eq:cond-disc-tors-FIQS-a}--\ref{eq:cond-disc-tors-FIQS-c})
  and a group homomorphism $p_0: \Gamma \rightarrow \Z/2\Z$. 
\end{itemize}
We do not require that $\hat{c}$ in (\ref{eq:cond-cpx-dim-Z}) is an
integer now. Let ${\cal S}_0$ denote the set of choices of
a data set $\{(k_i, R_i)\}$, $\Gamma$, $\epsilon$ and $p_0$;
there are only finitely many sets of data in ${\cal S}_0$ that
have a given $r$ and $\hat{c}$.

For one data set, $\{(k_i, R_i)\}$, $\Gamma$, $\epsilon$ and $p_0$, 
the Type 0 Gepner model SCFT is constructed by (a) tensoring the $r$ SCFTs,
(b) taking the $(\Z/2\Z)^{r-1}$-orbifold, where $(\Z/2\Z)^{r-1} \subset
(\Z/2\Z)^{r}_{\rm qnt}$, and further (c) taking the orbifold by $\Gamma$
with the discrete torsion phase given in (\ref{eq:def-4discTrs-byArf}). 
Its partition function is therefore 
\begin{align}
  Z_{\rm Gepn.0B}^{(3\hat{c},3\hat{c})} :=
  \sum_{\vec{y}}^{ \in \Gamma} \sum_{\lambda, \tilde{\lambda}}
  \frac{L_{\lambda, \tilde{\lambda}}}{2^{r+1}}  & \; 
 \left[   \sum_{\vec{\tilde{m}} \in}^{({\rm NS})}
\frac{1}{|\Gamma|} \sum_{\vec{x} \in}^{\Gamma}
\mathbb{E}\left[ - 2\vec{x} \bullet (\vec{\tilde{m}} + \vec{y}) \right]
 \right. 
    \label{eq:partFcn-Gepner-type0-SVOArepr} \\
& \qquad  \left( 
        [{\rm ch}_{\vec{\tilde{m}}+ 2\vec{y}}^{\lambda}]
        [\overline{{\rm ch}}^{\tilde{\lambda}}_{\vec{\tilde{m}}}]
        \epsilon^{00}(\vec{x},\vec{y})
        + [{\rm ch}_{\vec{\tilde{m}}+ 2\vec{y}}^{\lambda,-}]
        [\overline{{\rm ch}}^{\tilde{\lambda},-}_{\vec{\tilde{m}}}]
               \epsilon^{10}(\vec{x},\vec{y})
               \right)  \nonumber \\
 & \; +  \sum_{\vec{\tilde{m}} \in}^{({\rm R})}
\frac{1}{|\Gamma|} \sum_{\vec{x} \in}^{\Gamma}
\mathbb{E}\left[ - 2\vec{x} \bullet (\vec{\tilde{m}} + \vec{y}) \right]
   \nonumber \\
& \qquad  \left. \left( 
        [{\rm ch}_{\vec{\tilde{m}}+ 2\vec{y}}^{\lambda}]
        [\overline{{\rm ch}}^{\tilde{\lambda}}_{\vec{\tilde{m}}}]
        \epsilon^{01}(\vec{x},\vec{y})
        + [{\rm ch}_{\vec{\tilde{m}}+ 2\vec{y}}^{\lambda,-}]
        [\overline{{\rm ch}}^{\tilde{\lambda},-}_{\vec{\tilde{m}}}]
               \epsilon^{11}(\vec{x},\vec{y})
        \right) \right].  \nonumber
\end{align}
One can verify that it is modular invariant by using the properties
(\ref{eq:cond-disc-tors-S-c}--\ref{eq:cond-disc-tors-S-f}) of
$\{\epsilon^{00}, \epsilon^{10},\epsilon^{01}, \epsilon^{11}\}$.
Here, $[{\rm ch}^{\lambda}_{\vec{m}}]$ and $[{\rm ch}^{\lambda,-}_{\vec{m}}]$
are the characters ${\rm ch}^{\lambda}_{\vec{m}}$ and
${\rm ch}^{\lambda,-}_{\vec{m}}$ without the factor associated with the
fermionic part of the $\R^{8-2\hat{c}}$-target Type 0 SCFT. 

The properties (\ref{eq:cond-disc-tors-S-a}--\ref{eq:cond-disc-tors-S-g})
allow us to read out the spectrum of the Hilbert space
$({\cal H}_{\rm tot}^{\rm post})_{\rm Gepn.}$ of this CFT easily, and also 
of $({\cal H}_{\rm tot}^{\rm pre})_{\rm Gepn.}$; the Virasoro central
charge is $(c,\tilde{c}) = (3\hat{c}, 3\hat{c})$. The tensor product
${\cal A}^{\rm m.t.m.} := \otimes_{i=1}^r {\cal A}_i$
of the superchiral algebras of the $r$ internal SCFTs remains in the
untwisted sector ($\vec{y} = 0$) of $({\cal H}_{\rm tot}^{\rm pre})_{\rm Gepn.}$,
because of the property (\ref{eq:cond-disc-tors-S-b},
\ref{eq:cond-disc-tors-S-d}), so the spectrum in
$({\cal H}_{\rm tot}^{\rm dbl})_{\rm Gepn.}$ is described in terms of
the representations of ${\cal A}^{\rm m.t.m.}$ and its right-mover counter part 
$\widetilde{\cal A}^{\rm m.t.m.}$.
An irreducible representation of ${\cal A}^{\rm m.t.m.} \otimes
\widetilde{\cal A}^{\rm m.t.m.}$ is in $({\cal H}_{\rm tot}^{\rm dbl})_{\rm Gepn.}$
iff the pair $\lambda, \tilde{\lambda}$ is with
$L_{\lambda,\tilde{\lambda}} \neq 0$, and 
\begin{align}
NS-NS: & \qquad  -2 \vec{x} \bullet (\vec{\tilde{m}} + \vec{y} ) +
\frac{{\rm Arg}(\epsilon^{00}(\vec{x},\vec{y}))}{2\pi} \in \Z, \qquad {}^\forall \vec{x} \in \Gamma,
  \label{eq:cond-Gepner-survive-Z-w-d.t.} \\
R-R: & \qquad  -2 \vec{x} \bullet (\vec{\tilde{m}} + \vec{y} ) +
\frac{{\rm Arg}(\epsilon^{01}(\vec{x},\vec{y}))}{2\pi} \in \Z, \qquad {}^\forall \vec{x} \in \Gamma.
  \label{eq:cond-Gepner-survive-Z-w-d.t.for-R}
\end{align}
The states that remain in $({\cal H}_{\rm tot}^{\rm post})_{\rm Gepn.}$ than 
$({\cal H}_{\rm tot}^{\rm dbl})_{\rm Gepn.}$ are those with
(cf. $p_0(\vec{y}) \in \Z/2\Z$)
\begin{align}
  \sum_{i=1}^r (s_i-\tilde{s}_i)= 2p_0(\vec{y}) \in \Z/4\Z. 
 \label{eq:cond-Gepner-sTot-vs-sTilTot}
\end{align}

The left-moving superchiral algebra ${\cal A}^{\rm Gepn.}$ of such
a Type 0 Gepner model SCFT can be read out as follows. It contains
${\cal A}^{\rm m.t.m.}$, but also more representations
of ${\cal A}^{\rm m.t.m.}$ in general:
\begin{align}
  {\cal A}^{\rm Gepn.} & \; =
  \oplus_{y \in \Gamma_{\chi{\rm Alg}}}  
  \left( \otimes_{i=1}^r M_{\ell_i,2y_i}^{\oplus L_{\ell_i,0}^{R_i}} \right), \\
  &  \Gamma_{\chi {\rm Alg}} := \left\{ \vec{y} \in \Gamma \; \left| \;
  \sum_{i=1}^r \frac{x_iy_i}{\bar{k}_i} +
  \frac{ {\rm Arg}(\epsilon^{00}(\vec{x},\vec{y} )) } {2\pi} \in \Z
  \right. \;\; {}^\forall \vec{x} \in \Gamma \right\}. \nonumber 
\end{align}
The left-moving superchiral algebra ${\cal A}^{\rm Gepn.}$ always contains
an $N=2$ SCA with the central charge $3\hat{c}$;
not all the states in ${\cal A}^{\rm Gepn.}$ have integral charges under the 
U(1) symmetry in this $N=2$ SCA, however. 
The right-moving superchiral algebra $\widetilde{\cal A}^{\rm Gepn.}$ of
the Gepner model Type 0 SCFT can also be read out similarly:
\begin{align}
  \widetilde{\cal A}^{\rm Gepn.} = \oplus_{y \in \Gamma_{\chi {\rm alg}}}
  \left( \oplus_{i=1}^r
  \widetilde{M}_{\tilde{\ell}_i, -2y_i}^{\oplus L^{R_i}_{0,\tilde{\ell}_i}} \right)
  = \oplus_{y \in \Gamma_{\chi {\rm alg}}} \left( \oplus_{i=1}^r
  \widetilde{M}_{\tilde{\ell}_i, 2y_i}^{\oplus L^{R_i}_{0,\tilde{\ell}_i}} \right), 
\end{align}
where we used the bimultiplicative nature (\ref{eq:cond-disc-tors-S-a},
\ref{eq:cond-disc-tors-S-b}, \ref{eq:cond-disc-tors-S-d})
of $\epsilon^{00}$ at the equality in the middle. 

As a reminder, the Type 0 SCFTs formulated above are well-defined
regardless of whether the Virasoro central charges 
$(c,\tilde{c}) = (3\hat{c},3\hat{c})$ are with integral $\hat{c}$ or not, 
(or whether the U(1) charges of the $N=2$ SCAs are integral or not). 
They are formulated in such a way that, if $\hat{c} \in \{1,2,3,4\}$,  
tensoring it with the $\R^{8-2\hat{c}}$-target Type 0 SCFT and
orbifolding by the diagonal $\Z/2\Z$ symmetry of the quantum symmetries
$\Z/2\Z \times \Z/2\Z$ of the two Type 0 SCFTs reproduce the
Type 0 Gepner model critical SCFTs described in
section \ref{ssec:Gepn-crit-0SCFT}.

\vspace{5mm}
\label{pg:reintrp-AfrPhase-k=0-st}
{\bf Reinterpretation of the factor $(-1)^{{\rm Arf}[s+p_0]}(-1)^{{\rm Arf}[s]}$:}
In the rest of this section \ref{ssec:Gepn-Ncrit-0SCFT}, we explore the
relation among the phase factor $(-1)^{{\rm Arf}[s+p_0]}(-1)^{{\rm Arf}[s]}$
in (\ref{eq:def-4discTrs-byArf}), what one might call the
$(k,R) = (0, A_1)$ $N=2$ minimal model SCFT, and the Landau--Ginzburg orbifold
of a free massive chiral field. 

What one might call a level-0 $N=2$ minimal model SVOA
is with $c=0$, as we may extrapolate (\ref{eq:min-m-VirC}) to $k=0$. 
The vector space underlying any unitary SVOA with $c=0$  is
$\C {\bf 1}$, only with the vacuum state. The untwisted
(NS-type) representation of this trivial SVOA is $M_{0,0}$,
which may be split into ${\cal L}^0_{0,0}=\C{\bf 1}$ and
${\cal L}^0_{0,2}=\{ 0 \}$; this is compatible with 
(\ref{eq:N=2mm-list-repr}).  Although there is no convincing argument
to determine a twisted representation of such a trivial algebra, we
declare that there is just one irreducible Ramond-type representation of
this trivial SVOA, $M_{0,1}$ that splits into
\begin{align}
  {\cal L}^0_{1,1} \cong \C\ket{{\rm R}gs}, \quad {\rm and} \quad
  {\cal L}^0_{1,3}=\{0\};
  \label{eq:k=0-N=2mm-Rrepr}
\end{align}
certainly this is an acceptable extrapolation in that we still obey the list of irreducible representations (\ref{eq:N=2mm-list-repr}) and the property that $M_{\ell,\ell+1}$ has a single ground state (not a pair of degenerate ones),
which is in ${\cal L}^\ell_{\ell+1,1}$ (cf just below (\ref{eq:field-id})). 
It follows that ${\rm ch}^0_{0}={\rm ch}^0_{2}=1$,
${\rm ch}^{0,-}_{2b}= (-1)^b$ (for $2b \in 2\Z/4\Z$), 
and ${\rm ch}^0_1 = {\rm ch}^0_3=1$, ${\rm ch}^{0,-}_{1+2b}= (-1)^{b}$.
A more positive reason to adopt the extrapolation above is that the
orbifold interpretation below works well in fact. 
By inserting this $(k_0, R_0) = (0,A_1)$ minimal model SCFT
as the $i=0$-th one, and defining the characters $[{\rm ch}^\lambda_{\vec{m}}]$
and $[{\rm ch}^{\lambda,-}_{\vec{m}}]$ and the products
(\ref{eq:pairing-maxOrbGrp-charge-crtBos},
\ref{eq:inn-prod-on-maxOrbGrp-crtBos}) by including the $0$-th minimal model, 
\begin{align}
  Z_{\rm Gepn.0B}^{(3\hat{c},3\hat{c})} :=
  \sum_{\vec{y}}^{ \in \Gamma} \sum_{\lambda, \tilde{\lambda}}
  \frac{L_{\lambda, \tilde{\lambda}}}{2^{r+2}}  & \; 
    \sum_{\vec{\tilde{m}} \in}
\frac{1}{|\Gamma|} \sum_{\vec{x} \in}^{\Gamma}
\mathbb{E}\left[ - 2\vec{x} \bullet (\vec{\tilde{m}} + \vec{y}) \right]
  \epsilon(\vec{x},\vec{y})   
     \label{eq:partFcn-Gepner-type0-SVOArepr-w0th} \\
& \qquad  \left( 
        [{\rm ch}_{\vec{\tilde{m}}+ 2\vec{y}}^{\lambda}]
        [\overline{{\rm ch}}^{\tilde{\lambda}}_{\vec{\tilde{m}}}]
        + [{\rm ch}_{\vec{\tilde{m}}+ 2\vec{y}}^{\lambda,-}]
        [\overline{{\rm ch}}^{\tilde{\lambda},-}_{\vec{\tilde{m}}}]
               \right);  \nonumber 
\end{align}
the sign factors $(-1)^{p_0(\vec{x})p_0(\vec{y})}$ in the NS--NS sector
and $(-1)^{p_0(\vec{x})p_0(\vec{y})}(-1)^{p_0(\vec{x})}$ are now
a part of the phase $\mathbb{E}[-2\vec{x} \bullet (\vec{\tilde{m}}+\vec{y})]$,
and the sign factor $(-1)^{p_0(\vec{y})}$ is contained within
$[{\rm ch}^{\lambda,-}_{\vec{m}+2\vec{y}}][\overline{\rm ch}_{\vec{m}}^{\tilde{\lambda},-}]$. The equations
(\ref{eq:cond-Gepner-survive-Z-w-d.t.},
\ref{eq:cond-Gepner-survive-Z-w-d.t.for-R}) and
(\ref{eq:cond-Gepner-sTot-vs-sTilTot}) turn into
$-2 \vec{x} \bullet (\vec{\tilde{m}}+\vec{y}) +
    {\rm Arg}(\epsilon(\vec{x},\vec{y}))/(2\pi) \in \Z$ and
    $s_{\rm tot} = \tilde{s}_{\rm tot} \in \Z/4\Z$, respectively.  
If $p_0: \Gamma \rightarrow \Z/2\Z$
is trivial, the $i=0$-th minimal model SCFT does not do anything, and
is utterly harmless. 
Note that we do not need spacetime supersymmetry after compactification at all
(the conditions (iia, iib) in section \ref{ssec:from0B-toII}) in this argument. 

For $k \geq 1$, the $N=2$ minimal model SCFT $(k,R) = (k, A_{k+1})$ is regarded as the IR limit of the Landau--Ginzburg theory,
which is the $N=(2,2)$ supersymmetryic QFT with one chiral multiplet $\Phi$
and a superpotential $W = \Phi^{k+2}$
(see \cite{Witten:1993jg} and references therein). 
It is therefore natural to wonder if this relation extrapolates to the case $k=0$. 
The chiral superfield $\Phi$ consists of one complex scalar $\phi$,
a left-mover complex spinor field $\psi$ and a right-mover complex spinor $\tilde{\psi}$, and $W=\Phi^2$ is a mass term.  
As we study the IR spectrum of this theory, the massive scalar field $\phi$ is irrelevant.
In both the NS--NS sector ($\hat{\delta}_1=0$) and in the Ramond--Ramond sector ($\hat{\delta}_1=1$), one particle excitation  on a time-slice $S^1$ is always with a finite positive energy (no smaller than the mass),
so there is no vacuum degeneracy in each of those sectors in the IR limit of the $k=0$ Landau--Ginzburg theory. 
This is consistent with $M_{0,0} \otimes \widetilde{M}_{0,0} = \C {\bf 1}$
in the NS--NS sector and $M_{1,1}\otimes \widetilde{M}_{1,1} \cong \C$ in the R--R sector in the level-0 minimal model SVOA.
The present authors are not aware of a logic in the language of the Landau--Ginzburg theory, however, to determine the aciton of the orbifold group generator on the two ground states;
the orbifold genereator multiplies $(-1)^{p_0(\vec{x})}$ on the fields $\phi$, $\psi$ and $\tilde{\psi}$, so a state above a one-particle excitation must have an extra $(-1)^{p_0(\vec{x})}$ phase relatively to the state before the excitation,
but this argument does not determine the response of the NS--NS and R--R ground states under the generator. 
For this reason, the interpretation as the IR limit of the $k=0$ Landau--Ginzburg model has less information than (but is consistent with) the $k=0$ minimal model SCFT. Both (the latter, in particular) can replace the role of the Arf invariant in (\ref{eq:def-4discTrs-byArf}), or the freedom in the discrete torsion
in \cite[app.A]{Fuchs:1991vu} that is not visible when dealing only with
the $r$ minimal models. 
\label{pg:reintrp-AfrPhase-k=0-fin}

%%%%%%%%%%%%%%%%%%%%%%%%%%%%%%%%%%%%%%%%%%%%%%
\subsection{Type II Gepner Model SCFTs and Spacetime Supersymmetry}
\label{ssec:from0B-toII}
%%%%%%%%%%%%%%%%%%%%%%%%%%%%%%%%%%%%%%%%%%%%%%

It is colloquially said that a Type II SCFT is obtained from
a Type 0 SCFT, where the diagonal SVOA fermion parity $(-1)^{F_L-F_R}$
is gauged already, by gauging the chiral SVOA fermion parity
``$(-1)^{F_R}$'' further. This extra gauging is chiral, in that
only the orbifold projection is based only on the $\Z/2\Z$-grading 
$(-1)^{F_R}$ in the right moving sector; the twisted sectors
to be added are the representations of ${\cal A} \otimes \widetilde{\cal A}$
in the theory before the gauging
replaced only in the right-moving sector (NS-type representations
of $\widetilde{\cal A}$ replaced by R-type ones, and vice versa).
On the other hand, the gauging by $(\Z/2\Z)^r \times \Gamma \subset
{\cal G}^{\rm crt}_{\rm bos}$ in section \ref{ssec:Gepn-crit-0SCFT} 
is by a left-right diagonal symmetry action; indeed, only R--R sectors
are added or projected out in the gauging of $(\Z/2\Z)_{\rm frm.prty}$
or $(\Z/2\Z)_{\rm qnt.}$, and the projection
in the $(\Z/2\Z)^r_{\rm qnt.} \times \Gamma$ orbifold is based on
the charge $\mu_R + y$, which is the ``average''\footnote{
No such things as $\mu_R + y$ have been defined in this article, but
this is only meant as an intuitive support for $(\Z/2\Z)^r \times \Gamma$
to be diagonal. 
} %
of $\mu_L$ and $\mu_R$. 
Therefore, this extra $\Z/2\Z$ gauging leading to Type II SCFTs is quite
different in nature\footnote{
Lack of this observation, along with blurry distinction between
$G^{\rm crt}$ and ${\cal G}^{\rm crt}_{\rm bos}$, are the primary reasons
that literature has referred to the Gepner construction as
``orbifold-like'' process, $\beta$-prescription and the like and
has avoided describing it as an orbifold. 
} %
from the gauging of $(\Z/2\Z)^r \times \Gamma \subset
{\cal G}^{\rm crt}_{\rm bos}$ in
sections \ref{ssec:mtm-crit-SCFT}--\ref{ssec:Gepn-Ncrit-0SCFT}. 
So, we do not try to identify the $\Z/2\Z$ symmetry for
the extra orbifolding within the group ${\cal G}^{\rm crt}_{\rm bos}$
or ${\cal G}$.

The process of extra chiral gauging/orbifolding from a Type 0 SCFT
comes with further chocies; we have to specify what ``$(-1)^{F_R}$'' is.
We do not assume that a $\Z$-grading (such as fermion number) in the
right-mover representations has been given, or even exists, but we need
to require the existence of a structure explained in the following, and
fix a choice of it (choice of $(-1)^{F_R}$) to be gauged: we only consider
a Type 0 SCFT where (a) all the irreducible representations of
$\widetilde{\cal A}$
appearing in ${\cal H}_{\rm tot}^{\rm dbl}$ split into two irreducible
representations of the $\Z/2\Z$-even part $\widetilde{\cal A}^+$; for
such an irreducible representation $(\widetilde{M}_{\tilde{\gamma}},
\widetilde{Y}_{\tilde{\gamma}})$, there is a structure
$\widetilde{M}_{\tilde{\gamma}} = \oplus_{\tilde{s}_{\tilde{\gamma}} \in S_{\tilde{\gamma}}}
\widetilde{M}_{\tilde{\gamma} \tilde{s}_{\tilde{\gamma}}}$ where $S_{\tilde{\gamma}}$ is
a 2-element set on which the group
$S_{\tilde{0}} := \widetilde{\cal A}/\widetilde{\cal A}^+ \cong \Z/2\Z$
acts transitively; for all the triples $(\tilde{\alpha}; \tilde{\beta},
\tilde{\gamma})$ of irreducible representations of $\widetilde{\cal A}$
where the fusion constants are non-zero, there is
a set-theoretical map $f[\tilde{\alpha};\tilde{\beta},\tilde{\gamma}]:
S_{\tilde{\beta}} \times S_{\tilde{\gamma}} \rightarrow
S_{\tilde{\alpha}}$ that describes which triples of the irreducible
representations of $\widetilde{\cal A}^+$ have non-zero fusion constants.
Furthermore, within the class of Type 0 SCFTs with the property (a),
an additional structure for the chiral gauging is the existence of a universal
model of the fusion $f: S_{\tilde{\beta}} \times S_{\tilde{\gamma}}
\rightarrow S_{\tilde{\alpha}}$; namely, the universal model\footnote{
Imagine two or four Weyl and Majorana 2D fermions. The fusion rule among
the NS-type and R-type representations follows the group law of
$G_{D_1} \cong \Z/4\Z$ or
$G_{D_2} \cong \Z/2\Z \times \Z/2\Z$, respectively. 
} %
consists of the group $S_{NS} := \Z/2\Z$ and a 2-element $S_{NS}$-set
$S_{R}$ combined, $(S_{NS} \amalg S_R)$, with the fusion rule given by
either the group law of 
$\Z/2\Z \times \Z/2\Z$ or $\Z/4\Z$. An additional structure (and a choice)
(b) for the gauging of $(-1)^{F_R}$ is the existence (and a choice)
of identification $\tilde{\imath}_{\tilde{\gamma}}:
S_{\tilde{\gamma}} \rightarrow S_{NS}$
[resp. $\tilde{\imath}_{\tilde{\gamma}}: S_{\tilde{\gamma}} \rightarrow S_R$]
with the universal $S_{NS}$ [reps. $S_{R}$] as
$\Z/2\Z$-sets ($S_{\tilde{0}} \cong S_{NS}$) for all the
$\widetilde{\cal A}$-irreducible
NS-type [resp. R-type] representations appearing in
${\cal H}_{\rm tot}^{\rm dbl}$; a choice of those isomorphisms should be
such that the map $\tilde{\imath}_{\tilde{\alpha}} \cdot
f[\tilde{\alpha};\tilde{\beta},\tilde{\gamma}]$ from
$S_{\tilde{\beta}} \times S_{\tilde{\gamma}}$ to $S_{NS/R}$ agrees with
the group law $\Z/4\Z$ or $\Z/2\Z \times \Z/2\Z$ after the map 
$(\tilde{\imath}_{\tilde{\beta}} \times \tilde{\imath}_{\tilde{\gamma}})$
to $S_{NS/R} \times S_{NS/R}$.
A Type 0 SCFT with the property (a) does not necessarily have\footnote{
For the gauging of $(-1)^{F_L-F_R}$, such an identification with a universal
model of fusion is not necessary. What is used instead (implicitly
in this article) is a set of maps 
$\{ \sigma_\gamma: S_\gamma \rightarrow S_{\tilde{\gamma}}\}$
between $\Z/2\Z$ sets ($S_0 \cong S_{\tilde{0}} \cong \Z/2\Z$)
of a representation of ${\cal A}$ and
a representation of $\widetilde{\cal A}$ (that is compatible with the
left-mover and right-mover fusions $f[\alpha;\beta,\gamma]$ and
$f[\tilde{\alpha};\tilde{\beta},\tilde{\gamma}]$. 
} %
an identification $\{ \tilde{\imath}_{\tilde{\gamma}}\}$ with a universal model;
even when there is, such an identification may not be unique. 
$N=2$ minimal model Type 0 SCFTs (cf section \ref{ssec:N=2mm-SCFT}) are
good class of examples for exercise. We have a Type II SCFT for a given
choice of the additional structure ``$(-1)^{F_R}$'' in (b). 

For a given Type 0 SCFT, we have such questions as consistency conditions 
to gauge such a chiral $\Z/2\Z$ symmetry, and how many such
$\Z/2\Z$ symmetries there are. 
Setting aside those issues, however, there is one thing that is clear.

{\bf Construction and Additional Constraints on the Data:}
There is a class of $\Z/2\Z$ orbifolding (leading to Type II SCFTs) described
below where the string compactification on such SCFTs have spacetime
supersymmetry.  We begin
with presenting in terms of SCFTs with critical central charges
$(c,\tilde{c}) = (12, 12)$, so we are subject to the
condition (\ref{eq:cond-cpx-dim-Z})  (cf footnote \ref{fn:dim-Z}).
We choose
\begin{itemize}
\item [(iia)] the group structure on $S_{NS} \amalg S_R$ to be
  $\Z/2\Z \times \Z/2\Z$, where one of those $\Z/2\Z$ labels the distinction
  between untwisted (NS-type) vs twisted (R-type) representations of
  $\widetilde{\cal A}$, while the other $\Z/2\Z$ is due to whether the  
  charge under the U(1) current in the right-moving $N=2$ SCA is even or odd,
\item [(iib)] the diagonal $\Z/2\Z$ symmetry $(-1)^{F_L-F_R}$ gauged already
  in the Type 0 SCFT is due to the difference between the U(1) charge
  of the left-mover $N=2$ SCA and that of the right-mover $N=2$ SCA. 
\end{itemize}
It is customary to use
\begin{align}
  \beta_0 :=
%  (1; 1_{r \; {\rm elements}}; 1_{r \; {\rm elements}})
%      \in {\cal G}^{\rm crt}_{\rm bos}
  ({\rm sp}; 1_{r\; {\rm elements}}, \; 1_{r\; {\rm elements}})
  \in G^{\rm crt} 
  \label{eq:def-beta0-Gepnr-fermionNmbr-test}
\end{align}
to rewrite the conditions above as follows (cf \cite[(4.12)]{Gepner:1987qi}),
as we just have to deal with the U(1) charges in the representations of the
$N=2$ minimal model SVOAs reviewed around (\ref{eq:h&q-N=2mm-each-VOA-repr}):  
(iib) implies that all the states that survive in
$({\cal H}_{\rm tot}^{\rm dbl})^{\rm crt}_{\rm Gepn.}$ have
\begin{align}
  \beta_0 \bullet (\mu_L -\mu_R) \in \{ 0_{+\Z}, \; 1/2_{+\Z} \} \subset \Q/\Z,
  \label{eq:cond-in-type0-2a}
\end{align}
and the states in ${\cal H}_{\rm tot}^{\rm post}$
[resp. in $[({\cal H}_{\rm tot}^{\rm post})^\perp \subset
    {\cal H}_{\rm tot}^{\rm dbl}]$]
have $\mathbb{E}[\beta_0 \bullet (\mu_L-\mu_R)] = +1$
[resp. $-1$]. In other words, the left-mover U(1) charge and the right-mover
U(1) charge can differ only in $\Z$ for all of those states. 
The condition (iia) implies that all the
states in ${\cal H}_{\rm tot}^{\rm post}$ of the Type 0 SCFT with the
critical central charge have 
\begin{align}
  F_R/2 := \beta_0 \bullet \mu_R \in \{ 0_{+\Z}, \; 1/2_{+\Z} \} \subset
    \Q/\Z . 
\label{eq:def-Fr}
\end{align}
In other words, all the right-mover U(1) charges are integers. 

The conditions (iib) and (iia) also imply that the twisted sectors
of their respective $\Z/2\Z$ gauging are obtained from the spectrum of
$({\cal H}_{\rm tot}^{\rm post})^{\rm crt}_{\rm Gepn.}$ by the $\pm (1/2,1/2)$-unit
spectral flow and the $(0,\pm 1/2)$-unit flow, respectively, of the
left-mover and right-mover $N=2$ SCAs. The spectral flow of $(0,\pm 1)$-unit 
has to bring $({\cal H}_{\rm tot}^{\rm pre})^{\rm crt}_{\rm Gepn.}$ to itself---(**). 
Combined with the property that all the left-moving and right-moving
U(1) charges are integers (cf (\ref{eq:cond-in-type0-2a}, \ref{eq:def-Fr})), 
we see that the system described by this SCFT has spacetime supersymmetry 
(\cite{Banks:1987cy}, \cite{Banks:1988yz}). 

The conditions (\ref{eq:cond-in-type0-2a}), (\ref{eq:def-Fr}) and (**) can
be translated into those of the combinatorial data that only involve internal
$r$ SCFTs. First, 
the condition (\ref{eq:cond-in-type0-2a}) from (iib) is equivalent to
\begin{align}
  \beta_0 \bullet 2\gamma(\vec{y}) = \Z/\Z \in \Q/\Z, \qquad
  {}^\forall \vec{y} \in \Gamma, 
  \label{eq:cond-in-type0-2b}
\end{align}
as the condition (\ref{eq:cond-in-type0-2a}) is satisfied automatically
for all $y \in \Gamma_0^{\rm crt}\subset \Gamma^{\rm crt}$. 
The combinatorial data $\Gamma$ and $p_0$ of Type 0 Gepner construction is
therefore constrained when we impose (iib):
\begin{align}
  \sum_{i=1}^r \frac{y_i}{\bar{k}_i} \in 2^{-1} \Z/\Z \subset \Q/\Z,
  \qquad {}^\forall \vec{y} \in \Gamma, 
 \label{eq:cond-in-type0-2bb}  
\end{align}
and 
\begin{align}
  p_0: \vec{y} \longmapsto p_0(\vec{y}) = p_0^S(\vec{y}) := 
  2 \sum_{i=1}^r \frac{y_i}{\bar{k}_i} \in \Z/2\Z. 
    \label{eq:def-fcn-y0}
\end{align}
There is no freedom left for the homomorphism $p_0$. 

Secondly, the condition (**) is translated to the following:
\begin{align}
  \vec{y}_* \in \Gamma \subset {\cal G}, \qquad 
  \vec{y}_* := (1_{r \; {\rm elements}} ) \in {\cal G}
  \label{eq:cond-for-specFlow23-inGepner-A}
\end{align}
or equivalently,\footnote{
\label{fn:dim-Z}
The $\hat{c} \in \N$ condition (\ref{eq:cond-cpx-dim-Z}) is
satisfied automatically when we impose
(\ref{eq:cond-in-type0-2bb}--\ref{eq:cond-for-specFlow23-inGepner-A}),
because the condition (\ref{eq:def-fcn-y0}) implies that
$p_0^S(\vec{y}_*) = (r-\hat{c})_{+2\Z}$ is in $\Z/2\Z$. This value of
$p_0^S(\vec{y}_*)$ also indicates that $\gamma(\vec{y}_*) - \beta'_0$ is
contained in $\Gamma^{\rm crt}_0$. 
} %
\begin{align}
 \beta'_0 \in \Gamma^{\rm crt} \subset {\cal G}^{\rm crt}_{\rm bos}, \qquad  
 \beta'_0 := (\pm \hat{c};\vec{1};\vec{1}) \in {\cal G}^{\rm crt}_{\rm bos}. 
 \label{eq:def-betaPrm-0}
\end{align}
%
% because of the condition (**).
To make the case, we begin with
an observation (e.g., \cite[(5.12)]{Kazama:1988qp}, \cite[(4.8)]{Fuchs:1991vu})
that the spectral flow of $(-1/2)$-unit or $(-1)$-unit corresponds to 
\begin{align}
  \mu \mapsto \mu + \beta_0, \quad {\rm or} \quad
  \mu \mapsto \mu + 2\beta_0
  \label{eq:turn-into-spectFlow}
\end{align}
in $G^{\rm crt}$ 
regardless of whether $\hat{c}$ is odd or even. The condition (**)
implies that there must be (a) $\beta'_0 \in {\cal G}^{\rm crt}_{\rm bos}$
so that $2\beta'_0 = 2\beta_0 \in G^{\rm crt}$,
(b) $\vec{y}_* \in \Gamma$ so that
$\beta'_0 \sim \gamma(\vec{y}_*)$ mod $+\Gamma_0^{\rm crt}$ and
(c) $\vec{y}_* \in \Gamma_{\chi{\rm Alg}}$. The choice
$\beta'_0 \in {\cal G}^{\rm crt}_{\rm bos}$ in (\ref{eq:def-betaPrm-0}) is
the only solution to the condition (a); note that we should 
use (\ref{eq:def-embd-Gepn-orbGrp-2-lattDiscGrp}) for both even $\hat{c}$
and odd $\hat{c}$. 

Third, we argue that the condition (\ref{eq:def-Fr}) is translated
to 
\begin{align}
 \sum_{i=1}^r \frac{y_i(y_*)_i}{\bar{k}_i} + \frac{{\rm Arg}(\epsilon^{00}(\vec{y}_*, \vec{y}))}{2\pi} \in \Z, \qquad {}^\forall \vec{y} \in \Gamma.   
\label{eq:cond-Gepner-survive-specBy1-2}
\end{align}
To see this, remember that all the states in $Z^{\rm crt}_{\rm Gepn.0B}$ satisfy
\begin{align}
  2\beta'_0 \bullet (\mu_R + \gamma(\vec{y})) =
  \langle \beta'_0, \gamma(\vec{y}) \rangle \in \Q/\Z 
  \qquad {}^\forall \vec{y} \in \Gamma \subset {\cal G} 
\end{align}
because of the orbifold projection
condition (\ref{eq:cond-Gepner-proj-crt-type0-a},
\ref{eq:cond-Gepner-proj-crt-type0-b})
applied to $\beta'_0 \in \Gamma^{\rm crt}$. So, we can translate
the condition (\ref{eq:def-Fr}) as follows: 
\begin{align}
  \Z/\Z & \; = 2\beta_0 \bullet \mu_R = 2\beta'_0 \bullet \mu_R 
    = \langle \beta'_0, \gamma(\vec{y}) \rangle
  - 2\beta'_0\bullet \gamma(\vec{y}), \nonumber \\
  & \; = \langle \gamma(\vec{y}_*), \gamma(\vec{y}) \rangle
  -2 \gamma(\vec{y}_*) \bullet \gamma(\vec{y}) -2\sum_{i=1}^r 2\beta'_{(i)}\bullet \gamma(\vec{y}) \in \Q/\Z,  
\end{align}
where we used (\ref{eq:cond-in-type0-1b}) at the last equality.
The last term on the RHS
can be dropped because it is an integer. Now, (\ref{eq:def-4discTrs-a})
is used for the final step of translation. 

{\bf Summary:} We have described a construction of Type II SCFTs
with critical central charges $(c,\tilde{c})=(12,12)$ using
$r$ $N=2$ minimal model SCFTs so that the corresponding Type II string
vacua have supersymmetry on the spacetime $\R^{9-2\hat{c},1}$. 
A set of combinatorial data for such a construction is 
the same as what is listed in p. \pageref{pg:Gepner-cmb-daga-list},
but with a few more constraints as we impose the conditions (iia, iib).
The orbifold group $\Gamma$ has to
satisfy (\ref{eq:cond-for-specFlow23-inGepner-A}, \ref{eq:cond-in-type0-2bb}).
The condition (\ref{eq:cond-in-type0-2bb}) determines the maximal
subgroup $\Gamma_{\rm max}$ of ${\cal G}$ from which $\Gamma$ is chosen.
The group homomorphism $p_0$ in p. \pageref{pg:Gepner-cmb-daga-list}
should be $p_0^S$ in (\ref{eq:def-fcn-y0}) without a freedom left. 
The discrete torsion $\epsilon: \Gamma \times \Gamma \rightarrow S^1$
is determined by the condition (\ref{eq:cond-Gepner-survive-specBy1-2})
for any pair $(\vec{y}_*, \vec{y}) \in \Gamma\times \Gamma$
and $(\vec{y},\vec{y}_*)$. The discrete torsion $\epsilon$
for a pair from $\Gamma /\langle \vec{y}_*\rangle$ remains arbitrary
so long as it satisfies
(\ref{eq:cond-disc-tors-FIQS-a}--\ref{eq:cond-disc-tors-FIQS-c}). 
The collection of such combinatorial data sets is denoted by
${\cal S}_{\rm II}$; of course ${\cal S}_{\rm II} \subset {\cal S}_0$. 

The construction of a Type II SCFT for a data set in ${\cal S}_{\rm II}$
is unique. The data set in ${\cal S}_{\rm II}\subset {\cal S}_0$
determines a Type 0 Gepner model critical SCFT
($\hat{c}\in \N$ automatically; see footnote \ref{fn:dim-Z}).
As we consider only under the restriction (iia, iib), the choice of the chiral
$\Z/2\Z$ symmetry to be gauged is unique. 
The rest of the process is to tensor
the $\R^{1,1}$-target Type 0 SCFT and the $bc\beta\gamma$-ghost
sectors, and to gauge the chiral $\Z/2\Z$ symmetry. 

The spectrum of ${\cal A}^{\rm Gepn.} \otimes \widetilde{\cal A}^{\rm Gepn.}$
representations in the Type 0 Gepner model SCFT is computed by using 
(\ref{eq:cond-Gepner-survive-Z-w-d.t.}) for the NS--NS sector,
and (\ref{eq:cond-Gepner-survive-Z-w-d.t.for-R}) for the R--R sector.
The NS--R and R--NS sector states in the Type II Gepner model SCFT
are determined by using the chiral spectral flow of half-integral units.

It is possible to think of Type II SCFTs with a non-critical Virasoro
central charges $(c,\tilde{c})$. We have already exploited the
network of gauging $\Z/2\Z$ symmetry and its dual (quantum) symmetry
to extract a non-critical Type 0 SCFT with $(c,\tilde{c}) =
(3\hat{c},3\hat{c})$ from a Type 0 SCFT with the critical Virasoro
central charge in section \ref{ssec:Gepn-Ncrit-0SCFT}. The same can be
done for the chiral $\Z/2\Z$ gauging instead of the diagonal $\Z/2\Z$
gauging. When that is done, the non-critical Type II SCFT have
the universal structure of fusion on $S_{NS} \amalg S_R$ governed
by the group law $\Z/4\Z$ instead of $\Z/2\Z \times \Z/2\Z$, if  
the factored out $\R^{8-2\hat{c}}$-target SCFT has odd $(4-\hat{c})$. 

{\bf Notes:} \label{pg:notes}
Here, we explain subtle conceptual clarifications
that have been made in this article, and also how the choice of
combinatorial data for the Gepner construction has been generalized. 
Obviously the prescription in sections \ref{ssec:Gepn-crit-0SCFT}
and \ref{ssec:Gepn-Ncrit-0SCFT} without imposing (iia, iib) yields
a broader class of Type 0 SCFTs than those available in the literature,
but there is still a difference within the Type 0/II SCFTs subject
to the conditions (iia, iib). 

In many reviews on Gepner constructions that we have seen,
\label{pg:Gepnr-generator-beta0}
chosen
generators $\gamma\in \Gamma^{\rm crt} \subset {\cal G}^{\rm crt}_{\rm bos}$
are of the form of either (e.g., \cite[(4.11)]{Recknagel:1997sb} for odd $\hat{c}$, \cite[\S2]{Fuchs:1989yv})
\begin{align}
   \beta_* := (1;1_{r\; {\rm elements}}, \; 1_{r\; {\rm elements}})
   \in {\cal G}^{\rm crt}_{\rm bos};
   \label{eq:def-beta*-Gepnr-typical-genrtr}
\end{align}
or (e.g., \cite[(2.32)]{Font:1989gq}, \cite[(5.1)]{Fuchs:1991vu}, \cite[(2), (3)]{Font:1992uk})
\begin{align}
  \gamma = (0;\vec{y};\vec{0}) \in {\cal G}^{\rm crt}_{\rm bos}, \qquad
  \vec{y} \in {\cal G}.
  \label{eq:Gepner-generator-safe-ansatz}
\end{align}
One will notice the difference between $\beta_*$ and
$\beta'_0 \in {\cal G}^{\rm crt}_{\rm bos}$ when $\hat{c}$ is even.
This matters, because $\beta_0 \bullet 2 \beta'_0 = \Z/\Z \in \Q/\Z$ 
as we need in (\ref{eq:cond-in-type0-2a}, \ref{eq:cond-in-type0-2b}),
while $\beta_0 \bullet 2\beta_*=1/2+\Z$ for an even $\hat{c}$; this just
did not catch attention when the primary interest was in
particle phenomenology application (or mirror symmetry of Calabi--Yau
threefolds) with $\hat{c}=3$. This $\hat{c}$-dependence of the orbifold
group generator $\beta'_0$, combined with the $\hat{c}$-dependence of
$G_{D_{4-\hat{c}}} \subset G^{\rm crt}$ which the spectral-flow generator
$\beta_0$ is in, made the present authors realize that they are not much the same objects. The right way to think is
to start from $\beta_0 \in G^{\rm crt}$ for the $(-1/2)$-unit spectral flow
(and also a useful tool to measure $N=2$ SCA U(1) charge),
to think of $2\beta_0 \in G^{\rm crt}$ for the $(-1)$-unit spectral flow,
and finally to seek an appropriate orbifold group generator
$\beta'_0 \in {\cal G}^{\rm crt}_{\rm bos}$ for a Type 0 Gepner SCFT
so that the $\beta'_0$-twisted sector contains the $(-1,0)$-unit
spectral flow of the vacuum (i.e., $2\beta'_0 in G^{\rm crt}$ agrees with
$2\beta_0$), as we have presented in this article already. We are not
aware of a logic or reasoning that offers a shortcut here. 
See also footnote \ref{fn:improvements-in-extra-Gepner-review}. 

The class of choices of orbifold generators $\gamma(\vec{y})
\in \Gamma$ (with $p_0^S$ that may or may not be trivial)
is broader than that
in (\ref{eq:Gepner-generator-safe-ansatz}). 
This difference allows a broader class of the orbifold group $\Gamma$;
the condition (\ref{eq:cond-in-type0-2b})
would appear (e.g., \cite[(5.1)]{Fuchs:1991vu})
\begin{align}
  \sum_{i=1}^r \frac{y_i}{\bar{k}_i} = \Z/\Z \in \Q/\Z \qquad {}^\forall
  \vec{y} \in \Gamma,
  \label{eq:cond-in-type0-2bb-old}
\end{align}
when applied to $\gamma$ of the form of (\ref{eq:Gepner-generator-safe-ansatz});
in fact, it is enough to impose (\ref{eq:cond-in-type0-2bb})
for the condition (iib) due to the non-trivial homomorphism
$p_0^S: \Gamma \rightarrow \Z/2\Z$, and also the generalized form
of discrete torsion
$\{ \epsilon^{00}, \epsilon^{01}, \epsilon^{10}, \epsilon^{11}\}$. 
See an example below. 

\vspace{5mm}

{\bf Example ($\gamma(\vec{y})$ beyond (\ref{eq:Gepner-generator-safe-ansatz}),
  and $\epsilon^{00} \neq \epsilon$):}
\label{pg:Jacobi-EC}
It is often believed that the Landau--Ginzburg orbifold as
a 2D $N=(2,2)$ supersymmetric QFT with the superpotential 
\begin{align}
  W = y^2+x^4+z^4
  \label{eq:Jacobi-EC}
\end{align}
for the three chiral multiplets $y$, $x$, $z$, and the orbifold group
$\Z/4\Z$ generated by $y_*: (y,x,z) \mapsto (-y, i x, i z)$ flows
in the infrared to a 2D SCFT whose Type II version offers string
compactification over the Jacobi elliptic curve $W|_{(\ref{eq:Jacobi-EC})}=0$
with some complexified K\"{a}hler parameter. It is also believed
that this infrared 2D SCFT has something to do with the Gepner construction,
with the data $\{ (2,A_3), (2,A_3) \}$ and $\Gamma = \Z_4\vev{y_*} \subset
{\cal G}$.

Literature in the past often says that this $y_*$ does not satisfy
(\ref{eq:cond-in-type0-2bb-old}), so let us tensor, as a formality,
the $i=0$-th minimal model $(k_0, R_0) = (0, A_1)$
(or the $k_0=0$ Landau--Ginzburg theory) that is gapped and harmless in the infrared, along with $y_{i=0} = p_0^S(y_*)$.
More generally, it has been
said that the condition (\ref{eq:cond-in-type0-2bb-old}) may be violated
by $+1/2 +\Z$ in a Gepner construction
(so the condition is effectively (\ref{eq:cond-in-type0-2bb})),
because we can tensor the harmless 0-th minimal model with
$(k_0,R_0) = (0,A_1)$ as a formality, along with $y_0 = p_0^S(\vec{y})$ so
``the condition (\ref{eq:cond-in-type0-2bb-old}) for spacetime supersymmetry''
is formally satisfied when the $i=0$ contribution is included.  

As we have seen in the argument at the end of section
\ref{ssec:Gepn-Ncrit-0SCFT}, this $i=0$-th tensored minimal model of
level-$k_0=0$ is not just gapped and harmless, or a fudge factor
to cook the book and hide the mismatch $1/2+\Z$. It is not that
we can hide the mismatch, but there is such a freedom in the Gepner
construction (or orbifold theory of 2D SCFT in general); we have
seen in \ref{ssec:Gepn-Ncrit-0SCFT} that this freedom can be
captured in a couple of equivalent ways: one of which is
in tensoring the level-0 $N=2$ minimal tensor model SCFT, another
is as insertion of a phase in path-integral using the Arf invariants,
and the other is the conventional discrete torsion available in the
full critical Type 0 SCFT including the $R^{8-2\hat{c}}$-target SCFT,
all of which are within conformal field theories that allow treatment
purely in algebraic language. 
It should be remembered that this freedom has nothing to do with
spacetime supersymmetry (or the conditions (iia, iib)), or to do
with the gap between (\ref{eq:cond-in-type0-2bb-old})
and (\ref{eq:cond-in-type0-2bb-old}), but is much more general. 
This freedom has more to do with the worldsheet spin structure. 

Back in the specific example under consideration, the SCFT spectrum
can be computed purely algebraically, based on our Gepner construction. 
Because of the modification to the discrete torsion (\ref{eq:def-4discTrs-a},
\ref{eq:def-4discTrs-byArf}) with a non-trivial $p_0^S$, we have
$\epsilon^{00}(\vec{y}_*,\vec{y}_*)=-1$ in this case.
The spectrum of the Type II SCFT with this discrete torsion agrees
with that of a $T^2$ compactirification whose complex structure
is that of the Jacobi elliptic curve.

{\bf Example ($\hat{c}=2$):} for $\hat{c}$ to be 2 (complex 2-dimenions),
there are (\cite{Gepner:1987vz}, \cite{Lutken:1988hc}) 
17 different choices of $\{ k_i \}$, and 47 different choices of
$\{ (k_i, R_i) \}$. The full list of the 17 choices of $\{ k_i \}$
is found in \cite[Table 1]{Gepner:1987vz}
and \cite[Table 2]{Lutken:1988hc}.\footnote{
One choice, $\{k_i \} =\{ 1,1,1,2,2\}$, needs to be added to the
list in \cite{Gepner:1987vz}.
Ref. \cite{Lutken:1988hc} counted $(k,R) =(2,A_3)$ and
$(2, D_3)$ separately (whereas we do not here). 
} %

In the rest of this Example, we will be interested only in
the combinatorial data sets in ${\cal S}_{\rm II}$. 
For a choice $\{ (k_i, R_i) \} = \{ (1,A_2)^{\otimes 6} \}$,
there are 12 different choices of $\Gamma$ within $\Gamma_{\rm max}$.
The choice of a discrete torsion $\{ \epsilon \}$ is unique for
4 choices of $\Gamma$, while there is further variety for the 8 other
choices of $\Gamma$ consistent with (\ref{eq:def-fcn-y0},
\ref{eq:cond-Gepner-survive-specBy1-2}).
For $\{ (k_i, R_i) \}_{i=1}^r=\{
(1,A_2)^{\otimes 3}, \; (2, A_3)^{\otimes 2} \}$, there are four distinct
choices of $\Gamma$; the choice of a discrete torsion is unique
for any one of the four choices of $\Gamma$. 
It will be a doable task to complete the classification of
the choices of $\Gamma$ and $\{ \epsilon \}$ for the 15 other choices
of $\{ (k_i, R_i) \}$, but we have not done it yet.

As a reminder, the resulting Type II SCFTs may happen to be identical for
two diefferent Gepner model data sets, $\{ (k_i, R_i) \}$,
$\Gamma$ and $\{ \epsilon \}$. There are such pairs of Gepner model data
already with $\hat{c}=1$; there are also such pairs with $\hat{c}=2$. 

For some of the Gepner model data $\{ (k_i, R_i) \}_{i=1}^r$,
$\Gamma$ and $\{ \epsilon \}$, it is not hard to find an interpretation
as a special limit of the moduli space of the infrared limit of
$N=(2,2)$ supersymemtric abelian gauged linear sigma
models \cite{Witten:1993yc}.
It is not obvious whether such an interpretation is available for all
the Gepner model data, however. It is still a widely accepted belief that
the Type II SCFTs of all those Gepner model data with $\hat{c}=2$ allow
interpretation as compactification over a K3 surface or $T^4$. The Witten
index on the Ramond--Ramond sector Hilbert space indicates which
interpretation (K3 vs $T^4$) is appropriate. 

%%%%%%%%%%%%%%%%%%%%%%%%%%%%%%%%%%%%%%%%%%%%%%%%
\section{The Mirror Gepner Construction}
\label{sec:mirror}
%%%%%%%%%%%%%%%%%%%%%%%%%%%%%%%%%%%%%%%%%%%%%%%%

We should start in this section \ref{sec:mirror} by
stating the question we ask. Suppose that we have a Type 0 SCFT
(not necessarily with the critical central charge) where a copy of
the $N=2$ SCA is identified in each of the
left-moving and right-moving superchiral algebras;
this is done by indicating a special set of states $(\omega, g^c, g^a, j)$
in ${\cal A}$ corresponding to the currents, and also
$(\tilde{\omega}, \tilde{g}^c, \tilde{g}^a,
\tilde{\jmath})$ in $\widetilde{\cal A}$. 
We mean by its {\it mirror} in this article
a Type 0 SCFT that is identical to itself, but the identification
of ${\cal H}_{\rm tot}$, ${\cal A}$ and $\widetilde{\cal A}$ induces
$(\omega, g^{c/a},j) \leftrightarrow (\omega, g^{c/a},j)$ and
$(\tilde{\omega}, \tilde{g}^{c/a},\tilde{\jmath}) \leftrightarrow
(\tilde{\omega}, \tilde{g}^{a/c}, - \tilde{\jmath})$. 
This re-identification is always possible, and unique.

A non-trivial question arises when one specifies an ensemble ${\cal S}$
of sets of data (either combinatorial or geometric), and a dictionary
from the ensemble to the moduli space of Type 0 SCFTs along with an
identification of the $N=2$ SCA into ${\cal A}$ and
$\widetilde{\cal A}$. When one picks up a set of data in ${\cal S}$
and thinks of the corresponding Type 0 SCFT and identification,
does its mirror have some set of data in the same ensemble ${\cal S}$?
If there is, we call it a {\it mirror set of data}, but is
a mirror data in ${\cal S}$ unique?
What is an algorithm to find a mirror data, ${\cal S}\rightarrow {\cal S}$?
We apply this question here to the ensemble ${\cal S}={\cal S}_0$
introduced in section \ref{ssec:Gepn-Ncrit-0SCFT}, where
spacetime supersymmetry is absent in general.  
It is, in fact, better to set this question with one set of ensemble
${\cal S}$ on one hand, and another set of ensemble ${}^\circ {\cal S}$
on the mirror side, and seek a dictionary between ${\cal S}
\rightarrow {}^\circ {\cal S}$, as Berglund--H\"{u}bsch mirror
correspondence indicates explicitly \cite{Berglund:1991pp}. 
In the case we search for a mirror construction for Gepner
construction using the minimal tensor models, we do not have a
compelling evidence for a choice of ${}^\circ {\cal S}$ other than
${\cal S}$ itself, so we address this quesiton with ${}^\circ {\cal S} =
{\cal S}_0$ in this article. 
This question is obviously a continuation of the problem considered
in \cite{Greene:1990ud}, \cite[\S3]{Vafa:1994rv}, \cite{Kreuzer:1995yi} and
many other papers. 

{\bf Non-uniqueness:} The uniqueness/non-uniqueness question is in fact
not so much about the mirror process, but about the dictionary from
${\cal S}$ to the moduli space of Type 0 SCFTs and identification. 
There are known examples of multiple sets of combinatorial data of
Gepner construction that point to the same Type 0 SCFT and
identification.\footnote{
To name a few cases even within the ensemble ${\cal S}_{\rm II}$,
an identical Type 0/II SCFT and identification of a pair of $N=2$ SCA
is obtained from a common minimal tensor model
$\{ (k_i,R_i)\} = \{(2,A_3)^{\oplus 2}\}$ with two choices of
$(\Gamma,\epsilon)$ available in ${\cal S}_{\rm II}$; 
one is $\Gamma = \Z/4\Z$ and the other $\Gamma=\Z/4\Z \times \Z/4\Z$ 
(there is no freedom in $(\epsilon, p_0)$ for those $\Gamma$).

Another identical Type 0/II SCFT and an identification of a pair of $N=2$
SCA follows from three Gepner constructions in ${\cal S}_{\rm II}$;
one set of data is with $\{ (1,A_2), \; (4,A_5)\}$ and $\Gamma = \Z/6\Z$;
two others are with a common $\{ (1,A_2)^3 \}$, but with 
$\Gamma = \Z/3\Z$ and $\Gamma=\Z/3\Z \times \Z/3\Z$ (for all the three
sets of data, there is no freedom in $(\epsilon, p_0)$). 
The two specific Type 0 SCFTs here are elliptic-curve target SCFTs. 
% (cf \cite[footnoes 40, 41]{Kondo:2018mha}).
} %
So, this is enough to conclude that there is no chance finding a
one-to-one map ${\cal S} \rightarrow {\cal S}$ for ${\cal S}={\cal S}_0$
or ${\cal S}={\cal S}_{\rm II}$. 

{\bf How to find the mirror data:}
Here is a key observation in how to find a (combinatorial data of)
Gepner construction that is the mirror of a given Gepner construction
(called the {\it original theory}), with a set of data
$\{(k_i, R_i)\}_{i=1,\cdots, r}$, $\Gamma$ and
$\{ \epsilon^{00}, \epsilon^{01}, \epsilon^{10}, \epsilon^{11}\}$. 
When the vector space ${\cal H}_{\rm tot}^{\rm pre}$ of the original theory
(Type 0 SCFT) has the irreducible representation of
$\prod_{i=1}^r ({\cal A}_i \times \widetilde{\cal A}_i)$ labeled by 
$\lambda, \tilde{\lambda}, \vec{m}, \vec{\tilde{m}}$, the vector space
${\cal H}_{\rm tot}^{\rm pre}$ of the mirror has that
labeled by $\lambda, \tilde{\lambda}, \vec{m}, -\vec{\tilde{m}}$.
By using the parametrization $\vec{m} = \vec{\tilde{m}} + 2\vec{y}$
(with $\vec{y} \in \Gamma$), we see that the change in the sign of
$\vec{\tilde{m}}$ results in exchange of the followings: 
\begin{align}
  (\vec{m}+\vec{\tilde{m}}) = 2(\vec{\tilde{m}} + \vec{y}), \qquad
  (\vec{m}-\vec{\tilde{m}}) = 2\vec{y} \qquad
  \in \prod_{i=1}^r(2\Z/2\bar{k}_i\Z)
    \subset \prod_{i=1}^r (\Z/2\bar{k}_i\Z). 
\end{align}
So, if we are to find a combinatorial data $\{ (k_i, R_i)\}_{i=1,\cdots, r}$,
${}^\circ \Gamma$, $\{ {}^\circ \epsilon^{00}, {}^\circ \epsilon^{01}, {}^\circ \epsilon^{10}, {}^\circ \epsilon^{11} \}$ for the mirror, the candidate of
${}^\circ \Gamma$ is obtained by collecting $\vec{\tilde{m}} + \vec{y} =:
\vec{\eta}$ in ${\cal H}^{\rm pre}_{\rm tot}$ of the original theory. 
The candidate for the discrete torsion of the mirror,
$\{ {}^\circ \epsilon^{**}\}$, should be chosen so that the collection of
$2\vec{y}$ in the original theory survive
in the orbifold projection that leads to ${\cal H}_{\rm tot}^{\rm pre}$ of the
mirror.  Fortunately, the orbifold projection condition
(\ref{eq:cond-Gepner-survive-Z-w-d.t.},
\ref{eq:cond-Gepner-survive-Z-w-d.t.for-R}) is written in terms of
the two combinations above. 

{\bf Mirror orbifold group: }
%  $\tilde{\Gamma}^{NS}$ 
%
Following the observation above, we think of the following subset of
${\cal G}$ by collecting $\vec{\eta} := \vec{\tilde{m}} +\vec{y}$ satisfying
(\ref{eq:cond-Gepner-survive-Z-w-d.t.}):
\begin{align}
  {}^\circ \Gamma & \; = 
  {}^\circ \Gamma^{NS} := \cup_{\vec{y} \in \Gamma} {}^\circ \Gamma^{NS}_{\vec{y}},   \label{def:Union-of-Gamma-y-NS} \\
 &  {}^\circ \Gamma^{NS}_{\vec{y}} := \left\{ \vec{\eta} \in {\cal G} \; \left| \;
   -2 \vec{x} \bullet \vec{\eta} 
   + \frac{\text{Arg}(\epsilon^{00}(\vec{x},\vec{y}))}{2\pi i} = 0_{+\Z} \in \Q/\Z
   \qquad {}^\forall \vec{x} \in\Gamma \right. \right\}.
\end{align}
One can see that the subset ${}^\circ \Gamma^{NS} \subset {\cal G}$
forms a subgroup by using the additivity of
$\epsilon^{00}(\vec{x},\vec{y})$ with respect to
$\vec{y} \in \Gamma$ in the original theory. 

It is useful later to note the following structure in ${}^\circ \Gamma^{NS}$:
\begin{equation}
  {}^\circ \Gamma^{NS}=\amalg_{\bar{y}\in\Gamma/\Gamma^{NS*}}
     {}^\circ\Gamma_{\bar{y}}^{NS}, \qquad
     \Gamma^{NS*} := \{\vec{y}\in\Gamma|\epsilon^{00}(\vec{x},\vec{y})=1
     \ {}^\forall \vec{x} \in\Gamma\}.
\label{Mirror-Orbifold-Group-NS}
\end{equation}
The subset ${}^\circ \Gamma^{NS}_{\vec{y}} \subset {\cal G}$ depends
only on $\vec{y}$ mod $+\Gamma^{NS*}$, and the subsets
${}^\circ \Gamma^{NS}_{\vec{y}}$ for $\vec{y}$'s in different mod $+\Gamma^{NS*}$
equivalence classes are mutually disjoint. The abelian group structure
of ${}^\circ \Gamma^{NS}$ is compatible with the abelian group structure
of $\Gamma/\Gamma^{NS*}$. 

Similarly, think of the following subset of ${\cal G}$ by collecting
$\vec{\eta} := \vec{\tilde{m}} + \vec{y}$ satisfying
the orbifold projection condition (\ref{eq:cond-Gepner-survive-Z-w-d.t.for-R})
in the R-R sector: 
\begin{align}
 {}^\circ \Gamma^R & \; := \cup_{\vec{y} \in \Gamma} {}^\circ \Gamma^R_{\vec{y}}, \\ 
&  {}^\circ \Gamma^{R}_{\vec{y}} := \left\{ \vec{\eta} \in {\cal G} \; \left|
  -2 \vec{x} \bullet \vec{\eta} +\frac{\text{Arg}(\epsilon^{01}(\vec{x},\vec{y}))}{2\pi }= 0_{+\Z} \in\Q/\Z \qquad {}^\forall \vec{x} \in\Gamma
  \right. \right\}.
\end{align}
As one can see from (\ref{eq:def-4discTrs-a}, \ref{eq:def-4discTrs-c})
that $\epsilon^{01}(x,y)$ remains the same by changing $y$ by $\Gamma^{NS*}$, 
there is a structure 
\begin{equation}
{}^\circ \Gamma^{R} = \amalg_{\bar{y} \in \Gamma/\Gamma^{NS*}} {}^\circ \Gamma^R_{\bar{y}}.
    \label{def:Union-of-Gamma-y-R}
\end{equation}

We therefore encounter one necessary condition for a mirror Gepner
construction to exist, if we stick to the class of constructions\footnote{
One may think of a generalization by modifying the orbifold
projection in (\ref{eq:partFcn-Gepner-type0-SVOArepr}) as follows:
\begin{align}
  \sum_{\hat{\delta}_1}\sum_{\hat{\delta}_2}
  \sum_{\vec{y} \in}^{\Gamma} \sum_{\vec{x} \in}^{\Gamma}
  \mathbb{E}\left[-2\vec{x} \bullet (\vec{\tilde{m}} + \vec{y}) \right]
         \epsilon^{\hat{\delta}_2\hat{\delta}_1}(\vec{x},\vec{y})
 \Longrightarrow 
 \sum_{\hat{\delta}_1}\sum_{\hat{\delta}_2}
  \sum_{\vec{y} \in}^{\Gamma_{[\hat{\delta}_1]}} \sum_{\vec{x} \in}^{\Gamma_{[\hat{\delta}_2]}}
  \mathbb{E}\left[-2\vec{x} \bullet (\vec{\tilde{m}} + \vec{y}) \right]
       \epsilon^{\hat{\delta}_2\hat{\delta}_1}(\vec{x},\vec{y}), 
\end{align}
where $\hat{\delta}_1, \hat{\delta}_2 \in \Z/2\Z$ and
$\amalg_{\hat{\delta} \in \Z/2\Z} \Gamma_{[\hat{\delta}]}$ is a $\Z/2\Z$-graded
group not necessarily with $\Gamma_{[0]} = \Gamma_{[1]}$.
With the class of discrete torsion $\{ \epsilon^{\hat{\delta}_2\hat{\delta}_1} \}$
described in section \ref{ssec:Gepn-Ncrit-0SCFT}, however, the partition
function (\ref{eq:partFcn-Gepner-type0-SVOArepr}) would not be invariant
under the $T$-transformation when $\Gamma_{[1]} \neq \Gamma_{[0]}$.  
Such a generalization may still be possible by allowing discrete torsion
to be more general (cf \cite{Gaberdiel:2004vx}), but we do not explore
the possibility in this article. 
} %
described in section \ref{ssec:Gepn-Ncrit-0SCFT}.
The group ${}^\circ \Gamma^{NS} \subset {\cal G}$
and the subset ${}^\circ \Gamma^R \subset {\cal G}$ are not necessarily
the same, but they should agree if a mirror Gepner construction exists. 
To write down this necessary condition in terms of $\Gamma$, $p_0$ and
$\epsilon$ of the original Gepner construction, let us pause
for a moment for technical preparations. 

Note, first, that the group ${\cal G} = \prod_{i=1}^r \Z/\bar{k}_i\Z$ is
isomorphic to its dual group (the group ${\rm Char}({\cal G})$
of characters of ${\cal G}$);
\begin{align}
  {\cal G} \ni \vec{x} \mapsto
  \mathbb{E}[-2\vec{x} \bullet -] \in {\rm Char}({\cal G}),
  \label{eq:iso-G-charG}
\end{align}
or $(-2\vec{x}\bullet -) \in {\rm Hom}({\cal G}, \Q/\Z)$. 
Second, for any homomorphism $\psi: \Gamma_* \rightarrow \Q/\Z$
(or a character $\mathbb{E}[\psi(-)]$ of $\Gamma_*$) of a subgroup
$\Gamma_* \subset {\cal G}$, there is a lift
homorphism $\tilde{\psi}: {\cal G} \rightarrow \Q/\Z$
(a character $\mathbb{E}[\tilde{\psi}(-)]$ of ${\cal G}$),
because $\Q/\Z$ is injective. Combining these two facts, we see that
there is $\vec{y}_{\psi} \in {\cal G}$ s.t.
$\mathbb{E}[-2\vec{y}_\psi \bullet -]|_{\Gamma_*} = \mathbb{E}[\psi(-)]$
on $\Gamma_*$; such $\vec{y}_\psi \in {\cal G}$ for a given $\psi$
is with ambiguity
\begin{align}
  \Gamma_*^\perp := \left\{ \vec{z} \in {\cal G} \; | \;
    -2 \vec{x} \bullet \vec{z} = 0_{+\Z} \in \Q/\Z,
    \; {}^\forall \vec{x} \in \Gamma_*\right\}
    \cong {\rm Char}({\cal G}/\Gamma_*)
\end{align}
So, homomorphisms $\psi: \Gamma_* \rightarrow \Q/\Z$ of $\Gamma_*$ are
in one-to-one with ${\cal G}/\Gamma_*^\perp \cong {\rm Char}(\Gamma_*)$. 

One of applications of the technical observation above is to assign
\begin{align}
  [\vec{y}_0] \in {\cal G}/\Gamma^\perp, \qquad
  {\rm where~} \mathbb{E}[p_0(-)/2] =\mathbb{E}[ -2 \vec{y}_0 \bullet -]
  {\rm ~on~}\Gamma.
\end{align}
The property that $p_0/2:\Gamma \rightarrow \Q/\Z$ is in fact
$2^{-1}\Z/\Z$-valued is translated to the statement that
$2\vec{y}_0 \in \Gamma^\perp$. So, for a given group
$\Gamma \subset {\cal G}$, possible choice of $p_0$ is in one-to-one
with the 2-torsion elements of ${\cal G}/\Gamma^\perp$, or equivalently,
with the elements $(2\Gamma)^\perp/\Gamma^\perp$. 

As another application, think of $\epsilon^{00}(-,\vec{y})$ for each
$\vec{y} \in \Gamma$ as a character on $\Gamma$, so we may assign
\begin{align}
  \epsilon^{00\vee}: \Gamma \ni \vec{y} \mapsto [\vec{y}_{\epsilon00}^\vee]
  \in {\cal G}/\Gamma^\perp, \qquad
  {\rm where} \quad
  \epsilon^{00}(-,\vec{y}) = \mathbb{E}[-2\vec{y}_{\epsilon00}^\vee \bullet -]
  \quad {\rm in~}{\rm Char}(\Gamma); 
\end{align}
this assignment $\epsilon^{00\vee}$ is a group homomorphism.
In this notation,
\begin{align}
  {}^\circ \Gamma^{NS}_{\bar{y}}/\Gamma^\perp =  -[\vec{y}_{\epsilon00}^\vee] .
  \qquad 
  {}^\circ \Gamma^R_{\bar{y}}/\Gamma^\perp = - [\vec{y}_{\epsilon00}^\vee] + [\vec{y}_0].
   \label{eq:diff-mirror-Gns-Gr}   
\end{align}
and
\begin{align}
  {}^\circ \Gamma^{NS}/\Gamma^\perp
    = \epsilon^{00\vee}(\Gamma) \subset {\cal G}/\Gamma^\perp.
\end{align}

One more application is that ${}^\circ \Gamma^{NS}_{\vec{0}} =
{\rm Char}({\cal G}/\Gamma) = \Gamma^\perp \subset {\cal G}$, as is
known well.  
The mirror orbifold group ${}^\circ \Gamma = {}^\circ \Gamma^{NS}
\subset {\cal G}$ therefore always contains $\Gamma^\perp$,  
and may be larger if $\epsilon^{00}$ is non-trivial. 

Now, for a given $\Gamma$ and $p_0$ (or $[\vec{y}_0] \in
(2\Gamma)^\perp/\Gamma^\perp$), the necessary condition
${}^\circ \Gamma^{NS} = {}^\circ \Gamma^R$ for a mirror Gepner construction
to exist is stated in terms of the original theory. 
The condition that $[\vec{y}_0] \in {}^\circ \Gamma^{NS}$ implies that 
\begin{align}
  {}^\exists \overline{y_0^\vee} \in \Gamma/\Gamma^{NS*} \quad {\rm s.t.} \quad
   \mathbb{E}[-2\vec{x} \bullet \vec{y}_0]
   \epsilon^{00}(\vec{x},\overline{y_0^\vee}) = 1 \quad
           {}^\forall \vec{x} \in \Gamma.
  \label{eq:rltn-y0-y0vee}
\end{align}

The extra conditions in section \ref{ssec:from0B-toII} guarantees that
this necessary condition is satisfied. To see this, note first that
the choice $p_0 = p_0^S$ in (\ref{eq:cond-in-type0-2bb}, \ref{eq:def-fcn-y0})
implies that $[\vec{y}_0] = [\vec{y}_*] \in (2\Gamma)^\perp/\Gamma^\perp$.
The conditions (\ref{eq:cond-Gepner-survive-specBy1-2},
\ref{eq:cond-for-specFlow23-inGepner-A}, 
\ref{eq:cond-in-type0-2bb}) guarantee that this $[\vec{y}_0]$ is
in ${}^\circ \Gamma^{NS}_{\bar{y}_*} \subset {}^\circ \Gamma^{NS}$ indeed.  
$\overline{y_0^\vee} = y_* +\Gamma^{NS*}$. 
Incidentally, we have also seen that the mirror orbifold group
${}^\circ \Gamma = {}^\circ \Gamma^{NS}$ also satisfies a part of
the extra conditions, (\ref{eq:cond-for-specFlow23-inGepner-A}).
One also finds that the mirror orbifold group also satisfies the other
extra condition (\ref{eq:cond-in-type0-2bb}) by combining
(\ref{eq:cond-Gepner-survive-Z-w-d.t.}, \ref{eq:cond-Gepner-survive-specBy1-2},
\ref{eq:cond-in-type0-2bb}) in the original Gepner construction. 

\vspace{5mm}

{\bf Mirror discrete torsion:}
Once again, we do not necessarily assume that the extra
conditions in section \ref{ssec:from0B-toII} are satisfied;
we just assume that ${}^\circ \Gamma^{R} = {}^\circ \Gamma^{NS}$. 
The mirror discrete torsion ${}^\circ \epsilon^{00}$ and
${}^\circ \epsilon^{01}$ should be set so that
a mirror charge average ${}^\circ (\vec{\tilde{n}}+\vec{\eta})$
in the mirror $\vec{\eta}$-twisted sector in the Hilbert space
satisfies the orbifold projection condition 
(\ref{eq:cond-Gepner-survive-Z-w-d.t.},
\ref{eq:cond-Gepner-survive-Z-w-d.t.for-R})---(**2). In doing so, remember that 
${}^\circ (\vec{\tilde{n}}+\vec{\eta}) = \vec{y}$ and
$\vec{\eta} = (\vec{\tilde{m}}+\vec{y})$ in terms of the original theory.
Due to the mod $+\Gamma^{NS*}$ structure we have seen above, both the
set of charge averages $(\vec{\tilde{m}}+\vec{y})$ and the charge differences
$\vec{y}$ of the original theory can be split into the
subsets labeled by $\Gamma/\Gamma^{NS*}$. So, the mirror discrete torsion
can be determined for each of the cosets $\bar{y} \in \Gamma/\Gamma^{NS*}$. 

To start, pick one $\bar{y} \in \Gamma/\Gamma^{NS*}$ and think of
any $\vec{\eta} \in {}^\circ\Gamma^{NS}_{\bar{y}}$.
The algorithm (**2) determines a homomorphism 
${}^\circ \epsilon^{00}(-,\vec{\eta}):
{}^\circ \Gamma^{NS} \rightarrow S^1$ as follows: 
\begin{align}
  {}^\circ \Gamma^{NS} \ni \vec{\xi} \mapsto \frac{{\rm Arg}({}^\circ \epsilon^{00}(\vec{\xi}, \vec{\eta}))}{2\pi} & \; =
  2\vec{\xi} \bullet {}^\circ (\vec{\tilde{n}} + \vec{\eta})
  = 2\vec{\xi} \bullet \vec{y} \in \Q/\Z,  
\end{align}
where we picked one $\vec{y} \in \bar{y}$;
although the value $2\vec{\xi}\bullet \vec{y}$ in $\Q/\Z$
appears to depend on the $+\Gamma^{NS*}$ ambiguity in $\vec{y} \in \bar{y}$
it does not, in fact, because 
\begin{align}
  2\vec{\xi} \bullet \vec{y}
  = 2\vec{y} \bullet \vec{\xi}
  = \frac{{\rm Arg}(\epsilon^{00}(\vec{y}, \vec{x}))}{2\pi} \in \Q/\Z, 
\end{align}
where $\vec{\xi} \in {}^\circ \Gamma^{NS}_{\bar{x}}$ for an appropriate
$\bar{x} \in \Gamma/\Gamma^{NS*}$; the $\Q/\Z$-value in the last
expression does not depend on how we choose $\vec{y}$ from $\bar{y}$.
This also means that all the mirror charge averages
${}^\circ (\vec{\tilde{n}} + \vec{\eta}) = \vec{y}$ in the
$\vec{\eta}$-twisted sector survive the orbifold projection condition 
(\ref{eq:cond-Gepner-survive-Z-w-d.t.}) in the mirror theory. 
This reasoning can be applied to any $\bar{y} \in \Gamma/\Gamma^{NS*}$
for the decomposition of ${}^\circ \Gamma^{NS}$, so the algorithm (**2)
leads us to
\begin{align}
  {}^\circ \epsilon^{00}:
  {}^\circ \Gamma^{NS} \times {}^\circ \Gamma^{NS} \ni (\vec{\xi}, \vec{\eta})
  \mapsto \epsilon^{00}(\bar{y},\bar{x}) \in S^1,
\end{align}
where $\vec{\xi} \in {}^\circ \Gamma^{NS}_{\bar{x}}$ and
$\vec{\eta} \in {}^\circ \Gamma^{NS}_{\bar{y}}$. 

Next, pick one $\bar{y} \in \Gamma/\Gamma^{NS*}$, and think of any 
$\vec{\eta} \in {}^\circ \Gamma^R_{\bar{y}}$. The algorithm (**2) can be used
to determine a character 
${}^\circ \epsilon^{01}(-,\vec{\eta}): {}^\circ \Gamma^{NS} \rightarrow \Q/\Z$: 
\begin{align}
  {}^\circ \Gamma^{NS} \ni \vec{\xi} \mapsto 
  \frac{{\rm Arg}({}^\circ \epsilon^{01}(\vec{\xi}, \vec{\eta}))}{2\pi}
  = 2 \vec{\xi}\bullet {}^\circ (\vec{\tilde{n}} + \vec{\eta})
  = 2 \vec{\xi} \bullet \vec{y} \in \Q/\Z, 
\end{align}
where the $+\Gamma^{NS*}$ ambiguity in $\vec{y} \in \bar{y}$ still
remains at this moment. However, 
\begin{align}
  2\vec{\xi} \bullet \vec{y} = 2\vec{y} \bullet \vec{\xi}
  = \frac{{\rm Arg}(\epsilon^{00}(\vec{y}, \bar{x}))}{2\pi}
  = \frac{{\rm Arg}(\epsilon^{00}(\bar{y}, \bar{x}))}{2\pi} \in \Q/\Z,  
\end{align}
as we used the property (**2) for $\vec{\xi} \in {}^\circ \Gamma^{NS}_{\bar{x}}$
in the middle; now the value in $\Q/\Z$ is independent of the choice
of $\vec{y} \in \bar{y}$.
\begin{align}
  {}^\circ \epsilon^{01}: \Gamma^{NS}_{\bar{x}} \times \Gamma^R_{\bar{y}} \ni
  (\vec{\xi}, \vec{\eta}) \mapsto \epsilon^{00}(\bar{y}, \bar{x}) \in S^1. 
\end{align}
Equivalently (as we assume that $[\vec{y}_0] \in {}^\circ \Gamma^{NS}_{\overline{y_0^\vee}}$ for some ${}^\exists \overline{y_0^\vee} \in \Gamma/\Gamma^{NS*}$),
\begin{align}
  {}^\circ \epsilon^{01}: \Gamma^{NS}_{\bar{x}} \times \Gamma^{NS}_{\bar{y}} \ni
  (\vec{\xi}, \vec{\eta}) \mapsto
  \epsilon^{00}(\bar{y}- \overline{y_0^\vee}, \bar{x}),
\end{align}
where we used the relation between ${}^\circ \Gamma^R_{\bar{y}}$ and
${}^\circ \Gamma^{NS}_{\bar{y}}$ in (\ref{eq:diff-mirror-Gns-Gr}). 

The mirror discrete torsion ${}^\circ \epsilon^{10}$ and
${}^\circ \epsilon^{11}$ should be determined
as follows: as the orbifold projection operators for
$(\hat{\delta}_2,\hat{\delta}_1)
=(0,0)$ and $(0,1)$ in the original theory 
\begin{align*}
  \sum_{\vec{x} \in \Gamma} \epsilon^{00}(\vec{x},\vec{y})
  \frac{\mathbb{E}\left[ - 2\vec{x}\bullet (\vec{\tilde{m}}+\vec{y})\right]}
       {|\Gamma|} \quad {\rm and} \quad 
  \sum_{\vec{x} \in \Gamma} \epsilon^{01}(\vec{x},\vec{y})
  \frac{\mathbb{E}\left[ - 2\vec{x}\bullet (\vec{\tilde{m}}+\vec{y})\right]}
       {|\Gamma|}
\end{align*}
turn into
\begin{align*}
  \sum_{\vec{\xi} \in {}^\circ \Gamma} {}^\circ\epsilon^{00}(\vec{\xi},\vec{\eta})
  \frac{\mathbb{E}\left[ - 2\vec{\xi}\bullet {}^\circ (\vec{\tilde{n}}+\vec{\eta})\right]}
       {|{}^\circ \Gamma|} \quad {\rm and} \quad 
  \sum_{\vec{\xi} \in {}^\circ \Gamma} {}^\circ\epsilon^{01}(\vec{\xi},\vec{\eta})
  \frac{\mathbb{E}\left[ - 2\vec{\xi}\bullet {}^\circ (\vec{\tilde{n}}+\vec{\eta})\right]}
       {|{}^\circ \Gamma|} 
\end{align*}
in the mirror theory, the projection operators for $(\hat{\delta}_2,\hat{\delta}_1)=(1,0)$ and $(1,1)$ turn into
\begin{align*}
  \sum_{\vec{\xi} \in {}^\circ \Gamma} {}^\circ\epsilon^{00}(\vec{\xi},\vec{\eta})
  \frac{\mathbb{E}\left[ - 2(\vec{\xi}-\vec{y}_0)\bullet
      {}^\circ (\vec{\tilde{n}}+\vec{\eta})\right]}
       {|{}^\circ \Gamma|} \quad {\rm and} \quad 
  \sum_{\vec{\xi} \in {}^\circ \Gamma} {}^\circ\epsilon^{01}(\vec{\xi},\vec{\eta})
  \frac{\mathbb{E}\left[ - 2(\vec{\xi}-\vec{y}_0)\bullet
      {}^\circ (\vec{\tilde{n}}+\vec{\eta})\right]}  {|{}^\circ \Gamma|},  
\end{align*}
because $(-1)^{p_0(\vec{y})} = \mathbb{E}[2\vec{y}_0 \bullet \vec{y}]$
in the original theory is translated into $\mathbb{E}[2\vec{y}_0 \bullet
  {}^\circ (\vec{\tilde{n}}+\vec{\eta})]$ in the mirror. Therefore, 
\begin{align}
  {}^\circ \epsilon^{10}(\vec{\xi}, \vec{\eta}) & \; =
  {}^\circ \epsilon^{00}(\vec{\xi}+\vec{y}_0, \vec{\eta})
  = \epsilon^{00}(\bar{y}, \bar{x}+\overline{y_0^\vee}). \\
  {}^\circ \epsilon^{11}(\vec{\xi}, \vec{\eta}) & \; =
  {}^\circ \epsilon^{01}(\vec{\xi}+\vec{y}_0, \vec{\eta})
  = \epsilon^{00}(\bar{y}-\overline{y_0^\vee}, \bar{x} + \overline{y_0^\vee}) 
\end{align}
for $\vec{\xi} \in {}^\circ \Gamma_{\bar{x}}^{NS}$ and
$\vec{\eta} \in {}^\circ \Gamma_{\bar{y}}^{NS}$. 

To summarize, the four maps $\{ {}^\circ \epsilon^{00}, {}^\circ \epsilon^{10},
{}^\circ \epsilon^{01},  {}^\circ \epsilon^{11} \}$ from ${}^\circ \Gamma \times
{}^\circ \Gamma$ to $S^1$ factor through
\begin{align}
  {}^\circ \Gamma/ {}^\circ \Gamma_{\bar{0}}
  \cong \Gamma/\Gamma^{NS*},  
    \label{eq:isom-orbfd-grp-w-mirror}
\end{align}
and are given by
\begin{align}
  {}^\circ \epsilon^{00} : {}^\circ \Gamma_{\bar{x}}  \times {}^\circ \Gamma_{\bar{y}}
  \ni (\vec{\xi}, \vec{\eta}) \longmapsto & \; 
  {}^\circ \epsilon^{00} (\vec{\xi},\vec{\eta}) = \epsilon^{00}(\bar{y}, \bar{x}), \label{eq:mrr-discTrs-a} \\ 
    {}^\circ \epsilon^{10} : {}^\circ \Gamma_{\bar{x}}  \times {}^\circ \Gamma_{\bar{y}}
  \ni (\vec{\xi}, \vec{\eta}) \longmapsto & \; 
  {}^\circ \epsilon^{10} (\vec{\xi},\vec{\eta}) = \epsilon^{00}(\bar{y}, \bar{x}+\overline{y_0^\vee}), \\ 
    {}^\circ \epsilon^{01} : {}^\circ \Gamma_{\bar{x}}  \times {}^\circ \Gamma_{\bar{y}}
  \ni (\vec{\xi}, \vec{\eta}) \longmapsto & \; 
  {}^\circ \epsilon^{01} (\vec{\xi},\vec{\eta}) = \epsilon^{00}(\bar{y}-\overline{y_0^\vee}, \bar{x}), \\ 
    {}^\circ \epsilon^{11} : {}^\circ \Gamma_{\bar{x}}  \times {}^\circ \Gamma_{\bar{y}}
  \ni (\vec{\xi}, \vec{\eta}) \longmapsto & \; 
      {}^\circ \epsilon^{11} (\vec{\xi},\vec{\eta}) = \epsilon^{00}(\bar{y}-\overline{y_0^\vee}, \bar{x}+\overline{y_0^\vee}).
      \label{eq:mrr-discTrs-d}
\end{align}

Remember that four maps $\{{}^\circ \epsilon \}$ have been determined
uniquely as above, but that was under an assumption that a mirror Gepner
construction exists. We see below that they satisfy the conditions
(\ref{eq:cond-disc-tors-S-a}--\ref{eq:cond-disc-tors-S-f}) for the
group ${}^\circ \Gamma \subset {\cal G}$; this means
that the condition ${}^\circ \Gamma^R = {}^\circ \Gamma$ is the only one
on a Gepner construction data ${\cal S}_0$ in
section \ref{ssec:Gepn-Ncrit-0SCFT} for
a mirror Gepner construction to exist. 

The first step in doing so is to note---by using the properties of $\epsilon^{00}$---that
${}^\circ \epsilon(\vec{\xi}_1+\vec{\xi}_2,\vec{\eta}) = {}^\circ \epsilon^{00}(\vec{\xi}_1,\vec{\eta}) {}^\circ \epsilon^{00}(\vec{\xi}_2,\vec{\eta})$,
${}^\circ \epsilon^{00}(\vec{\xi},\vec{\eta}) = {}^\circ \epsilon^{00}(-\vec{\eta},\vec{\xi})$ and ${}^\circ \epsilon^{00}(\vec{\xi}+2\vec{\eta},\vec{\eta})$
for ${}^\forall \vec{\xi}, \vec{\eta} \in {}^\circ\Gamma$. This means that
the map ${}^\circ \epsilon^{00}:{}^\circ \Gamma \times {}^\circ \Gamma \rightarrow
S^1$ has a presentation of the form (\ref{eq:def-4discTrs-a}) for some
${}^\circ \epsilon$ of the group ${}^\circ \Gamma$ satisfying the
conditions (\ref{eq:cond-disc-tors-FIQS-a}--\ref{eq:cond-disc-tors-FIQS-c})
and the group homomorphism ${}^\circ p_0: {}^\circ \Gamma \rightarrow \Z/2\Z$
given by 
\begin{align}
  {}^\circ \Gamma \subset {}^\circ \Gamma_{\bar{x}} \ni \vec{\xi}
  \longmapsto (-1)^{{}^\circ p_0(\vec{\xi})} =
  {}^\circ \epsilon^{00}(\vec{\xi},\vec{\xi})
  = \epsilon^{00}(\bar{x},\bar{x}) = (-1)^{p_0(\bar{x})} \in \{ \pm 1\}.      
\end{align}
A map ${}^\circ \epsilon: {}^\circ \Gamma \times {}^\circ \Gamma \rightarrow
S^1$ satisfying (\ref{eq:cond-disc-tors-FIQS-a}--\ref{eq:cond-disc-tors-FIQS-c})
can be extracted from ${}^\circ \epsilon^{00}$ and ${}^\circ p_0$ by
using the relation (\ref{eq:def-4discTrs-a}). 
Next, we can use the additivity of $\epsilon^{00}(x,y)$ with respect to $x$
and also $y$, and also the relation (\ref{eq:rltn-y0-y0vee}) to see
that the two maps ${}^\circ \epsilon^{01}$ and ${}^\circ \epsilon^{10}$ also
have the parametrization by ${}^\circ \epsilon$ and ${}^\circ p_0$
(that have already been specified); indeed, 
\begin{align}
 \epsilon^{00}(- \overline{y_0^\vee}, \bar{x}) = \mathbb{E}[2\vec{x}\bullet \vec{y}_0] = (-1)^{p_0(\bar{x})} = (-1)^{{}^\circ p_0(\vec{\xi})}.  
\end{align}
Finally, the same argument can be repeated for the map ${}^\circ \epsilon^{11}$
to see that ${}^\circ \epsilon^{11}$ has either the parametrization
(\ref{eq:def-4discTrs-d}) or possibly modified from it by a sign factor
$(-1)^{p_0(y_0^\vee)}$. An overall sign that is independent of
$\vec{\xi}, \vec{\eta}\in {}^\circ \Gamma$ is the freedom allowed by
the conditions (\ref{eq:cond-disc-tors-S-a}--\ref{eq:cond-disc-tors-S-f}). 

A Gepner construction in section \ref{ssec:from0B-toII}, which is
subject to the extra conditions (\ref{eq:cond-in-type0-2bb},
\ref{eq:def-fcn-y0},  \ref{eq:cond-for-specFlow23-inGepner-A},
\ref{eq:cond-Gepner-survive-specBy1-2}) has its mirror within this
narrower class ${\cal S}_{\rm II}$ of Gepner constructions.
Indeed, we have already seen earlier
that the four extra conditions guarantee that ${}^\circ \Gamma^R =
{}^\circ \Gamma$ (so, a mirror Gepner construction exists), and that
the mirror Gepner data also satisfy (\ref{eq:cond-in-type0-2bb},
\ref{eq:cond-for-specFlow23-inGepner-A}). To see that the mirror
Gepner data also satisfy the remaining two extra conditions
(\ref{eq:def-fcn-y0}, \ref{eq:cond-Gepner-survive-specBy1-2}), note
that
\begin{align}
  \mathbb{E}[2\vec{\eta}\bullet \vec{y}_* ] = \epsilon^{00}(\vec{y}_*,\bar{y})
  = \mathbb{E}[2\vec{y}_* \bullet \vec{y}]  = (-1)^{p_0(\bar{y})};
\end{align}
the orbifold projection condition (\ref{eq:cond-Gepner-survive-Z-w-d.t.})
for $\vec{y}_* = \vec{y}_0 \in \Gamma$ (cf (\ref{eq:def-fcn-y0},
\ref{eq:cond-for-specFlow23-inGepner-A} in the original theory)
is used in the first equality, and the
condition (\ref{eq:cond-Gepner-survive-specBy1-2}) in the original theory
is used in the second equality. Now, the condition (\ref{eq:def-fcn-y0})
is satisfied in the mirror theory, because $(-1)^{{}^\circ p_0(\vec{\eta})} = \mathbb{E}[2\vec{\eta} \bullet \vec{y}_*]$.
The condition (\ref{eq:cond-Gepner-survive-specBy1-2}) in the mirror theory
is also satisfied because ${}^\circ \epsilon(\vec{y}_*, \vec{\eta}) =
\mathbb{E}[2\vec{y}_* \bullet \vec{y}]$ (the orbifold projection
condition (\ref{eq:cond-Gepner-survive-Z-w-d.t.}) in the mirror thoery,
with $\vec{y}_* \in {}^\circ \Gamma$) is equal to $\mathbb{E}[2\vec{y}_* \bullet \vec{\eta}]$ as we have seen above. 

\newpage

%%%%%%%%%%%%%%%%%%%%%%%%%%%%%%%%%%%%%%%%%%%%%%%%%%%%%
\section{SCFT Vertex Operators of Gepner Models}
\label{sec:CFTop}
%%%%%%%%%%%%%%%%%%%%%%%%%%%%%%%%%%%%%%%%%%%%%%%%%%%%

A minimal set of algebraic data of a 2D CFT consists not just of
the representation of the left-moving and right-moving chiral
algebras ${\cal A}$ and $\widetilde{\cal A}$ on the Hilbert space
${\cal H}_{\rm tot}$, but also the CFT vertex operators $Y_{\rm tot}(x,z)$
acting on ${\cal H}_{\rm tot}$ of all the states $x \in {\cal H}_{\rm tot}$.
Earlier sections were about the representation of $x \in {\cal A}^{\rm m.t.m.}
\otimes \widetilde{\cal A}^{\rm m.t.m.} \subset
({\cal H}_{\rm tot}^{\rm dbl})_{\rm Gepn.}$,
not of all the states in $({\cal H}_{\rm tot}^{\rm dbl})_{\rm Gepn.}$. 

Given the fact that Gepner model SCFTs are obtained by modifying
minimal tensor model SCFTs, it is most convenient if we can specify/describe 
the SCFT vertex operator $YY_{\rm tot}^{\rm Gepn.}$ of a Gepner model
by using the SCFT vertex operator $YY_{\rm tot}^{\rm m.t.m.}$ of
the minimal tensor model it came from.  A large fraction of articles reviewing
Gepner model SCFTs does not offer information about
SCFT vertex operators $YY_{\rm tot}^{\rm Gepn.}$ more than saying\footnote{
\label{fn:spc-flw}
It is worth pointing out that the term {\it spectral flow} (or spectral flow
operator) has been used in a couple of different meanings. That has been so
even when we restrict our attention to the one associated with the U(1) current
of an $N=2$ superconformal algebra (SCA).

In one meaning, it is associated with an automorphism
${\cal P}_\theta$ ($\theta \in \R$) of the $N=2$ SCA;
${\cal P}_\theta: J_n \mapsto J_n + \theta \hat{c} \delta_n$,
$L_n \mapsto L_n + \theta J_n + \delta_n \theta^2 \hat{c}/2$ and
$G^{c}_\nu \mapsto G^{c}_{\nu + \theta}$ (see \cite{Schwimmer:1986mf}).
For any representation $(M_1,\rho_1)$
of the algebra where $G^c(e^{2\pi i}z) = G^c(z) e^{-2\pi i (\nu_1-1/2)}$,
there is a representation $(M_2,\rho_2)$ where $G^c(e^{2\pi i} z) = G^c(z)e^{-2\pi i (\nu_1+\theta-1/2)}$, and a vector-space isomorphism
$P_\theta: M_1 \rightarrow M_2$ so that $\rho_2(x) \cdot P_\theta =
P_\theta \cdot \rho_1(x)$, and $\rho_2(x) = \rho_1({\cal P}_\theta(x))$;
this is true for any $\theta \in \R$ (e.g. \cite[\S3.4]{Greene:1996cy}). 
There is a minor variation of ${\cal P}_\theta$ and $P_\theta$
when applied to an SVOA that contains the $N=2$ SCA, but is larger.
This is a statement about an SVOA and its representation theory, but has
nothing to do with a 2D SCFT.

It also refers to a state
$st_{\theta,\tilde{\theta}} \in {\cal H}_{\rm tot}^{\rm dbl.}$ of a 2D
non-holomorphic SCFT whose conformal weights are
$(h(st_{\theta,\tilde{\theta}}), \tilde{h}(st_{\theta,\tilde{\theta}}))
= (\theta^2 \hat{c}/2, \tilde{\theta}^2 \hat{c}/2)$ and the $N=2$ SCA
charges $(q(st_{\theta,\tilde{\theta}}), \tilde{q}(st_{\theta,\tilde{\theta}}))
= (\theta \hat{c}, \tilde{\theta}\hat{c})$. The spectrum of
$(\theta,\tilde{\theta})\in \R^2$ for such states is determined by
the Hilbert space of the 2D SCFT. The SCFT vertex operator
$YY_{\rm tot}(st_{\theta,\tilde{\theta}},z)$, often denoted by
$e^{i \theta \sqrt{\hat{c}}\varphi} e^{i\tilde{\theta}\sqrt{\hat{c}}\tilde{\varphi}}$
in physics literature, depends on a point in 2D (Riemann surface).

A third meaning is a part of the mode operators of
$YY_{\rm tot}(st_{\theta,\tilde{\theta}},z) = \sum_{\nu,\tilde{\nu}}
[st_{\theta,\tilde{\theta}}]_{\nu,\tilde{\nu}}
[z^{h(st)+\nu} \bar{z}^{\tilde{h}(st)+\tilde{\nu}}]^{-1}$.
When $YY_{\rm tot}(st_{\theta,\tilde{\theta}},z)$ acts on a subspace of
${\cal H}_{\rm tot}^{\rm dbl}$ with $J_0 = q$ and $\tilde{J}_0 = \tilde{q}$, 
the linear operator ${\cal U}_{\theta,\tilde{\theta}} :=
[st_{\theta,\tilde{\theta}}]_{-h(st)-\theta q, -\tilde{h}(st) -\tilde{\theta}\tilde{q}}$
has the property $L_0 {\cal U}_{\theta,\tilde{\theta}} =
{\cal U}_{\theta,\tilde{\theta}} (L_0 + \theta q + \theta^2 \hat{c}/2)$
and $J_0 {\cal U}_{\theta,\tilde{\theta}} = {\cal U}_{\theta,\tilde{\theta}} \cdot
(J_0 + \theta \hat{c})$, similarly to
$P_\theta \otimes \widetilde{P}_{\tilde{\theta}}$. The linear operator
${\cal U}_{\theta,\tilde{\theta}}$ does not have the property of
$P_\theta \otimes \widetilde{P}_{\tilde{\theta}}$ for $n\neq 0$, however.

The difference between $P\otimes \widetilde{P}$ and ${\cal U}$ is relevant
in a description of boundary states, but irrelevant when one only deals with
character formulas, or correspondence between representations
(rather than that between states). Most of the usage of this term in this
article is in the latter category.
} %
``because of the spectral flow'', or ``by using the spectral flow operator''.  
This section goes beyond this slogan.

This slogan in physics literature can be implemented into a four-step
procedure to determine the SVOA vertex operator\footnote{
An SVOA ${\cal A}$ has a decomposition ${\cal A}^+ \oplus {\cal A}^- =
\oplus_{s \in \Z/2\Z} {\cal A}_{s_0}$,
by the conformal weights mod $\Z$; the SVOA vertex operator $YY$ of
${\cal A}$ consists of four intertwining operators $\{ Y_{s_0, s_1} \; | \;
s_0, s_1 \in \Z/2\Z\}$ of the VOA ${\cal A}^+$, where $Y_{s_0, s_1}:
{\cal A}_{s_0} \rightarrow {\rm Hom}({\cal A}_{s_1}, {\cal A}_{s_0 \cdot s_1})
[[z,z^{-1}]]$. Similarly, ${\cal YY}$ (intertwining {\it super}operator),
${\cal VV}$ (the vector space of intertwining {\it super}operators),
${\cal II}$ (a set of irreducible representations of an {\it S}VOA),
and $YY_{\rm tot}$ (an {\it S}CFT vertex operator) are multiplets of
non-super versions. We should have used this notation in
footnote \ref{fn:SCFT-min}, though. 
} %
$YY^{\rm Gepn.}$ of the
left-mover superchiral algebra ${\cal A}^{\rm Gepn.} =
\oplus_{ \vec{y} \in \Gamma_{\chi {\rm Alg}}} M_{\vec{0},2\vec{y}}$
(and also the right-mover counterpart).
\begin{itemize}
\item (Step 1) The subspace $M_{\vec{0},2\vec{y}} \subset {\cal A}^{\rm Gepn.}$
for $\vec{y} \in \Gamma_{\chi {\rm Alg}}$ is regarded as ${\cal A}^{\rm m.t.m.} \cdot
v_\eta = M_{\vec{0},\vec{0}} \cdot v_\eta$ for a spectral flow state $v_\eta \in
M_{\vec{0},2\vec{y}}$ (i.e., generated by $\{ [u]_m \cdot v_\eta \; |
u \in M_{\vec{0},\vec{0}}\}$, where $YY^{\rm Gepn.}(u,z) =:
\sum_m [u]_m /z^{h(u)+m}$); here,
$\eta = (\eta_{1,\cdots, r}) \in \Z^{\oplus r}$ is a lift of
$\vec{y}= (y_{i=1,\cdots,r}) \in \prod_i \Z/\bar{k}_i\Z$. One can establish
a dictionary between this description of the vector space $M_{\vec{0},2\vec{y}}$
with another description, ${\cal U}_{-\eta} \cdot M_{\vec{0},\vec{0}}$
($(-\eta)$-unit spectral flow of all the states in $M_{\vec{0},\vec{0}}$),
by exploiting \underline{the skewsymmetry invariance} that
$YY^{\rm Gepn.}$ is supposed to have.
\item (Step 2) The dictionary in Step 1 can be used to compute
$YY^{\rm Gepn.}(u, z_1) \cdot YY^{\rm Gepn.}(v_\eta,z_2)$ on $M_{\vec{0},\vec{0}}$
and also $YY^{\rm Gepn.}(v_\eta,z_1) \cdot YY^{\rm Gepn.}(u,z_2)$ on
$M_{\vec{0},\vec{0}}$ and compare them, for any $u \in M_{\vec{0},\vec{0}}$ and
a spectral flow state $v_\eta \in M_{\vec{0},2\vec{y}}$ for
$y \in \Gamma_{\chi {\rm Alg}}$.
Because $YY^{\rm Gepn.}$ is supposed to satisfy \underline{the super-Jacobi
  identity},
we can compute the action of $YY^{\rm Gepn.}([u]_m \cdot v_\eta,z)$ on
$M_{\vec{0},\vec{0}}$.  This means that
we can determine $YY^{\rm Gepn.}(x,z)$ on $M_{\vec{0},\vec{0}}$ for any
$x \in M_{\vec{0},2\vec{y}}$. 
\item (Step 3)
Let $L_0$ be the free abelian group of all the possible
lifts $\eta$ of all $\vec{y} \in \Gamma_{\chi {\rm Alg}}$. It is an integral
lattice by $(\eta,\eta')_{L_0} := \sum_i \eta_i \eta'_i \hat{c}_i$;
it is integral indeed because 
\begin{align}
\sum_i \frac{\eta_i \eta'_ik_i}{\bar{k}_i} & \; \equiv
-2 \sum_i \frac{\eta_i \eta'_i}{\bar{k}_i} \equiv
\sum_i \frac{\eta_i^2+\eta_i^{'2} -(\eta_i+\eta'_i)^2}{\bar{k}_i}
\label{eq:def-lattice-4left-bilinF} \\
& \equiv \frac{{\rm Arg}\left[ \epsilon^{00}(\vec{y}+\vec{y}',\vec{y}+\vec{y}')/\epsilon^{00}(\vec{y},\vec{y})\epsilon^{00}(\vec{y}',\vec{y}') \right]}{2\pi}
\equiv 0 \quad ({\rm mod~}+\Z).  \nonumber 
\end{align}
Let ${\cal A}_J$ be the vertex operator subalgebra of ${\cal A}^{\rm m.t.m.}$
of the rank-$r$ chiral bosons, each of which is in the coset
VOA (\ref{eq:coset-VOA}). The vector subspace ${\cal A}_J \cdot L_0 \subset
\oplus_{\vec{y} \in \Gamma_{\chi {\rm Alg}}} M_{\vec{0},2\vec{y}}$ admits a structure
of an SVOA,  which is denoted by $SVOA(L_0)$ here;
$YY^{\rm Gepn.}$ on this subspace $SVOA(L_0)$ is that of the stadnard 
  lattice SVOA; the cocycle factor is determined from the bilinear form
  $(\eta,\eta')_{L_0}$ above.  Now, as the next step, the action of
  $YY^{\rm Gepn.}(x,z)$ of $x \in SVOA(L_0)$ on
  the whole space ${\cal A} = \oplus_{\vec{y}_2 \in \Gamma_{\chi {\rm Alg}}}
  M_{\vec{0},2\vec{y}_2}$ is determined by (i) translating the states in
  $M_{\vec{0},2\vec{y}_2}$ by the dictionary in Step 1 into the form of
  ${\cal U}_{-\eta_2} \cdot u$ for some $u \in M_{\vec{0},\vec{0}}$ and then
  by (ii) using the $x$--$v_\eta$ OPE of $SVOA(L_0)$. All the states
  that appear in the OPE are somewhere on $M_{\vec{0},2(\vec{y}+\vec{y}_2)}$,
  so we know from Step 2 (iii) how they act on $M_{\vec{0},\vec{0}}$.
\item (Step 4) Now, we can compute both
  $YY^{\rm Gepn.}(u,z_1) \cdot YY^{\rm Gepn.}(v,z_2)$ and
  $YY^{\rm Gepn.}(v,z_1) \cdot YY^{\rm Gepn.}(u,z_2)$ on the whole space for
  $u \in M_{\vec{0},\vec{0}}$ and $v \in SVOA(L_0)$. By comparing them,
  and using \underline{the super-Jacobi identity} that $YY^{\rm Gepn.}$ is
  supposed to satisfy, we can determine $YY^{\rm Gepn.}([u]_m \cdot v,z)$ on
  the whole space. Because the whole space is generated
  by the states of the form $[u]_m \cdot v$, we can determined all
  of $YY^{\rm Gepn.}(x,z)$ acting on the whole space ${\cal A}^{\rm Gepn.}$
  for any $x \in {\cal A}^{\rm Gepn.}$. 
\end{itemize}
This implementation seems to work, although it is not straightforward. 
The present authors prepared an exposition that fills in details of the wordy description above, but we do not write it down here, because
we need more than that. We need $YY^{\rm Gepn.}(x,z)$---the
SCFT vertex operator---acting on $({\cal H}_{\rm tot}^{\rm dbl})_{\rm Gepn.}$
(with non-holomorphic dependence on $z$) for all
$x \in ({\cal H}_{\rm tot}^{\rm dbl})_{\rm Gepn.}$.

It may be possible that a story similar
to the implementation above works (in a downstairs approach), by making
an analogy between VOA and CFT. $M_{\vec{0},\vec{0}}$ 
is replaced by the part of $({\cal H}_{\rm tot}^{\rm dbl})_{\rm m.t.m.}$ that
survives the orbifold projection, and $SVOA(L_0)$ is replaced by the lattice
CFT that involves both the left-mover and right-mover charges.
A concern at the technical level in this downstair approach,\footnote{
The Gepner construction is an orbifold by an automorphism of the minimal tensor
model (non-holomorphic) {\it SCFT}; the symmetry $\Gamma$ is not an automorphism
of the left-mover (holomorphic) {\it superchiral algebra}
${\cal A}^{\rm m.t.m.}$. So, an upstairs approach would have to deal
with $\Gamma$-twisted representations of the whole states
$({\cal H}_{\rm tot}^{\rm dbl})_{\rm m.t.m.}$, and also the mutually
non-local operators of the states in those twisted representations 
(and apply projection operators to restrict the class of states to which
we pay attention to). So we would need a non-holomorphic CFT version (than
a VOA version) of twisted representations and intertwining algebra of states
in those representations. 
} %
however, is that
there is, in general, $(\lambda,\vec{\tilde{m}})$ where $\vec{\tilde{m}}$
satisfies the orbifold projection
condition (\ref{eq:cond-Gepner-survive-Z-w-d.t.},
\ref{eq:cond-Gepner-survive-Z-w-d.t.for-R}) only with $\vec{y} \in \Gamma$
but not in $\Gamma_{\chi {\rm Alg}}$. Then such a $\vec{y}$-twisted sector state
is not generated from the orbifold-surviving untwisted sector states
by the action of the lattice CFT. A more theoretical issue is to prepare
a theory of non-holomorphic 2D CFT at the level of solidness that we see
in the theory of VOA today, so we can treat with confidence the complex
phases and branch cuts of the SCFT vertex operators. 
For those reasons, the present authors do not rely on this naive VOA-CFT
analogy in this article. 

Instead, this section provides two thinking
frameworks for the construction of the SCFT vertex operators
$YY_{\rm tot}^{\rm Gepn.}$. One in section \ref{ssec:CFTop-by-sc-ext} is more
abstract, based on the theory of simple
current extension and representation theory of the minimal tensor model
superchiral algebras
${\cal A}^{\rm m.t.m.} \otimes \widetilde{\cal A}^{\rm m.t.m.}$.
In this approach, we can avoid the multi-step procedure like the
``implementation'' above, due to the power of the theory of simple current
extension; we manage to go beyond SVOA to deal with SCFT as a whole by
exploiting the theory of intertwining superoperators of both
${\cal A}^{\rm m.t.m.}$ and $\widetilde{\cal A}^{\rm m.t.m.}$. 
The other approach in section \ref{ssec:CFTop-by-bos-lattice} is more
concrete, using the decomposition of the coset VOA (\ref{eq:coset-VOA})
into the U(1) VOA and the parafermion VOA. This approach avoids the
multi-steps in the ``implementation'' going back and forth
between the representations of ${\cal A}^{\rm m.t.m.}$ and the lattice/U(1)
VOA, by woking on the U(1)-parafermion decomposition completely, and
by giving up manifest action of the worldsheet $N=2$ SCA. The non-holomorphic
aspects of CFT are not difficult at all when we deal with lattice
CFTs.

The discussions in this section apply to the Gepner constructions
whose data is in ${\cal S}_0$, not necessarily in ${\cal S}_{\rm II}$; we
do not use the conditions (iia, iib),
or equivalently (\ref{eq:cond-in-type0-2bb},
\ref{eq:def-fcn-y0}, \ref{eq:cond-for-specFlow23-inGepner-A},
\ref{eq:cond-Gepner-survive-specBy1-2}). 

%%%%%%%%%%%%%%%%%%%%%%%%%%%%%%%%%%%%%%%%%%%%%%%%%
\subsection{Through the Theory of Simple Current Extension}
\label{ssec:CFTop-by-sc-ext}
%%%%%%%%%%%%%%%%%%%%%%%%%%%%%%%%%%%%%%%%%%%%%%%%%%

Here we articulate a thinking framework of how to build the SCFT vertex
operator $YY_{\rm tot}^{\rm Gepn.}$ of the Gepner model SCFTs, from
that $YY_{\rm tot}^{\rm m.t.m.}$ of the minimal tensor model SCFT,
using the theory of simple current extension. 
As our goal is to lay a solid foundation to be able to deal with complex
phases and branch cuts of SCFT vertex operators, we use the language
of SVOA heavily in this section, along with a theory of intertwining
{\it super}operators (supermultiplets of intertwining operators). 
While a minimum review on those theoretical background is inserted at a couple
of places in this section \ref{ssec:CFTop-by-sc-ext}, we keep it to be
almost minimal in this article in order not to distract readers from
the main story. Readers interested in further details of (or skeptical of)
the theory of intertwining superoperators are referred
to \cite[\S4]{Watari:2026scft}.

\vspace{5mm}

  Given a rational VOA $(V, Y)$ whose all the irreducible representations are
unitary, an irreducible representation $s$ in the set of all the irreducible
representations ${\cal I} = {\cal I}((V,Y))$ is said to be a {\it simple
  current} \cite{Schellekens:1989am} if
its fusion $(s \cdot ): {\cal I} \rightarrow \Z[{\cal I}]$ is only
a permutation on the set ${\cal I}$; in other words, there is a
set-theoretical map $f(s,-): {\cal I} \ni s' \mapsto f(s,s') \in {\cal I}$
so $s \cdot s' = f(s,s') \in \Z[{\cal I}]$;
put differently, $\dim_\C {\cal V}_{s,s'}^{f(s,s')} = 1$ and
$\dim_\C {\cal V}_{s,s'}^{s''} =0$ for ${}^\forall s'' \in {\cal I}
\backslash f(s,s')$, where ${\cal V}_{b,c}^a$ for $a,b,c\in {\cal I}$ is
the vector space of intertwining operators of the VOA $(V, Y)$.
To take the coset VOA (\ref{eq:coset-VOA}) as an example,
${\cal L}^\ell_{m,s}$ with $\ell = 0, k$ are simple currents. 

The notion of simple current can also be extended
to, and its theorey fully developed for, a rational SVOA $(V, YY)$ with
a $\Z/2\Z$-grading
$V = \oplus_{s_0 \in S_0} V_{s_0}$ ($S_0 = \Z/2\Z$) such that each
one of either NS- or R-type irreducible representation $M$ is unitary, and has
a decomposition into two irreducible representations
$M = \oplus_{s \in S_M} M_s$ labeled by a 2-element set $S_M$ on which
$S_0 = \Z/2\Z$ acts non-trivially. 
In the case of the $N=2$ minimal model SVOA of level-$k$,
$\{ s=0,2\} \subset \Z/4\Z$ or $\{ s=1,3\} \subset \Z/4\Z$ is the 2-element
set  on which $S_0 = \Z/2\Z = 2\Z/4\Z$ acts. The SVOA representations
$M_{\ell,m}$ with $\ell = 0, k$ are simple currents. 
\vspace{3mm}

Relevant to this present article are simple currents of the left-mover 
superchiral algebra ${\cal A}^{\rm m.t.m.}$ of the
minimal tensor model SCFT with $\{ (k_i, R_i)\}$;
we will restrict our attention to the cases all $R_{i=1,\cdots,}$ are
the $A_{k_i+1}$ type, to keep the story simple in the rest of this
section \ref{ssec:CFTop-by-sc-ext}. The irreducible representations
of ${\cal A}^{\rm m.t.m.}$ of the form $\otimes_{i=1}^r M_{\ell_i,m_i} =
M_{\lambda,\vec{m}}$ with $\ell_i =0$ ($i=1,\cdots, r$) are simple currents; 
$\ell_i = k_i$ is acceptable, but those representations are not different
from those with $\ell_i = 0$ because of (\ref{eq:field-id}). 

Remember that the goal of this section \ref{ssec:CFTop-by-sc-ext} is to
use the SCFT vertex operator
$YY_{\rm tot}^{\rm m.t.m.}$ as a collection of elements 
$(YY_{\rm tot}^{\rm m.t.m.})^{((\lambda_3,\vec{m}_3);(\lambda_1,\vec{m}_1),(\lambda_2,\vec{m}_2))} \in
{\cal VV}_{(\lambda_1,\vec{m}_1), \; (\lambda_2,\vec{m}_2)}^{(\lambda_3,\vec{m}_3)} \otimes \widetilde{\cal VV}_{(\lambda_1,\vec{m}_1), \; (\lambda_2,\vec{m}_2)}^{(\lambda_3,\vec{m}_3)}$ of the minimal tensor model SCFT to 
determine the SCFT vertex operator\footnote{
All the representation lebels $(\lambda,\vec{m})\otimes (\tilde{\lambda},\vec{\tilde{m}})$ in this section \ref{ssec:CFTop-by-sc-ext} show up in the form
of $(\vec{m},\vec{\tilde{m}}) = (\vec{\tilde{m}},\vec{\tilde{m}})$ or
$(\vec{\tilde{m}}+2\vec{y}, \vec{\tilde{m}})$, we find it more convenient to
write $\vec{\tilde{m}}$ elsewhere in this article as $\vec{m}$ in
section \ref{ssec:CFTop-by-sc-ext}. 
} %
$YY_{\rm tot}^{\rm Gepn.}$ as a collection of elements in
${\cal VV}_{(\lambda_1,\vec{m}_1+2\vec{y}_1), \; (\lambda_2,\vec{m}_2+2\vec{y}_2)}^{(\lambda_3,\vec{m}_3+2\vec{y}_3)} \otimes \widetilde{\cal VV}_{(\lambda_1,\vec{m}_1), \; (\lambda_2,\vec{m}_2)}^{(\lambda_3,\vec{m}_3)}$, or equivalently to determine $YY_{\rm tot}^{\rm Gepn.}(x, z)$
acting on the Gepner model Hilbert space for 
$x \in M_{\lambda_1, \vec{m}_1+2\vec{y}_1} \otimes \widetilde{M}_{\lambda_1, \vec{m}_1}$. 
Here, and also in the rest of this section \ref{ssec:CFTop-by-sc-ext},
${\cal VV}_{R_1,R_2}^{R_3}$ [resp. $\widetilde{\cal VV}_{\tilde{R}_1,\tilde{R}_2}^{\tilde{R}_3}$] stands for the vector space of intertwining superoperators
of the SVOA ${\cal A}^{\rm m.t.m.}$ [resp. $\widetilde{\cal A}^{\rm m.t.m.}$]
 (to be elaborated more in the following).
An idea for this problem in the literature is to use the
isomorphism between the tensor product of the vector spaces of intertwining
superoperators
\begin{align}
&  {\cal VV}_{(\vec{0},2\vec{y}_1), \; (\vec{0}, 2\vec{y}_2)}^{(\vec{0},2\vec{y}_3)} \otimes
  {\cal VV}_{(\vec{0},2\vec{y}_3), \; (\lambda_3,\vec{m}_3)}^{(\lambda_3,\vec{m}_3+2\vec{y}_3)} \otimes
  {\cal VV}_{(\lambda_1,\vec{m}_1), \; (\lambda_2,\vec{m}_2)}^{(\lambda_3,\vec{m}_3)}
  \label{eq:smpl-crrnt-keyIso-temp} \\
& \qquad  \cong
  {\cal VV}_{(\vec{0},2\vec{y}_1), (\lambda_1,\vec{m}_1)}^{(\lambda_1,\vec{m}_1+2\vec{y}_1)} \otimes
  {\cal VV}_{(\vec{0},2\vec{y}_2), \; (\lambda_2,\vec{m}_2)}^{(\lambda_2,\vec{m}_2+2\vec{y}_2)} \otimes
  {\cal VV}_{(\lambda_1,\vec{m}_1+2\vec{y}_1), \; (\lambda_2,\vec{m}_2+2\vec{y}_2)}^{(\lambda_3,\vec{m}_3+2\vec{y}_3)} \nonumber 
\end{align}
to map $YY_{\rm tot}^{\rm m.t.m.}$ into $YY_{\rm tot}^{\rm Gepn.}$; here, because 
all of $M_{(\vec{0}, 2\vec{y}_1)}$, $M_{(\vec{0},2\vec{y}_2)}$ and
$M_{(\vec{0},2\vec{y}_3)}$
are simple currents of the SVOA ${\cal A}^{\rm m.t.m.}$, four out of the
six vector spaces of the intertwining superoperators in the isomorphism
above are of 1-dimension, so the isomorphism above indeed turns into\footnote{
\label{fn:boxtimes-univ} 
In a modern math language (cf \cite{MR1338873}, \cite[\S5]{MR2691082}), the construction method is stated as follows.
In writing it down, the following properties are useful: for a module
$M_\beta$ and $M_\gamma$ of a VOA $(V, Y)$, (i) there is one module of 
the VOA denoted by $M_\beta \boxtimes M_\gamma$, along with an
intertwining operator $\widetilde{\cal Y}$ of type
$(\beta \boxtimes \gamma; \beta,\gamma)$, and (ii) an intertwining
operator ${\cal Y}$ of type $(\alpha;\beta,\gamma)$ is in one-to-one
with a homomorphism of $(V,Y)$-module $\Phi_{\cal Y}: M_\beta \boxtimes M_\gamma
\rightarrow M_{\alpha}$; the relation is ${\cal Y} = \Phi_{\cal Y} \cdot
\widetilde{\cal Y}$.

For a given intertwining operator ${\cal YY}^{\rm m.t.m.}$ of type
$((\lambda_3,\vec{m}_3);(\lambda_1,\vec{m}_1),(\lambda_2,\vec{m}_2))$,
we have a homomorphism 
\[
(M_{\vec{0},2\vec{y}_1} \boxtimes M_{\vec{0},2\vec{y}_2}) \boxtimes (M_{\lambda,\vec{m}_1} \boxtimes M_{\lambda_2,\vec{m}_2}) \rightarrow M_{\vec{0},2\vec{y}_{1+2}} \boxtimes M_{\lambda_3,\vec{m}_3} \rightarrow M_{\lambda_3,\vec{m}_3+2\vec{y}_{1+2}}
\]
of ${\cal A}^{\rm m.t.m.}$ modules, given by $\Phi_{YY^{[\vec{y}_{1+2},\vec{m}_3]}} \cdot
(\Phi_{YY^{(\vec{y}_1,\vec{y}_2)}} \otimes \Phi_{{\cal YY}^{\rm m.t.m.}})$. 
There is also another homomorphism
\[
(M_{\vec{0},2\vec{y}_1} \boxtimes M_{\vec{0},2\vec{y}_2}) \boxtimes
(M_{\lambda,\vec{m}_1} \boxtimes M_{\lambda_2,\vec{m}_2}) \rightarrow
(M_{\vec{0},2\vec{y}_1} \boxtimes M_{\lambda,\vec{m}_1}) \boxtimes
(M_{\vec{0},2\vec{y}_2} \boxtimes M_{\lambda_2,\vec{m}_2}) \rightarrow
M_{\lambda_1,\vec{m}_1+2\vec{y}_1} \boxtimes M_{\lambda_2, \vec{m}_2+2\vec{y}_2}
\]
 of $\underline{\cal A}$ modules, 
given by $\alpha \cdot c^{-1}_\rho \cdot \alpha \cdot (c_\rho\alpha^{-1})$ with
the associativity morphism and commutativity morphisms $\alpha$ and $c_\rho$
as in $(*)_\rho$, followed by $\Phi_{YY^{[\vec{y}_1,\vec{m}_1]}} \otimes
\Phi_{YY^{[\vec{y}_2,\vec{m}_2]}}$. One homomorphism $\Phi: M_{\lambda_1,\vec{m}_1+2\vec{y}_1}
\boxtimes M_{\lambda_2, \vec{m}_2+2\vec{y}_2} \rightarrow
M_{\lambda_3,\vec{m}_3+2\vec{y}_{1+2}}$
of ${\cal A}^{\rm m.t.m.}$ modules is determined then so that the two routes
agree; this assignment $\Phi_{{\cal YY}^{\rm m.t.m.}} \mapsto \Phi$ is in one-to-one
with the construction ${\cal YY}^{\rm m.t.m.} \mapsto {\cal YY}^{\rm Gepn.}$
in the main text.
} %
an isomorphism ${\cal VV}_{(\lambda_1,\vec{m}_1), \; (\lambda_2,\vec{m}_2)}^{(\lambda_3,\vec{m}_3)} \cong {\cal VV}_{(\lambda_1,\vec{m}_1+2\vec{y}_1), \; (\lambda_2,\vec{m}_2+2\vec{y}_2)}^{(\lambda_3,\vec{m}_3+2\vec{y}_3)}$ as soon as we make a choice of a basis element
$YY^{(\vec{y}_1,\vec{y}_2)}$ [resp. $YY^{[\vec{y}_a,(\lambda_a,\vec{m}_a)]}$ with $a=1,2,3$] of the
1-dimensional vector space ${\cal VV}_{(\vec{0},2\vec{y}_1), \; (\vec{0},2\vec{y}_2)}^{(\vec{0},2\vec{y}_3)}$ [resp. ${\cal V}_{(\vec{0},2\vec{y}_a), \; (\lambda_a, \vec{m}_a)}^{
    (\lambda_a, \vec{m}_a+2\vec{y}_a)}$].

The remaining task of this
section \ref{ssec:CFTop-by-sc-ext} is to provide necessary details 
of the isomorphism (\ref{eq:smpl-crrnt-keyIso-temp}), and spell out
the conditions on the choice of basis elements $YY^{(\vec{y}_1,\vec{y}_2)}$ for
$\vec{y}_{1,2} \in \Gamma$ and  $YY^{[\vec{y},(\lambda,\vec{m})]}$ for each pair
$\vec{y}$ and $(\lambda,\vec{m})$
that survive the orbifold projection condition
(\ref{eq:cond-Gepner-survive-Z-w-d.t.},
\ref{eq:cond-Gepner-survive-Z-w-d.t.for-R}).
What has been said in the literature is to choose $YY^{[\vec{y},(\lambda,\vec{m})]}$
so that $Y^{[\vec{y},(\lambda,\vec{m})]}(v_{\eta})$ of a properly normalized
spectral flow state $v_{\eta} \in M_{\vec{0},2\vec{y}}$ of $(-2\eta,0)$-unit flow
($\eta = (\eta_{i=1,\cdots,r}) \in \Z^{\oplus r}$ so that
$\eta_i + \bar{k}_i\Z = y_i$) is the ``plane wave vertex operator''
that is denoted often by $e^{-i \eta_i \sqrt{\hat{c}_i} \varphi_i}$; $\varphi_i$
for each of $i=1,\cdots, r$ is a left-moving chiral boson, with the property
that
\begin{align}
  \varphi_{(i)}(z) \varphi_{(j)}(0) \sim -  \delta_{i,j} \; \ln (z), \qquad
  J^{N=2}_{(i)} = i \sqrt{\hat{c}_i} \partial \varphi_{(i)}. 
\end{align}
The basis element $YY^{(\vec{y}_1,\vec{y}_2)}$ should reproduce the fusion algebra
of the chiral boson sectors including the spectral flow states
$v_{\eta_1}$, $v_{\eta_2}$ and $v_{\eta_3}$. The literature rarely talks about
the right choice of complex phases of those $YY^{[\vec{y}_a,(\lambda_a,\vec{m}_a)]}$
(or $e^{-i\eta_i \sqrt{\hat{c}_i} \varphi_{(i)}}$) and $YY^{(\vec{y}_1,\vec{y}_2)}$,
however. In fact, there is no unique well-motivated
isomorphism (\ref{eq:smpl-crrnt-keyIso-temp}), but there are multiple
well-motivated ones that differ from one another by complex phases. 
So, we cannot even talk about the right conditions on the basis elements
$YY^{[\vec{y}_a,(\lambda_a,\vec{m}_a)]}$ and $YY^{(\vec{y}_1,\vec{y}_2)}$ before
specifying which isomorphism of the form (\ref{eq:smpl-crrnt-keyIso-temp}) is
used to set-up a dictionary between $YY_{\rm tot}^{\rm m.t.m.}$ and
$YY_{\rm tot}^{\rm Gepn.}$.  

% Here is a technical reminder at the end of this opening statement
% in this section \ref{ssec:CFTop-by-sc-ext}. An intertwining superoperator
% $YY \in {\cal VV}_{(\lambda_1,\vec{m}_1), \; (\lambda_2,\vec{m}_2)}^{(\lambda_ 3,\vec{m}_3)}$
% consists of $2^2$ intertwining operators of $\underline{\cal A}^+$
% (the $\Z/2\Z$-even part of the minimal tensor model superchiral algebra
% $\underline{\cal A}$), and of $(2^r)^2$ intertwining operators of
% $\otimes_i {\cal A}_{(i)}^+$. way to keep track of correlation
% among the $2^2$ intertwining operators is reviewed in our separate paper. 
% We will drop reference to $\lambda$ to save space,
% $Y$ instead of $YY$, ${\cal V}$ instead of ${\cal VV}$,
% ${\cal I}$ instead of ${\cal II}$ to save space. 

\vspace{5mm}

There are multiple different ways to build an isomorphism
of the form (\ref{eq:smpl-crrnt-keyIso-temp}) as a combination
of fusion isomorphisms and skewsymmetry isomorphisms of intertwining
superoperators of the minimal tensor model ${\cal A}^{\rm m.t.m.}$. 
Think of this: 
\begin{equation}
  (*)_\rho: 
  \xymatrix{
   {}_{[ini]} \; 
    {\cal VV}_{2\vec{y}_1,2\vec{y}_2}^{2\vec{y}_3}{\cal VV}_{2\vec{y}_3,\vec{m}_3}^{\vec{m}_3+2\vec{y}_3}
  {\cal VV}_{\vec{m}_1,\vec{m}_2}^{\vec{m}_3} \ar[r]^{{\cal F}_{12}^{-1}} &
  {\cal VV}_{2\vec{y}_1,\vec{m}_3+2\vec{y}_2}^{\vec{m}_3+2\vec{y}_3}{\cal VV}_{2\vec{y}_2,\vec{m}_3}^{\vec{m}_3+2\vec{y}_2}
  {\cal VV}_{\vec{m}_1,\vec{m}_2}^{\vec{m}_3} \ar[d]^{\Omega_3} \; {}_{[1]} \\
  {}_{[3]} \; {\cal VV}_{2\vec{y}_1,\vec{m}_3+2\vec{y}_2}^{\vec{m}_3+2\vec{y}_3}{\cal VV}_{2\vec{y}_2,\vec{m}_2}^{\vec{m}_2+2\vec{y}_2}
  {\cal VV}_{\vec{m}_2+2\vec{y}_2,\vec{m}_1}^{\vec{m}_3+2\vec{y}_2} \ar[d]^{\Omega_3^{-1}} &  
  {\cal VV}_{2\vec{y}_1,\vec{m}_3+2\vec{y}_2}^{\vec{m}_3+2\vec{y}_3} {\cal VV}_{2\vec{y}_2,\vec{m}_3}^{\vec{m}_3+2\vec{y}_2}
  {\cal VV}_{\vec{m}_2,\vec{m}_1}^{\vec{m}_3} \ar[l]^{{\cal F}_{23}} \; {}_{[2]} \\
  {}_{[4]} \; {\cal VV}_{2\vec{y}_1,\vec{m}_3+2\vec{y}_2}^{\vec{m}_3+2\vec{y}_3} {\cal VV}_{2\vec{y}_2,\vec{m}_2}^{\vec{m}_2+2\vec{y}_2}
  {\cal VV}_{\vec{m}_1,\vec{m}_2+2\vec{y}_2}^{\vec{m}_3+2\vec{y}_2} \ar[r]^{{\cal F}_{13}} &
  {\cal VV}_{2\vec{y}_1,\vec{m}_1}^{\vec{m}_1+2\vec{y}_1}{\cal VV}_{2\vec{y}_2,\vec{m}_2}^{\vec{m}_2+2\vec{y}_2}
  {\cal VV}_{\vec{m}_1+2\vec{y}_1,\vec{m}_2+2\vec{y}_2}^{\vec{m}_3+2\vec{y}_3}  \; {}_{[fin]}
}, \label{eq:isom-4-mtm-Gepn-optr-cnstr}
\end{equation}
where representation labels $(\lambda,\vec{m})$ and $(\vec{0},2\vec{y})$
are replaced by $\vec{m}$ and $2\vec{y}$, and the symbol $\otimes$
mitted to save space in (\ref{eq:isom-4-mtm-Gepn-optr-cnstr});
harmless labels such as [1], [2] are assigned to each term in the middle
of this isomorphism for easier reference in the discussion later. 
In the each step of the isomorphism, ${\cal F}$ is the fusion isomorphism
$\oplus_{s_5 \in {\cal II}} ({\cal VV}_{s_1,s_5}^{s_4} \otimes
{\cal VV}_{s_2,s_3}^{s_5}) \rightarrow \oplus_{s_6 \in {\cal II}}
({\cal VV}_{s_1,s_2}^{s_6} \otimes {\cal VV}_{s_6,s_3}^{s_4})$
(or $(s_1 \cdot (s_2 \cdot s_3))^{s_4} \rightarrow ((s_1 \cdot s_2)
\cdot s_3)^{s_4}$ in a space-saving notation using {\it parenthesized words});
in all the three fusion isomorphisms in $(*)_\rho$ above, the intermediate
representations $s_5$ and $s_6$ are unique for the given choices of
$s_1, s_2, s_3, s_4$, because the representation $(\vec{0},2\vec{y})$
are simple currents. We use a skewsymmetry isomorphism
$\Omega_\rho: {\cal VV}_{s_1,s_2}^{s_3} \rightarrow {\cal VV}_{s_2,s_1}^{s_3}$
in the second map ($[1]\rightarrow[2]$) in the diagram above,
and $\Omega_\rho^{-1} = \Omega_{-1-\rho}$ in the fourth map
($[3]\rightarrow [4]$) in the diagram to define $(*)_\rho$.
The skewsymmetry isomorphisms intertwining operators are labeled by
$\rho \in \Z$, because $\Omega_\rho$ assigns a new intertwinig operator
$\Omega_\rho (YY)(v, \underline{z})$ that reproduces matrix elements of the
original intertwinig operator $YY(v,z)$ analytically continued to
$z= e^{\pi i (2\rho+1)}\underline{z}$; we need to keep track of the branch cuts.
Already our choice of the fourth map as $\Omega_{-1-\rho}$ correlated with
the second map $\Omega_\rho$ restricts a class of isomorphisms of the
form (\ref{eq:smpl-crrnt-keyIso-temp}), and moreover, we will be interested
only in the isomorphisms $(*)_\rho$ with $\rho = 0, -1$ (corresponding
to the analytic continuation $z \rightarrow e^{ \pm \pi i} \underline{z}$). 
%%%%%%%%%%%%%%%%%%%%%%%%%%%%%%%%%%%%%%%%%%%%%%%%%%%%
\begin{figure}[tbp]
\begin{center}
  \includegraphics[scale=0.7]{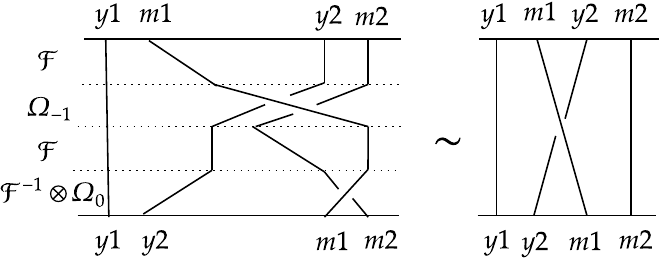}  
\caption{\label{fig:mtm2Gep-CFTop-cnstr} The graphical presentation of
 the chain of isomorphisms $(*)_{\rho=0}$ that converts
 ${\cal YY} \in {\cal VV}_{(\lambda_1,\vec{m}_1),(\lambda_2,\vec{m}_2)}^{(\lambda_3,\vec{m_3})}$ to the one in ${\cal VV}_{(\lambda_1,\vec{m}_1+2\vec{y}_1),(\lambda_2,\vec{m}_2+2\vec{y}_2)}^{(\lambda_3,\vec{m}_3+2\vec{y}_3)}$. The isomorphism
 ${\cal F}^{-1} \otimes \Omega_\rho: [ini]\rightarrow [2]$ at the bottom
 of the graph is followed by ${\cal F}: [2]\rightarrow [3]$ a little above
 in the graph, and further up in the graph.  
Due to the pentagon and 
 hexagon identities, only the corresponding braid-group element is relevant. 
The graphical presentation of $(*)_{-1}$ is obtained by exchaning overcrossing
and undercrossing (the skewysymmetry isomorphisms $\Omega_0 \leftrightarrow \Omega_{-1}$). }
\end{center}
\end{figure}
%%%%%%%%%%%%%%%%%%%%%%%%%%%%%%%%%%%%%%%%%%%%%%%%%%%%

By using the pentagon identity and hexagon identities (available
for $\Omega_{0}$ and $\Omega_{-1}$), the isomorphisms $(*)_{\rho}$ with
$\rho = 0, -1$ can be rewritten as a sequence of fusion and skew-symmetry
isomorphisms in many different ways; 
so long as the graphical presentation of such a sequence is homotopically
the same as the one in Fig. \ref{fig:mtm2Gep-CFTop-cnstr}
(i.e., the braiding group element is the same), we have one common
isomorphism $(*)_\rho$, with either $\rho = 0$
or $-1$ \cite[app. B]{Moore:1988qv}. 

\vspace{5mm}

{\bf The superchiral algebra ${\cal A}^{\rm Gepn.}$ of the Gepner
  model SCFT:} we construct the state-operator correspondence $YY^{\rm Gepn.}$
of the superchiral algebra ${\cal A}^{\rm Gepn.}$ of the Gepner model SCFT,
using intertwining superoperators of ${\cal A}^{\rm m.t.m.}$; this is therefore
nothing more than an example of the simple current extension
theory \cite{MR1408523}, except that we do so with manifest SVOA structure. 
This
$YY^{\rm Gepn.}$ is the same as the CFT vertex operator $YY_{\rm tot}^{\rm Gepn.}$
for states in ${\cal A} = \oplus_{\vec{y} \in \Gamma_{\chi {\rm Alg}}}
(M_{\vec{0},2\vec{y}} \otimes
\tilde{\bf 1}) \subset ({\cal H}_{\rm tot}^{\rm pre})_{\rm Gepn.}$, acting on
${\cal A}^{\rm Gepn.} \subset ({\cal H}_{\rm tot}^{\rm pre})_{\rm Gepn.}$ and
ending in ${\cal A}^{\rm Gepn.} \subset ({\cal H}_{\rm tot}^{\rm pre})_{\rm Gepn.}$.
So, we use the isomorphism
$(*)_\rho$ for $(\lambda_a,\vec{m}_a) = (\vec{0}, \vec{0})$ for $a=1,2,3$,
and $\vec{y}_1, \vec{y}_2 \in \Gamma_{\chi {\rm Alg}} \subset \Gamma$.
The isomorphism $(*)_\rho$ is used to convert the state-operator correspondence
of the minimal tensor model left-mover superchiral algebra,
$(YY^{\rm m.t.m.}\otimes {\rm id}_{\widetilde{\cal A}^{\rm m.t.m.}}) \in
({\cal VV}_{(\vec{0},\vec{0}), (\vec{0},\vec{0})}^{(\vec{0},\vec{0})}
\otimes \widetilde{\cal VV}_{(\vec{0},\vec{0}), (\vec{0},\vec{0})}^{(\vec{0},\vec{0})})$.

We impose the following conditions on the basis elements
$YY^{(\vec{y}_1,\vec{y}_2)}$ and $YY^{[\vec{y},(\lambda,\vec{m})]}$:
\begin{itemize}
\item (C1$_0$)  $YY^{(\vec{0}, \;\vec{y})} := YY_{\vec{0},2\vec{y}}^{\rm m.t.m.}$
  (the representation of ${\cal A}^{\rm m.t.m.}$ on $M_{\vec{0},2\vec{y}}$)
  for $\vec{y} \in \Gamma_{\chi {\rm Alg}}$
\item (C2${}_0$) the skewsymmetry isomorphism $\Omega$ maps
  $YY^{(\vec{y}_1,\vec{y}_2)}$ to $YY^{(\vec{y}_2,\vec{y}_1)}$ for
  $\vec{y}_{1,2} \in \Gamma_{\chi {\rm Alg}}$, 
\item (C3${}_0$) the fusion isomorphism ${\cal F}$
 maps $(YY^{(y_1,y_{2+3})} \otimes YY^{(y_2,y_3)})$ in 
 $({\cal VV}_{(\vec{0},2\vec{y}_1), (\vec{0},2\vec{y}_{2+3})}^{(\vec{0},2\vec{y}_{1+2+3})}
 \otimes {\cal VV}_{(\vec{0},2\vec{y}_2),(\vec{0},2\vec{y}_3)}^{(\vec{0},2\vec{y}_{2+3})})$
 to $(YY^{(\vec{y}_1,\vec{y}_2)} \otimes YY^{(\vec{y}_{1+2},\vec{y}_3)})$ in 
 $({\cal VV}_{(\vec{0},2\vec{y}_1), (\vec{0},2\vec{y}_2)}^{(\vec{0},2\vec{y}_{1+2})}
 \otimes {\cal VV}_{(\vec{0},2\vec{y}_{1+2}), (\vec{0},2\vec{y}_3)}^{(\vec{0},2\vec{y}_{1+2+3})})$ 
  for  $\vec{y}_{1,2,3} \in \Gamma_{\chi {\rm Alg}}$,    
\end{itemize}
and 
\begin{itemize}
\item (D1) $YY^{[\vec{y}, (\vec{0},\vec{0})]} = \Omega YY^{\rm m.t.m.}_{\vec{0},2\vec{y}}$ for
  $\vec{y} \in \Gamma_{\chi {\rm Alg}}$.  
\end{itemize}
We do not specify the subscript $\rho' \in \Z$ of
$\Omega_{\rho'}(YY^{(\vec{y}_1,\vec{y}_2)})
= YY^{(\vec{y}_2,\vec{y}_1)}$ in C2${}_0$, because
$\Omega_{\rho'}YY^{(\vec{y}_1,\vec{y}_2)}$ is
independent of $\rho'\in \Z$ when $\vec{y}_1, \vec{y}_2 \in
\Gamma_{\chi {\rm Alg}}$
(so $M_{\vec{0},2\vec{y}_{1,2}}$ are in ${\cal A}^{\rm Gepn.}$). 
It follows then that the state-operator correspondence $YY^{\rm Gepn.}:
{\cal A}^{\rm Gepn.} \rightarrow {\rm End}({\cal A}^{\rm Gepn.})[[z,z^{-1}]]$ of
the Gepner model
superchiral algebra ${\cal A}^{\rm Gepn.} = \oplus_{\vec{y} \in \Gamma_{\chi {\rm Alg}}}
M_{\vec{0},2\vec{y}}$ is given by
\[
YY^{\rm Gepn.} = \{ YY^{(\vec{y}_1,\vec{y}_2)} \; | \;
\vec{y}_1, \vec{y}_2 \in \Gamma_{\chi {\rm Alg}}\}. 
\]
This statement follows from tracking the map $(*)_\rho$;
$(YY^{(\vec{y}_1,\vec{y}_2)}\otimes YY^{[\vec{y}_{1+2},(\vec{0},\vec{0})]} \otimes
YY^{\rm m.t.m.})$ in the vector space at $[ini]$ is mapped to
$(YY^{(\vec{y}_1,\vec{y}_2)} \otimes \Omega YY^{\rm m.t.m.}_{(\vec{0},2\vec{y}_2)}
\otimes YY^{\rm m.t.m.})$ in the vector space at [2], to
$(YY^{(\vec{y}_1,\vec{y}_2)} \otimes \Omega Y^{\rm m.t.m.}_{(\vec{0},2\vec{y}_2)} \otimes
\Omega YY^{\rm m.t.m.}_{(\vec{0},2\vec{y}_2)})$ at [3], to
$(YY^{(\vec{y}_1,\vec{y}_2)} \otimes \Omega YY^{\rm m.t.m.}_{(\vec{0},2\vec{y}_2)}
\otimes YY^{\rm m.t.m.}_{(\vec{0},2\vec{y}_2)})$ at [4], and finally to
$(\Omega YY^{\rm m.t.m.}_{(\vec{0},2\vec{y}_1)} \otimes
\Omega YY^{\rm m.t.m.}_{(\vec{0},2\vec{y}_2)} \otimes YY^{(\vec{y}_1,\vec{y}_2)})$.
We obtain $YY^{(\vec{y}_1,\vec{y}_2)} \in {\cal VV}_{(\lambda_1,\vec{m}_1+2\vec{y}_1), (\lambda_2,\vec{m}_2+2\vec{y}_2)}^{(\lambda_3,\vec{m}_3+2\vec{y}_{1+2})}$ for
$\vec{m}_{1,2,3} = \vec{0}$ by stripping off $YY^{[\vec{y}_1,(\vec{0},\vec{0})]} =
\Omega YY^{\rm m.t.m.}_{(\vec{0},2\vec{y}_1)}$
and $YY^{[\vec{y}_2,(\vec{0},\vec{0})]} = \Omega YY^{\rm m.t.m.}_{(\vec{0},2\vec{y}_2)}$. 

One subtlety that we have not mentioned so far is the fact that
the vector space of intertwining superoperators ${\cal VV}_{\beta,\gamma}^\alpha$
of an SVOA $(V, YY)$, with $\alpha,\beta,\gamma \in {\cal II}$, is defined
with two auxiliary parameters \cite[4.1.7, 4.1.8, 4.1.12]{Watari:2026scft} 
one of which is relevant here already,
and the other in a discussion later (only when Ramond-type representations
are relevant). The notion of an intertwining \underline{super}operator
${\cal YY} \in
{\cal VV}_{\beta,\gamma}^\alpha$ is to keep track of correlation among four
intertwining operators ${\cal YY} = \{ Y_{\beta s_\beta, \gamma s_\gamma} \in {\cal V}_{\beta s_\beta, \gamma s_\gamma}^{\alpha f(s_\beta,s_\gamma)}  \; |
\; s_\beta \in S_\beta, \; s_\gamma \in S_\gamma \}$ by fusion and braiding
isomorphisms; we need to specify how to choose a reference point in
measuring the correlation, and how to deal with branch cuts in braiding, and
hence we need auxiliary parameters for those choices.  
In practice, the first auxiliary parameter is a choice of identification
of the 2-element $\Z_2$-set $i_\beta: S_\beta \cong \Z/2\Z$ of the
representaiton $M_\beta$ of the SVOA $(V, YY)$; there are SCFTs with
a natural choice (those for ${\cal S}_{\rm II}$, as in
section \ref{ssec:from0B-toII}), but not all the SCFTs have a natural or
canonical choice, so we have to choose in that case. 
The skewsymmetry isomorphisms on intertwining \underline{super}operatros
have been defined for $\Omega_{\rho'}: {\cal VV}_{\beta,\gamma}^\alpha(i_\beta)
\rightarrow {\cal VV}_{\gamma,\beta}^\alpha(i_\gamma)$, and the fusion isomorphism
has been defined for ${\cal VV}_{\beta_1,\delta}^\alpha(i_{\beta_1}) \otimes
{\cal VV}_{\beta_2,\gamma}^\delta(i_{\beta_2}) \rightarrow
{\cal VV}_{\beta_1,\beta_2}^{\epsilon}(i_{\beta_1}) \otimes
{\cal VV}_{\epsilon,\gamma}^\alpha(i_{\beta_1} + i_{\beta_2})$;
see \cite[4.1.9+14]{Watari:2026scft} and \cite[4.1.10+15]{Watari:2026scft},
respectively, and references therein for further information;
here, $i_{\beta_1} + i_{\beta_2}$ is the identification of 2-element $\Z_2$-set 
$S_\epsilon \rightarrow \Z/2\Z$ so that the fusion
$(\beta_1 s_{\beta_1}) \cdot (\beta_2 s_{\beta_2}) \in S_\epsilon$ followed by
the map ``$(i_{\beta_1} + i_{\beta_2})$'' agrees with the map
$(i_{\beta_1} \times i_{\beta_2}):
  S_{\beta_1} \times S_{\beta_2} \rightarrow \Z/2\Z \times \Z/2\Z$ followed by
  the group addition law $\Z/2\Z \times \Z/2\Z \rightarrow \Z/2\Z$. 
For more information, see \cite[\S4]{Watari:2026scft} and references therein.
  
Back in our context, the background information above is applied
to the SVOA ${\cal A}^{\rm m.t.m.}$; the 2-element $\Z_2$ set
of the representation $(\lambda, \vec{m})$ is
$S_{(\lambda,\vec{m})} =\{ \sum_i s_i \in 2\Z/4\Z + \sum_i (\ell_i - m_i)\}$, so a choice of
$i_{(\lambda,\vec{m})}$ is in between whether $i_{(\lambda,\vec{m})}$ assigns 
$\sum_i s_i = \sum_i (\ell_i-m_i)+4\Z$ in $S_{(\lambda, \vec{m})}$ to
$0 \in \Z/2\Z$, or
$\sum_i s_i = 2+ \sum_i (\ell_i-m_i)+4\Z$ to $0 \in \Z/2\Z$. 
We should have specified a choice of
$i_{(\vec{0},2\vec{y})}: S_{(\vec{0},2\vec{y})} \cong \Z/2\Z$, before talking about
a choice a basis element $YY^{(\vec{y}_1,\vec{y}_2)}$ of the 1-dimensional
vector space ${\cal VV}_{(\vec{0},2\vec{y}_1), \; (\vec{0},2\vec{y}_2)}^{(\vec{0},2\vec{y}_{1+2})}(i_{(\vec{0},2\vec{y}_1)})$. The condition C2${}_0$ demands that
a chosen basis element $YY^{(\vec{y}_1,\vec{y}_2)}$ of
${\cal VV}_{(\vec{0},2\vec{y}_1), (\vec{0},2\vec{y}_2)}^{(\vec{0},2\vec{y}_{1+2})}
(i_{(\vec{0},2\vec{y}_1)})$ should be mapped to
the chosen basis element of ${\cal VV}_{(\vec{0},2\vec{y}_2), (\vec{0},2\vec{y}_1)}^{(\vec{0},2\vec{y}_{1+2})} (i_{(\vec{0},2\vec{y}_2)})$. The condition C3${}_0$
has an extra subtlety, because it makes sense right away, only if  
\begin{align}
i_{(\vec{0},2\vec{y}_1)} + i_{(\vec{0},2\vec{y}_2)}
= i_{(\vec{0},2\vec{y}_{1+2})}, \qquad
{}^\forall \vec{y}_1, \vec{y}_2 \in \Gamma_{\chi {\rm Alg}}.  
\end{align}
We use the following identification for $\vec{y} \in \Gamma_{\chi {\rm Alg}}$,
\begin{align}
  i_{(\vec{0},2\vec{y})} : S_{(\vec{0},2\vec{y})} = \{ \sum_i s_i \in 2\Z/4\Z \} \ni
  \sum_i s_i \longmapsto \frac{\sum_i s_i}{2} - p_0(\vec{y}) \in \Z/2\Z,
  \label{eq:S2-id-inGamma}
\end{align}
and there is nothing wrong in extending this to
$\vec{y} \in \Gamma$; because $p_0: \Gamma \rightarrow \Z/2\Z$ is
a homomorphism, we have the desired property above, and the condition
C3${}_0$ makes sense. 

To claim that the vector space ${\cal A}^{\rm Gepn.}$ is combined
with $YY^{\rm Gepn.}$ so constructed to be an SVOA, one has to make sure
that $YY^{\rm Gepn.}$ is invariant under the skewsymmetry isomorphism
of intertwining superoperators, and has the associativity.
The conditions C2${}_0$ and C3${}_0$ on the choice of the basis elements
immediately guarantee that, because $YY^{\rm Gepn.}$ is given literally by
$YY^{(\vec{y}_1,\vec{y}_2)}$.

In the construction of the state-operator correspondence within the
superchiral algebra of the Gepner model SCFT, we did not specify
whether we use $\Omega_\rho$ with $\rho = 0$ or $\rho = -1$, because
the dummy/coordinate variable $z$ shows up only in integral powers.

What is being done here is equivalent to the four-step procedure at the
beginning of this section \ref{sec:CFTop}. In physics intuition,
we have a lattice SVOA associated with the positive definite rank-$r$
lattice $L_{0}$, where 
\begin{align}
  L_{00}^{\vec{y}} & \; := \left\{ \eta = (\eta_{i=1,\cdots, r}) \in \Z^{\oplus r} \; | \;
  \eta_i +\bar{k}_i\Z = y_i \in \Z/\bar{k}_i\Z \right\},
  \label{eq:def-lattice-4left-00y} \\
  L_0 &\; = \oplus_{\vec{y} \in \Gamma_{\chi {\rm Alg}}} L_{00}^{\vec{y}}, \qquad
  L := \oplus_{\vec{y} \in \Gamma} L_{00}^{\vec{y}}. 
\end{align}
For each $\eta \in L_0$, the spectral flow state $v_{\eta}$
(the $(-\eta,0)$-unit flow of the vacuum state of the minimal tensor model
SCFT) is with $q = - \hat{c}_i \eta_i$ and $h= \hat{c}_i \eta_i^2/2$; when
all the representations of the minimal tensor model SVOA are split into
the U(1) sector and the rest, $v_\eta$ is purely in the U(1) sector tensored
with the ground state of the remaining sector. A choice of basis elements
$\{ YY^{(\vec{y}_1,\vec{y}_2)} \; | \; \vec{y}_2 \in \Gamma_{\chi {\rm Alg}}\}$
subject to (C1--C3) determines the vertex operators of the states $v_\eta$
with $\eta \in L^{\vec{y}_1}_{00}$.
Conversely, there is no freedom left for $YY^{(\vec{y}_1,\vec{y}_2)}$ in the
1-dimensional vector space ${\cal VV}_{(\vec{0},2\vec{y}_1), \;
  (\vec{0},2\vec{y}_2)}^{(\vec{0},2\vec{y}_{1+2})}(i_{(\vec{0},2\vec{y}_1)})$ when
the vertex operators of the states
$\{ v_\eta \; | \; \eta \in L^{\vec{y}_1}_{00} \}$ on the U(1) excitations
of the states $\{ v_{\eta} \; | \; \eta \in L^{\vec{y}_2}_{00} \}$. This
converse statement is a key feature of the theory of simple current
extension that allows us to skip/hide the non-straightforward four steps
at the beginning of section \ref{sec:CFTop}.

\vspace{5mm}

% \label{pg:Gep-repr-cnstr-2-beg}
Let us construct {\bf the representation
 $YY^{\rm Gepn.}_{(\lambda, [\vec{m}_0+2\vec{y}_0])}$ of the Gepner model superchiral
  algebra ${\cal A}^{\rm Gepn.}$} associated with some $\lambda$ and one pair
$(\vec{y}_0, \vec{m}_0)$ that survives the orbifold projection
(\ref{eq:cond-Gepner-survive-Z-w-d.t.},
\ref{eq:cond-Gepner-survive-Z-w-d.t.for-R});
a representation is in fact for $\lambda$, the coset
$\vec{y}_0 + \Gamma_{\chi {\rm Alg}} \in \Gamma/\Gamma_{\chi {\rm Alg}}$
and the orbit $\vec{m}_0 + 2\Gamma_{\chi {\rm Alg}} \subset
\prod_i \Z/2\bar{k}_i\Z$. 
We do not pay attention to whether a representation to be constructed
here is irreducible.\footnote{
We expect that this representation is reducible when the group
${\rm Ker}\left[ (\prod_{i=1}^r \Z/\bar{k}_i\Z)
  \rightarrow ( \prod_{i=1}^{r_{k/2}}\Z/\bar{k}_i\Z) \right] \cap \Gamma_{\chi {\rm Alg}}$
has non-trivial 2-torsion elements; here, $r_{k/2}$ is the number of
the tensored factors (the number of $i \in \{1,\cdots, r\}$)
where $\ell_i = k_i/2$ in $\lambda = (\ell_1,\cdots, \ell_r)$;
an element $g$ of an abelian group is a {\it 2-torsion} if $2g=0$
in the group. When this group is isomorphic to $(\Z/2\Z)^{r_{shrt}}$, 
the vector space $\oplus_{\Delta \vec{y} \in \Gamma_{\chi {\rm Alg}}}
M_{\lambda, \vec{m}_0+ 2(\vec{y}_0 + \Delta \vec{y})}$ contains $2^{r_{shrt}}$ identical
representation spaces of ${\cal A}^{\rm m.t.m.}$. The present authors expect that
the representation of ${\cal A}^{\rm Gepn.}$ on this vector space splits
into $2^{r_{shrt}}$ distinct irreducible representations; this question
is known as the fixed-point resolution in the representation theory of
simple current extensions, and there is an extensive literature
(see e.g., \cite{Schellekens:1990xy}, \cite{Fuchs:1996dd},
\cite{MR1935493}, \cite{MR2038578} % , \cite{MR4720880}
and references therein).
The authors apologize for not spending enough time to go through the
literature to be able to write down statements with confidence. 
} %

A representation $YY^{\rm Gepn.}_{(\lambda,[\vec{m}_0+2\vec{y}_0])}$ of the algebra 
${\cal A}^{\rm Gepn.} \cong \oplus_{\vec{y}_1 \in \Gamma_{\chi {\rm Alg}}} M_{\vec{0},2\vec{y}_1}$ on
the vector space
$\oplus_{\Delta \vec{y} \in \Gamma_{\chi {\rm Alg}}}
M_{\lambda, \vec{m}_0 + 2(\vec{y}_0+\Delta \vec{y})}$ consists of
$(\# \Gamma_{\chi {\rm Alg}})^2$
intertwining superoperators of ${\cal A}^{\rm m.t.m.}$, 
$\{ YY_{(\vec{0},2\vec{y}_1), \; (\lambda, \vec{m}_0 + 2(\vec{y}_0+\Delta \vec{y}) )} \}$,
labeled
by $\vec{y}_1, \Delta \vec{y} \in \Gamma_{\chi {\rm Alg}}$. There are multiple 
different ways labeled by $\delta \vec{y} \in \Gamma_{\chi {\rm Alg}}$ in
constructing 
$YY_{(\vec{0},2\vec{y}_1), \; (\lambda, \vec{m}_0 + 2(\vec{y}_0 + \Delta \vec{y}))}$: 
that is to use $(*)_\rho$ with $\vec{y}_2 = \vec{y}_0 +\Delta \vec{y}
- \delta \vec{y}$ and
$\vec{m}_2 = \vec{m}_0 + 2\delta \vec{y})$; for any $\delta \vec{y} \in
\Gamma_{\chi {\rm Alg}}$, we have the same desired $\vec{m}_2+2\vec{y}_2 =
\vec{m}_0 + 2(\vec{y}_0 + \Delta \vec{y})$; this issue was present also 
in the construction of the defining representation $YY^{\rm Gepn.}$, but we
just focused on one construction with $\delta \vec{y} = 0$  
(and $\vec{m}_0 = \vec{0}$, $\vec{y}_0 = \vec{0}$,
$\vec{y}_2 = \Delta \vec{y}$).
We use the isomorphism $(*)_\rho$
(for each $\delta \vec{y} \in \Gamma_{\chi {\rm Alg}}$) to translate
the representation $YY^{\rm m.t.m.}_{(\lambda, \vec{m}_0 + 2\delta \vec{y})}
\in {\cal VV}_{(\vec{0},\vec{0}), \; (\lambda, \vec{m}_0 + 2\delta \vec{y})}^{
  (\lambda, \vec{m}_0 + 2\delta \vec{y})}$
of the minimal tensor model superchiral algebra ${\cal A}^{\rm m.t.m.}$,
along with the choice of basis elements
$YY^{(\vec{y}',\vec{y}'')}$ with $\vec{y}', \vec{y}''$ in
$\Gamma_{\chi {\rm Alg}}$ or $\vec{y}_0 + \Gamma_{\chi {\rm Alg}}$ and
$YY^{[\vec{y},(\lambda,\vec{m})]}$ with $\vec{y} \in \vec{y}_0 +
\Gamma_{\chi {\rm Alg}}$ and $\vec{m} \in \vec{m}_0 + 2\Gamma_{\chi {\rm Alg}}$.  

We impose the following conditions on the basis elements
$YY^{(\vec{y}_1,\vec{y}_2)}$ and $YY^{[\vec{y},(\lambda,\vec{m})]}$:
\begin{itemize}
\item (C1$_1$): extend (C1$_0$) for $\vec{y} \in \Gamma$, 
\item (C3$_1$): extend (C3$_0$) for $\vec{y}_1 \in \Gamma_{\chi {\rm Alg}}$ and
  $\vec{y}_{2,3} \in \Gamma$,
\end{itemize}
and 
\begin{itemize}
\item (D2) there exists a collection
  $\{ YY^{[\delta \vec{y}, (\lambda,\vec{m}_0)]} \in
  {\cal VV}_{(\vec{0},2\delta \vec{y}),(\lambda,\vec{m}_0)}^{(\lambda,\vec{m}_0+2\delta \vec{y})} \;
  | \;  \delta \vec{y} \in \Gamma_{\chi {\rm Alg}} \}$
  so that\footnote{
As a special case of (D2) and (\ref{eq:temp-aux-choice-fM0-vs-fM}), where
$(\lambda,\vec{m}_0)=(\vec{0},\vec{0})$, it follos that 
$YY^{[\vec{y},(\vec{0},2\vec{y}')]} = YY^{(\vec{y},\vec{y}')}$.
  } % 
\begin{align}
  {\cal F}: YY^{[\vec{y}_0 + \Delta \vec{y}-\delta \vec{y}, (\lambda,
      \vec{m}_0 + 2\delta \vec{y})]}
  \otimes YY^{[\delta \vec{y},(\lambda,\vec{m}_0)]} \mapsto
  YY^{(\vec{y}_0+\Delta \vec{y}-\delta \vec{y}, \delta \vec{y})} \otimes Y^{[\vec{y}_0 + \Delta \vec{y},(\lambda,\vec{m}_0)]}
  \label{eq:temp-aux-choice-fM0-vs-fM}
\end{align}
for $\Delta \vec{y}, \delta \vec{y} \in \Gamma_{\chi {\rm Alg}}$. 
\end{itemize}
The condition (D2) is intended to make sure that the constructed
$YY_{(\vec{0},2\vec{y}_1), \; (\lambda, \vec{m}_0+2(\vec{y}_0+\Delta \vec{y}))}$ does not
depend on the choice of $\delta \vec{y} \in \Gamma_{\chi {\rm Alg}}$, as
we see below. Before doing so, we explain what is being done
in the choice of basis elements and choice of $\delta \vec{y}$ in plain
language. 

[intuition]: A choice of
$\{ YY^{[\vec{y}, (\lambda,\vec{m}_0)]} \in {\cal VV}_{(\vec{0},2\vec{y}), \; (\lambda, \vec{m}_0)}^{(\lambda, \vec{m}_0 + 2\vec{y})} \; | \; \vec{y} = \vec{y}_0 + \Delta \vec{y} \in \vec{y}_0 + \Gamma_{\chi {\rm Alg}} \}$ is,
in the language that is often used in physics literature, to introduce
a reference frame in $M_{\lambda,\vec{m}_0+2(\vec{y}_0+\Delta \vec{y})}$ by using
the states in $M_{\lambda,\vec{m}_0}$ as the anchor (spectral flow).\footnote{
What is written here in ``physics language'' is neither rigorous nor 
well-defined (see footnote \ref{fn:spc-flw}). Furthermore, there are
infinitely many spectral flow states $v_\eta \in M_{\vec{0},2(\vec{y}_0+\Delta \vec{y})}$ and corresponding ${\cal U}_{-\eta}$'s; the reference frames with those
${\cal U}_{-\eta}$ are tightly constrained to each other, and highly redundant. 
What is determined precisely by $YY^{[\vec{y},(\lambda,\vec{m})]}$ for
$\vec{y} \in \vec{y}_0 + \Gamma_{\chi {\rm Alg}}$ and
$\vec{m} \in \vec{m}_0+2\Gamma_{\chi {\rm Alg}}$
is $\Phi_{YY^{[\vec{y},(\lambda,\vec{m})]}}$ in footnote \ref{fn:boxtimes-univ},
and is free from this redundancy. 
} %
It is just as natural, however, to use the states in
$M_{\lambda, \vec{m}_0+2\delta \vec{y}}$ with any $\delta \vec{y} \in \Gamma_{\chi {\rm Alg}}$ 
to introduce another reference frame in $M_{\lambda,\vec{m}_0+2(\vec{y}_0+\Delta \vec{y})}$; 
because the idea in the construction using $(*)_\rho$ is to
set up a dictionary between the states in $M_{\lambda, \vec{m}_0+2(\vec{y}_0+\Delta \vec{y})}
\otimes \widetilde{M}_{\tilde{\lambda},\vec{m}_0}$ of the Gepner model CFT
with the states 
$M_{\lambda, \vec{m}} \otimes \widetilde{M}_{\tilde{\lambda},\vec{m}}$ for some 
$\vec{m} \in \vec{m}_0 + 2\Gamma_{\chi {\rm Alg}}$ 
available in the minimal tensor model CFT, it is fine to use any one of
$M_{\lambda, \vec{m}}\otimes \widetilde{M}_{\tilde{\lambda},\vec{m}}$ as a
the anchor so long as $\vec{m} \in \vec{m}_0 + 2\Gamma_{\chi {\rm Alg}}$.
We only need to make sure consistency among
the multiple different ways to introduce a reference frame to
$M_{\lambda, \vec{m}_0 + 2(\vec{y}_0 + \Delta \vec{y})}$. 
For this to be the case, there should exist a dictionary (correspondence
between the states) among the candidates of the anchor,
$M_{\lambda, \vec{m}}$ with $\vec{m} \in \vec{m}_0 +
2\Gamma_{\chi {\rm Alg}}$; the condition (D2) requires that there exist 
$YY^{[\delta \vec{y}, (\lambda, \vec{m}_0)]}$
for $\delta \vec{y} \in \Gamma_{\chi {\rm Alg}}$ because of that.
The condition (\ref{eq:temp-aux-choice-fM0-vs-fM}) implies\footnote{
A network of reference system $\{ YY^{[\vec{y},(\lambda, \vec{m})]} \; | \; \vec{y} \in \vec{y}_0 + \Gamma_{\chi {\rm Alg}}, \; \vec{m} \in \vec{m}_0 + 2\Gamma_{\chi {\rm Alg}}\}$ is
therefore built out of a choice of $\{ YY^{[\vec{y},(\lambda, \vec{m}_0)]} \; | \; \vec{y} \in \vec{y}_0 + \Gamma_{\chi {\rm Alg}}\}$ and
$\{ YY^{[\delta \vec{y}, (\lambda, \vec{m}_0)]} \; | \;
\delta \vec{y} \in \Gamma_{\chi {\rm Alg}} \}$, by using
(\ref{eq:temp-aux-choice-fM0-vs-fM}). Once a network is built, then
the consistency relation in a more general form follows:
$\delta \vec{y}', \vec{y}-\vec{y}_0 \in \Gamma_{\chi {\rm Alg}}$,
$\vec{m}-\vec{m}_0 \in 2\Gamma_{\chi {\rm Alg}}$, 
\begin{align}
  {\cal F}: YY^{[\vec{y}, (\lambda,\vec{m} + 2\delta \vec{y}')]} \otimes
  YY^{[\delta \vec{y}',(\lambda,\vec{m})]} \mapsto
  YY^{(\vec{y}, \delta \vec{y}')} \otimes YY^{[\vec{y}+\delta \vec{y}',(\lambda,\vec{m})]},
  \qquad 
  \label{eq:temp-aux-choice-fM0-vs-fM-gen}
\end{align}
which uses a more general version of the dictionary among the anchors 
$\{ YY^{[\delta \vec{y}',(\lambda,\vec{m})]} \; | \; \delta \vec{y}'
\in \Gamma_{\chi {\rm Alg}}\}$
for general $\vec{m} \in \vec{m}_0 + 2\Gamma_{\chi {\rm Alg}}$ determined
by the fusion isomorphism 
\begin{align}
  {\cal F}: YY^{[\delta \vec{y}', (\lambda,\vec{m}_0+2\delta \vec{y})]} \otimes
  YY^{[\delta \vec{y},(\lambda,\vec{m}_0)]}
  \mapsto YY^{(\delta \vec{y}',\delta \vec{y})} \otimes YY^{[\delta \vec{y}' + \delta \vec{y}, (\lambda,\vec{m}_0)]},
  \qquad \delta \vec{y}, \delta \vec{y}' \in \Gamma_{\chi {\rm Alg}}.
  \label{eq:temp-aux-dict-amng-anchr-spcFlw}
\end{align}
To derive (\ref{eq:temp-aux-choice-fM0-vs-fM-gen}), use the pentagon identity
for the four representations
$(\vec{0},2\vec{y})$, $(\vec{0}, 2\delta \vec{y}')$, $(\vec{0},2\delta \vec{y})$
$(\lambda, \vec{m}_0)$, along with (\ref{eq:temp-aux-choice-fM0-vs-fM},
\ref{eq:temp-aux-dict-amng-anchr-spcFlw}) and C3${}_1$. 
} %
that
the reference system introduced in $M_{\lambda, \vec{m}_0 + 2(\vec{y}+\Delta \vec{y})}$
from $M_{\lambda, \vec{m}_0}$ (by $YY^{[\vec{y} + \Delta \vec{y},(\lambda, \vec{m}_0)]}$)
is the same as the reference system using the dictionary
(by $YY^{[\Delta \vec{y}, \vec{m}_0]}$) to $M_{\lambda, \vec{m}_0 + 2\Delta \vec{y}}$
first, and then by $YY^{[\vec{y},(\lambda, \vec{m}_0+2\Delta \vec{y})]}$. 

\begin{figure}[tbp]
\begin{center}
   \includegraphics[scale=0.7]{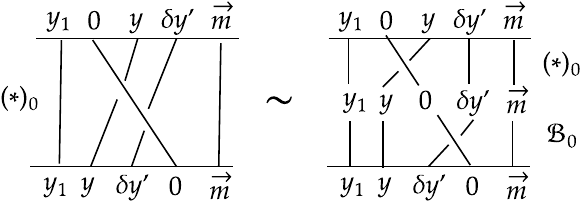}  
\caption{\label{fig:cmm-dgrm-spcFlw-cnsstnt}The graphical presentation
of the isomorphisms along the two routes in the
diagram (\ref{eq:diagr-spcFlw-cnsistency}), starting from the up-middle term
and ending at the right two terms; the graph on the left is for the route
going directly to the upper-right corner of (\ref{eq:diagr-spcFlw-cnsistency}), while the graph on the right for the route through the left two terms and the bottom row. The corresponding braid-group elements are the same, so the diagram commutes. }
\end{center}
\end{figure}

To see that the constructed
$YY_{(\vec{0},2 \vec{y}_1),(\lambda,\vec{m}_0+2(\vec{y}_0+\Delta \vec{y}))}$ is
independent of $\delta \vec{y} \in \Gamma_{\chi {\rm Alg}}$
we use the fact that the following diagram commutes
(see Fig. \ref{fig:cmm-dgrm-spcFlw-cnsstnt}): 
\begin{equation}
  \xymatrix{
    (((\vec{y}_1 \cdot \vec{y})\cdot \delta \vec{y}') \cdot (0 \cdot \vec{m})) & 
    ((\vec{y}_1 \cdot (\vec{y} \cdot \delta \vec{y}')) \cdot (0 \cdot \vec{m})) \ar[l]^{\cal F}
    \ar[r]^{(*)_\rho}
    % \ar[d]^{\cal F}
    &
    ((\vec{y}_1 \cdot 0)\cdot ((\vec{y} \cdot \delta \vec{y}') \cdot \vec{m})) \\
    ((\vec{y}_1 \cdot \vec{y}) \cdot (\delta \vec{y}' \cdot (0 \cdot \vec{m})))
    \ar[u]^{\cal F} \ar[r]^{{\cal B}_\rho} &
    ((\vec{y}_1 \cdot \vec{y}) \cdot (0 \cdot (\delta \vec{y}' \cdot \vec{m}))) \ar[r]^{(*)_\rho} &
    ((\vec{y}_1 \cdot 0)\cdot (\vec{y} \cdot (\delta \vec{y}' \cdot \vec{m}))) \ar[u]^{\cal F}
  }; \label{eq:diagr-spcFlw-cnsistency}
\end{equation}
here, the irreducible representations $(\vec{0},2\vec{y}_1)$,
$(\vec{0}, 2(y_0+\Delta \vec{y} - \delta \vec{y} -\delta \vec{y}'))$,
$(\lambda, \vec{m}_0+2\delta \vec{y})$ of ${\cal A}^{\rm m.t.m.}$ are
abbreviated by $\vec{y}_1$, $\vec{y}$, $\vec{m}$, respectively, to save space.
In the diagram above, each term in the form of a parenthesized word with
letters in $\vec{y}_1$, $\vec{y}$, $\vec{m}$ is an alternative expression
for the tensor product of vector spaces of intertwining superoperators,
as we have introduced already below (\ref{eq:isom-4-mtm-Gepn-optr-cnstr}).  
To keep track of the intertwining superoperators in this diagram, use the
condition (D2) in (\ref{eq:temp-aux-choice-fM0-vs-fM},
\ref{eq:temp-aux-choice-fM0-vs-fM-gen}) in the two vertical ${\cal F}$'s,
and (C3${}_1$ with $\vec{y}\in \Gamma$ as $\vec{y}_2$) at the upper-left
${\cal F}$. The condition (D2) guarantees that the input
(chosen basis elements) for the $(*)_\rho$ construction in the
top row is compatible with the input for the $(*)_\rho$ construction
in the bottom row; the $\delta y$-independence follows from the
commutativity of the diagram then. 
%%%%%%%%%%%%%%%%%%%%%%%%%%%%%%%%%%%%%%%%%
%%%%%%%%%%%%%%%%%%%%%%%%%%%%%%%%%%%%%%%%%

The constructed
$\{YY^{\rm Gepn.}_{(\vec{0},2\vec{y}_1),(\lambda, \vec{m}_0+2(\vec{y}_0+\Delta \vec{y}))}\}$
satisfies the associativity because of (C3${}_1$) with
$\vec{y}_3 \in \vec{y}_0 + \Gamma_{\chi {\rm Alg}}$, 
and the commutativity because of (C2) along with the associativity. So,
$\{ YY^{\rm Gepn.}_{(\vec{0},2\vec{y}_1),(\lambda,\vec{m}_0+2(\vec{y}_0+\Delta \vec{y}))}\}$
is a representation of ${\cal A}^{\rm Gepn.}$. This representation restricted
to ${\cal A}^{\rm m.t.m.}$ is that of the minimal tensor model, because of
(C1${}_1$) with $\vec{y} \in \vec{y}_0+\Gamma_{\chi {\rm Alg}}$.

\vspace{5mm}

{\bf SCFT vertex operator:} Let us proceed to the construction of
the whole SCFT vertex operator $YY_{\rm tot}^{\rm Gepn.}$. It should be 
a choice of tensor product of intertwining superoperators
of ${\cal A}^{\rm Gepn.}$ and $\widetilde{\cal A}^{\rm Gepn.}$
that is invariant under the skewsymmetry and fusion isomorphisms. 
To construct an SCFT vertex operator of the Gepner model SCFT,
we convert that of the minimal tensor model
\begin{align}
  YY_{\rm tot}^{\rm m.t.m.} & \; =
\left\{ \left. {\cal YY}_{(\lambda_1,\vec{m}_1),(\lambda_2,\vec{m}_2)}^{(\lambda_3,\vec{m}_3)} \otimes
\widetilde{\cal YY}_{(\lambda_1,\vec{m}_1), \; (\lambda_2,\vec{m}_2)}^{(\lambda_3,\vec{m}_3)} \; \right| \; \right. \\
& \; \left. (\lambda_a,\vec{m}_a) \; a=1,2, \; |\lambda_1-\lambda_2| \leq
\lambda_3 \leq {\rm Min}(\lambda_1+\lambda_2, 2\vec{k}-\lambda_1-\lambda_2)
\right\} \nonumber 
\end{align}
by using the isomorphism\footnote{
We will specify later (at p. \pageref{pg:set-rho=s})
which of $\rho \in \{ \pm 1\}$ to be used. 
} %
 $(*)_\rho$ along with a choice of basis elements  
$\{ YY^{(\vec{y}_1,\vec{y}_2)} \; | \; \vec{y}_1,\vec{y}_2 \in \Gamma\}$ and
$\{ YY^{[\vec{y},(\lambda, \vec{\tilde{m}})]} \; | (\vec{y},\vec{\tilde{m}})~{\rm surviving~}
(\ref{eq:cond-Gepner-survive-Z-w-d.t.},
\ref{eq:cond-Gepner-survive-Z-w-d.t.for-R}) \}$. 
Remember that ${\cal YY}_{(\lambda_1,\vec{m}_1),(\lambda_2,\vec{m}_2)}^{(\lambda_3,\vec{m}_3)} \widetilde{\cal YY}_{(\lambda_1,\vec{m}_1), \; (\lambda_2,\vec{m}_2)}^{(\lambda_3,\vec{m}_3)}$ above is not the tensor of general intertwining superoperators of
${\cal A}^{\rm m.t.m.}$ and $\widetilde{\cal A}^{\rm m.t.m.}$, 
but is a special choice from ${\cal VV}_{(\lambda_1,\vec{m}_1),(\lambda_2,\vec{m}_2)}^{(\lambda_3,\vec{m}_3)} \otimes
\widetilde{\cal VV}_{(\lambda_1,\vec{m}_1), \; (\lambda_2,\vec{m}_2)}^{(\lambda_3,\vec{m}_3)}$ so that the collection $YY_{\rm tot}^{\rm m.t.m.}$ as a whole is invariant
under the skewsymmetry and fusion isomorphisms.

Here are two subtle details to pay attention to before talking about
the conditions to impose on the choice of the basis elements; both are
about the auxiliary parameters in the definition of the (vector space of)
intertwining superoperators. In the input $YY_{\rm tot}^{\rm m.t.m.}$,
$\widetilde{\cal YY}_{(\lambda_1,\vec{m}_1), \; (\lambda_2,\vec{m}_2)}^{(\lambda_3,\vec{m}_3)}$ is from $\widetilde{\cal VV}_{(\lambda_1,\vec{m}_1), \; (\lambda_2,\vec{m}_2)}^{(\lambda_3,\vec{m}_3)}(\tilde{\imath}_{(\lambda_1,\vec{m}_1)})$, where an identification
$\tilde{\imath}_{(\lambda,\vec{m})}$ of 2-element $\Z_2$ sets maps
two possible choices of $\sum_{i=1}^r \tilde{s}_i \in \Z/4\Z$ to $\Z/2\Z$, 
and
${\cal YY}_{(\lambda_1,\vec{m}_1), \; (\lambda_2,\vec{m}_2)}^{(\lambda_3,\vec{m}_3)}$
from ${\cal VV}_{(\lambda_1,\vec{m}_1), \; (\lambda_2,\vec{m}_2)}^{(\lambda_3,\vec{m}_3)}
(i_{(\lambda_1,\vec{m}_1)})$, also with a choice of the auxiliary parameter
$i_{(\lambda,\vec{m})}$ mapping two possible values of $\sum_i s_i \in \Z/4\Z$
to $\Z/2\Z$. We make it a rule to choose $i_{(\lambda,\vec{m})}$ for the
left-mover to be identical to $\tilde{\imath}_{(\lambda, \vec{m})}$ for the
right-mover, when replacing $\sum_i \tilde{s}_i$ by $\sum_i s_i$.
With this rule, $i_{(\lambda,\vec{m})}(s_{\rm tot}) = \tilde{\imath}_{(\lambda,\vec{m})}(\tilde{s}_{\rm tot})\in \Z/2\Z$ is equivalent to
whether the states in ${\cal L}^{\lambda}_{\vec{m},\vec{s}} \otimes \widetilde{\cal L}^{\lambda}_{\vec{m},\vec{\tilde{s}}}$ remain in
$({\cal H}_{\rm tot}^{\rm post})_{\rm m.t.m.}$. The sequence of isomorphisms
$(*)_\rho$ starting from ${\cal VV}_{(\vec{0},2\vec{y}_1), \;
  (\vec{0},2\vec{y}_2)}^{(\vec{0},2\vec{y}_{1+2})}(i_{(\vec{0},2\vec{y}_1)})$,
${\cal VV}_{(\lambda_1,\vec{m}_1),\; (\lambda_2,\vec{m}_2)}^{(\lambda_3,\vec{m}_3)}(i_{(\lambda_1,\vec{m}_1)})$ and ${\cal VV}_{(\vec{0},2\vec{y}_{1+2}), \; (\lambda_3,\vec{m}_3)}^{(\lambda_3,\vec{m}_3+2\vec{y}_{1+2})}(i_{(\vec{0},2\vec{y}_{1+2})})$ ends up in
${\cal VV}_{(\vec{0},2\vec{y}_1), \; (\lambda_1,\vec{m}_1)}^{(\lambda_1,\vec{m}_1+2\vec{y}_1)}(i_{(\vec{0},2\vec{y}_1)})$, 
${\cal VV}_{(\vec{0},2\vec{y}_2), \; (\lambda_2,\vec{m}_2)}^{(\lambda_2,\vec{m}_2+2\vec{y}_2)}(i_{(\vec{0},2\vec{y}_2)})$ and
${\cal VV}_{(\lambda,\vec{m}_1+2\vec{y}_1), \; (\lambda_2,\vec{m}_2+2\vec{y}_2)}^{(\lambda_3,\vec{m}_3+2\vec{y}_{1+2})}(i_{(\vec{0},2\vec{y}_1)} + i_{(\lambda_1,\vec{m}_1)})$.
So, the SCFT vertex operator
$YY_{\rm tot}^{\rm Gepn.}$ constructed through $(*)_\rho$ is naturally with
the auxiliary parameter
\begin{align}
  i_{(\vec{0},2\vec{y}_1)}+ i_{(\lambda_1,\vec{m}_1)} =:
  i^{\rm Gepn.}_{(\lambda_1,\vec{m}_1+2\vec{y}_1)}
  \label{eq:diff-S2-id-mtm-vs-Gep-a}
\end{align}
for the left mover and $\tilde{\imath}_{(\lambda_1,\vec{m}_1)}$ for the right-mover,
a combination where 
\begin{align}
  i^{\rm Gepn.}_{(\lambda_1,\vec{m}_1+2\vec{y}_1)}(s_{\rm tot}^{\rm Gepn.}) =
  \tilde{\imath}_{(\lambda_1,\vec{m}_1)}(\tilde{s}_{\rm tot})
  \Leftrightarrow
  s_{\rm tot}^{\rm Gepn.} = \tilde{s}_{\rm tot} + p_0(\vec{y}_1) \in \Z/4\Z,
  \label{eq:diff-S2-id-mtm-vs-Gep-b}
\end{align}
i.e., when the states are in $({\cal H}_{\rm tot}^{\rm post})_{\rm Gepn.}$
rather than $[({\cal H}_{\rm tot}^{\rm post})^\perp_{\rm Gepn.} \subset
  ({\cal H}_{\rm tot}^{\rm dbl})_{\rm Gepn.}]$.

% Although there is nothing wrong in using either one of two possible
% choices of $\tilde{\imath}_{(\lambda,\vec{m})}$ for each of the
% irreducible representaitons $(\lambda,\vec{m})$ of the minimal
% tensor model (and $\widetilde{\cal YY}_{(\lambda_1,\vec{m}_1),\;
% (\lambda_2,\vec{m}_2)}^{(\lambda_3,\vec{m}_3)}$ should be translated
% properly) so long as one is
% interested only in the minimal tensor model CFT, it is more convenient
% if we choose the identifications $\tilde{\imath}_{(\lambda,\vec{m})}$ so that
% %
% \begin{align}
%   \tilde{\imath}_{(\lambda,\vec{m} + 2y)} = \tilde{\imath}_{(\lambda,\vec{m})}
%   + p_0(y), \qquad y \in \Gamma_{\chi {\rm Alg}} 
% \end{align}
% %
% when we use $YY_{\rm tot}^{\rm m.t.m.}$ to build $YY_{\rm tot}^{\rm Gepn.}$. 
% Then $i^{\rm Gepn.}_{(\lambda,\vec{m}+2y)} = i_{(\lambda,\vec{m}+2y)}$. 
% Such a correlated choice of the identification is possible and compatible
% with (\ref{eq:field-id}) so long as $\lambda = (\ell_1,\cdots, \ell_r)$
% do not contain $\ell_i = k_i/2$.

One more auxiliary parameter $r \in \Z/2\Z$ is necessary in formulating
(the vector space of) intertwining superoperators of type
$(\alpha;\beta,\gamma)$ of an SVOA when the representation $\beta$ of
the SVOA $(V_0 \oplus V_1, YY)$ is Ramond-type. 
An intertwining superoperator ${\cal YY}$ of type $(\alpha;\beta,\gamma)$
consists of four intertwining operators of the underlying VOA $V_0$,
$\{ {\cal Y}_{s_\beta, s_\gamma} \; | \; s_\beta \in S_\beta, \; s_\gamma \in S_\gamma
\}$, where ${\cal Y}_{s_\beta, s_\gamma}$ and ${\cal Y}_{s_\beta,s'_\gamma}$ with
the same $s_\beta$ and opposite $s_\gamma, s'_\gamma \in S_\gamma$
are correlated under the braiding isomorphism ${\cal B}_{-1-r}: (V_1 \cdot (M_{\beta s_\beta} \cdot M_{\gamma s_\gamma})) \rightarrow (M_{\beta s_\beta} \cdot
(V_1 \cdot M_{\gamma s_\gamma}))$. The parameter $r \in \Z/2\Z$ chooses which
one of the braiding isomorphisms is used to set the correlation
among the four component operators ${\cal Y}_{s_\beta,s_\gamma}$; when
$M_{\beta}$ is a Ramond-type representation, the angle $+\pi$ monodromy
of $V_1 \subset V$ around $M_{\beta}$ and the angle $-\pi$ monodromy
are not the same.

Here, we quote a few statements from \cite{Watari:2026scft} on the
$r\in \Z/2\Z$-dependence that we need in this article; 
readers interested in systematic exposition of the idea
of vector space of intertwining superoperators are referred to
\cite[\S4]{Watari:2026scft}. First, the skewsymmetry isomorphism is defined as 
$\Omega_\rho: {\cal VV}_{\beta,\gamma}(i_\beta, r)|^{{\rm if~}\beta{\rm ~Rmnd}}_{r=-1-\rho}
\rightarrow {\cal VV}_{\gamma,\beta}^\alpha(i_\gamma, r')|^{{\rm if~}\gamma{\rm ~Rmnd}}_{r'=\rho}$; if the representation $\beta$ [resp. $\gamma$] of the SVOA
is NS-type, the vector space ${\cal VV}_{\beta,\gamma}^\alpha(r)$
[resp. ${\cal VV}_{\gamma,\beta}^\alpha(r')$] is independent of the
auxiliary parameter $r$ [resp. $r'$] in $\Z/2\Z$, so we can ignore
the comments on $r$ [resp. $r'$]. Second, the fusion isomorphism is defined
as ${\cal F}: \oplus_{\delta} ({\cal VV}_{\beta_1,\delta}^\alpha(r_1)\otimes
{\cal VV}_{\beta_2,\gamma}^\delta(r_2)) \rightarrow \oplus_{\epsilon}
({\cal VV}_{\beta_1,\beta_2}^\epsilon(r_3)\otimes {\cal VV}_{\epsilon,\gamma}^\alpha(r_4))$ if (i) $r_1=r_3 \in \Z/2\Z$ if $\beta_1$ is Ramond type,
(ii) $r_2 = r_4 \in \Z/2\Z$ if $\beta_2$ is Ramond type but $\beta_1$
is NS type, and (iii) $r_1=r_2=r_3 \in \Z/2\Z$ if both $\beta_1$ and $\beta_2$
are Ramond type. 

Back in the context of this article, 
we use the isomorphism $(*)_\rho$ with either $\rho = 0$ or $\rho = -1$
in converting the SCFT vertex operator of the minimal tensor model
to that of the Gepner models, so we should use the SCFT vertex operator
of the minimal tensor model $YY_{(\lambda,\vec{m}_1), \; (\lambda_2,\vec{m}_2)}^{(\lambda_3,\vec{m}_3)} \widetilde{YY}_{(\lambda,\vec{m}_1), \; (\lambda_2,\vec{m}_2)}^{(\lambda_3,\vec{m}_3)}$ in ${\cal VV}_{(\lambda,\vec{m}_1), \; (\lambda_2,\vec{m}_2)}^{(\lambda_3,\vec{m}_3)}(i_{(\lambda_1,\vec{m}_0)}, -1-\rho) \otimes
\widetilde{\cal VV}_{(\lambda,\vec{m}_1), \; (\lambda_2,\vec{m}_2)}^{(\lambda_3,\vec{m}_3)}(\tilde{\imath}_{(\lambda_1,\vec{m}_0)}, \rho)$, if $M_{\lambda_1,\vec{m}_1}$ is
a Ramond representation of ${\cal A}^{\rm m.t.m.}$. At the end of the
sequence of isomorphisms in $(*)_\rho$, the SCFT vertex operator so constructed
is in
\[
{\cal VV}_{(\lambda_1,\vec{m}_1+2\vec{y}_1), \; (\lambda_2,\vec{m}_2+2\vec{y}_2)}^{(\lambda_3,\vec{m}_3+2\vec{y}_3)}(i_{(\lambda_1,\vec{m}_1)} + i_{(\vec{0},2\vec{y}_1)}, -1-\rho) \otimes
\widetilde{\cal VV}_{(\lambda,\vec{m}_1), \; (\lambda_2,\vec{m}_2)}^{(\lambda_3,\vec{m}_3)}(\tilde{\imath}_{(\lambda_1,\vec{m}_1)}, \rho);
\]
indeed, the isomorphism $\Omega_\rho$ from [1] to [2] in $(*)_\rho$
requires $r= \rho$ in ${\cal VV}_{(\lambda_2,\vec{m}_2),
  (\lambda_1,\vec{m}_1)}^{(\lambda_3,\vec{m}_3)}(r)$ at [2]; the fusion
isomorphism ${\cal F}$ from [2] to [3] requires, if $(\lambda_2,\vec{m}_2)$
is Ramond type, that $r= \rho$ in ${\cal VV}_{(\lambda_2,\vec{m}_2+2\vec{y}_2),\; (\lambda_1,\vec{m}_1)}^{(\lambda_3,\vec{m}_3+2\vec{y}_2)}(r)$ at [3] because of the
pattern (ii) above; $\Omega_{-1-\rho}$ from [3] to [4] implies that
$r = -1-\rho$ in ${\cal VV}_{(\lambda_1,\vec{m}_1), \; (\lambda_2,\vec{m}_2+2\vec{y}_2)}^{(\lambda_3,\vec{m}_3+2\vec{y}_2)}(r)$ at [4]; the last ${\cal F}$ from [4] to [fin]
in $(*)_\rho$
implies---if $(\lambda_1,\vec{m}_1)$ is Ramond type---that we end up with
$r=-1-\rho$ in ${\cal VV}_{(\lambda_1,\vec{m}_1+2\vec{y}_1), \;(\lambda_2,\vec{m}_2+2\vec{y}_2)}^{(\lambda_3,\vec{m}_3+2\vec{y}_{1+2})}(r)$ at [fin]---as claimed above---because of the pattern (ii) above. 

% 
% All the intertwining operators here are super-quartets. When the representation
% $(\ell_1,m_1+2y_1)$ is a Ramond type, we do have a canonical set-theoretical
% id $S_{(\ell,m+2y)} \cong \Z/2\Z$ when $\hat{c}$ is even, while we have to
% make a choice (charge = -1/2 mod $+2\Z$ as even ones). 
% The construction $(*)_\rho$ uses $\Omega_\rho$ first, and then
% $\Omega_{\rho}^{-1}$ later. So, we may start from
% $Y_{\rm tot} = {\cal Y} \otimes \widetilde{\cal Y}$ in 
% ${\cal V}_{m_1,m_2}^{m_3}(r=-1-\rho)\otimes
% \widetilde{\cal V}_{m_1,m_2}^{m_3}(r=\rho)$
% of the minimal tensor model, and end up with
% ${\cal V}_{m_1+2y_1,m_2+2y_2}^{m_3+2y_3}(-1-\rho)\otimes
% \widetilde{\cal V}_{m_1,m_2}^{m_3}(\rho)$.
% 

Now, here is the extra conditions we impose on the choice of basis elements:
\begin{itemize}
\item (C2${}_2$) extend C2${}_0$ for $\vec{y}_1, \vec{y}_2 \in \Gamma$, with 
\begin{align}
  \Omega_{\rho}: YY^{(\vec{y}_1,\vec{y}_2)} \longmapsto
  YY^{(\vec{y}_2,\vec{y}_1)} \left( \mathbb{E} \left[  \sum_i \frac{1}{\bar{k}_i} y^i_1 y^i_2 
    \right] \epsilon^{00}(\vec{y}_1, \vec{y}_2) \right)^{-(2\rho+1)}
  , \quad \vec{y}_{1,2} \in \Gamma,  
\end{align}
from ${\cal VV}_{(\vec{0},2\vec{y}_1), \; (\vec{0},2\vec{y}_2)}^{(\vec{0},2\vec{y}_{1+2})}(r=-1-\rho)$
to ${\cal VV}_{(\vec{0},2\vec{y}_2), \; (\vec{0},2\vec{y}_1)}^{(\vec{0},2\vec{y}_{1+2})}(r=\rho)$. 
\item (C3${}_2$) extend C3${}_1$ for $\vec{y}_{1,2,3} \in \Gamma$. 
\end{itemize}

\begin{figure}[tbp]
 \begin{center}
    \includegraphics[scale=0.7]{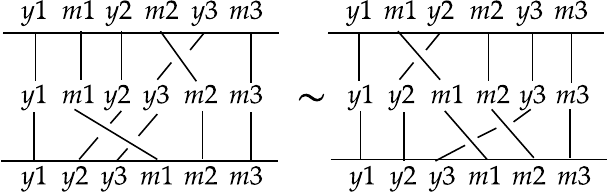}  
 \caption{\label{fig:cmm-dgrm-4-Finv}The graphical presentations of the
   the isomorphisms along the two routes in the diagram (\ref{eq:diagr-Finv}),
   starting form the [UL] corner and ending at the [DR] corner. 
   The graph on the left is for the route down the left column in the diagram,
   while the one on the right for the route down the right column.
   The corresponding braid-group elements are the same, hence the diagram
   commutes. 
 }
 \end{center}
\end{figure}

Let us first verify that the SCFT vertex operator of the Gepner model SCFT
constructed in this way is invariant under the fusion isomorphism
${\cal F}_{\rm tot} = {\cal F} \otimes \widetilde{\cal F}$, by using
the fact that the SCFT vertex operator $YY_{\rm tot}^{\rm m.t.m.}$
of the minimal tensor model is invariant under ${\cal F}_{\rm tot}$,
and also the condition (C3${}_2$) on the choice of basis elements.
To this end, we use the fact that the isomorphisms in the following
diagram 
\begin{equation}
  \xymatrix{  
    {}_{[UL]} \;\;
    ((\vec{y}_1 \cdot (\vec{y}_2 \cdot \vec{y}_3)) \cdot (\vec{m}_1 \cdot
    (\vec{m}_2 \cdot \vec{m}_3))) \ar[r]^{{\cal F} \otimes {\cal F}}
    \ar[d]^{(*)_\rho} &
    (((\vec{y}_1 \cdot \vec{y}_2) \cdot \vec{y}_3) \cdot ((\vec{m}_1 \cdot
    \vec{m}_2) \cdot \vec{m}_3)) \ar[d]^{(*)_\rho} \;\; {}_{[UR]} \\
    {}_{[ML]} \;\;
    ((\vec{y}_1 \cdot \vec{m}_1) \cdot ((\vec{y}_2 \cdot \vec{y}_3) \cdot
    (\vec{m}_2 \cdot \vec{m}_3))) \ar[d]^{(*)_\rho} &
    (((\vec{y}_1 \cdot \vec{y}_2) \cdot (\vec{m}_1 \cdot \vec{m}_2)) \cdot
    (\vec{y}_3 \cdot \vec{m}_3)) \ar[d]^{(*)_\rho} \;\;
    {}_{[MR]} \\
    {}_{[DL]}\;\;
    ((\vec{y}_1 \cdot \vec{m}_1) \cdot ((\vec{y}_2 \cdot \vec{m}_2) \cdot
    (\vec{y}_3 \cdot \vec{m}_3))) \ar[r]^{{\cal F}} &
    (((\vec{y}_1 \cdot \vec{m}_1) \cdot (\vec{y}_2 \cdot \vec{m}_2)) \cdot
    (\vec{y}_3 \cdot \vec{m}_3)) \;\; {}_{[DR]}
  }
 \label{eq:diagr-Finv}
\end{equation}
are well-defined and commute; here, the symbols $\vec{y}_{1,2,3}$ stand for
the representations $(\vec{0},2\vec{y}_{1,2,3})$ of ${\cal A}^{\rm m.t.m.}$,
and the symbols $\vec{m}_{a}$ with $a=1,2,3$ for the representations
$(\lambda_{a}, \vec{m}_{a})$; each of those parenthesized words made up
of those symbols above stands for the tensor product of the vector spaces
of intertwining superoperators, as before. When any one
of the representations $(\lambda_a, \vec{m}_a)$ is Ramond-type, use
the auxiliary parameter $r=-1-\rho$ for
${\cal VV}_{(\lambda_a,\vec{m}_a), *}^{*}(r)$; then we have the pattern of
auxiliar parameters necessary for $(*)_\rho$ and the fusion isomorphisms. 
The diagram above commutes because the graphic presentation of
the two vertical sequences of $(*)_\rho \cdot (*)_\rho$ are identical
in Fig. \ref{fig:cmm-dgrm-4-Finv}. 
%%%%%%%%%%%%%%%%%%%%%%%%%%%%%%%%%%%%%%%%%
%%%%%%%%%%%%%%%%%%%%%%%%%%%%%%%%%%%%%%%%%

The commuting diagram above is used to see the invariance of
$YY_{\rm tot}^{\rm Gepn.}$ under ${\cal F}_{\rm tot}$, by tracking
\begin{align}
 & YY^{(\vec{y}_1,\vec{y}_{2+3})} \otimes YY^{(\vec{y}_2,\vec{y}_3)} \otimes YY^{[\vec{y}_{1+2+3}, (\lambda_4,\vec{m}_4)]} \otimes  \label{eq:temp-Finv-Gepn-CFTop-cntrs-UL} \\
 & \left\{ \left(
  YY_{(\lambda_1,\vec{m}_1), (\lambda_5,\vec{m}_{2+3})}^{(\lambda_4,\vec{m}_{1+2+3})} \otimes
  \widetilde{YY}_{(\lambda_1,\vec{m}_1), (\lambda_5,\vec{m}_{2+3})}^{(\lambda_4,\vec{m}_{1+2+3})} \right) \otimes \left( 
    YY_{(\lambda_2,\vec{m}_2), \; (\lambda_3,\vec{m}_3)}^{(\lambda_5,\vec{m}_{2+3})} \otimes \widetilde{YY}_{(\lambda_2,\vec{m}_2), \; (\lambda_3,\vec{m}_3)}^{(\lambda_5,\vec{m}_{2+3})} \right) \right\}_{\lambda_5}  \nonumber 
\end{align}
at the [UL] corner of the diagram; the first three intertwining
superoperators and the 4th and 6th ones
of (\ref{eq:temp-Finv-Gepn-CFTop-cntrs-UL})---all of which are for the
left-mover
${\cal A}^{\rm m.t.m.}$---are in the tensored vector space
of intertwining superoperators in the diagram above; the 5th and 7th are 
intertwining superoperators of the right-mover
$\widetilde{\cal A}^{\rm m.t.m.}$ omitted from the diagram just to save space.
The 4th-5th and 6th--7th intertwining superoperators in the left-right
combination are the SCFT vertex operators of the minimal tensor model.
By following the isomorphisms $(*)_\rho \cdot (*)_\rho$ along the left column
in the diagram, the SCFT vertex operator of the Gepner model SCFT is obtained:
\begin{align}
&  YY^{[\vec{y}_1,(\lambda_1,\vec{m}_1)]} \otimes YY^{[\vec{y}_2,(\lambda_2,\vec{m}_2)]} \otimes
  YY^{[\vec{y}_3,(\lambda_3,\vec{m}_3)]} \otimes \nonumber \\
&  \left\{ \left( YY_{(\lambda_1,\vec{m}_1+2\vec{y}_1), \; (\lambda_5,\vec{m}_{2+3}+2\vec{y}_{2+3})}^{{\rm Gepn.}(\lambda_4,\vec{m}_{1+2+3}+2\vec{y}_{1+2+3})} \otimes
  \widetilde{YY}_{(\lambda_1,\vec{m}_1), (\lambda_5,\vec{m}_{2+3})}^{(\lambda_4,\vec{m}_{1+2+3})} \right)  \right. \label{eq:temp-Finv-Gepn-CFTop-cntrs-LL} \\
&  \qquad \otimes \left. 
  \left( YY_{(\lambda_2,\vec{m}_2+2\vec{y}_2), \; (\lambda_3,\vec{m}_{3}+2\vec{y}_{3})}^{{\rm Gepn.}(\lambda_5,\vec{m}_{2+3}+2\vec{y}_{2+3})} \otimes
  \widetilde{YY}_{(\lambda_2,\vec{m}_2), (\lambda_3,\vec{m}_{3})}^{(\lambda_5,\vec{m}_{2+3})} \right) \right\}_{\lambda_5}. \nonumber 
\end{align}
The fusion isomorphism at the bottom of the commuting diagram
acts on the 4th and 6th intertwining operators
of (\ref{eq:temp-Finv-Gepn-CFTop-cntrs-LL}), so this is the left-mover
part of ${\cal F}_{\rm tot} = {\cal F} \otimes \widetilde{\cal F}$ applied
on the Gepner model SCFT operators. To test whether
${\cal F}_{\rm tot}$ maps
$\{YY_{\rm tot}^{\rm Gepn.} \otimes YY_{\rm tot}^{\rm Gepn.}\}_{\lambda_5}$
with $|\lambda_2-\lambda_3| \leq \lambda_5 \leq {\rm Min}(\lambda_2+\lambda_3, 2\vec{k}-\lambda_2-\lambda_3)$ to
$YY_{\rm tot}^{\rm Gepn.} \otimes YY_{\rm tot}^{\rm Gepn.}$
in the fused channel, we can just apply $\widetilde{\cal F}$ on the 5th and 7th
intertwining operators of (\ref{eq:temp-Finv-Gepn-CFTop-cntrs-LL}) along with
the isomorphism ${\cal F}$ at the bottom of the diagram. The image
of this ${\cal F}_{\rm tot}$ in the Gepner model CFT can be studied by
using the other route in the commuting diagram. 

Now, let us track (\ref{eq:temp-Finv-Gepn-CFTop-cntrs-UL}) along the other
route in the diagram. The isomorphism ${\cal F} \otimes {\cal F}$ along
with $\widetilde{\cal F}$ on the 5th and 7th intertwining superoperators
of (\ref{eq:temp-Finv-Gepn-CFTop-cntrs-UL}) maps (\ref{eq:temp-Finv-Gepn-CFTop-cntrs-UL}) into
\begin{align}
 & YY^{(\vec{y}_1,\vec{y}_{2})} \otimes YY^{(\vec{y}_{1+2},\vec{y}_3)} \otimes YY^{[\vec{y}_{1+2+3}, (\lambda_4,\vec{m}_4)]} \otimes  \label{eq:temp-Finv-Gepn-CFTop-cntrs-UR} \\
 & \left\{ \left(
  YY_{(\lambda_1,\vec{m}_1), (\lambda_2,\vec{m}_{2})}^{(\lambda_6,\vec{m}_{1+2})} \otimes
  \widetilde{YY}_{(\lambda_1,\vec{m}_1), (\lambda_2,\vec{m}_{2})}^{(\lambda_6,\vec{m}_{1+2})} \right) \otimes \left( 
    YY_{(\lambda_6,\vec{m}_{1+2}), \; (\lambda_3,\vec{m}_3)}^{(\lambda_4,\vec{m}_{1+2+3})} \otimes \widetilde{YY}_{(\lambda_6,\vec{m}_{1+2}), \; (\lambda_3,\vec{m}_3)}^{(\lambda_4,\vec{m}_{1+2+3})} \right) \right\}_{\lambda_6},   \nonumber 
\end{align}
because of the condition C3${}_2$ on the choice of basis elements,
and also of the ${\cal F}_{\rm tot} = {\cal F} \otimes \widetilde{\cal F}$
invariance of the SCFT vertex operator of the minimal tensor model. 
By tracking this (\ref{eq:temp-Finv-Gepn-CFTop-cntrs-UR}) along the
isomorphism $(*)_\rho \cdot (*)_\rho$ along the right column of the
diagram, we obtain
\begin{align}
&  YY^{[\vec{y}_1,(\lambda_1,\vec{m}_1)]} \otimes YY^{[\vec{y}_2,(\lambda_2,\vec{m}_2)]} \otimes
  YY^{[\vec{y}_3,(\lambda_3,\vec{m}_3)]} \otimes \nonumber \\
&  \left\{ \left( YY_{(\lambda_1,\vec{m}_1+2\vec{y}_1), \; (\lambda_2,\vec{m}_{2}+2\vec{y}_{2})}^{{\rm Gepn.}(\lambda_6,\vec{m}_{1+2}+2\vec{y}_{1+2})} \otimes
  \widetilde{YY}_{(\lambda_1,\vec{m}_1), (\lambda_2,\vec{m}_{2})}^{(\lambda_6,\vec{m}_{1+2})} \right)  \right. \label{eq:temp-Finv-Gepn-CFTop-cntrs-LR} \\
&  \qquad \otimes \left. 
  \left( YY_{(\lambda_6,\vec{m}_{1+2}+2\vec{y}_{1+2}), \; (\lambda_3,\vec{m}_{3}+2\vec{y}_{3})}^{{\rm Gepn.}(\lambda_4,\vec{m}_{1+2+3}+2\vec{y}_{1+2+3})} \otimes
  \widetilde{YY}_{(\lambda_6,\vec{m}_{1+2}), (\lambda_3,\vec{m}_{3})}^{(\lambda_4,\vec{m}_{1+2+3})} \right) \right\}_{\lambda_6}. \nonumber 
\end{align}
So, ${\cal F}_{\rm tot}={\cal F} \otimes \widetilde{\cal F}$ in the
Gepner model SCFT maps the SCFT vertex operator in
(\ref{eq:temp-Finv-Gepn-CFTop-cntrs-LL}) to the SCFT vertex operator in 
(\ref{eq:temp-Finv-Gepn-CFTop-cntrs-LR}) indeed. 

Secondly, let us verify that the SCFT vertex operator $YY_{\rm tot}^{\rm Gepn.}$
constructed so far is invariant under the skewsymmetry isomorphism
$\Omega_{\rm tot} = \Omega_r \otimes \widetilde{\Omega}_{-1-r}$. Although
it is fine to use any $r \in \Z$ to test this, we use
$r=\rho$ when $YY_{\rm tot}^{\rm Gepn.}$ is contructed by using the isomorphism
$(*)_\rho$ because that is the easiest.

\begin{figure}[tbp]
\begin{center}
    \includegraphics[scale=0.7]{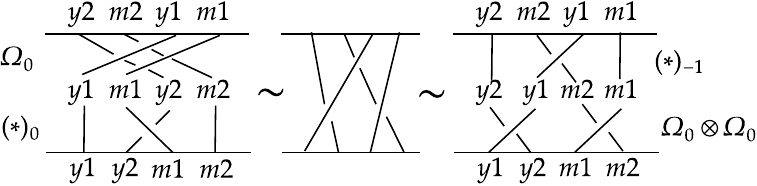}  
\caption{\label{fig:cmm-dgrm-4-OmegInv}The graphical presentations of the
  isomorphisms along the two routes in the diagram (\ref{eq:diagr-OmegInv}),
  starting from the [UL] corner and ending at the [DR] corner. The graph
  on the left is for the route through the [UR] corner, while the one on the
  right for the route through the [DL] corner. The diagram therefore commutes. 
  }
\end{center}
\end{figure}

Two observations will be used for this purpose. One is the fact that the
following diagram of isomorphisms is well-defined and commutes:
\begin{equation}
  \xymatrix{
  {}_{[UL]} \;\;  ((\vec{y}_1 \cdot \vec{y}_2) \cdot (\vec{m}_1 \cdot \vec{m}_2))\otimes
    (\vec{m}_1\cdot \vec{m}_2) \ar[r]^{(*)_\rho \otimes {\rm id}.}
    \ar[d]^{\Omega_\rho \otimes {\rm id}. \otimes (\Omega_\rho \otimes \widetilde{\Omega}_{-1-\rho})} &
    ((\vec{y}_1 \cdot \vec{m}_1) \cdot (\vec{y}_2 \cdot \vec{m}_2)) \otimes
    (\vec{m}_1 \cdot \vec{m}_2) \ar[d]^{{\rm id}\otimes {\rm id}\otimes (\Omega_\rho \otimes \widetilde{\Omega}_{-1-\rho})} \;\; {}_{[UR]} \\
    {}_{[LL]} \;\; ((\vec{y}_2 \cdot \vec{y}_1) \cdot (\vec{m}_2 \cdot \vec{m}_1)) \otimes
    (\vec{m}_2 \cdot \vec{m}_1) \ar[r]^{(*)_{-1-\rho} \otimes {\rm id}.}  &
    ((\vec{y}_2 \cdot \vec{m}_2) \cdot (\vec{y}_1 \cdot \vec{m}_1)) \otimes
    (\vec{m}_2 \cdot \vec{m}_1) \;\; {}_{[LR]} 
  } .
  \label{eq:diagr-OmegInv}
\end{equation}
Once again, we used abbreviated symbols for irreducible representations
of ${\cal A}^{\rm m.t.m.}$ and $\widetilde{\cal A}^{\rm m.t.m.}$
($\vec{y}$ for $(\vec{0},2\vec{y})$ and $\vec{m}$ for $(\lambda,\vec{m})$);
the symbols to the left [resp. right] of $\otimes$ are
for the representations of ${\cal A}^{\rm m.t.m.}$ [resp.
  $\widetilde{\cal A}^{\rm m.t.m.}$]; the parentehsized words of those
representation symbols stand for the tensor product of the vector spaces
of intertwining superoperators fused in the order designated by the
parentheses. We have added tags such as ${}_{[UL]}$, ${}_{[LL]}$ to refer
to each term in this diagram in later discussions. 
To see that this diagram commutes, it is enough to compare the graphical
presentations of two routes, as in Fig. \ref{fig:cmm-dgrm-4-OmegInv}. 
%%%%%%%%%%%%%%%%%%%%%%%%%%%%%%%%%%%%%%%%%
%%%%%%%%%%%%%%%%%%%%%%%%%%%%%%%%%%%%%%%%%

The other observation is the difference between the isomorphisms
$(*)_\rho$ with $\rho \in \{0,-1\}$; although the isomorphisms $(*)_\rho$
with different $\rho$ should be used---if the representation
$(\lambda_1,\vec{m}_1)$ is Ramond type---for different vector spaces
${\cal VV}_{(\lambda_1,\vec{m}_1), \; (\lambda_2,\vec{m}_2)}^{(\lambda_3,\vec{m}_{1+2})}(-1-\rho)$ in the domain and different vector spaces
${\cal VV}_{(\lambda_1,\vec{m}_1+2\vec{y}_1), \; (\lambda_2,\vec{m}_2+2\vec{y}_2)}^{(\lambda_3,\vec{m}_{1+2}+2\vec{y}_{1+2})}(-1-\rho)$ in the target, we focus on the difference in the
isomorphisms $(*)_\rho$ here, and will come back to the conversion
between the vector spaces of intertwining superoperators with different
auxiliary parameters later (p. \pageref{pg:conv-Rmnd-intOp-diff-r}).
We claim that the isomorphisms $(*)_\rho$ with
$\rho \in \{0,-1\}$ are related by 
\begin{align}
  (*)_{\rho + \Delta \rho} = (*)_\rho \cdot \mathbb{E}\left[ (\Delta \rho)
    \left( - \sum {}_i \frac{y^{(2)}_i (m_{2,i} - m_{3,i})}{\bar{k}_i} \right)
    \right] = (*)_\rho \cdot \mathbb{E}\left[
    (\Delta \rho) \sum {}_i \frac{y^{(2)}_i m_{1,i}}{\bar{k}_i} \right].
 \label{eq:temp-Gepn-CFTop-cnstr-rhoDpndnc-4NSNS}
\end{align}
Indeed, the isomorphism $\Omega_{\rho + \Delta \rho}$ from [1] to [2] in
$(*)_{\rho + \Delta \rho}$ is
$\mathbb{E}[(\Delta \rho)(-h_1-h_2+h_3)]$ times $\Omega_\rho$ in $(*)_\rho$;
here,
$\Delta \rho = -(1+2\rho)$, and $h_a$ with $a=1,2,3$ are the $L_0$-eigenvalues
mod $+\Z$ on $M_{\lambda_a,\vec{m}_a}$.
The isomorphism $\Omega_{\rho+\Delta \rho}^{-1}$
from [3] to [4] in $(*)_{\rho + \Delta \rho}$ is $\mathbb{E}[(-\Delta \rho)(-h'_2-h_1+h'_3)]$ times $\Omega_\rho^{-1}$ in $(*)_\rho$, where $h'_a$ with $a=2,3$
are the $L_0$-eigenvalues mod $+\Z$ on $M_{\lambda_a,\vec{m}_a+2\vec{y}_2}$.
Using the
fact that the representations $M_{\lambda_a,\vec{m}_a+2\vec{y}_2}$ with $a=2,3$
are obtained by the spectral flow of $(-y_2)$ unit from $M_{\lambda_a,\vec{m}_a}$,
we see that $(h'_2-h_2) - (h'_3-h_3) = - \sum {}_i
y^{(2)}_i(m_{2,i}-m_{3,i})/\bar{k}_i$ mod $+\Z$, so the
relation (\ref{eq:temp-Gepn-CFTop-cnstr-rhoDpndnc-4NSNS}) follows.\footnote{
The $L_0$-eigenvalue mod $+\Z$ on $M_{(\lambda_a,\vec{m}_a)}$ or
$M_{(\lambda_a,\vec{m}_a+2\vec{y}_a)}$ changes by $1/2 +\Z$ if the representations
are NS-type, and pick up states with different $\sum_i s_i$. This difference
cancels in $(h_1-h_1)$, $(h'_2-h_2)$ and $(h'_3-h_3)$, however.  
} % 

Now, let us use the two observations above to see that the Gepner model
SCFT vertex operator is invariant under $\Omega_{\rm tot} =
\Omega_\rho \otimes \widetilde{\Omega}_{-1-\rho}$; we do so by tracking
\begin{align}
  YY^{(\vec{y}_1,\vec{y}_2)} \otimes YY^{[\vec{y}_{1+2},(\lambda_3,\vec{m}_{1+2})]} \otimes \left(
  YY_{(\lambda_1,\vec{m}_1),\; (\lambda_2,\vec{m}_2)}^{(\lambda_3,\vec{m}_{1+2})} \otimes
  \widetilde{YY}_{(\lambda_1,\vec{m}_1),\; (\lambda_2,\vec{m}_2)}^{(\lambda_3,\vec{m}_{1+2})} \right)
  \label{eq:temp-OmegInv-Gepn-CFTop-UL}
\end{align}
in the upper-left corner of the diagram in two different routes. 
By tracking this product of operators (\ref{eq:temp-OmegInv-Gepn-CFTop-UL})
in [UL]$\rightarrow$[UR], first, we have the SCFT vertex operator of the Gepner
model in the [UR] corner of the diagram, and then
$YY^{[\vec{y}_1,(\lambda_1,\vec{m}_1)]} \otimes YY^{[\vec{y}_2,(\lambda_2,\vec{m}_2)]}
\otimes \Omega_{\rm tot}(YY_{\rm tot}^{\rm Gepn.})$ at the [LR] corner. 
So we can read out $\Omega_{\rm tot}(YY_{\rm tot}^{\rm Gepn.})$ by tracking
the same product of operators in the downstairs route in this commuting
diagram. The product of operators (\ref{eq:temp-OmegInv-Gepn-CFTop-UL})
is sent to
\begin{align}
  \left( \mathbb{E}\left[ \sum_i \frac{y^{(1)}_i y^{(2)}_i}{\bar{k}_i} \right]
  \epsilon^{00}(y_1,y_2) \right)^{-(2\rho+1)} \; 
  YY^{(y_2,y_1)} \otimes YY^{[y_{1+2},(\lambda_3,\vec{m}_{1+2})]} \otimes YY_{\rm tot}^{\rm m.t.m.}
  \label{eq:temp-OmegInv-Gepn-CFTop-LL}
\end{align}
in the [LL] corner of the diagram, because of the condition (C4) on
the basis elements $\{ YY^{(\vec{y},\vec{y}')} \; | \;
\vec{y},\vec{y}' \in \Gamma \}$, and $\Omega_{\rm tot}(YY_{\rm tot}^{\rm m.t.m.})=
YY_{\rm tot}^{\rm m.t.m.}$. To work on the downstairs route of the diagram
from the [LL] corner to the [LR] corner, however,
we need different arguments depending on whether the representation
$(\lambda_2,\vec{m}_2)$ is NS-type, or Ramond type. 

Think of the NS-type case, first. Then the vector space of intertwining
superoperators of type $((\lambda_3,\vec{m}_3); \; (\lambda_2,\vec{m}_2),
(\lambda_1,\vec{m}_1))$ is independent of the auxiliary parameter $r$,
so that is not a concern. Still we have, in the [LL]
corner (\ref{eq:temp-OmegInv-Gepn-CFTop-LL}), 
the chosen of basis elements $YY^{(\vec{y}_2,\vec{y}_1)}$ and
$YY^{[\vec{y}_{1+2},(\lambda_3,\vec{m}_3)]}$ 
prepared for the construction using $(*)_\rho$ on one hand, while
we need to use $(*)_{-1-\rho} = (*)_{\rho -(1+2\rho)}$ to go to the [LR]
corner of the diagram. This product of operators
(\ref{eq:temp-OmegInv-Gepn-CFTop-LL}) would naturally
yield $YY_{\rm tot}^{\rm Gepn.}$ if we were to use $(*)_\rho$ downstair.
The difference between $(*)_{\rho - (1+2\rho)}$ and $(*)_\rho$ is
found already in 
(\ref{eq:temp-Gepn-CFTop-cnstr-rhoDpndnc-4NSNS}). So, the image
in the [LR] corner is
\begin{align}
  \left( \mathbb{E}\left[ \sum_i \frac{y^{(1)}_i (y^{(2)}_i + m_{2,i})}{\bar{k}_i} \right]
  \epsilon^{00}(\vec{y}_1,\vec{y}_2) \right)^{-(2\rho+1)} \!\!\!\!\! 
  YY^{[\vec{y}_1,(\lambda_1,\vec{m}_1)]} \otimes YY^{[\vec{y}_2,(\lambda_2,\vec{m}_2)]} \otimes
  YY_{\rm tot}^{\rm Gepn.}.
\end{align}
The expression within the large parenthesis is 1 because of the
orbifold projection condition (\ref{eq:cond-Gepner-survive-Z-w-d.t.})
on $(\vec{y}_2, \vec{m}_2)$ in the NS--NS sector. 
So, the Gepner model SCFT vertex operator is invariant under $\Omega_{\rm tot}$,
when the representation $(\lambda_2,\vec{m}_2)$ is NS-type. 

Let us now work on the case the representation $(\lambda_2,\vec{m}_2)$ is
Ramond-type. We should note then that the product of operators
(\ref{eq:temp-OmegInv-Gepn-CFTop-LL}) is in
${\cal VV}_{(\vec{0},2y_2), \; (\vec{0},2y_1)}^{(\vec{0},2y_{1+2})} \otimes
{\cal VV}_{(\vec{0},2y_{1+2}), \; (\lambda_3,\vec{m}_{1+2})}^{(\lambda_3,\vec{m}_{1+2}+2y_{1+2})}
\otimes {\cal VV}_{(\lambda_2,\vec{m}_2), \; (\lambda_1,\vec{m}_1)}^{(\lambda_3,\vec{m}_{1+2})}(\rho)\otimes \widetilde{\cal VV}_{(\lambda_2,\vec{m}_2), \; (\lambda_1,\vec{m}_1)}^{(\lambda_3,\vec{m}_{1+2})}(-1-\rho)$; the isomorphism $(*)_{-1-\rho} \otimes {\rm id}:
        [LL] \rightarrow [LR]$
is from this vector space. To compare with the isomorphism $(*)_\rho$ for
the construction of the Gepner model SCFT vertex operator, we need
this diagram that does NOT commute:
\label{pg:conv-Rmnd-intOp-diff-r}
\begin{equation}
  \xymatrix{
 {}_{[LL]} \;\; \otimes {\cal VV}_{\vec{m}_2,\vec{m}_1}^{\vec{m}_{1+2}}(\rho) \otimes
 \widetilde{\cal VV}_{\vec{m}_2,\vec{m}_1}^{\vec{m}_{1+2}}(-1-\rho) 
 \ar[r]^{(*)_{-1-\rho}} \ar[d]^{{\rm can.}({\rm m.t.m.})}
 &
 \otimes {\cal VV}_{\vec{m}_2+2y_2,\vec{m}_1+2y_1}^{\vec{m}_{1+2}+2y_{1+2}}(\rho) \otimes
 \widetilde{\cal VV}_{\vec{m}_2,\vec{m}_1}^{\vec{m}_{1+2}}(-1-\rho) \; {}_{[LR]}
 \ar[d]^{{\rm can.}({\rm Gepn.})} \\
 {}_{[XL]} \; \otimes {\cal VV}_{\vec{m}_2,\vec{m}_1}^{\vec{m}_{1+2}}(-1-\rho) \otimes
 \widetilde{\cal VV}_{\vec{m}_2,\vec{m}_1}^{\vec{m}_{1+2}}(\rho) \ar[r]^{(*)_\rho} & 
 \otimes {\cal VV}_{\vec{m}_2+2y_2,\vec{m}_1+2y_1}^{\vec{m}_{1+2}+2y_{1+2}}(-1-\rho) \otimes
 \widetilde{\cal VV}_{\vec{m}_2,\vec{m}_1}^{\vec{m}_{1+2}}(\rho) \; {}_{[XR]}
  };
  \label{eq:diagr-OmegInv-ext}
\end{equation}
the $[LL]$ and $[LR]$ terms and the isomorphism $(*)_{-1-\rho} \otimes {\rm id}$
between them are the same as that of the diagram (\ref{eq:diagr-OmegInv}); 
to save space, $\lambda$ in the representation labels and the vector space
${\cal VV}_{\vec{y}_2,\vec{y}_1}^{\vec{y}_{1+2}} \otimes
{\cal VV}_{\vec{y}_{1+2},\vec{m}_{1+2}}^{\vec{m}_{1+2}+2\vec{y}_{1+2}} \otimes \cdots $ to which
$YY^{(\vec{y}_2,\vec{y}_1)} \otimes YY^{[\vec{y}_{1+2},\vec{m}_{1+2}]}$ belongs to is omitted
in the term $[LL]$, and similarly in the three other terms in this diagram. 
Our strategy is to compute $(*)_{-1-\rho}:[LL] \rightarrow [LR]$
by going through $[XL]$ and $[XR]$, and implementing the phase factors
that account for the non-commutativity of the diagram. 

Here, we need a little more quote from \cite[\S4]{Watari:2026scft}.
Isomorphisms from ${\cal VV}_{R_1,R_2}^{R_3}(\rho) \otimes \widetilde{\cal VV}_{\tilde{R}_1,\tilde{R}_2}^{\tilde{R}_3}(-1-\rho)$ of a left-mover SVOA ${\cal A}_0$ and
a right-mover SVOA $\widetilde{\cal A}_0$ to 
${\cal VV}_{R_1,R_2}^{R_3}(\rho + \Delta \rho) \otimes \widetilde{\cal VV}_{\tilde{R}_1,\tilde{R}_2}^{\tilde{R}_3}(-1-\rho-\Delta \rho)$ are of the form of
multiplying $\mathbb{E}[(\Delta \rho)(i_{R_2}(s_2) -\tilde{\imath}_{\tilde{R}_2}(\tilde{s}_2))/2]$ for some identification of 2-element $\Z/2\Z$-sets
$i_{R_2}: S_{R_2} \rightarrow \Z/2\Z$ and $\tilde{\imath}_{\tilde{R}_2}:
\widetilde{S}_{\tilde{R}_2} \rightarrow \Z/2\Z$
if the representations $R_1$ and $\tilde{R}_1$ are Ramond-type.
The trivial identity isomorphism is fine if $R_1$ and $\tilde{R}_1$ are
NS-type; that was why we ignored this when $R_1=(\lambda_2,\vec{m}_2)$
is NS-type. Moreover, not all the isomorphisms with arbitrarily chosen pairs
of $i_{R_2}$ and $\tilde{\imath}_{\tilde{R}_2}$ have a desirable property
when dealing with SCFT vertex operators of a Type 0 SCFT; suppose that
the superchiral algebras of a given Type 0 SCFT contain ${\cal A}_0$
and $\widetilde{\cal A}_0$; when we choose $i_{R_2}$ and
$\tilde{\imath}_{\tilde{R}_2}$ so that $i_{R_2}=\tilde{\imath}_{\tilde{R}_2}$
is equivalent to be in ${\cal H}_{\rm tot}^{\rm post}$ but not
$[({\cal H}_{\rm tot}^{\rm post})^\perp \subset {\cal H}_{\rm tot}^{\rm dbl}]$,
the isomorphism above that uses such a pair $i_{R_2}$ and
$\tilde{\imath}_{\tilde{R}_2}$ keeps the CFT vertex operator acting
on ${\cal H}_{\rm tot}^{\rm post} \subset {\cal H}_{\rm tot}^{\rm dbl}$. 
We call such isomorphisms a canonical isomorphism of that given
Type 0 SCFT \cite[4.1.12, 4.2.4]{Watari:2026scft}. 

Back in the present context, where
$R_1 = (\lambda_2,\vec{m}_2)$ is Ramond type, we have
$\Delta \rho = -(1+2\rho)$, ${\cal A}_0 = {\cal A}^{\rm m.t.m.}$ and
$R_2 = (\lambda_1, \vec{m}_1)$. We hope to use the canonical isomorphism
of the minimal tensor model SCFT from $[LL]$ to $[XL]$ to translate
$YY_{\rm tot}^{\rm m.t.m.}$ with $r=\rho$ to $r= -1-\rho$, while
we use the canonical isomorphism of the Gepner model SCFT from $[LR]$
to $[XR]$ to translate $YY_{\rm tot}^{\rm Gepn.}$ with $r=\rho$ and $-1-\rho$.
\begin{align}
  {\rm can.}({\rm m.t.m.}) & \; = \mathbb{E}[-(\Delta \rho)
    (i_{(\lambda_1,\vec{m}_1)}(s_{1,{\rm tot}})-\tilde{\imath}_{(\lambda_1,\vec{m}_1)}(\tilde{s}_{1,{\rm tot}}))], \\
  {\rm can.}({\rm Gepn.}) & \; =
  \mathbb{E}[-(\Delta \rho)
    (i_{(\lambda_1,\vec{m}_1+2\vec{y}_1)}^{\rm Gepn.}(s_{1,{\rm tot}})
    -\tilde{\imath}_{(\lambda_1,\vec{m}_1)}(\tilde{s}_{1,{\rm tot}}))].
\end{align}
The route from $[LL]$ through $[XL]$ and $[XR]$ to reach $[LR]$ is designed
to map the product of operators in (\ref{eq:temp-OmegInv-Gepn-CFTop-LL})
except the phase factor to
$YY^{[\vec{y}_1,(\lambda_1,\vec{m}_1)]} \otimes YY^{[\vec{y}_2,(\lambda_2,\vec{m}_2)]}
\otimes YY_{\rm tot}^{\rm Gepn.}$. 

The extra complex phase in the route through $[XL]$ and $[XR]$ compared
with the direct route to $[LR]$ has two sources. One is from the difference
between the isomorphism $(*)_{-1-\rho}$ in $[XL] \rightarrow [XR]$
from $(*)_\rho$ in $[LL] \rightarrow [LR]$, for which we can use
(\ref{eq:temp-Gepn-CFTop-cnstr-rhoDpndnc-4NSNS}) as we already did
in the case $(\lambda_2,\vec{m}_2)$ is NS type. The other is from
the difference between ${\rm can.}({\rm m.t.m.})$ from
${\rm can.}({\rm Gepn.})$, which we can evaluate by using
$i^{\rm Gepn.}_{(\lambda_1,\vec{m}_1+2\vec{y}_1)}= i_{(\lambda_1,\vec{m}_1)}
+ p_0(\vec{y}_1)$
(cf (\ref{eq:diff-S2-id-mtm-vs-Gep-a}, \ref{eq:diff-S2-id-mtm-vs-Gep-b})
and (\ref{eq:S2-id-inGamma})). 
By combining
the remaining complex phase in (\ref{eq:temp-OmegInv-Gepn-CFTop-LL})
with those two complex phases for the non-commutativity of the diagram
(\ref{eq:diagr-OmegInv-ext}), we see that
\begin{align}
  \left( \mathbb{E}\left[ \sum_i \frac{y^{(1)}_i(y^{(2)}_i+m_{2,i})}{\bar{k}_i} \right] \epsilon^{00}(\vec{y}_1,\vec{y}_2)
  (-1)^{p_0(\vec{y}_1)} \right)^{-(2\rho+1)} = 1
\end{align}
for a pair $(\vec{y}_2,\vec{\tilde{m}}_2)$ for Ramond--Ramond sector states
in the Gepner model SCFT, because of the orbifold projection condition
(\ref{eq:cond-Gepner-survive-Z-w-d.t.for-R}). This completes the proof
that the conditions C2${}_2$ and C3${}_2$ on the choice of basis elements
in the construction via $(*)_\rho$ are sufficient for the invariance
of the CFT vertex operator $YY_{\rm tot}^{\rm Gepn.}$ under ${\cal F}_{\rm tot}
={\cal F} \otimes \widetilde{\cal F}$ and $\Omega_{\rm tot} =
\Omega_r \otimes \widetilde{\Omega}_{-1-r}$. 

\vspace{5mm}

{\bf Unitarity structure and crossing symmetry:}  
It is better to impose a few more conditions on the choice of basis elements
$YY^{(\vec{y}_1, \vec{y}_2)}$ and $YY^{[\vec{y},(\lambda,\vec{m})]}$.
To state what they are for, we need to insert a brief review
on unitarity structure and crossing symmetry on a general 2D CFT and
Type 0 SCFT. 

\label{pg:Moore-natural-rev-4chiral}
Given a 2D modular invariant rational CFT whose left-mover and right-mover
chiral algebras are ${\cal A}$ and $\widetilde{\cal A}$, there must be a set of
irreducible representations ${\cal I}$ and $\widetilde{\cal I}$
of ${\cal A}$ and $\widetilde{\cal A}$, respectively, and a
one-to-one map $\pi: {\cal I} \rightarrow \widetilde{\cal I}$
so that the Hilbert space of the CFT has the structure
${\cal H}_{\rm tot} \cong \oplus_{\alpha \in {\cal I}} (M_\alpha \otimes
\widetilde{M}_{\pi(\alpha)})$ (cf \cite[\S4]{Moore:1988ss}).
There should also exist a set-theoretical
map $\Upsilon^{\cal I}: {\cal I} \rightarrow {\cal I}$ so that
the contragradient representation $(M_\alpha^\vee, \Theta Y_\alpha)$
of one representation $(M_\alpha, Y_\alpha)$ for $\alpha \in {\cal I}$
is equivalent to $(M_{\Upsilon^{\cal I}(\alpha)},
Y_{\Upsilon^{\cal I}(\alpha)})$ for one $\Upsilon^{\cal I}(\alpha) \in {\cal I}$.
In other words, there must be an isomorphism $\lambda_\alpha: M_\alpha^\vee
\rightarrow M_{\Upsilon^{\cal I}(\alpha)}$ so 
\begin{align}
  \lambda_\alpha \cdot (\Theta Y_\alpha)(v,z) \cdot \lambda_\alpha^{-1}
  = Y_{\Upsilon(\alpha)}(v,z) \qquad {}^\forall v \in {\cal A},
\end{align}
Similarly, there must be a map $\widetilde{\Upsilon}^{\cal I}:
\widetilde{\cal I} \rightarrow \widetilde{\cal I}$ that is uniquely
determined by $\widetilde{\Upsilon}^{\cal I} \cdot \pi =
\pi \cdot \Upsilon^{\cal I}$, and an isomorphism
$\tilde{\lambda}_{\tilde{\alpha}}: \widetilde{M}_{\tilde{\alpha}}^\vee
\rightarrow \widetilde{M}_{\widetilde{\Upsilon}^{\cal I}(\tilde{\alpha})}$
of $\widetilde{\cal A}$ modules for each $\tilde{\alpha}
\in \widetilde{\cal I}$. When the 2D CFT is unitary, each
of the modules $M_\alpha$ and $\widetilde{M}_{\tilde{\alpha}}$ for
$\alpha \in {\cal I}$ and $\tilde{\alpha} \in \widetilde{\cal I}$
should have a positive definite Hermitian inner product, and the
isomorphisms $\lambda_\alpha$ and $\tilde{\lambda}_{\tilde{\alpha}}$
should be unitary with respect to the Hermitian inner products.
The unitary maps $\lambda_\alpha$ and $\tilde{\lambda}_{\tilde{\alpha}}$
are used to construct an anti-linear CPT map on ${\cal H}_{\rm tot}$,
although details of the CPT map do not concern us in this article. 
A {\it unitarity structure} of a 2D CFT refers to those $\{ \lambda_\alpha\}$,
$\{ \tilde{\lambda}_{\tilde{\alpha}} \}$ and the Hermitian inner
products.

There is another role of the isomorphisms $\{ \lambda_\alpha \}$ and
$\{ \tilde{\lambda}_{\tilde{\alpha}}\}$ other than the CPT maps; 
we say that a CFT vertex operator $Y_{\rm tot}$ of a given CFT is invariant
under the crossing symmetry, when
\begin{align}
  \lambda_\gamma \tilde{\lambda}_{\tilde{\gamma}} \cdot
  (\Theta_s \widetilde{\Theta}_{-1-s} Y_{\rm tot}^{(\alpha;\beta,\gamma)(\tilde{\alpha};\tilde{\beta},\tilde{\gamma})}) \cdot
  (\lambda_\alpha\tilde{\lambda}_{\tilde{\alpha}})^{-1}
  = Y_{\rm tot}^{(\Upsilon^{\cal I}(\gamma),\beta,\Upsilon^{\cal I}(\alpha))(\widetilde{\Upsilon}^{\cal I}(\tilde{\gamma});\tilde{\beta},\widetilde{\Upsilon}^{\cal I}(\tilde{\alpha}))}
  \label{eq:xing-sym-rltn-bos}
\end{align}
for all the possible types $(\alpha;\beta,\gamma)$ and
$(\tilde{\alpha};\tilde{\beta},\tilde{\gamma})$ available in the given CFT. 
It is necessary to insert the $\C$-linear maps $\lambda_\alpha$,
$\lambda_\gamma$, $\tilde{\lambda}_{\tilde{\alpha}}$ and
$\tilde{\lambda}_{\tilde{\gamma}}$ even to compare the objects on both sides,
before we ask if they agree. When both sides agree 
(if (\ref{eq:xing-sym-rltn-bos}) is satisfied for one $s \in \Z$, then 
it also is for all other $s \in \Z$ in a modular invariant CFT), 
we say that the given CFT
has crossing symmetry (or that $Y_{\rm tot}$ is invariant under the
crossing symmetry isomorphism $\Theta_{\rm tot} = \Theta_s \otimes
\widetilde{\Theta}_{-1-s}$).

In practice, it is not necessary to test (\ref{eq:xing-sym-rltn-bos})
for {\it all} of those different types. Once the
relation (\ref{eq:xing-sym-rltn-bos})
is confirmed for the type $(0;\beta,\Upsilon^{\cal I}(\beta))$
and $(\tilde{0};\tilde{\beta},\widetilde{\Upsilon}^{\cal I}(\tilde{\beta}))$
({\it pair-annihilation type}),
then the relation (\ref{eq:xing-sym-rltn-bos}) for all other types
follows. This shortcut
comes from the fact that the crossing symmetry isomorphism $\Theta_{s}$ on
a general intertwining operator ${\cal Y}_{\beta,\gamma}^\alpha \in
{\cal V}_{\beta,\gamma}^\alpha$ of a VOA ${\cal A}$ is implemented by
(\cite[(3.8b), (4.1d)]{Moore:1988qv}, \cite[(4.7)]{MR2387861}, cf
also \cite[2.3.12]{Watari:2026scft})
\begin{align}
  \Omega_{s} \cdot {\cal F}: {\cal V}_{\Upsilon^{\cal I}(\alpha),\alpha}^0 \otimes
   {\cal V}_{\beta,\gamma}^\alpha & \; \rightarrow
   {\cal V}_{\beta,\Upsilon^{\cal I}(\alpha)}^{\Upsilon^{\cal I}(\gamma)} \otimes
   {\cal V}_{\Upsilon^{\cal I}(\gamma),\gamma}^0  \label{eq:Theta-by-OmegaF} \\
  \left( \lambda_0 (\Theta_{s}\Omega Y_{\Upsilon^{\cal I}(\alpha)})
  \lambda_{\Upsilon^{\cal I}(\alpha)}^{-1} \right) \otimes {\cal Y}_{\beta,\gamma}^\alpha & \; \longmapsto
  \Theta_{-1-s} \left( \lambda_{\Upsilon^{\cal I}(\alpha)}^{-1}
        {\cal Y}_{\beta,\gamma}^\alpha
  \lambda_{\Upsilon^{\cal I}(\gamma)} \right) 
  \otimes
  \left( \lambda_0 (\Theta_{s}\Omega Y_{\Upsilon^{\cal I}(\gamma)})
  \lambda_{\Upsilon^{\cal I}(\gamma)}^{-1} \right).   \nonumber 
\end{align}
Once we manage to confirm that the CFT vertex operator $Y_{\rm tot}$
in the pair annihilation type (i.e., $\gamma = 0$, $\tilde{\gamma} = 0$,
$\alpha = \beta$ and $\tilde{\alpha} = \tilde{\beta}$), i.e.,
\begin{align}
  Y_{\rm tot}^{{\rm pair~ann.}} = \lambda_0\tilde{\lambda}_0 \cdot
  \left( (\Theta_s \Omega Y_\beta) \otimes (\widetilde{\Theta}_{-1-s}
  \widetilde{\Omega} \widetilde{Y}_{\tilde{\beta}}) \right) \cdot
  \lambda_{\beta}^{-1}\tilde{\lambda}_{\tilde{\beta}}^{-1}, 
\end{align}
then the invariance of $Y_{\rm tot}$ under the crossing symmetry isomorphism
$\Theta_{\rm tot}$ follows from the invariance of $Y_{\rm tot} \otimes Y_{\rm tot}$
under ${\cal F}_{\rm tot}$ and the invariance of $Y_{\rm tot}$ under
$\Omega_{\rm tot}$. ---(**3)

In a Type 0 CFT, there are small twists to the story reviewed above, when
we hope to deal with the unitarity structure and crossing symmetry invariance
while maintaining the representation theory of the {\it super}chiral algebras 
${\cal A}$ and $\widetilde{\cal A}$ manifest.
First, the superchiral algebra
${\cal A}$ contains states with conformal weights in $1/2+\Z$, so
the crossing symmetry isomorphism $\Theta$ on $YY_\alpha$ of a representation
$(M_\alpha, YY_\alpha)$ has to come with a parameter $\Theta_s$ with
$s \in \Z/2\Z$ ($s \in \Z$, but $\Theta_{s+2}=\Theta_s$ when conformal
weights in ${\cal A}$ are in $2^{-1}\Z$). Second, similarly to the fact
that the definition of a skewysmmetry isomorphism on intertwining
{\it super}operators $\Omega_r: {\cal VV}_{\beta,\gamma}^\alpha(i_\beta)
\rightarrow {\cal VV}_{\gamma,\beta}^\alpha(i_\gamma)$ involves a pair of
identifications $i_\beta$ and $i_\gamma$ of 2-element $\Z/2\Z$-sets, 
definition of a crossing symmetry isomorphism on the vector space of
intertwining {\it super}operators ${\cal VV}_{\beta,\gamma}^\alpha(i_\beta)$ also
depends on $i_\gamma$. The isomorphisms $\lambda_\alpha: M_\alpha^\vee \rightarrow
M_{\Upsilon^{\cal I}(\alpha)}$ of SVOA representations are therefore characterized by
\begin{align}
  \lambda_{\alpha; \epsilon_s,i_\alpha} \cdot (\Theta_s YY_\alpha)(v,z) \cdot
  \lambda_{\alpha; \epsilon_s,i_\alpha}^{-1} = YY_{\Upsilon^{\cal I}(\alpha)}(v,z) \qquad
         {}^\forall v \in {\cal A}
         \label{eq:equiv-rltn-4-cntrGrepr-SVOA}
\end{align}
instead; $\epsilon_s := 2s+1 \in \{ \pm 1\}$ for $s \in \{ 0,-1\}$ that 
represents $s \in \Z/2\Z$. The crossing symmetry invariance condition
on the SCFT vertex operator is still 
\begin{align}
  (\lambda_{\gamma;\epsilon_s,i_\gamma}
  \tilde{\lambda}_{\tilde{\gamma}; \epsilon_{-1-s},\tilde{\imath}_{\tilde{\gamma}}}) \cdot
  (\Theta_s \widetilde{\Theta}_{-1-s} YY_{\rm tot}^{(\alpha;\beta,\gamma)(\tilde{\alpha};\tilde{\beta},\tilde{\gamma})}) & \; \cdot
  (\lambda_{\alpha; \epsilon_s,i_\alpha}
  \tilde{\lambda}_{\tilde{\alpha}; \epsilon_{-1-s},\tilde{\imath}_{\tilde{\alpha}}})^{-1}
  \label{eq:xing-sym-inv-SVOA}   \\
 &\;  = YY_{\rm tot}^{(\Upsilon^{\cal I}(\gamma);\beta,\Upsilon^{\cal I}(\alpha))(\widetilde{\Upsilon}^{\cal I}(\tilde{\gamma});\tilde{\beta},\widetilde{\Upsilon}^{\cal I}(\tilde{\alpha}))}  \nonumber 
\end{align}
for $s \in \{ 0,-1\}$. 
Just like in the case of bosonic CFT, it is enough to test this condition
only for the pair annihilation type $(0;\beta,\Upsilon^{\cal I}(\beta))$, and
then the crossing-symmetry invariance of the CFT vertex operator $YY_{\rm tot}$
in all the other types follows from the invairnance under ${\cal F}_{\rm tot}$
and $\Omega_{\rm tot}$. Further details, if necessary, will be
found in \cite[\S4]{Watari:2026scft}.

\vspace{5mm}
Let us now go back to the context in this article,
as we have a minimal review on the unitarity structure and crossing
symmetry on a general CFT and SCFT already.
We have a unitarity structure for the minimal tensor
model SCFT, with $\{ \lambda_{(\lambda,\vec{m}); \epsilon_s, i_{(\lambda,\vec{m})}} \}$
% and Hermitian inner products for the representations of
for ${\cal A}^{\rm m.t.m.}$,
$\{ \tilde{\lambda}_{(\lambda,\vec{m}); \epsilon_{-1-s},\tilde{\imath}_{(\lambda,\vec{m})}} \}$
% and Hermitian inner products for the representations of
for $\widetilde{\cal A}^{\rm m.t.m.}$, and Hermitian inner products on those
representation spaces. Because the Hilbert space of a Gepner model
SCFT is built out of the same set of representation spaces of
${\cal A}^{\rm m.t.m.}$ and $\widetilde{\cal A}^{\rm m.t.m.}$, we may ask if
the unitarity structure of the original minimal tensor model (say, of
$s = 0$ or $s=-1$) is compatible with the representation of
${\cal A}^{\rm Gepn.}$ and $\widetilde{\cal A}^{\rm Gepn.}$, and if
the CFT vertex operator $YY_{\rm tot}^{\rm Gepn.}$ is invariant under the
crossing symmetry when we use the isomorphisms $\{ \lambda \}$ and
$\{ \tilde{\lambda}\}$ of either $s=0$ or $s=-1$ of the
${\cal A}^{\rm m.t.m.}$ and $\widetilde{\cal A}^{\rm m.t.m.}$ modules. 

For this to be realized, we choose to construct the SCFT vertex operator
$YY_{\rm tot}^{\rm Gepn.}$ by using the isomorphism $(*)_{\rho}|_{\rho =s}$,
to recycle the unitairity structure $\{\lambda_{(\lambda,\vec{m}); \epsilon_s}\}$
and $\{ \tilde{\lambda}_{(\lambda,\vec{m});\epsilon_{-1-s}} \}$ of
the minimal tensor model SCFT. Because the condition C2${}_2$ depends
on $\rho \in \{0,-1\}$, we should now read it for the version with
$\rho = s \in \{ 0,-1\}$ correlated with the choice of a unitarity structure.
\label{pg:set-rho=s}
On top of this, we impose the following conditions: 
\begin{itemize}
\item (C4) for $\vec{y} \in \Gamma$ (the expression on the RHS is independent
  of $s$),
\begin{align}
 YY^{(-\vec{y},\vec{y})} = \lambda_{(\vec{0},\vec{0});\epsilon_s} \cdot (\Theta_s \Omega YY^{\rm m.t.m.}_{(\vec{0},-2\vec{y})} ) \cdot \lambda_{(\vec{0},-2\vec{y});\epsilon_s}^{-1}, 
\end{align}
\item (D3) for any pair of $\vec{y}$ and $\vec{m}$ subject to the orbifold projection condition (\ref{eq:cond-Gepner-survive-Z-w-d.t.},
  \ref{eq:cond-Gepner-survive-Z-w-d.t.for-R}), the isomorphism $(*)_s$
  from ${\cal VV}_{(\vec{0},-2\vec{y}), \; (\vec{0},2\vec{y})}^{(\vec{0},\vec{0})}
  \otimes {\cal VV}_{(\vec{0},\vec{0}), \; (\vec{0},\vec{0})}^{(\vec{0},\vec{0})} \otimes
          {\cal VV}_{(\lambda,-\vec{m}), \; (\lambda,\vec{m})}^{(\vec{0},\vec{0})}$
  is given by  
\begin{align}
&  \left[
    \lambda_{(\vec{0},\vec{0});\epsilon_s}
    (\Theta_s\Omega YY_{(\vec{0},-2\vec{y})}^{\rm m.t.m.})
    \lambda_{(\vec{0},-2\vec{y});\epsilon_s}^{-1} \right] \otimes YY^{\rm m.t.m.} \otimes
  \left[ \lambda_{(\vec{0},\vec{0});\epsilon_s}
    (\Theta_s\Omega YY_{(\lambda,-\vec{m}}^{\rm m.t.m.})
    \lambda_{(\lambda,-\vec{m});\epsilon_s}^{-1} \right] \nonumber \\
&  \mapsto YY^{[-\vec{y},(\lambda,-\vec{m})]} \otimes YY^{[\vec{y},(\lambda,\vec{m})1]} \otimes
  \left( \lambda_{(\vec{0},\vec{0});\epsilon_s}
  \cdot (\Theta_s\Omega YY_{(\lambda,-\vec{m}-2\vec{y})}^{\rm m.t.m.}) \cdot
  \lambda_{(\lambda,-\vec{m}-2\vec{y});\epsilon_s}^{-1}\right).  
\end{align}
\end{itemize}
The rest is to verify that those conditions are sufficient for
the relation (\ref{eq:xing-sym-inv-SVOA}) in the pair annihilation type and 
the relation (\ref{eq:equiv-rltn-4-cntrGrepr-SVOA}) both hold
for $YY_{\rm tot}^{\rm Gepn.}$ along with the isomorphisms
$\{ \lambda_{(\lambda,\vec{m}); \epsilon_s} \}$
and $\{ \tilde{\lambda}_{(\lambda,\vec{m});\epsilon_{-1-s}} \}$ of
the minimal tensor model. 

The combination of the conditions (C4) and (D3) implies immediately
that $YY_{\rm tot}^{\rm Gepn.}$ is
\begin{align}
  \left( \lambda_{(\vec{0},\vec{0});\epsilon_s}
  \cdot (\Theta_s\Omega YY_{(\lambda,-\vec{m}-2\vec{y})}^{\rm m.t.m.}) \cdot
  \lambda_{(\lambda,-\vec{m}-2\vec{y});\epsilon_s}^{-1}\right) \otimes 
  \left( \tilde{\lambda}_{(\vec{0},\vec{0});\epsilon_s}
  \cdot (\widetilde{\Theta}_{-1-s}\widetilde{\Omega}
  \widetilde{YY}_{(\lambda,-\vec{m}}^{\rm m.t.m.}) \cdot
  \tilde{\lambda}_{(\lambda,-\vec{m});\epsilon_s}^{-1}\right)
  \label{eq:temp-GepnCFTvo-pairAnnType}
\end{align}
in the pair annihilation types
$(0;(\lambda, -\vec{m}-2\vec{y}), (\lambda, \vec{m}+2\vec{y})) \otimes
(0;(\lambda,-\vec{m}), (\lambda,\vec{m}))$; remember that
the basis element $YY^{[\vec{0},(\vec{0},\vec{0})]}$ should be $YY^{\rm m.t.m.}$
because of the condition C1${}_0$, and
$YY_{\rm tot}^{\rm m.t.m.}$ is
$(\lambda_{(\vec{0},\vec{0});\epsilon_s} \tilde{\lambda}_{\vec{0};\epsilon_{-1-s}}) \cdot
\left( (\Theta_s \Omega YY^{\rm m.t.m.}_{(\lambda,-\vec{m})}) \otimes
(\widetilde{\Theta}_{-1-s}\widetilde{\Omega}
\widetilde{YY}^{\rm m.t.m.}_{(\lambda,-\vec{m})}) \right) \cdot
(\lambda_{(\lambda,-\vec{m});\epsilon_s}
\tilde{\lambda}_{(\lambda,-\vec{m}); \epsilon_{-1-s}})^{-1}$ in the pair
annihilation type
$(0;(\lambda,-\vec{m}),(\lambda,\vec{m})) \otimes
(0; (\lambda,-\vec{m}),(\lambda, \vec{m}))$
because of the crossing-symmetry invariance of $YY_{\rm tot}^{\rm m.t.m.}$.
On the other hand, we know already that $YY_{\rm tot}^{\rm Gepn.}$ is
\begin{align*}
 \left\{  (\Omega YY^{\rm m.t.m.}_{(\lambda, -\vec{m}-2\vec{y})}) \otimes
 (\widetilde{\Omega} \widetilde{YY}^{\rm m.t.m.}_{(\lambda,-\vec{m})})
 \; | \; \vec{m} \in \vec{m}_0 + 2\Gamma_{\chi {\rm Alg}} ,
  \vec{y} \in \vec{y}_0 + \Gamma_{\chi {\rm Alg}} \right\}
\end{align*}
in the types $( (\lambda,[-\vec{m}-2\vec{y}]); (\lambda, [-\vec{m}-2\vec{y}]), 0) \otimes
((\lambda, [-\vec{m}]); (\lambda, [-\vec{m}]), \tilde{0})$,
because we have seen already that $YY_{\rm tot}^{\rm Gepn.}$ is invariant
under $\Omega_{\rm tot} = \Omega_r \otimes \widetilde{\Omega}_{-1-r}$. 
So, the crossing-symmetry condition is evaluated on both sides independently
for the pair annihilation type of ${\cal A}^{\rm Gepn.} \otimes
\widetilde{\cal A}^{\rm Gepn.}$, and the relation (\ref{eq:xing-sym-inv-SVOA})
has been confirmed in the pair annihilation type. 

\vspace{3mm}

Let us pause for a moment to make two remarks before moving further. 
First, the condition (D3) restricts the normalization of the basis
elements $YY^{[\vec{y},(\lambda,\vec{m})]}$ for projection-surviving pairs $\vec{y}$ and
$\vec{m}$ for the first time; apart from special cases of $\lambda$ such
as $\lambda = \vec{0}$ and $\vec{m} \in 2\Gamma_{\chi {\rm Alg}}$, the
conditions (D1) and (D2) allow simultaneous rescaling of
$YY^{[\vec{y},(\lambda,\vec{m})]}$ by a non-zero constant within each class of
$\vec{m} \in \vec{m}_0 + 2\Gamma_{\chi {\rm Alg}}$ and $\vec{y} \in
\vec{y}_0+\Gamma_{\chi {\rm Alg}}$. If we hope to
use the same isomorphisms $\{ \lambda_{(\lambda,\vec{m});\epsilon_s}\}$
for the unitary structure of the Gepner model SCFT, however, the dictionary between
$M_{\lambda,\vec{m}_0}$ and $M_{\lambda, \vec{m}_0 + 2\vec{y}}$ set by
$YY^{[\vec{y},(\lambda,\vec{m_0})]}$ has no chance to retain a free rescaling. So
it was obvious that the conditions (D1, D2) are not enough.\footnote{
In an intuitive language, the condition (D3) is like using a {\it
  unit-normalized} 
spectral flow state $v_{\eta}$ ($\eta  \in L^{\vec{y}}_{00}$) to set up
a dictionary, instead of a state $v_\eta$ with arbitrary normalization. 
} %

The other is on the condition (C4). Although this condition refers only
to the correspondence between $YY^{(-\vec{y},\vec{y})}$ and  $YY^{(-\vec{y},0)}$ with $\vec{y} \in \Gamma$
under the crossing symmetry isomorphism, it is possible to derive the
relation for all the other $YY^{(\vec{y}_1,\vec{y}_2)}$ with $\vec{y}_1,\vec{y}_2 \in \Gamma$. 
We use the fact (**3) reviewed earlier, with the representations
$\alpha\Rightarrow (\vec{0},2\vec{y}_3)$, $\beta \Rightarrow (\vec{0},2\vec{y}_1)$ and
$\gamma \Rightarrow(\vec{0},2\vec{y}_2)$.
The condition (C4) indicates that $YY^{(-\vec{y}_3,\vec{y}_3)}$ is already
of the form appropriate for the operator in
${\cal V}_{\Upsilon^{\cal I}(\alpha),\alpha}^0 \Rightarrow
{\cal VV}_{(\vec{0},-2\vec{y}_3), (\vec{0},2\vec{y}_3)}^{(\vec{0},\vec{0})}$ in
(\ref{eq:Theta-by-OmegaF}), and we apply (\ref{eq:Theta-by-OmegaF}) for
the case ${\cal Y}_{\beta,\gamma}^\alpha \Rightarrow YY^{(\vec{y}_1,\vec{y}_2)}$. 
The fusion isomorphism brings $YY^{(-\vec{y}_3,\vec{y}_3)} \otimes
YY^{(\vec{y}_1,\vec{y}_2)}$ to $YY^{(-\vec{y}_3,\vec{y}_1)} \otimes
YY^{(-\vec{y}_2,\vec{y}_2)}$ (cf. (C3${}_1$)). Here, $YY^{(-\vec{y}_2,\vec{y}_2)}$
is in the form appropriate for the operator in
${\cal V}_{\Upsilon^{\cal I}(\gamma),\gamma}^0\Rightarrow
{\cal VV}_{(\vec{0},-2\vec{y}_2), (\vec{0},2\vec{y}_2)}^{(\vec{0},\vec{0})}$
in (\ref{eq:Theta-by-OmegaF}) (cf (C4)).  
The remaining skewsymmetry isomorphism $\Omega_{s}$
on $YY^{(-\vec{y}_3,\vec{y}_1)}$ brings it to 
\begin{align}
  YY^{(\vec{y}_1,-\vec{y}_3)} \left( \mathbb{E}\left[ - \sum_i
    \frac{y^{(1)}_i y^{(3)}_i}{\bar{k}_i} \right] \epsilon^{00}(\vec{y}_1,\vec{y}_3) \right)^{-(2s+1)} \!\!\!\! = \Theta_{-1-s}(\lambda_{-2\vec{y}_3;\epsilon_s}^{-1} 
  YY^{(\vec{y}_1,\vec{y}_2)} \lambda_{-2\vec{y}_2;\epsilon_s} )
  \label{eq:Theta-rltn-4basisElmnts-a}
\end{align}
as we may use C2${}_2$, or equivalently,
\begin{align}
  \lambda_{-2\vec{y}_3;\epsilon_s} (\Theta_s YY^{(\vec{y}_1,-\vec{y}_3)}) \lambda_{-2\vec{y}_2;\epsilon_s}^{-1}
  \left( \mathbb{E}\left[ - \sum_i \frac{y^{(1)}_iy^{(3)}_i}{\bar{k}_i} \right]
  \epsilon^{00}(\vec{y}_1,\vec{y}_3) \right)^{-(2s+1)}
   = YY^{(\vec{y}_1,\vec{y}_2)}. 
\end{align}

\vspace{3mm}

Let us go back to the main story, and move on to verify that the unitary
structure $\{ \lambda_{(\lambda,\vec{m});\epsilon_s} \}$ and
$\{ \tilde{\lambda}_{(\lambda,\vec{m});\epsilon_{-1-s}} \}$
of the minimal tensor model can also be used for the Gepner model Type 0
SCFT, by using the fact (\ref{eq:temp-GepnCFTvo-pairAnnType}) that has
been established. 
The proof for this is, in fact, a special case
of showing the crossing symmetry invariance of $YY_{\rm tot}^{\rm Gepn.}$
for all possible types; the special case is a little easier technically,
however, and a proof for the special case is enough for all the general
types because of the fact (**3) reviewed earlier. 

\begin{figure}[tbp]
\begin{center}
   \includegraphics[scale=0.7]{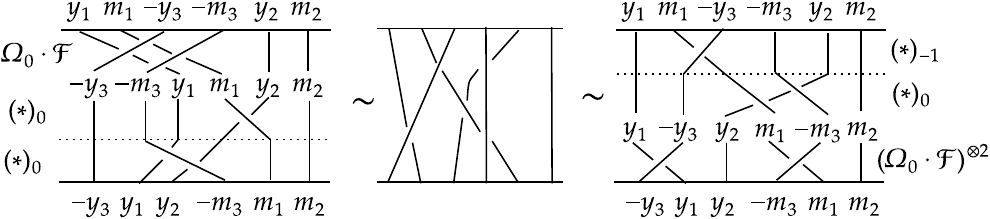}  
\caption{\label{fig:cmm-dgrm-4-ThetaInv}The graphical presentation of the
isomorphisms along the two routes in the diagram (\ref{eq:diagr-ThetaInv}),
starting from the [UL] corner and ending at the [DR] corner.
The graph on the left is for the route down the left column in the
diagram (\ref{eq:diagr-ThetaInv}), while the one on the right for the
other route. The diagram commutes because the braiding group elements
of the two graphs are the same. 
}
\end{center}
\end{figure}

To do this, we use the following diagram 
\begin{equation}
  \xymatrix{
    {}_{[UL]} \; ((-\vec{y}_3 \cdot (\vec{y}_1 \cdot \vec{y}_2)) \cdot
    (-\vec{m}_3 \cdot (\vec{m}_1 \cdot \vec{m}_2)))
    \ar[r]^{\Omega_s{\cal F}\otimes \Omega_s {\cal F}} \ar[d]^{(*)_s} & 
   (((\vec{y}_1 \cdot -\vec{y}_3) \cdot \vec{y}_2) \cdot
    ((\vec{m}_1 \cdot -\vec{m}_3) \cdot \vec{m}_2)) \ar[d]^{(*)_s} \; {}_{[UR]}
    \\
    {}_{[ML]} \; ((-\vec{y}_3 \cdot -\vec{m}_3) \cdot ((\vec{y}_1 \cdot \vec{y}_2) \cdot
    (\vec{m}_1 \cdot \vec{m}_2))) \ar[d]^{(*)_s} &
    ( ((\vec{y}_1 \cdot -\vec{y}_3) \cdot (\vec{m}_1 \cdot -\vec{m}_3)) \cdot
    (\vec{y}_2 \cdot \vec{m}_2)) \ar[d]^{(*)_{-1-s}} \; {}_{[MR]} \\
    {}_{[DL]} \; ((-\vec{y}_3 \cdot -\vec{m}_3) \cdot ((\vec{y}_1 \cdot \vec{m}_1) \cdot
    (\vec{y}_2 \cdot \vec{m}_2))) \ar[r]^{\Omega_s {\cal F}} &
    ( ((\vec{y}_1 \cdot \vec{m}_1) \cdot (-\vec{y}_3 \cdot -\vec{m}_3)) \cdot
    (\vec{y}_2 \cdot \vec{m}_2)) \; {}_{[DR]}
  };
  \label{eq:diagr-ThetaInv}
\end{equation}
it commutes, as one can see from Fig. \ref{fig:cmm-dgrm-4-ThetaInv}.
%%%%%%%%%%%%%%%%%%%%%%%%%%%%%%%%%%%%%%%%%
%%%%%%%%%%%%%%%%%%%%%%%%%%%%%%%%%%%%%%%%%
%%%%%%%%%%%%%%%%%%%%%%%%%%%%%%%%%%%%%%%%%%%%%%%%%%%%%%
$\vec{y}_a$ and $\vec{m}_a$ for $a=1,2,3$ are the symbols for the
${\cal A}^{\rm m.t.m.}$-representations $(\vec{0},2\vec{y}_a)$ and
$(\lambda_a,\vec{m}_a)$, repsectively, and each term stands
for the tensor product of the vector spaces of intertwining superoperators of
${\cal A}^{\rm m.t.m.}$ where the parentheses indicate the order
of fusions, as we also did in the diagrams
(\ref{eq:diagr-Finv}, \ref{eq:diagr-OmegInv}) to save space.  
Each pair $\vec{y}_a$ and $\vec{m}_a$ satisfies the the orbifold projection
condition (\ref{eq:cond-Gepner-survive-Z-w-d.t.},
\ref{eq:cond-Gepner-survive-Z-w-d.t.for-R}).
We are interested in the case $\vec{y}_3= \vec{y}_1+\vec{y}_2$,
$\vec{m}_3 = \vec{m}_1+\vec{m}_2$ this time, whereas $\vec{y}_3$ and
$\vec{m}_3$ were
independent from $\vec{y}_{1,2}$ and $\vec{m}_{1,2}$ in the diagram
(\ref{eq:diagr-Finv}). When we focus on the proof in the special case as
explained already, we will use $\vec{y}_1 \in \Gamma_{\chi {\rm Alg}} \subset
\Gamma$
and $(\lambda_1,\vec{m}_1) = (\vec{0},\vec{0})$ (so $(\lambda_3,\vec{m}_3)=
(\lambda_2,\vec{m}_2)$). 

The diagram is used to track the product of intertwining superoperator
\begin{align}
  & \left( YY^{(-\vec{y}_3,\vec{y}_{1+2})} \otimes
  YY^{(\vec{y}_1,\vec{y}_2)} \right) \otimes
       \label{eq:temp-prdOp-2track4-xingSymInv} \\
  &\qquad \qquad\qquad
  \left( ( \lambda_{(\vec{0},\vec{0}); \epsilon_s}
  (\Theta_s \Omega YY^{\rm m.t.m.}_{(\lambda_3,-\vec{m}_3)})
  \lambda_{(\lambda_3,-\vec{m}_3); \epsilon_s}^{-1} ) 
  \otimes {\cal YY}_{(\lambda_1,\vec{m}_1), (\lambda_2,\vec{m}_2)}^{(\lambda_3,\vec{m}_{1+2})} \right) \otimes Y^{\rm m.t.m.} \nonumber 
\end{align}
at the upper-left corner [UL] of the diagram in two different routes
ending at the down-right corner [DR]. We will keep an intertwining
superoperator ${\cal YY} \in
{\cal VV}_{(\lambda_1,\vec{m}_1), (\lambda_2,\vec{m}_2)}^{(\lambda_3,\vec{m}_{1+2})}$
to be general for the moment, and will consider a special choice for it later. 
The sequence of isomorphisms $(*)_s \cdot (*)_s$ maps this
product of operators to 
\begin{align}
  YY^{[-\vec{y}_3,(\lambda_3,-\vec{m}_3)]}
  \otimes
  \left( ({\cal YY}^{\rm Gepn.})_{(\lambda_1,\vec{m}_1+2\vec{y}_1),(\lambda_2,\vec{m}_2+2\vec{y}_2)}^{(\lambda_3,\vec{m}_3+2\vec{y}_3)} \otimes YY^{[\vec{y}_1,(\lambda_1,\vec{m}_1)]} \otimes YY^{[\vec{y}_2,(\lambda_2,\vec{m}_2)]} \right) & \label{eq:temp-prdOp-2track4-xingSymInv-atDL} \\
  \otimes \left( \lambda_{(\vec{0},\vec{0});\epsilon_s}
  (\Theta_s \Omega YY_{(\lambda_3,-\vec{m}_3-2\vec{y}_3)}^{\rm m.t.m.})
  \lambda_{(\lambda_3,-\vec{m}_3-2\vec{y}_3); \epsilon_s}^{-1} \right) & \; \in [DL];
  \nonumber 
\end{align}
the last factor of (\ref{eq:temp-prdOp-2track4-xingSymInv-atDL})
in ${\cal VV}^{(\vec{0},\vec{0})}_{(\lambda_3,-\vec{m}_3-2\vec{y}_3),
  (\lambda_3,\vec{m}_3+2\vec{y}_3)}$
arises already in [ML] after the first $(*)_s$ isomorphism,
as we have already seen at (\ref{eq:temp-GepnCFTvo-pairAnnType}).
The isomorphism $\Omega_s \cdot {\cal F}$ from [DL] to [DR] can be
evaluated by using (**3), because the last factor of
(\ref{eq:temp-prdOp-2track4-xingSymInv-atDL}) is in the appropriate form. 
The operator (\ref{eq:temp-prdOp-2track4-xingSymInv}) is therefore
mapped into 
\begin{align}
  &  YY^{[-\vec{y}_3,(\lambda_3,-\vec{m}_3)]} \otimes YY^{[\vec{y}_1,(\lambda_1,\vec{m}_1)]} \otimes YY^{[\vec{y}_2,(\lambda_2,\vec{m}_2)]} 
  \label{eq:temp-prdOp-2track4-xingSymInv-atDR} \\
  &  \qquad \qquad \otimes \left( \lambda_{(\vec{0},\vec{0});\epsilon_s}
  (\Theta_s \Omega YY_{(\lambda_2,-\vec{m}_2-2\vec{y}_2)}^{\rm m.t.m.})
  \lambda_{(\lambda_2,-\vec{m}_2-2\vec{y}_2);\epsilon_s }^{-1} \right)  \nonumber \\
  &  \qquad \qquad \otimes
  \Theta_{-1-s}\left( \lambda_{(\lambda_3,-\vec{m}_3-2\vec{y}_3);\epsilon_s}^{-1} 
  ({\cal YY}^{\rm Gepn.})_{\vec{m}_1+2\vec{y}_1, \vec{m}_2+2\vec{y}_2}^{\vec{m}_3+2\vec{y}_3}
  \lambda_{(\lambda_2,-\vec{m}_2-2\vec{y}_2); \epsilon_s} \right) \in [DR]. \nonumber
\end{align}
In this route through [DL], intertwining superoperators of the Gepner model
CFT are constructed, and then the crossing-symmetry isomorphism is applied.
So, we may compute what it is by tracking the same object
(\ref{eq:temp-prdOp-2track4-xingSymInv}) in the other route of the diagram
(\ref{eq:diagr-ThetaInv}). 

In the isomorphism from [UL] to [UR], 
one combination $(\Omega_s \cdot {\cal F})$ acts on the 3rd and 4th factors
of (\ref{eq:temp-prdOp-2track4-xingSymInv}), and the other combination
$(\Omega_s \cdot {\cal F})$ on the first two factors of
(\ref{eq:temp-prdOp-2track4-xingSymInv}); to evaluate them, we can use
(**3) (cf. the condition (C4) on $YY^{(-\vec{y}_3,\vec{y}_3)}$): 
\begin{align}
  &  \left( \Theta_{-1-s}(\lambda_{(\vec{0},-2\vec{y}_3);\epsilon_s}^{-1}
  YY^{(\vec{y}_1,\vec{y}_2)} \lambda_{(\vec{0},-2\vec{y}_2);\epsilon_s}) \otimes
  YY^{(-\vec{y}_2,\vec{y}_2)} \right) \otimes YY^{\rm m.t.m.}
  \label{eq:temp-prdOp-2track4-xingSymInv-UR-gen} \\
& \qquad \otimes \Theta_{-1-s} \left(\lambda_{(\lambda_3,-\vec{m}_3);\epsilon_s}^{-1}
 {\cal YY} \lambda^{-1}_{(\lambda_2,-\vec{m}_2);\epsilon_s} \right) \otimes
(\lambda_{(\vec{0},\vec{0});\epsilon_s}
(\Theta_s\Omega YY_{(\lambda_2,-\vec{m}_2)})
\lambda_{(\lambda_2,-\vec{m}_2);\epsilon_s}^{-1}) \in [UR] \nonumber 
\end{align}
The first factor of (\ref{eq:temp-prdOp-2track4-xingSymInv-UR-gen})
can be rewritten by using the consequence (\ref{eq:Theta-rltn-4basisElmnts-a})
of (C4, C3${}_1$, C2${}_2$). 
From here on, we focus on a special case that we announced earlier;
as we set $\vec{y}_1 \in \Gamma_{\chi {\rm Alg}}$ and
$(\lambda_1,\vec{m}_1)= (\vec{0},\vec{0})$, the phase factor in
(\ref{eq:Theta-rltn-4basisElmnts-a}) is trivial because of
the $\vec{y}_1$--$\vec{m}_1$ relation (\ref{eq:cond-Gepner-survive-Z-w-d.t.})
then. As we also
focus on the special choice ${\cal YY}^{\rm m.t.m.} \Rightarrow
YY^{\rm m.t.m.}_{(\lambda_2,\vec{m}_2)}$ (the representation of
${\cal A}^{\rm m.t.m.}$) in
${\cal VV}_{(\lambda_1,\vec{m}_1), (\lambda_2,\vec{m}_2)}^{(\lambda_3,\vec{m}_3)}
\Rightarrow {\cal VV}_{(\vec{0},\vec{0}), (\lambda_2,\vec{m}_2)}^{(\lambda_2,\vec{m}_2)}$,
we can use the defining property (\ref{eq:equiv-rltn-4-cntrGrepr-SVOA})
of $\lambda_{(\lambda_2,-\vec{m}_2);\epsilon_s}$. So, for the special case of
interst, the operator (\ref{eq:temp-prdOp-2track4-xingSymInv-UR-gen}) is
\begin{align}
&  YY^{(\vec{y}_1,-\vec{y}_3)}\otimes YY^{(-\vec{y}_2,\vec{y}_2)}  \otimes YY^{\rm m.t.m.}
  \label{eq:temp-prdOp-2track4-xingSymInv-UR-spc} \\
  & \qquad YY^{\rm m.t.m.}_{(\lambda_2,-\vec{m}_2)} \otimes 
  (\lambda_{(\vec{0},\vec{0});\epsilon_s}
(\Theta_s\Omega YY_{(\lambda_2,-\vec{m}_2)})
\lambda_{(\lambda_2,-\vec{m}_2);\epsilon_s}^{-1}) \in [UR]. \nonumber 
\end{align}

The sequence isomorphisms $(*)_{-1-s} \cdot (*)_{s}$ from [UR] to [DR]
maps this (\ref{eq:temp-prdOp-2track4-xingSymInv-UR-spc}) to 
\begin{align}
  \left[ \lambda_{(\vec{0},\vec{0});\epsilon_s}
    (\Theta_s\Omega YY_{(\lambda_2,-\vec{m}_2-2\vec{y}_2)})
    \lambda_{(\lambda_2,-\vec{m}_2-2\vec{y}_2);\epsilon_s}^{-1} \right] \otimes
  YY^{[-\vec{y}_2,(\lambda_2,-\vec{m}_2)]} \otimes YY^{[\vec{y}_2,(\lambda_2,\vec{m}_2)]} &   \\
  \otimes YY^{(\vec{y}_1,-\vec{y}_3)} \otimes  YY^{\rm m.t.m.}_{(\lambda_2,-\vec{m}_2)} & \in [MR]
  \nonumber 
\end{align}
because of (C4, D3) and (\ref{eq:temp-GepnCFTvo-pairAnnType}), and 
then to
\begin{align}
 & \left[ \lambda_{(\vec{0},\vec{0});\epsilon_s}
    (\Theta_s\Omega YY_{(\lambda_2,-\vec{m}_2-2\vec{y}_2)})
    \lambda_{(\lambda_2,-\vec{m}_2-2\vec{y}_2);\epsilon_s}^{-1} \right]
  \otimes YY^{[\vec{y}_2,(\lambda_2,\vec{m}_2)]} \otimes
  \label{eq:temp-prdOp-2track4-xingSymInv-DR-spc2} \\
  & \qquad
  YY^{[\vec{y}_1,(\vec{0},\vec{0})]} \otimes YY^{[-\vec{y}_3,(\lambda_2,-\vec{m}_2)]} \otimes
  (YY^{\rm Gepn.})_{(\vec{0},\vec{0}+2\vec{y}_1), (\lambda_2,-\vec{m}_2-2\vec{y}_3)}^{(\lambda_2,-\vec{m}_2-2\vec{y}_2)} \in [DR], \nonumber
\end{align}
where the last factor is the representation $(\lambda_2,[-\vec{m}_2])$
of ${\cal A}^{\rm Gepn.}$ on $M_{\vec{0},2\vec{y}_1} \subset {\cal A}^{\rm Gepn.}$. 
The difference between $(*)_{-1-s}$ and $(*)_s$ from [MR] to [DR]
is irrelevant in the special case treated here (as only integral powers
of dummy variables $z$ are involved in representations rather than
intertwining operators). 

Now, the relation (\ref{eq:equiv-rltn-4-cntrGrepr-SVOA}) is established
for the representation $(\lambda_2,[\vec{m}_2])$ of ${\cal A}^{\rm Gepn.}$
while using the data $\{ \lambda_{(\lambda,\vec{m});\epsilon_s} \}$
of the unitarity structure of the minimal tensor model, by comparing
(\ref{eq:temp-prdOp-2track4-xingSymInv-atDR})
and (\ref{eq:temp-prdOp-2track4-xingSymInv-DR-spc2}): 
\begin{align}
 &  \Theta_{-1-s}\left( \lambda_{(\lambda_2,-\vec{m}_2-2\vec{y}_3);\epsilon_s}^{-1}
  (YY^{\rm Gepn.})_{(\vec{0},2\vec{y}_1), (\lambda_2,\vec{m}_2+2\vec{y}_2)}^{(\lambda_2,\vec{m}_2+2\vec{y}_3)}
  \lambda_{(\lambda_2,-\vec{m}_2-2\vec{y}_2);\epsilon_s} \right) \\
  & \qquad \qquad \qquad \qquad \qquad \qquad \qquad \qquad \qquad 
  =
  (YY^{\rm Gepn.})_{(\vec{0},2\vec{y}_1), (\lambda_2,-\vec{m}_2-2\vec{y}_3)}^{(\lambda_2,-\vec{m}_2-2\vec{y}_2)}.  \nonumber 
\end{align}
It is possible to push the analysis using the diagram (\ref{eq:diagr-ThetaInv})
further for the general case, by retaining the phase factor at [UR],
and paying attention to the difference between $(*)_{-1-s}$ and $(*)_s$
in the isomorphism from [MR] to [DR]. We expect that those two effects
cancel against each other, just like we saw in the analysis for
the invariance of $YY_{\rm tot}^{\rm Gepn.}$ under $\Omega_{\rm tot}$, and
one will be able to establish the invariance of $YY_{\rm tot}^{\rm Gepn.}$
under $\Theta_{\rm tot}$ through this direct computation. We do not do that
in this article, however, because the $\Theta_{\rm tot}$-invariance in all types
follow from the $\Theta_{\rm tot}$-invariance in the pair annihilation
type, along with the confirmation that the unitary maps $\{ \lambda \}$
and $\{ \tilde{\lambda} \}$ do set up the equivalence between the
contragradient representations and the conjugate representations. 

\vspace{5mm}

To summarize, suppose that we are given
the SCFT vertex operator $YY_{\rm tot}^{\rm m.t.m.}$
and unitary maps $\{ \lambda_{(\lambda,\vec{m}); \epsilon_s} \}$ and
$\{ \tilde{\lambda}_{(\lambda,\vec{m});\epsilon_s} \}$ for a unitarity structore
of the minimal tensor model,
where $s$ is one of $\{ 0, -1\}$, and $\epsilon_s := 2s+1$. Suppose
that those data satisfy $\Omega_{\rm tot}$-invariance,
${\cal F}_{\rm tot}$-invariance and $\Theta_{\rm tot}$-invariance. Then
one can use the same set of unitary maps to give a unitarity structure
on the Hilbert space of a Gepner model SCFT for $(\Gamma,\epsilon,p_0)$,
and also construct the SCFT vertex operator $YY_{\rm tot}^{\rm Gepn.}$
through the isomorphism $(*)_\rho|_{\rho =s}$
in (\ref{eq:isom-4-mtm-Gepn-optr-cnstr}) so that
the costructed $YY_{\rm tot}^{\rm Gepn.}$ and the unitarity structure is
invariant under $\Omega_{\rm tot}$, ${\cal F}_{\rm tot}$ and $\Theta_{\rm tot}$. 
The condition on the choice of basis elements
$\{ YY^{(\vec{y},\vec{y}')} \; | \; \vec{y},\vec{y}' \in \Gamma\}$ and
$\{ YY^{[\vec{y},(\lambda,\vec{m})]} \; | \; (\ref{eq:cond-Gepner-survive-Z-w-d.t.},
\ref{eq:cond-Gepner-survive-Z-w-d.t.for-R}) \}$ are
(C1${}_1$, C2${}_2$, C3${}_2$, C4) and (D1--D3). The story itself
is precisely the expected one, and the conditions on the choice of
basis elements are also very reasonable; the primary task
of this section \ref{ssec:CFTop-by-sc-ext} has been to stamp out
phase ambiguities, to bring branch cuts under control and to
lay down the logic and algorithm for the construction of SCFT vertex
operators in terms of the simple current extension theory.  

\newpage 

%%%%%%%%%%%%%%%%%%%%%%%%%%%%%%%%%%%%%%%%%%%%%%%%%%
\subsection{Via the Lattice--Parafermion Decomposition}
\label{ssec:CFTop-by-bos-lattice}
%%%%%%%%%%%%%%%%%%%%%%%%%%%%%%%%%%%%%%%%%%%%%%%%%

For a Gepner Type 0 SCFT, it is often said that the Gepner model as an
orbifold theory of the minimal tensor model can be constructed (including
its SCFT vertex operators) by spectral flow of U(1) currents.
The coset VOA (\ref{eq:coset-VOA})---the $\Z/2\Z$-even part ${\cal A}_i^+$
of an $N=2$ minimal models---contains the U(1)${}^{N=2}$ current of the
$N=2$ SCA. So, a possible strategy is to try to implement the orbifold
projection and construction of twisted sectors in terms of the lattice CFT
of those U(1)${}^{N=2}$ currents from the left-mover $N=2$ minimal models and
$\widetilde{\U(1)}^{N=2}$ currents from the right-mover $N=2$ minimal models. 
In this approach, the action of neither the superconformal algebra nor the
minimal tensor model SVOA is manifest, while the advantage is that only
the lattice CFT is involved (hopefully) in the orbifold process. 

\vspace{5mm}

{\bf Preparation: U(1)-parafermion decomposition of the coset VOA:}
We begin with the decomposition of the coset VOA (\ref{eq:coset-VOA})
with $c = 3k_i/\bar{k}_i$ into the chiral boson VOA 
$\U(1)^{N=2}$ with the central charge $1$, and the $\Z_{k_i}$ parafermion
VOA with the central charge $3k_i/\bar{k}_i -1$.
Representations ${\cal L}^\ell_{m,s}$ of the coset
VOA (\ref{eq:coset-VOA}) also need to be decomposed into representations of
the $\U(1)^{N=2}$ and the parafermion VOAs. 
It is then possible to modify only the U(1) part of the Hilbert space 
$({\cal H}_{\rm tot}^{\rm post})_{\rm m.t.m.}$ in the hope of obtaining
$({\cal H}_{\rm top}^{\rm post})_{\rm Gepn.}$. 

The first step is to work on the Affine $\hat{\mathfrak{su}}(2)_k$ VOA
used in the coset VOA (\ref{eq:coset-VOA}).
The spin-$\ell/2$ representation of the $\hat{\mathfrak{su}}(2)_k$ VOA
decomposes as \cite{Kac:1984mq}, \cite{Fateev:1985mm}, \cite{Gepner:1986hr}
\begin{align}
  M^{\SU(2)_k}_{\ell} \cong \oplus_{sh = -\ell/2}^{-\ell + k-1} \left(
  W^{\U(1)_k}_{2sh} \otimes  P^k_{sh,\ell/2} \right),
  \label{eq:dcmp-SU(2)k-2-U1+paraF}
\end{align}
when the VOA is restricted to its vertex operator subalgebra
$\U(1)_k \otimes ({\rm parafermion}_k)$. Here, 
$({\rm parafermion}_k)$ stands for a VOA called the $\Z_k$ parafermion
theory. Among the $2k$ irreducible representations $W^{\U(1)_k}_{2sh}$ of the
$\U(1)_k$ VOA (lattice $\langle +2k \rangle$ VOA) labeled by
$2sh \in \Z/2k\Z$, only those
with $2sh \equiv \ell$ mod 2 show up in the decomposition.
The list of irreducible representations of the $\Z_k$ parafermion VOA
consists of $P^k_{sh,sp}$ with $2sp \in \{ 0,1,\cdots, k\}$, $2sh \in \Z/2k\Z$,
subject to the identification  
\begin{align}
P^k_{sh, sp} = P^k_{sh', sp'} \qquad {\rm when} \quad 
2sp' = k - 2 sp, {\rm ~and~} 2sh' = 2 sh + k \in \Z/2k\Z.
\label{eq:field-id-para}
\end{align}

Secondly, we want to plug in the decomposition (\ref{eq:dcmp-SU(2)k-2-U1+paraF})
into the LHS of (\ref{eq:dcmp-N=2-minimalM-SVOA-repr-2-bosCst-reprs}), 
and reorganize the irreducible representation
$W^{\U(1)_k}_{2sh} \otimes W^{\U(1)_2}_{-s}$ of the VOA $\U(1)_k \otimes \U(1)_2$
into the representations of $\U(1)^{N=2} \otimes \U(1)_{\bar{k}}$ on the RHS
of (\ref{eq:dcmp-N=2-minimalM-SVOA-repr-2-bosCst-reprs}). This allows us
to find out the decomposition of the representation ${\cal L}^\ell_{m,s}$
of the coset VOA $(\hat{\mathfrak{su}}(2)_k \times \hat{\mathfrak{so}}(2)_1)
/\U(1)_{\bar{k}}$ into the irreducible representations of the vertex operator
subalgebra $\U(1)^{N=2} \otimes ({\rm parafermion}_k)$. 
% on the RHS of (\ref{eq:dcmp-N=2-minimalM-SVOA-repr-2-bosCst-reprs}), which
To do so, however, we need to understand the spectrum of the
VOA $\U(1)^{N=2}$ better. 

The $\U(1)_{\bar{k}}$ symmetry of the coset construction (\ref{eq:coset-VOA})
is embedded into $\U(1)_k \times \U(1)_2$ through $-J^m = J^{2sh} - J^s$;
here, $J^s$ and $J^{2sh} = 2J^3$ are the U(1) currents of $\U(1)_2 =
\hat{\mathfrak{so}}(2)_1$ and the Cartan $\U(1)_k$ of
$\hat{\mathfrak{su}}(2)_k$, respectively, normalized
so that their charges fill $\Z$ but not elsewhere. 
The current $J^{N=2}$ of $\U(1)^{N=2}$ in the $N=2$ SCA has to be orthogonal
to $J^m$, so it is proportional to $(2 J^{2sh} + k J^s)$; the U(1)
charge (\ref{eq:h&q-N=2mm-each-VOA-repr}) is $1/(-2\bar{k})$ times the
eigenvalue of $(2J^{2sh} + kJ^s) = (\bar{k}J^s - 2 J^m)$. 

When $k$ is odd, all the possible irreducible representations of
$\U(1)_k \times \U(1)_2$ labeled by $(2sh, s) \in \Z/2k\Z \times \Z/4\Z$
give rise to the charges $q^\perp$ of
$J^\perp := -(2J^{2sh} + kJ^s)$ filling $\Z$;  
\begin{align}
  \left( \begin{array}{c} q_\perp \\ - m \end{array} \right) =
  \left( \begin{array}{cc} -2 & k \\ 1 & 1  \end{array}
  \right) \left( \begin{array}{c} 2sh \\ -s \end{array} \right), \quad 
  \left( \begin{array}{c} 2sh/(2k) \\ -s/4 \end{array} \right)
  = \left( \begin{array}{cc} -2 & 1 \\ k & 1 \end{array} \right)
  \left( \begin{array}{c} q_\perp/(4k\bar{k}) \\ -m/2\bar{k}
  \end{array} \right).    
\end{align}
The VOA for $J^\perp$
is of level $2k\bar{k}$ (i.e., $\U(1)_{2k\bar{k}}$); this is because $J^s$ is
of level $2$ and $J^{2sh}$ of level $k$ (so,
$k^2 \cdot 2 + 2^2 \cdot k = 2k\bar{k}$). A change in the charge $q_\perp$ by
$\Delta q_\perp = 4k\bar{k}$ results in $\Delta 2sh$ and $\Delta s$ divisible
by $2k$ and $4$, respectively (i.e., in the charge lattice of the VOA
$\U(1)_{k}\times \U(1)_2$), so the VOA
$\U(1)_{2k\bar{k}} \times \U(1)_{\bar{k}}$ by
$J^\perp$ and $J^m$ is indeed a vertex operator subalgebra of
$\U(1)_{k} \times \U(1)_2$; the charge lattice of $\U(1)_{2k\bar{k}} \times
\U(1)_{\bar{k}}$ is an index-$\bar{k}$ sublattice of that of
$\U(1)_{k} \times \U(1)_2$; an irreducible representation of
the VOA $\U(1)_k\times \U(1)_2$ therefore splits into $\bar{k}$ irreducible
representations of $\U(1)_{2k\bar{k}} \times \U(1)_{\bar{k}}$. Here is the
formula:
\begin{align}
  W^{\U(1)_k}_{2sh} \otimes W^{\U(1)_2}_{-s} \cong \oplus_{t \in}^{\Z/\bar{k}\Z} \left( 
  W^{\U(1)_{2k\bar{k}}}_{-(4sh + s k) + 4k t} \otimes W^{\U(1)_{\bar{k}}}_{2sh - s + 4t}
  \right), 
  \label{eq:dcmp-U1U1-to-U1U1-kOdd}
\end{align}
because the integral points of the charge lattice of the VOA
$\U(1)_k\times \U(1)_2$ are regarded as rational points of that of the
VOA $\U(1)_{2k\bar{k}} \times \U(1)_{\bar{k}}$ as in
\begin{align*}
  \left( \begin{array}{c} q_\perp/(4k\bar{k}) \\ -m/2\bar{k}
  \end{array} \right) = \left( \begin{array}{cc} -1/\bar{k} & 1/\bar{k} \\
    k/\bar{k} & 2/\bar{k} \end{array} \right)
  \left( \begin{array}{c} 2sh/(2k) \\ -s/4 \end{array} \right)
  = \left( \begin{array}{cc} & 1/\bar{k} \\ 1 & 2/\bar{k} \end{array} \right)
  \left( \begin{array}{c}2sh/(2k) \\ -2sh/(2k)-s/4 \end{array} \right). 
\end{align*}
The RHS of (\ref{eq:dcmp-U1U1-to-U1U1-kOdd}) as a whole remains the same
when we change $2sh$ by $+2k$ [resp. $s$ by $+4$] and replace the summation
variable $t$ by $t+1$ [resp. $t+1$]. The following form is also equivalent
to (\ref{eq:dcmp-U1U1-to-U1U1-kOdd}) because
$4k(k+3)t \equiv 4kt$ in $\Z/4k\bar{k}\Z$,
and we will use this later: 
\begin{align}
  W^{\U(1)_k}_{2sh} \otimes W^{\U(1)_2}_{-s} \cong \oplus_{t \in}^{\Z/\bar{k}\Z} \left( 
  W^{\U(1)_{2k\bar{k}}}_{-(4sh + s k) + 4k(k+3)t} \otimes W^{\U(1)_{\bar{k}}}_{2sh - s + 4t}
  \right).   \label{eq:dcmp-U1U1-to-U1U1-kOdd-2}
\end{align}

When $k$ is even, we set $J^\perp$ to be $-(2J^{2sh} + k J^s)/2$, so that
there are states with $J^\perp = \pm 1$ (or any integers) in some
representations of $\U(1)_{k} \times \U(1)_2$.
\begin{align}
  \left( \begin{array}{c} q^\perp \\ -m \end{array} \right) =
  \left( \begin{array}{cc} - 1 & k/2 \\ 1 & 1  \end{array}
  \right) \left( \begin{array}{c} 2sh \\ -s \end{array} \right),
  \;\;
  \left( \begin{array}{c} 2sh/(2k) \\ -s/4 \end{array} \right) = 
  \left( \begin{array}{cc} -1 & 1 \\ k/2 & 1 \end{array}
  \right) \left( \begin{array}{c} q_\perp/(k\bar{k}) \\ - m/(2\bar{k})
  \end{array} \right). 
\end{align}
Now, the VOA for $J^\perp$ is $\U(1)_{k\bar{k}/2}$; the charge lattice
of the VOA $\U(1)_{k\bar{k}/2} \times \U(1)_{\bar{k}}$ fits within that
of the VOA $\U(1)_k \times \U(1)_2$ with index $\bar{k}/2$. 
An irreducible representation of the latter VOA splits into those of the
former (vertex operator subalgebra) as follows: 
\begin{align}
  W^{\U(1)_k}_{2sh} \otimes W^{\U(1)_2}_{-s} \cong \oplus_{t \in }^{\Z/(\bar{k}/2)\Z}
  \left( W^{\U(1)_{k\bar{k}/2}} _{-(2sh + sk/2) + 2k t} \otimes
  W^{\U(1)_{\bar{k}}}_{2sh -s + 4t} \right);
  \label{eq:dcmp-U1U1-to-U1U1-kEven}
\end{align}
we read off how the integral points of the charge lattice of
$\U(1)_k\times \U(1)_2$ show up as rational points in that of
$\U(1)_{k\bar{k}/2}\times \U(1)_{\bar{k}}$ through 
\begin{align*}
  \left( \begin{array}{c} q^\perp/(k\bar{k}) \\ -m/(2\bar{k}) \end{array} \right)
  = \left(\begin{array}{cc} & 2/\bar{k} \\ 1 & 2/\bar{k} \end{array} \right)
  \left( \begin{array}{c} 2sh/(2k) \\ -2 sh/(2k) - s/4 \end{array} \right). 
\end{align*}
The RHS of (\ref{eq:dcmp-U1U1-to-U1U1-kEven}) as a whole remains the same
when we change $2sh$ by $+2k$ [resp. $s$ by $+4$] and replace $t$ by
$t+\bar{k}/2$ [resp. $t+1$]. 

To complete the second step, we read $W^{\U(1)_{\bar{k}}}_{2sh-s+4t}$
in (\ref{eq:dcmp-U1U1-to-U1U1-kOdd}, \ref{eq:dcmp-U1U1-to-U1U1-kOdd-2})
as $W^{\U(1)_{\bar{k}}}_{-m}$, where
$4t \in 4\Z/4\bar{k}\Z$ is set $-2sh + s-m$ by choosing an appropriate
lift $m\in \Z/4\bar{k}\Z$ of $m \in \Z/2\bar{k}\Z$; this difference
in $-m$ by $+2\bar{k}$ does not matter when substituted to
the combination $+4t \; k(k+3)$ mod $+4k\bar{k}\Z$
in (\ref{eq:dcmp-U1U1-to-U1U1-kOdd-2}) because $(k+3)$ is even. 
So, when $k$ is odd, the following result
\begin{align}
  {\cal L}^\ell_{m,s} \cong \oplus_{sh \in }^{-\ell/2 + \Z/k\Z}
  \left( W^{\U(1)_{2k\bar{k}}}_{-mk(k+3)-2sh (k+1)(k+2) + sk\bar{k}}
  \otimes P^k_{sh,\ell/2}, \right)
  \label{eq:dcmp-coset-to-U1-para-kOdd}
\end{align}
in \cite[(2.18)]{Gepner:1987qi} (cf also \cite[p.189]{Blumenhagen:2009zz}, \cite[(7.81)]{Eguchi:2015book}) is reproduced. 
When $k$ is even, on the other hand, $t \in \Z/(\bar{k}/2)\Z$
(and $4t \in 4\Z/2\bar{k}\Z$) is read off by setting
$W^{\U(1)_{\bar{k}}}_{2sh-s+4t}$ in (\ref{eq:dcmp-U1U1-to-U1U1-kEven})
as $W^{\U(1)_{\bar{k}}}_{-m}$ in
(\ref{eq:dcmp-N=2-minimalM-SVOA-repr-2-bosCst-reprs});
\begin{align}
  &  {\cal L}^\ell_{m,s} \cong \oplus_{sh \in }^{\frac{s}{2}-\frac{m}{2} + 2\Z/k\Z}
  \left( W^{\U(1)_{k\bar{k}/2}}_{ -mk/2 -sh \bar{k}} \otimes P^k_{sh,\ell/2}\right) .
  \label{eq:dcmp-coset-to-U1-para-kEven}
\end{align}
The expressions on the RHS of (\ref{eq:dcmp-coset-to-U1-para-kOdd},
\ref{eq:dcmp-coset-to-U1-para-kEven}) remain the same when we change
$\ell \rightarrow k-\ell$, $m \rightarrow m+\bar{k}$ and $s \rightarrow s+2$;
this works as a consistency check with the relation (\ref{eq:field-id}). 

At the end of this preparation, we use the result
(\ref{eq:dcmp-coset-to-U1-para-kOdd}, \ref{eq:dcmp-coset-to-U1-para-kEven})
to solve the following inverse problem: when the decomposition
of ${\cal L}^{\ell}_{m,s}$ involves $W^{\U(1)_{2k\bar{k}}}_{q^\perp} \otimes
P^k_{sh,\ell/2}$ [resp. $W^{\U(1)_{k\bar{k}/2}}_{q^\perp} \otimes P^k_{sh,\ell/2}$]
when $k$ is odd [resp. even], find out other coset-VOA representations 
${\cal L}^{\ell + \Delta \ell}_{m+\Delta m, s+\Delta s}$ that yield 
$W^{\U(1)_{2k\bar{k}}}_{q^\perp}$ [resp. $W^{\U(1)_{k\bar{k}/2}}_{q^\perp}$] 
along with some $P^k_{sh+\Delta sh, \ell'/2}$. When $k$ is odd, solutions
are parametrized by $\Delta \sigma \in \Z/2\Z$:  
\begin{align}
  \Delta (2sh) = k (\Delta \sigma) \in \Z/2k\Z, \quad
  \Delta m = \bar{k} (\Delta \sigma) \in \Z/2\bar{k}\Z,  \quad
  \Delta s = 2 (\Delta \sigma) \in \Z/4\Z, 
 \label{eq:temp-cosVOAreps-4-q=0-kOdd}
\end{align}
so $\ell'$ mod 2 is equal to $\Delta \sigma + \ell$ mod 2; odd $\ell$'s
and even $\ell$'s are exchanged when $\Delta \sigma$ is odd. 
The pair of ${\cal L}^\ell_{m,s}$'s (and the corresponding pair of
$P^k_{sh,\ell/2}$'s) sharing $W^{\U(1)_{2k\bar{k}}}_{q^\perp}$, parametrized by
$\Delta \sigma \in \Z/2\Z$, is in fact one and the same representation
of the coset VOA (\ref{eq:coset-VOA}) (of the $\Z_k$ parafermion VOA)
because of (\ref{eq:field-id}) and (\ref{eq:field-id-para}). 
So, the charge $q^\perp$ of $\U(1)_{2k\bar{k}}$ uniquely specifies
the parafermion representation $P^k_{sh,\ell/2}$ it is coupled to,
and also the coset VOA representation ${\cal L}^\ell_{m,s}$ from which it arises.

For an even $k$, solutions are parametrized by
$\Delta \sigma \in \Z/4\Z$:
\begin{align}
 \Delta (2sh) = \frac{k}{2} (\Delta \sigma)  \in \Z/2k\Z, \quad 
 \Delta m = - \frac{\bar{k}}{2} (\Delta \sigma) \in \Z/2\bar{k}\Z, \quad
 \Delta s = - \Delta \sigma \in \Z/4\Z.
 \label{eq:temp-cosVOAreps-4-q=0-kEven}
\end{align}
%
% which is with $(\Delta s) +2\Z = (\Delta \sigma) +2\Z \in \Z/2\Z$.
When $\Delta \sigma$ is even, both $\Delta m$ and $\Delta \ell$
are even; when $\Delta \sigma$ is odd, however,
\begin{align*}
  \left\{ \begin{array}{lll}
    4|k &  \Delta \ell \quad {\rm even}, & \Delta m \quad {\rm odd}, \\
    4|\bar{k} & \Delta \ell \quad {\rm odd}, & \Delta m \quad {\rm even}.
    \end{array} \right.
\end{align*}
The variation by $\Delta \sigma \in 2\Z/4\Z$ corresponds to
the identical coset VOA representation and parafermion representation
due to (\ref{eq:field-id}, \ref{eq:field-id-para}), similarly to the
case $k$ is odd. 

\vspace{5mm}

{\bf Structure of the Hilbert spaces:}
The left-mover chiral algebras ${\cal A}^+$ of both the minimal tensor model
and the Gepner models contain the VOA
\begin{align}
  {\cal A}^+_{\rm U(1)-paraF} := 
  \otimes_{i=1}^r \left( \U(1)_{2^{2[k_i]-1}k_i\bar{k}_i} \otimes
  ({\rm parafermion}_{k_i}) \right),
\end{align}
and their right-mover chiral algebras also contain a copy of this VOA. 
Here, $[k_i] := 1$ if $k_i$ is odd and $[k_i]:= 0$ if $k_i$ is even. 
Instead of regarding the Hilbert spaces ${\cal H}_{\rm tot}^{\rm post}$ of those
two CFTs as representation spaces of ${\cal A}_{\U(1){\rm -paraF}}^+$ and
$\widetilde{\cal A}_{\U(1){\rm -paraF}}^+$, we may regard them as representation
spaces of the two modular-covariant CFTs on  
\begin{align*}
  {\cal H}_{\U(1)} = \otimes_{i=1}^r \left(\U(1)_{2^{2[k_i]-1}k_i\bar{k}_i} \otimes \widetilde{\U(1)}_{2^{2[k_i]-1}k_i\bar{k}_i}\right), \quad
  {\cal H}_{\rm paraF} = (P^{\vec{k}}_{\vec{0},\lambda/2} \otimes
  \widetilde{P}^{\vec{k}}_{\vec{\tilde{0}},\tilde{\lambda}/2})^{\oplus L_{\lambda,\tilde{\lambda}}}.  
\end{align*}
Both the minimal tensor model and the Gepner models should have the structure
\begin{align}
 {\cal H}_{\rm tot}^{\rm post} \cong \oplus \left( \left( {\rm repr~or~}{\cal H}_{\U(1)} \right) \otimes \left( {\rm repr~of~}{\cal H}_{\rm paraF} \right) \right). 
 \label{eq:dcmp-HtotPst-LattCFTrep-paraFcftRep}
\end{align}

Given the Hilbert space ${\cal H}_{\rm tot}^{\rm post}$ of a CFT, one can
define the subspaces ${\cal H}_{\U(1)}^{\rm max}$
[resp. ${\cal H}_{\rm paraF}^{\rm max}$] by collecting all the states in
${\cal H}_{\rm tot}^{\rm post}$ that are coupled to the ground state
of ${\cal H}_{\rm paraF}$) [resp. to the ground state of ${\cal H}_{\U(1)}$]. 
This is similar to the way left-moving and right-moving chiral algebras
are determined from the Hilbert space of a 2D CFT. 
It is plausible that
the structure theory in \cite[\S4]{Moore:1988ss}---for a CFT with rational
left-mover and
right-mover chiral algebras there (reviewed in
p. \pageref{pg:Moore-natural-rev-4chiral})---can be recycled to the present
situation for ${\cal H}_{\U(1)}^{\rm max}$ and ${\cal H}_{\rm paraF}^{\rm max}$;
this means that there is a one-to-one correspondence
$\pi: I_{\U(1)}^{\rm max} \rightarrow I_{{\rm paraF}}^{\rm max}$ of the set
of all the distinct irreducible representations of ${\cal H}_{\U(1)}^{\rm max}$
and ${\cal H}_{\rm paraF}^{\rm max}$, and the Hilbert space has the
structure
\begin{align}
  {\cal H}_{\rm tot}^{\rm post} \cong ({\cal H}_{\U(1)}^{\rm max} \otimes
  {\cal H}_{\rm paraF}^{\rm max}) \oplus
  \oplus_{a \in I_{\U(1)}^{\rm max} \backslash 0} ({\cal H}_{\U(1)}^{\rm max}{\rm repr}_a)
  \otimes ({\cal H}_{\rm paraF}^{\rm max}{\rm repr}_{\pi(a)}). 
  \label{eq:dcmp-HtotPst-LattCFTrep-paraFcftRep-byMax}
\end{align}
The maximally extended ${\cal H}_{\U(1)}^{\rm max}$ is still a lattice CFT,
which is determined by the signature $(r,r)$ lattice
$LL[\vec{0},\vec{\tilde{0}}]$ of charges $(\vec{q}^\perp, \vec{\tilde{q}}^\perp)$
of $W_{\vec{q}^\perp}^{\U(1)} \otimes \widetilde{W}_{\vec{\tilde{q}}^\perp}^{\U(1)}$
coupled to ${\cal H}_{\rm paraF}$ in ${\cal H}_{\rm tot}^{\rm post}$.   

So the study of the structure (\ref{eq:dcmp-HtotPst-LattCFTrep-paraFcftRep},
\ref{eq:dcmp-HtotPst-LattCFTrep-paraFcftRep-byMax}) is reduced to the study
of (a) the full charge lattice $LL[\vec{0},\vec{\tilde{0}}]$, (b) of the list
of $(\vec{sh}, \vec{\widetilde{sh}})$ of the parafermion representations
coupled to $W_{\vec{0}}^{\U(1)} \otimes \widetilde{W}_{\vec{\tilde{0}}}^{\U(1)}$
in ${\cal H}_{\rm tot}^{\rm post}$, and (c) of the list of
all the $(\vec{sh}, \vec{\widetilde{sh}})$ that appear in
${\cal H}_{\rm tot}^{\rm post}$. The study of (a) and (b) is similar, in spirit,
to the study of the left-mover and right-mover chiral algebras ({\it maximally
  extended} VOAs) of a given CFT. The list of irreducible representations
$I_{\U(1)}^{\rm max}$ is the same as the discriminant group $LL[\vec{0},\vec{\tilde{0}}]^\vee/LL[\vec{0},\vec{\tilde{0}}]$ of the lattice
$LL[\vec{0},\vec{\tilde{0}}]$.
The one-to-one map $\pi$ is not hard to find in the present context,  
because the parameter $(\vec{sh},\vec{\widetilde{sh}})$ is used to label
the representations on both sides. 

%
% Whenever we know that a ${\cal H}_{\rm paraF}$-representation
% $(P^{\vec{k}}_{\vec{sh},\lambda/2}\otimes
% \widetilde{P}^{\vec{k}}_{\vec{\widetilde{sh}},\tilde{\lambda}/2})^{\oplus L_{\lambda, \tilde{\lambda}}}$ ever survives in
% ${\cal H}_{\rm tot}^{\rm post}$, the list of $(\vec{q}^\perp,
% \vec{\tilde{q}}^\perp)$ coupled to this parafermion representation in
% ${\cal H}_{\rm tot}^{\rm post}$ must be 
% %
% \begin{align}
%   LL[\vec{sh}, \vec{\widetilde{sh}}] = LL[\vec{0},\vec{\tilde{0}}] +
%   \left( - \bar{k}_i sh_i \left\{ \begin{array}{ll} 2(k_i+1) & {\rm odd}, \\
%     1 & {\rm even} \end{array} \right\} ,
%   - \bar{k}_i \widetilde{sh}_i \left\{ \begin{array}{ll} 2(k_i+1) & {\rm odd},
%     \\  1 & {\rm even} \end{array} \right\} \right).
%   \label{eq:charge-latt-shift-law}
% \end{align}
% %

\vspace{5mm}

{\bf When it comes to the minimal tensor model} with
$\{ (k_i, R_i)_{i=1,\cdots, r} \}$, we have a complete answer to the
questions (a--c).  
Let us begin with the description of the charge lattice
$LL[\vec{0},\vec{\tilde{0}}]_{\rm m.t.m.}$.  Here are some preparations.
The lattice of charges
$(\vec{q}^\perp,\vec{\tilde{q}}^\perp)$ coupled to ${\cal H}_{\rm paraF}$
in $({\cal H}_{\rm tot}^{\rm post})_{\rm m.t.m.}$ of course contains
that of ${\cal H}_{\U(1)}$, i.e., the signature $(r,r)$ lattice 
\begin{align}
  L_{00}' \oplus L'_{00}[-1], \qquad
  L_{00}' := \oplus_{i=1}^r \langle +2^{2[k_i]}k_i\bar{k}_i \rangle.  
\end{align}
The symbol $L[-1]$ for a lattice $L$ is the same as $L$ as a free abelian
group, while the bilinear form is multiplied by $-1$.  
The positive definite lattice $L'_{00}$ here is an index-$2^{[k_i]}$
sublattice\footnote{
It was fine to deal with an integral lattice in
section \ref{ssec:CFTop-by-sc-ext}, as we dealt with {\it S}VOAs there;
we need an {\it even} lattice here, however, because we deal with VOAs. 
} 
of $L_{00}^{\vec{0}}$ in (\ref{eq:def-lattice-4left-00y},
\ref{eq:def-lattice-4left-bilinF}). The generator of the rank-1 lattice
$\langle + 2^{2[k_i]}k_i\bar{k}_i \rangle$ [resp.
  $\langle - 2^{2[k_i]}k_i\bar{k}_i \rangle$] in the left-mover
[resp. right-mover] VOA is denoted by $e_i$ [resp. $\tilde{e}_i$];
the generator of the dual lattice $\langle \pm 2^{2[k_i]}k_i\bar{k}_i
\rangle^\vee$ dual to $e_i$ [resp. $\tilde{e}_i$] is denoted by
$e_i^\vee$ [resp. $\tilde{e}_i^\vee$]. 

The charge lattice $LL[\vec{0},\vec{\tilde{0}}]_{\rm m.t.m.}$
of the minimal tensor model with $\{ (k_i, R_i)\}$ contains more elements
in $(L'_{00} \oplus L'_{00}[-1])^\vee$ beyond $L'_{00} \oplus L'_{00}[-1]$;
the additional charges $(\vec{q}^\perp, \vec{\tilde{q}}^\perp)$ include 
\begin{align}
  \left(\sum_j 2^{[k_j]-1} k_j(e_j^\vee + \tilde{e}_j^\vee)\right), \quad
  2^{[k_i]} k_i(e_i^\vee + \tilde{e}_i^\vee)_{i=1,\cdots, r},
  \label{eq:charge-latt-mtm-commonGen}
\end{align}
where the 1st element is for $m_j = \bar{m}_j = s_j = \tilde{s}_j = -1$
for all $j$ and with even $\ell_j$ and $\tilde{\ell}_j$; the
remaining family of elements above labeled by $i \in \{1,\cdots, r\}$
are for $m_j=\bar{m}_j= s_j = \tilde{s}_j = -2 \delta_{j,i}$ with even
$\ell_j$ and $\tilde{\ell}_j$; they are all in the untwisted sector
of the $(\Z/2\Z)^{r-1}_{\rm qnt.}$-orbifold from the ``original theory''
in section \ref{ssec:mtm-crit-SCFT} (the tensor product of the $r$
minimal model Type 0 SCFTs).
In the $(\Z/2\Z)^{r-1}_{\rm qnt.}$-twisted sectors, we should still have
$m_i = \tilde{m}_i$ for all $i=1,\cdots, r$, but $s_i$ and $\tilde{s}_i$
can differ by $2\Z/4\Z \subset \Z/4\Z$ at each $i$, so long as
$s_{\rm tot}= \tilde{s}_{\rm tot} \in \Z/4\Z$. At $i \in \{1,\cdots, r\}$
where $k_i$ is even, however, $sh_i = \widetilde{sh}_i$ (and $m_i=\tilde{m}_i$)
implies that $s_i = \tilde{s}_i \in \Z/4\Z$. Therefore,
$LL[\vec{0},\vec{\tilde{0}}]_{\rm m.t.m.}$ also contains elements of the form 
\begin{align}
  2k_i \bar{k}_i (e_i^\vee - e_{i+1}^\vee)_{i=1,2,\cdots, r_{\rm odd}-1}
  \label{eq:charge-latt-mtm-GenIfOdd}
\end{align}
where we sorted $\{1,\cdots, r\}$ so that $k_j$ are odd for the first
$r_{\rm odd}$ of those; in particular, there is none of this kind if all $k_i$'s
are even.

The maximally extended charge lattice $LL[\vec{0},\vec{\tilde{0}}]_{\rm m.t.m.}$
of the U(1) CFT computed above is larger by 
\begin{align}
  [LL[\vec{0},\vec{\tilde{0}}]_{\rm m.t.m.} : L'_{00} \oplus L'_{00}[-1] ]
  = \left\{ \begin{array}{ll}
    \prod_{i=1}^r (2^{2[k_i]}\bar{k}_i), & {}^\exists k \; k_k {\rm ~is~odd}, \\
    2\prod_{i=1}^r \bar{k}_i, & {}^\forall i \; k_i {\rm ~is~even}.
    \end{array} \right. 
\end{align}
Now we know the maximally extended U(1) CFT,
$({\cal H}_{\U(1)}^{\rm max})_{\rm m.t.m.}$. 
The fact that $[(L'_{00} \oplus L'_{00}[-1])^\vee :
  (L'_{00}\oplus L'_{00}[-1])] = \prod_{i=1}^r (2^{2[k_i]}k_i\bar{k}_i)^2$ is
used to find that there are
\begin{align}
  [LL[\vec{0},\vec{\tilde{0}}]_{\rm m.t.m.}^\vee :
    LL[\vec{0},\vec{\tilde{0}}]_{\rm m.t.m.} ] = \left\{ \begin{array}{ll}
    \prod_{i=1}^r k_i^2 & {}^\exists k, \; k_k {\rm ~is~odd}, \\
    2^{-2} \prod_{i=1}^r k_i^2 & {}^\forall i, \; k_i {\rm ~is~even}
  \end{array} \right.
  \label{eq:temp-mtm-Nmbr-irrep-LattCFT}
\end{align}
distinct irreducible representations of the lattice CFT
$({\cal H}_{\U(1)}^{\rm max})_{\rm m.t.m.}$.

The parafermion representations coupled to
$(W_{\vec{0}}^{\U(1)} \otimes \widetilde{W}_{\vec{\tilde{0}}}^{\U(1)})$
in the minimal tensor model Hilbert space (the question (b)) are
\begin{align}
  \left\{ \begin{array}{ll}
    \otimes_{i=1}^r (P^{k_i}_{0,\ell_i/2} \otimes
    \widetilde{P}^{k_i}_{0,\tilde{\ell}_i/2})^{\oplus L_{\ell_i,\tilde{\ell}_i}}  &
              {}^\exists j \; k_j {\rm ~is~odd}, \\
    \otimes_{i=1}^r \left[ \left(\oplus_{\sigma_i \in}^{\Z/4\Z} (
      P^{k_i}_{\sigma_i k_i/2,\ell_i/2} \otimes \widetilde{P}^{k_i}_{\sigma_i k_i/2, \tilde{\ell}_i/2} )^{\oplus L_{\ell_i,\tilde{\ell}_i}} \right)_{/2} \right]_{\sigma_i-\sigma_j \in 2\Z/4\Z} & {}^\forall i \; k_i {\rm ~is~even}. \end{array} \right.
  \label{eq:mtm-paraF-max}
\end{align}
The result above is obtained by exploiting
(\ref{eq:temp-cosVOAreps-4-q=0-kOdd}, \ref{eq:temp-cosVOAreps-4-q=0-kEven}).
Indeed, we first note from (\ref{eq:temp-cosVOAreps-4-q=0-kOdd}, \ref{eq:temp-cosVOAreps-4-q=0-kEven}) that $\sigma_i = \tilde{\sigma}_i$ in $\Z/2\Z$
when $k_i$ is odd and $\sigma_i = \tilde{\sigma}_i$ in $\Z/4\Z$
when $k_i$ is even (for each of $i =1,\cdots, r$), because
only the coset-VOA representations ${\cal L}^{\ell_i}_{m_i,s_i}\otimes
\widetilde{\cal L}^{\tilde{\ell}_i}_{\bar{m}_i,\tilde{s}_i}$ with $m_i = \bar{m}_i
\in \Z/2\bar{k}_i\Z$ show up in the Hilbert space of the minimal tensor model. 
All the parafermion representions coupled to $W^{\U(1)}_{\vec{0}}\otimes
\widetilde{W}_{\vec{\tilde{0}}}^{\U(1)}$ therefore originate from the untwisted
sector of the $(\Z/2\Z)^{r-1}_{\rm qnt.}$-orbifold ($s_i = \tilde{s}_i$ for
$i=1,\cdots r$). The subscript $\sigma_i-\sigma_j \in 2\Z/4\Z$ in the case
all $k_i$ are even is a reminder that all the states in the minimal tensor
model are in the NS--NS representations in all the $r$ tensored SCFTs,
or in the R--R representations in all the $r$ tensored SCFTs, not mixed up.
The subscript $/2$ is a reminder that one and the same pair of left-mover
and right-mover coset VOA representations show up $2^r$ times
in (\ref{eq:partFcn-m.t.m.+Mkwsk-type0-1},
\ref{eq:partFcn-m.t.m.+Mkwsk-type0-2}) accompanied by the factor $2^{-r}$
because of the field identification (\ref{eq:field-id}); this
redundant description still remains as the field identification
(\ref{eq:field-id-para}) among the left-mover and right-mover parafermion
representations. So, we have the factor $2^{-1}$ at each of the $r$ 
minimal model SCFTs. The same reasoning applied to a minimal tensor
model simplifies when one of $k_i$'s (say, $k_j$) is odd: 
in that $j$-th minimal model, the coset-VOA representations
that contain the charge $(q_j^\perp, \tilde{q}_j^\perp ) = (0,0)$
are NS--NS representations, so we can restrict our attention to 
$\sigma_i \in 2\Z/4\Z \subset \Z/4\Z$ even in the $i$-th minimal model
where $k_i$ is even. 
We can then remove the redundancy $/2$ completely by choosing one
representative from the equivalent expressions of one representation
(\ref{eq:field-id-para}). The maximally extended parafermion CFT
$({\cal H}_{\rm paraF}^{\rm max})_{\rm m.t.m.}$
in the minimal tensor model Hilbert space agrees with ${\cal H}_{\rm paraF}$
when at least one of $k_i$'s is odd; on the other hand, it is larger than
${\cal H}_{\rm paraF}$ when all $k_i$'s are even. 

Finally, we state the result of the list of
${\cal H}_{\rm paraF}$-representations that ever show up in
$({\cal H}_{\rm tot}^{\rm post})_{\rm m.t.m.}$ coupled to some
$(W^{\U(1)}_{\vec{q}^\perp} \otimes \widetilde{W}^{\U(1)}_{\vec{\tilde{q}}^\perp})$;
this is the question (c). If at least one of $k_i$'s is odd,
the orbits of such parafermion representations are grouped into
representations of $({\cal H}_{\rm paraF}^{\rm max})_{\rm m.t.m.} =
{\cal H}_{\rm paraF}$ labeled by
\begin{align}
  \prod_{i=1}^r \left\{ \begin{array}{ll}
    {\rm odd~}k_i: &  \{(sh_i, \widetilde{sh}_i) \in (\Z/k_i\Z)^2 \} \quad
    [(\ell_i,\tilde{\ell}_i) \; {\rm both~even}], \\
    {\rm even~}k_i: & \{ (sh_i,\widetilde{sh}_i) \in (\Z/k_i\Z)^2 \}_{/2}
    \amalg \{ (sh_i,\widetilde{sh}_i) \in (1/2,1/2)+(\Z/k_i\Z)^2 \}_{/2}.
  \end{array} \right.
  \label{eq:mtm-paraF-reprs-someKiOdd}
\end{align}
The subscript $/2$ in the $i$-th factor for an even $k_i$ is a reminder
that the representation of $(P^{k_i}_{0,\ell_i/2}\otimes
\widetilde{P}^{k_i}_{0,\tilde{\ell}_i/2})^{\oplus L_{\ell_i,\tilde{\ell}_i}}$ for
$(sh_i,\widetilde{sh}_i)$ and that of $(sh_i,\widetilde{sh}_i) + (k_i/2,k_i/2)$
are the same, so we should avoid double counting. This double counting
is avoided already for the $i$-th factor with an odd $k_i$
by restricting attention to the representations with integral $sh_i$, 
$\widetilde{sh}_i$, $\ell_i/2$ and $\tilde{\ell}_i/2$. All of those parafermion
representations show up in the Hilbert space of the minimal tensor model
as a part of the structure (\ref{eq:dcmp-HtotPst-LattCFTrep-paraFcftRep}).
So, there are $\prod_i k_i^2$ representations of
$({\cal H}_{\rm paraF}^{\rm max})_{\rm m.t.m.}$.

In a minimal tensor model where all $k_i$ are even, on the other hand, 
the representations of ${\cal H}_{\rm paraF}$ (of the form of the block 
$\otimes_i (P^{k_i}_{sh_i,\ell_i/2}\otimes
\widetilde{P}^{k_i}_{\widetilde{sh}_i,\tilde{\ell}_i/2})^{\oplus L_{\ell_i,\tilde{\ell}_i}}$)
that show up in the minimal tensor model Hilbert space are with
$(\vec{sh},\vec{\widetilde{sh}})$ in 
\begin{align}
 \left( \prod_i \left[(\Z/k_i\Z)^2 \amalg \left((1/2,1/2) + (\Z/k_i\Z)^2\right)\right] \right)_{\sum_i (sh_i - \widetilde{sh}_i) \in 2\Z}; 
\end{align}
the constraint that $\sum_i (sh_i-\widetilde{sh}_i)$ is even in
$({\cal H}_{\rm tot}^{\rm post})_{\rm m.t.m.}$ is due to the fact that
$sh_i - \widetilde{sh}_i \equiv (s_i/2-m_i/2)-(\tilde{s}_i/2-\tilde{m}_i/2)
= (s_i-\tilde{s}_i)/2$ mod $+2\Z$ for each $i$ in the minimal tensor model
with even $k_i$, and because of the constraint $s_{\rm tot} = \tilde{s}_{\rm tot}$
in $\Z/4\Z$ (the latter constraint $s_{\rm tot} = \tilde{s}_{\rm tot}$
does not restrict the variety of combinations of $\vec{sh}$ and
$\vec{\widetilde{sh}}$ when at least one of $k_i$ is odd).
Those $2^{-1} \prod_{i=1}^r (2k_i^2)$ blocks of representations of
${\cal H}_{\rm paraF}$ are grouped into the representations
of $({\cal H}_{\rm paraF}^{\rm max})_{\rm m.t.m.}$, which consists of
$2 \cdot \prod_i 2$ blocks. So, overall $2^{-2} \prod_i k_i^2$ representations
of $({\cal H}_{\rm paraF}^{\rm max})_{\rm m.t.m.}$ show up in
the Hilbert space of the minimal tensor model
$({\cal H}_{\rm tot}^{\rm post})_{\rm m.t.m.}$. 

To recap, the maximally extended U(1) CFT in
$({\cal H}_{\rm tot}^{\rm post})_{\rm m.t.m.}$ is with the charge lattice
$LL[\vec{0},\vec{\tilde{0}}]_{\rm m.t.m.}$
in (\ref{eq:charge-latt-mtm-commonGen}, \ref{eq:charge-latt-mtm-GenIfOdd}),
and the maximally extended     
parafermion CFT is $({\cal H}_{\rm paraF}^{\rm max})_{\rm m.t.m.}$; 
both are modular covariant but not necessarily invariant.
The number of distinct representations of the
$LL[\vec{0},\vec{\tilde{0}}]_{\rm m.t.m.}$ CFT and the number of distinct
representations of $({\cal H}_{\rm paraF}^{\rm max})_{\rm m.t.m.}$ have
been studied independently; the
former (\ref{eq:temp-mtm-Nmbr-irrep-LattCFT}) agrees with the latter. 
That they agree is consistent with the expected structure 
(\ref{eq:dcmp-HtotPst-LattCFTrep-paraFcftRep-byMax}).

\vspace{5mm}

{\bf A Gepner model SCFT} of a given minimal tensor model with $\{ (k_i,R_i)\}$
comes with additional data $(\Gamma, \epsilon, p_0)$. We have a complete
answer to the questions (a--c) on the Gepner model Type 0 SCFTs (with any
$\Gamma, \epsilon, p_0$) for any minimal tensor model where all $k_i$'s are
odd. For a minimal tensor model with at least one even $k_i$, we do not have
a systematic solution to the question (b, c).
% 
% and hence the structure
% (\ref{eq:dcmp-HtotPst-LattCFTrep-paraFcftRep}) of the Gepner model Hilbert
% spaces $({\cal H}_{\rm tot}^{\rm post})_{\rm Gepn.}$ under the maximally
% extended lattice- and parafermion CFTs.  
% of the Gepner model $({\cal H}_{\U(1)}^{\rm max})_{\rm Gepn.}$ and
% $({\cal H}_{\rm paraF}^{\rm max})_{\rm Gepn.}$.

Let us begin with {\bf the question (b)}. In addressing the question (b),
where we list up $(\vec{sh}, \vec{\widetilde{sh}})$
that are compatible with $(\vec{q}^\perp, \vec{\tilde{q}}^\perp) =
(\vec{0},\vec{0})$, note that there are $\otimes_i
({\cal L}^{\ell_i}_{m_i,s_i}\otimes
\widetilde{\cal L}^{\tilde{\lambda}_i}_{\tilde{m}_i,\tilde{s}_i})$ with
$m_i \neq \tilde{m}_i \in \Z/2\bar{k}_i\Z$ in Gepner model CFTs
(any $s_i - \tilde{s}_i \in 2\Z/4\Z$ is acceptable
already in a minimal tensor model). Therefore, $\sigma_i$ and
$\tilde{\sigma}_i$ in $\Z/2\Z$ [resp. $\Z/4\Z$] (if $k_i$ is odd [resp. even])
in the $i$-th minimal model can be chosen almost independently. Because
$m_i - \tilde{m}_i = 2y_i \in \Z/2\bar{k}_i\Z$ in the Gepner models on
one hand, and $m_i - \tilde{m}_i$ is tied with $\sigma_i - \tilde{\sigma}_i$
through (\ref{eq:temp-cosVOAreps-4-q=0-kOdd},
\ref{eq:temp-cosVOAreps-4-q=0-kEven}) on the other hand, we can find out
which twisted sector has a chance to yield a solution
$(\vec{sh},\vec{\widetilde{sh}})$ with $\sigma_i \neq \tilde{\sigma}_i$.

Think of Gepner models of a minimal tensor model {\bf where all $k_i$ are odd},
first. 
All the solutions to $(\vec{sh}, \vec{\widetilde{sh}})$ come from the
untwisted sector then; to see this, note that 
$m_i-\tilde{m}_i$ should be even because $2y_i \in 2\Z/2\bar{k}_i\Z$, while
$m_i-\tilde{m}_i$ is $-\bar{k}_i (\sigma_i-\tilde{\sigma}_i)$ for
$\sigma_i, \tilde{\sigma}_i \in \Z/2\Z$; because $\bar{k}_i$ is odd,
$\sigma_i = \tilde{\sigma}_i$ and $y_i = 0\in \Z/\bar{k}_i\Z$. 
Within the untwisted sector, all the solutions
$(\vec{sh}, \vec{\widetilde{sh}})$ in the minimal tensor model
% and hence $({\cal H}_{\rm paraF}^{\rm max})_{\rm m.t.m.}=
% {\cal H}_{\rm paraF}$, 
survive the orbifold projection, % as they are the minimum ${\cal H}_{\rm paraF}$.
so, $({\cal H}_{\rm paraF}^{\rm max})_{\rm Gepn.} = {\cal H}_{\rm paraF} =
({\cal H}_{\rm paraF}^{\rm max})_{\rm m.t.m.}$ for all possible
$(\Gamma, \epsilon, p_0)$ so long as all $k_i$ are odd. 

Let us get the question (c) done in the case all $k_i$ are odd. 
As $({\cal H}_{\rm paraF}^{\rm max})_{\rm Gepn.} = {\cal H}_{\rm paraF} =
(P^{\vec{k}}_{\vec{0},\lambda/2}\otimes
\widetilde{P}^{\vec{k}}_{\vec{\tilde{0}},\tilde{\lambda}/2})^{
  \oplus L_{\lambda,\tilde{\lambda}}}$, it is enough to
keep in mind the representations of $({\cal H}_{\rm paraF}^{\rm max})_{\rm Gepn.}$
of the form $(P^{\vec{k}}_{\vec{sh},\lambda/2}\otimes
\widetilde{P}^{\vec{k}}_{\vec{\widetilde{sh}},\tilde{\lambda}/2})^{\oplus L_{\lambda,\tilde{\lambda}}}$ for $(\vec{sh}, \vec{\widetilde{sh}})$ in
$\prod_{i=1}^r \{ (\Z/k_i\Z)^2 \amalg (1/2,1/2)+(\Z/k_i\Z)^2 \}$. 
In fact, it is enough to keep in mind only those with
$\prod_{i=1}^r (\Z/k_i\Z)^2$ when all $k_i$ are odd because
of (\ref{eq:field-id-para}). All of those representations of
$({\cal H}_{\rm paraF}^{\rm max})_{\rm Gepn.}$---for any one in
$\prod_i (\Z/k_i\Z)^2$---show up in
$({\cal H}_{\rm tot}^{\rm post})_{\rm Gepn.}$ in combination with some
appropriate $W^{\U(1)}_{\vec{q}^\perp}\otimes
\widetilde{W}^{\U(1)}_{\vec{\tilde{q}}^\perp}$, because one
${\cal L}^{\lambda}_{\vec{m},\vec{s}}\otimes \widetilde{\cal L}^{\tilde{\lambda}}_{\vec{\tilde{m}},\vec{\tilde{s}}}$ with even $\ell_i$ and $\tilde{\ell}_i$ 
give rise to $(P^{\vec{k}}_{\vec{sh},\lambda/2} \otimes \widetilde{P}^{\vec{k}}_{\vec{\widetilde{sh}},\tilde{\lambda}/2})$ in $({\cal H}_{\rm tot}^{\rm post})_{\rm Gepn.}$
indiscriminately in $(\vec{sh}, \vec{\widetilde{sh}}) \in \prod_i (\Z/k_i\Z)^2$. 
To summarize, $({\cal H}_{\rm paraF}^{\rm max})_{\rm Gepn.}$ is the same
as $({\cal H}_{\rm paraF}^{\rm max})_{\rm m.t.m.}$, and all the $\prod_i k_i^2$
different representations of
$({\cal H}_{\rm paraF}^{\rm max})_{\rm Gepn.}=
({\cal H}_{\rm paraF}^{\rm max})_{\rm m.t.m.}$ do show up in the Gepner model
CFTs $({\cal H}_{\rm tot}^{\rm post})_{\rm Gepn.}$---for any
$(\Gamma,\epsilon, p_0)$---as well as in the minimal tensor model 
$({\cal H}_{\rm tot}^{\rm post})_{\rm m.t.m.}$. The orbifolds converting the
minimal tensor model to the Gepner models with $(\Gamma, \epsilon, p_0)$
keep the parafermion CFT of the minimal tensor model intact. 

For a minimal tensor model {\bf where at least one $k_i$ is even},
neither is this conclusion generally true, nor do we have a comprehensive
solution to the questions (b, c). Even in the question (b), some
representations of ${\cal H}_{\rm paraF}$ in $({\cal H}_{\rm paraF}^{\rm max}
)_{\rm m.t.m.}$ do not necessarily remain in
$({\cal H}_{\rm paraF}^{\rm max})_{\rm Gepn.}$, some representations of
${\cal H}_{\rm paraF}$ that did not show up in
$({\cal H}_{\rm tot}^{\rm post})_{\rm m.t.m.}$ may show up in
$({\cal H}_{\rm paraF}^{\rm max})_{\rm Gepn.}$, and even the number of
representations of $({\cal H}_{\rm paraF}^{\rm max})_{\rm Gepn.}$ appearing
in $({\cal H}_{\rm tot}^{\rm post})_{\rm Gepn.}$ is not necessarily the
same as the number of representations of
$({\cal H}_{\rm paraF}^{\rm max})_{\rm m.t.m.}$ appearing in
$({\cal H}_{\rm tot}^{\rm post})_{\rm m.t.m.}$. It is enough to look at
a few simplest Gepner constructions that result in supersymmetric
compactifications (i.e., with data in ${\cal S}_{\rm II}$)
: $\{ (k_i, R_i)\} = \{ (2,A_3)^{\oplus 2} \}$
with $\Gamma = \Z/4\Z$ generaged by $\vec{y}_*$ (for the Jacobi
elliptic curve) and
$\{ (k_i,R_i)\} = \{ (2,A_3)^{\oplus 4} \}$ with $\Gamma = \Z/4\Z$ generated
by $\vec{y}_*$ (for the Fermat quartic K3 surface). The orbifold processes
from a minimal tensor model to Gepner model CFTs modify the parafermion
sector in general, not just the U(1) sector, if at least one of $k_i$ is even. 

To see where the difference is from, let us try to address the question (b)
a little further. For the $i$-th minimal model where $k_i$ is odd,
we do have $\sigma_i = \tilde{\sigma}_i \in \Z/2\Z$,
and this freedom in $\Z/2\Z$ common to both the left-mover and right-mover
is essentially the uninteresting one
(the field identification (\ref{eq:field-id}, \ref{eq:field-id-para}), as we
have seen already). For the $i$-th minimal model where $k_i$ is even,
on the other hand, it is possible that $\sigma_i -\tilde{\sigma}_i =
2 \epsilon_i \in 2\Z/4\Z$ for $\epsilon_i \in \Z/2\Z$, when
$m_i - \tilde{m}_i = -\bar{k}_i \epsilon_i \in \Z/2\bar{k}_i\Z$ should
be equal to $2y_i \in 2\Z/2\bar{k}_i\Z$, and $sh_i - \widetilde{sh}_i
= \epsilon_i k_i/2  \in \Z/k_i\Z$. Therefore, a representation of
${\cal H}_{\rm paraF}$ with such $(\vec{sh}, \vec{\widetilde{sh}})$ may be
contained in $({\cal H}_{\rm paraF}^{\rm max})_{\rm Gepn.}$, when the orbifold group
$\Gamma$ contains a 2-torsion element ($\vec{y} \in \Gamma$ s.t.
$2\vec{y} = 0 \in \Gamma$). Whether those states do survive in
$({\cal H}_{\rm paraF}^{\rm max})_{\rm Gepn.}$ depends on details of orbifold
projection (\ref{eq:cond-Gepner-survive-Z-w-d.t.},
\ref{eq:cond-Gepner-survive-Z-w-d.t.for-R}) and also
on (\ref{eq:cond-Gepner-sTot-vs-sTilTot}) if $p_0 \neq p_0^S$. 

Let us move on to {\bf the question (a)}, which is to determine the list
$LL[\vec{0},\vec{\tilde{0}}]_{\rm Gepn.}$ of U(1) charges
$(\vec{q}^\perp, \vec{\tilde{q}}^\perp)$ of
$(W^{\U(1)}_{\vec{q}^\perp} \otimes \widetilde{W}^{\U(1)}_{\vec{\tilde{q}}})$
coupled to ${\cal H}_{\rm paraF}$ in the Hilbert space $({\cal H}_{\rm tot}^{\rm post})_{\rm Gepn.}$. The set $LL[\vec{0},\vec{\tilde{0}}]_{\rm Gepn.}$ of U(1)
charges becomes a lattice of signature $(r,r)$ under the bilinear form
on $L_{00}^{'\vee} \oplus L'_{00}[-1]^\vee$.  It is useful to note that
the homomorphism $p_0: \Gamma \rightarrow \Z/2\Z$ is trivial (hence
$\epsilon^{00} = \epsilon^{01} = \epsilon$) when all $k_i$ are odd;
this is because all the elements in $\Gamma$ would be of odd order, while
$0\in \Z/2\Z$ is the only element of that kind. 

Before writing down $LL[\vec{0}, \vec{\tilde{0}}]_{\rm Gepn.}$ directly,
let us list up the U(1)-charges $Q\widetilde{Q}_y^\perp \subset
L_{00}^{'\vee} \oplus L'_{00}[-1]^\vee$ that show up
in combination with $(P^{\vec{k}}_{\vec{0},\lambda/2}\otimes \widetilde{P}^{\vec{k}}_{\vec{\tilde{0}},\tilde{\lambda}/2})^{\oplus L_{\lambda,\tilde{\lambda}}}$
in the decomposition of 
${\cal L}^\lambda_{\vec{\tilde{m}}+2y, \vec{s}} \otimes \widetilde{\cal L}^{\tilde{\lambda}}_{\vec{\tilde{m}},\vec{\tilde{s}}}$ with non-zero
$L_{\lambda,\tilde{\lambda}}$, $s_{\rm tot}-\tilde{s}_{\rm tot} = 2p_0(y) \in \Z/4\Z$,
$\tilde{s}_i-\tilde{s}_j \in 2\Z/4\Z$ for any pair $i,j$. It follows
from this that all $\ell_i$ and $\tilde{\ell}_i$ are even; 
for the $i$-th minimal model where $k_i$ is even, 
$s_i = [m_i] \in \Z/4\Z$ and $\tilde{s}_i = [\tilde{m}_i] \in \Z/4\Z$.
As a trial expression for
$Q\widetilde{Q}^\perp_{\vec{y}}$, think of $Q\widetilde{Q}^{\perp {\rm temp}}_{\vec{y}}$
for each $\vec{y} \in \Gamma$ given by 
\begin{align}
  \left\{ \left.
  \zeta := \delta \zeta - \sum_i e_i^\vee y_i k_i
  \left\{ \begin{array}{l} 2(k_i+3) \\ 1 \end{array} \right\}
  + e^\vee_{{}^\exists {\rm odd}} p'_0(\vec{y}) 2k_{\rm odd}\bar{k}_{\rm odd}
  \; \right| \;  {}^\forall \delta \zeta \in LL[\vec{0},\vec{\tilde{0}}]_{\rm m.t.m.} \right\}; 
  \label{eq:temp-charges-upstair}
\end{align}
in the second term of $\zeta$, $2(k_i+3)$ shoud be used when $k_i$ is odd
while $1$ be used when $k_i$ is even; the 3rd term of $\zeta$
is meant to pick up one of $j \in \{1,\cdots, r\}$ where $k_j$ is odd
and add just $e^\vee_j p'_0(\vec{y}) 2k_j\bar{k}_j$ without a sum over $j$, 
unless all $k_i$ are even. Here, 
\[
p'_0(\vec{y}) := p_0(\vec{y}) - (\sum_i^{(k_i {\rm ~even})} y_i) \in \Z/2\Z
\]
for $\vec{y} \in \Gamma$ (here, $y_i$ mod 2 is well-defined for
$y_i +\bar{k}_i\Z$ when $k_i$ is even), whose motivation is explained below. 
The ambiguity in $y_i +\bar{k}_i\Z \subset \Z$ or choice $j \in
\{1,\cdots, r_{\rm odd}\}$ only results in the ambiguity of $\zeta$ in
$LL[\vec{0},\vec{\tilde{0}}]_{\rm m.t.m.}$, so the set
of charges $Q\widetilde{Q}_{\vec{y}}^{\perp {\rm temp}}$ is well-defined for
$y \in \Gamma$. The second term in $\zeta$ is meant to capture
$m_i - \tilde{m}_i = 2y_i$.

When the underlying minimal model has only even $k_i$'s,
the contributions to $(\vec{q}^\perp, \vec{\tilde{q}}^\perp)$ with
$(\vec{sh},\vec{\widetilde{sh}}) = (\vec{0},\vec{\tilde{0}})$ should come
from the coset-VOA representations with $s_{\rm tot}-\tilde{s}_{\rm tot} = \sum_i
([m_i]-[\tilde{m}_i]) = 2 (\sum_{i=1}^r y_i)$ mod $+4\Z$,
so the condition $s_{\rm tot}-\tilde{s}_{\rm tot} = 2p_0(\vec{y}) \in \Z/4\Z$ 
implies that $Q\widetilde{Q}^{\perp}_{\vec{y}}$
is $Q\widetilde{Q}^{\perp {\rm temp}}_{\vec{y}}$ (without the 3rd term of $\zeta$)
indeed for $\vec{y} \in \Gamma$ in the kernel of the homomorphism $p'_0:
\Gamma \rightarrow \Z/2\Z$, while $Q\widetilde{Q}^{\perp}_{\vec{y}} =0$ for
all other $\vec{y} \in \Gamma$. 

When at least one of $k_i$'s are odd in the underlying minimal tensor model,
the contributions with $(\vec{sh}, \vec{\widetilde{sh}}) =
(\vec{0}, \vec{\tilde{0}})$ should be from the coset-VOA representations
with $\sum_i^{\rm even} (s_i-\tilde{s}_i) = 2 \sum_i^{\rm even}y_i$, and
$\sum_i^{\rm odd}(s_i-\tilde{s}_i) = 2p'_0(\vec{y})$ in $2\Z/4\Z$. 
Because the charge lattice $LL[\vec{0},\vec{\tilde{0}}]_{\rm m.t.m.}$ of
the minimal tensor model is determined by contributions with
$\sum_i^{\rm odd}(s_i - \tilde{s}_i) = 0 \in \Z/4\Z$ 
[and $\sum_i^{\rm even}(s_i-\tilde{s}_i) = 0\in \Z/4\Z$], 
the third term in $\zeta$ is meant to adjust
$\sum_i^{\rm odd}(s_i - \tilde{s}_i)$ by $+2p'_0(\vec{y})+4\Z$,
along with the corresponding charges
$(\vec{q}^{\perp}, \vec{\tilde{q}}^{\perp})$. 

The charge lattices $LL[\vec{0},\vec{\tilde{0}}]_{\rm Gepn.}$ of the Gepner
model CFTs are the subsets of $\oplus_{\vec{y} \in \Gamma} Q\widetilde{Q}^{\perp}_{\vec{y}}$
that correspond to $\vec{\tilde{m}}$ and $\vec{y}$ (with $\vec{m}
=\vec{\tilde{m}}+2\vec{y}$) subject to the orbifold projection condition
(\ref{eq:cond-Gepner-survive-Z-w-d.t.},
\ref{eq:cond-Gepner-survive-Z-w-d.t.for-R}). We claim that
\begin{align}
  LL[\vec{0},\vec{\tilde{0}}]_{\rm Gepn.} & = \oplus_{y\in \Gamma} \left\{ \left. 
  \zeta \in Q\widetilde{Q}^\perp_{\vec{y}} \; \right|
     {}^\forall \vec{x} \in \Gamma,   \;\;      
     (X_{\vec{x}} , \zeta) =_{+\Z} \sum_i \frac{x_iy_i}{\bar{k}_i} + 
     \frac{{\rm Arg}(\epsilon^{00}(\vec{x},\vec{y}))}{2\pi}  \right\};
  \label{eq:charge-latt-Gepn}
\end{align}
here, for $\vec{x} \in \Gamma \subset \prod_{i=1}^r \Z/\bar{k}_i\Z$,
we set $X_{\vec{x}} \in (L_{00}^{'\vee} \oplus L'_{00}[-1]^\vee)$ by 
\begin{align}
  X_{x} := \sum_i  \tilde{e}_i^\vee x_i k_i \left\{ \begin{array}{l}
    2(k_i+3) \\ 1 \end{array} \right\}  \quad
  + \tilde{e}_{{}^\exists {\rm even}}^\vee \bar{k}_{\rm even} p'_0(\vec{x}); 
\end{align}
in the first term, the coefficient $2(k_i+3)$ should be used when $k_i$ is odd,
and $1$ when $k_i$ is even; the ambiguity in $x_i +\bar{k}_i\Z$
changes $X_{\vec{x}}$ in $(L'_{00} \oplus L'_{00}[-1])$, so
$(X_{\vec{x}}, \zeta) +\Z$ remains unchanged (well-defined).
The second term is meant to pick one of $i \in \{1,\cdots, r\}$ where
$k_i$ is even if there is any, and add that term for that $i$ without
a sum over $i$; its role
is to detect $\delta \zeta \in LL[\vec{0},\vec{\tilde{0}}]_{\rm m.t.m.}$
associated with the Ramond--Ramond representation, because
\begin{align}
  \left( \tilde{e}_{\rm even}^\vee \bar{k}_{\rm even}, \; \sum_j 2^{[k_j]-1} k_j
  (e^\vee_j+\tilde{e}_j) \right) = - \frac{1}{2} ;  
\end{align}
if all $k_i$ are odd in a minimal tensor model, we just forget about
the second term of $X_{\vec{x}}$; $p'_0(\vec{x}) = p_0(\vec{x})
- \sum_i^{\rm even}x_i = 0$ anyway. 

To see that the condition on $\zeta$ is equivalent to the orbifold
projection conditions (\ref{eq:cond-Gepner-survive-Z-w-d.t.},
\ref{eq:cond-Gepner-survive-Z-w-d.t.for-R}), let us compute
$(X_x,\delta \zeta)$. The contributions to $(X_{\vec{x}}, \delta \zeta) + \Z$
where $k_i$ is odd are 
\begin{align*}
(X_{\vec{x}}, \delta \zeta)^{\rm odd} & \; = 
  \sum_i^{(k_i {\rm ~odd})} \left( 2 x_i k_i(k_i+3) \right) \frac{-1}{4k_i\bar{k}_i}
  \left( -\bar{m}_i k_i(k_i+3)
  + \tilde{s}_i k_i\bar{k}_i \right), \nonumber \\
  & \; = \sum_i^{\rm odd} x_i k_i(k_i+3)\frac{1}{2\bar{k}_i}  \bar{m}_i (k_i+3)
    \nonumber \\
  & \; = \sum_i^{\rm odd} x_i \bar{m}_i \frac{\bar{k}_i-2}{2\bar{k}_i}
    (k_i+3)^2 = - \sum_i^{\rm odd} x_i \bar{m}_i  \frac{(k_i+3)^2}{\bar{k}_i}
  = - \sum_i^{\rm odd} \frac{x_i \bar{m}_i}{\bar{k}_i},
        \nonumber 
\end{align*}
all of those equalities are mod $+\Z$; 
we used $4|2(k_i+3)$ in the 2nd equality to drop the
$\tilde{s}_i$-dependent term, and also $2|(k_i+3)$ in the 4th equality
to drop $\bar{k}_i$ from $k_i = \bar{k}_i-2$. 
The contributions from $i \in \{1,\cdots, r\}$ where $k_i$ is even is 
\begin{align*}
  (X_{\vec{x}}, \delta \zeta)^{\rm even} & \; = 
  - \frac{p'_0(\vec{x})}{2}_{\rm if~RR} +  \sum_i^{(k_i {\rm ~even})} k_i x_i
  \frac{-1}{k_i\bar{k}_i} \left( -\bar{m}_i \frac{k_i}{2} \right) =
   - \frac{p'_0(\vec{x})}{2}_{\rm if~RR} +  \sum_i^{\rm even} 
   \frac{x_i\tilde{m}_ik_i}{2\bar{k}_i} \\
   & = - \frac{p'_0(\vec{x})}{2}_{\rm if~RR} 
   + \frac{\sum_i^{\rm even} x_i \tilde{m}_i}{2}
   - \sum_i^{\rm even} \frac{x_i\tilde{m}_i}{\bar{k}_i}; 
\end{align*}
because $\tilde{m}_i$ (and $\tilde{s}_i$) are even [resp. odd]
for an even $\tilde{\ell}_i$ (and $\widetilde{sh}_i=0$) in the NS--NS
[resp. RR] sector,
\begin{align*}
  (X_{\vec{x}}, \delta \zeta)^{\rm even} = -\frac{p_0(\vec{x})}{2}_{\rm if~RR}
  - \sum_i^{\rm even}\frac{x_i \tilde{m}_i}{\bar{k}_i}. 
\end{align*}
Now, the orbifold projection codintion (\ref{eq:cond-Gepner-survive-Z-w-d.t.},
\ref{eq:cond-Gepner-survive-Z-w-d.t.for-R}) is reproduced by substituting
$(X_{\vec{x}},\delta \zeta)^{\rm odd} + (X_{\vec{x}},\delta \zeta)^{\rm even}$ to
the condition in (\ref{eq:charge-latt-Gepn}). 

Now, we have all the necessary information to describe the maximally
extended U(1) CFT within $({\cal H}_{\rm tot}^{\rm post})_{\rm Gepn.}$: 
the charge lattice $LL[\vec{0},\vec{\tilde{0}}]_{\rm Gepn.}$.
The set of charges
$LL[\vec{0},\vec{\tilde{0}}]_{\rm Gepn.}$ is a free abelian group indeed,
because the shift in $\zeta$---the 2nd and 3rd terms
in (\ref{eq:temp-charges-upstair})---is additive
(mod $L'_{00} \oplus L'_{00}[-1]$) with respect to $\vec{y} \in \Gamma$, and
the inhomogeneous term in the projection condition
in (\ref{eq:charge-latt-Gepn}) is also additive (mod $\Z$) with
respect to $\vec{y} \in \Gamma$. Once an even lattice $\Lambda$ is given, then
the lattice CFT is given uniquely; a cocycle factor in $H^2(\Lambda,S^1)$
necessary for the construction of the vertex operator of a lattice CFT is
determined by the bilinear form on $\Lambda$; this statement
from \cite[\S5.4, \S5.4]{MR1417941} is valid not just for a (holomorphic) VOA,
but also for a (non-holomorphic) CFT \cite[\S8]{Polchinski:1998rq}. 

\vspace{5mm}

{\bf Comparison between $({\cal H}_{\rm tot}^{\rm post})_{\rm m.t.m.}$ and
  $({\cal H}_{\rm tot}^{\rm post})_{\rm Gepn.}$:}
The charge lattice $LL[\vec{0},\vec{\tilde{0}}]_{\rm Gepn.}$ of
a Gepner model CFT contains the lattice $(L'_{00} \oplus L'_{00}[-1])$,
and is contained in $(L'_{00} \oplus L'_{00}[-1])^\vee$;
neither the lattices $LL[\vec{0},\vec{\tilde{0}}]_{\rm Gepn.}$ of
Gepner models derived from a minimal tensor model
contain the lattice $LL[\vec{0},\vec{\tilde{0}}]_{\rm m.t.m.}$ nor
in the other way around in general. 

{\bf If all $k_i$'s are odd}, there is one thing in common among the charge
lattices $LL[\vec{0},\vec{\tilde{0}}]_{\rm Gepn.}$ of Gepner models
derived from one minimal tensor model with $\{ (k_i, R_i)\}$, 
despite the variations among $LL[\vec{0},\vec{\tilde{0}}]_{\rm Gepn.}$ for
different choices of $(\Gamma,\epsilon)$. That is, 
\begin{align}
  [LL[00]_{\rm m.t.m.}: L'_{00}\oplus L'_{00}[-1]] =
  [LL[00]_{\rm Gepn.}:  L'_{00}\oplus L'_{00}[-1]]
  \label{eq:temp-GepnChrgLatt-indx-kOdd}
\end{align}
as we verify below, so $[LL[00]_{\rm m.t.m.}^\vee : LL[00]_{\rm m.t.m.}] =
[LL[00]_{\rm Gepn.}^\vee : LL[00]_{\rm Gepn.}]$, and the number of distinct
irreducible representations $\# I_{\U(1)}^{\rm max}$ are the same among all
of those lattice CFTs, and is also equal to $\#I_{\rm paraF}^{\rm max}$
of $({\cal H}_{\rm paraF}^{\rm max})_{\rm m.t.m.}$ and
$({\cal H}_{\rm paraF}^{\rm max})_{\rm Gepn.}$ (as expected in the
discussion around (\ref{eq:dcmp-HtotPst-LattCFTrep-paraFcftRep-byMax})). 
As we have already seen that 
$\#I_{\rm paraF}^{\rm max}$ is common for
a minimal tensor model and Gepner models derived from it (when all $k_i$ are
odd), and that $\#I_{\rm paraF}^{\rm max} = \#I_{\U(1)}^{\rm max}$ in a minimal
tensor model, the rest is to compare $\#I_{\U(1)}^{\rm max}$ between
a minimal tensor model and Gepner models derived from it.

The consistency check (\ref{eq:temp-GepnChrgLatt-indx-kOdd}) can be done
by combining two observations. The first of the two is that
the untwisted sector ($\vec{y} = 0\in \Gamma$) contribution to
$LL[\vec{0},\vec{\tilde{0}}]_{\rm Gepn.}$ is an index-$\#\Gamma$ 
sublattice of 
$LL[\vec{0},\vec{\tilde{0}}]_{\rm m.t.m.} = Q\widetilde{Q}^\perp_{\vec{y}=0}$,
and the other is that the $\vec{y}$-twisted sector contribution to
$LL[\vec{0},\vec{\tilde{0}}]_{\rm Gepn.}$ is a constant shift of
that of the untwisted sector; so, we lose and win the same amount
when comparing $LL[\vec{0},\vec{\tilde{0}}]_{\rm Gepn.}$ against 
$LL[\vec{0},\vec{\tilde{0}}]_{\rm m.t.m.}$, to
have (\ref{eq:temp-GepnChrgLatt-indx-kOdd}) and then the
common value of $\#I_{\U(1)}^{\rm max}$. 

To verify the first observation claimed above, note first that the
pairing between $\delta \zeta \in LL[\vec{0},\vec{\tilde{0}}]_{\rm m.t.m.}$
and $X_{\vec{x}}$ passes through $LL[\vec{0},\vec{\tilde{0}}]_{\rm m.t.m.} \rightarrow
L'_{00}[-1]^\vee$, and also that the image of the projection to
$L'_{00}[-1]^{\vee}$
is spanned by $\{ A (\sum_j \tilde{e}_j^\vee k_j ), \;
2B_i k_i \tilde{e}_i^\vee \; | i=(1), 2,\cdots, r, A, B_i \in \Z\}$
(cf (\ref{eq:charge-latt-mtm-commonGen})).  The pairing (mod $\Z$) is
\begin{align*}
  \sum_i 2x_i k_i (k_i+3) & \; \frac{-1}{4k_i\bar{k}_i} (A + 2B_i)k_i
  = - \sum_i \frac{x_i(A+2B_i)k_i(k_i+3)}{2\bar{k}_i} \\
 & =_{+\Z} \sum_i \frac{x_i(A+2B_i)(k_i+3)}{\bar{k}_i}
  =_{+\Z} \sum_I \frac{x_i(A+2B_i)}{\bar{k}_i} \in \Q/\Z, 
\end{align*}
so the fraction of $\delta \zeta \in LL[\vec{0},\vec{\tilde{0}}]_{\rm m.t.m.}$
that survive in $LL[\vec{0},\vec{\tilde{0}}]_{\rm Gepn.}$ is
the fraction of characters of the group ${\cal G}$ that vanish on the subgroup
$\Gamma  \subset {\cal G}$. The first observation now follows.  

To verify the second observation claimed above, it is enough to
show that there is $\delta \zeta_{\vec{y}}
\in LL[\vec{0},\vec{\tilde{0}}]_{\rm m.t.m.}$
for each $\vec{y} \in \Gamma$ so that the character
$\Gamma \ni \vec{x} \mapsto (X_{\vec{x}}, \delta \zeta_{\vec{y}}) = -
\sum_i x_i \tilde{m}_i/\bar{k}_i \in \Q/\Z$ agrees with the character
$\Gamma \ni x \mapsto \sum_i (x_iy_i/\bar{k}_i) +
    {\rm Arg}(\epsilon^{00}(\vec{x},\vec{y}))/(2\pi) + \Z$.
 Because any character of $\Gamma \subset {\cal G}$ is available as
 restriction of a character of ${\cal G}$, we can find an appropriate
 character $- \sum_i \tilde{m}_ix_i/\bar{k}_i$ with
 $\tilde{m}_i \in \Z/\bar{k}_i\Z$ chosen appropriately for each
 $\vec{y} \in \Gamma$.
 By choosing an appropriate lift $\tilde{m}_i \in \Z/2\bar{k}_i\Z$,
 it is possible to set all $\{\tilde{m}_{i=1,\cdots, r}\}$ even
 (or all odd), so we can find $\delta \zeta_{\vec{y}} \in
 LL[\vec{0},\vec{\tilde{0}}]_{\rm m.t.m.}$ in this way.  

{\bf To recap,} we have managed to determine the maximally extended
lattice subCFT $({\cal H}_{\U(1)}^{\rm max})_{\rm Gepn.} \subset
({\cal H}_{\rm tot}^{\rm post})_{\rm Gepn.}$ for all the Gepner models;
the charge lattice $LL[\vec{0},\vec{\widetilde{0}}]_{\rm Gepn.}$
is determined by (\ref{eq:charge-latt-Gepn}). Its set of irreducible
representations are $I_{\U(1)}^{\rm max} = LL[\vec{0},\vec{\widetilde{0}}]_{\rm Geon.}^\vee/LL[\vec{0},\vec{\widetilde{0}}]_{\rm Gepn.}$, given a structure of
an abelian group, and the fusion algebra is its group ring
$\Z[I_{\U(1)}^{\rm max}]$. The CFT vertex operator within
$({\cal H}_{\U(1)}^{\rm max})_{\rm Gepn.}$ is determined uniquely
by useing the 2-cocycle in $H^2(LL[\vec{0},\vec{\tilde{0}}]_{\rm Gepn.},S^1)$
corresponding to the intersection form of the lattice
$LL[\vec{0},\vec{\tilde{0}}]_{\rm Gepn.}$.

In the case the original minimal tensor model $\{ (k_i, R_i)\}$ is only
with odd $k_i$'s, we have also determined the maximally extended parafermion
subCFT $({\cal H}_{\rm paraF}^{\rm Gepn.}) \subset
({\cal H}_{\rm tot}^{\rm post})_{\rm Gepn.}$, and also identified a complete list
of irreducible representations of $({\cal H}_{\rm paraF}^{\rm max})_{\rm Gepn.}$.
The whole structure (\ref{eq:dcmp-HtotPst-LattCFTrep-paraFcftRep-byMax})
has been identified. We have further confirmed that
($\star$1)
$({\cal H}_{\rm paraF}^{\rm max})_{\rm Gepn.}$ and its irreducible representations
remain the same as those of the original minimal tensor model
$({\cal H}_{\rm paraF}^{\rm max})_{\rm m.t.m.} = {\cal H}_{\rm paraF}$
with $I_{\rm paraF}^{\rm max}$ in 1-to-1 with
$(\vec{sh},\vec{\widetilde{sh}}) \in \prod_{i=1}^r (\Z/k_i\Z)^2$
(see (\ref{eq:mtm-paraF-max}, \ref{eq:mtm-paraF-reprs-someKiOdd})), ($\star$2)
the irreducible representations of $({\cal H}_{\U(1)}^{\rm max})_{\rm Gepn.}$
are also labeled by the same set of $(\vec{sh},\vec{\widetilde{sh}})$,
whose charges are 
\begin{align}
   LL[\vec{sh}, \vec{\widetilde{sh}}] = LL[\vec{0},\vec{\tilde{0}}] +
   \left( - \bar{k}_i sh_i  2(k_i+1)  ,
   - \bar{k}_i \widetilde{sh}_i  2(k_i+1)  \right), 
 %  \label{eq:charge-latt-shift-law}
\end{align}
simply shifted from $LL[\vec{0},\vec{\tilde{0}}]_{\rm Gepn.}$, 
and ($\star$3) each of the charge weight
lattice $LL[\vec{sh},\vec{\widetilde{sh}}]_{\rm Gepn.}$ of a Gepner model
SCFT (for any $(\Gamma,\epsilon)$) and
$LL[\vec{sh},\vec{\widetilde{sh}}]_{\rm m.t.m.}$ of the original
minimal tensor model with the same $(\vec{sh},\vec{\widetilde{sh}})$
share an index-$\#\Gamma$ sublattice of both. Overall, ($\star$4)
the orbifold process from $({\cal H}_{\rm tot}^{\rm post})_{\rm m.t.m.}$ to
$({\cal H}_{\rm tot}^{\rm post})_{\rm Gepn.}$ only touches the U(1) charge
weight lattices, while keeping the parafermion part intact. 
When one of the $k_i$ of the minimal tensor model is even,
the properties ($\star$1--4) are not generally true. 

\vspace{5mm}

{\bf The CFT vertex operator:}
The main question in this section \ref{ssec:CFTop-by-bos-lattice} is
how to determine the CFT vertex operator on the whole space
$({\cal H}_{\rm tot}^{\rm max})_{\rm Gepn.}$ acting on itself. 
In the presence of a structure
(\ref{eq:dcmp-HtotPst-LattCFTrep-paraFcftRep-byMax}) in
$({\cal H}_{\rm tot}^{\rm post})_{\rm m.t.m.}$ and
$({\cal H}_{\rm tot}^{\rm max})_{\rm Gepn.}$, the full CFT vertex operator 
\begin{align}
  Y_{\rm tot} = \left\{ Y_{\rm tot}^{(b+c;b,c)} \; | \;
  b, c \in I_{\U(1)}^{\rm max} \right\}
\end{align}
is of the form $Y_{\rm tot}^{(b+c;b,c)} =
  (Y_{{\rm tot}.\U(1)}^{{\rm max}})_{b,c}^{b+c} \otimes
  (Y_{{\rm tot}.{\rm paraF}}^{\rm max})_{\pi(b),\pi(c)}^{\pi(b+c)}$ for
the minimal tensor model and the Gepner models alike. Note that
the fusion constants are either 1 or 0 in the fusion algebra
$\Z[I_{\U(1)}^{\rm max}]$ of the lattice CFT, and the fusion algebra of
$({\cal H}_{\rm paraF}^{\rm max})$ must be isomorphic to it through $\pi$.
So, the direct tensor-product form above does not lose generality. 
What we referred to as the CFT vertex operator of the subCFT
$({\cal H}_{\U(1)}^{\rm max})_{\rm Gepn./m.t.m.}$ corresponds to
$(Y_{{\rm tot.}\U(1)}^{\rm max})_{0,0}^0$ in the notation here. 

For Gepner model CFTs derived from a minimal tensor model with
only odd $k_i$'s, we have identified the structure
(\ref{eq:dcmp-HtotPst-LattCFTrep-paraFcftRep-byMax}) for the Gepner models
and for the original minimal tensor model already.
Because the orbifold process does not touch the parafermion sector,
and also because we have one common parametrization
$(\vec{sh},\vec{\widetilde{sh}})$ for the sets
$(I_{\rm paraF}^{\rm max})_{\rm m.t.m.} \cong
({\cal I}_{\rm paraF}^{\rm max})_{\rm Gepn.}$, 
the operators $(Y_{{\rm tot}.{\rm paraF}}^{\rm max})_{\pi(b),\pi(c)}^{\pi(b+c)}$
for the minimal tensor model should be used as they are
in the Gepner model SCFTs (regardless of $(\Gamma,\epsilon)$). 
For a given $b, c \in (I_{\U(1)}^{\rm max})_{\rm Gepn.} \cong
(I_{\rm paraF}^{\rm max})_{\rm Gepn.}$, the remaining quesetion is to determine
an appropriate choice of $(Y_{{\rm tot}.\U(1)}^{\rm max})_{b,c}^{b+c}$
from the 1-dimensional vector space of intertwining operators
of CFT($LL[\vec{0},\vec{\tilde{0}}]_{\rm Gepn.}$), when we already know
the choice of intertwining operators
$(Y_{{\rm tot}.\U(1)}^{\rm max})_{b,c}^{b+c}$ of the
$LL[\vec{0},\vec{\tilde{0}}]_{\rm m.t.m.}$-CFT so that
$Y_{\rm tot}^{\rm m.t.m.} = Y_{{\rm tot.}\U(1)}^{\rm m.t.m.} \otimes
Y_{{\rm tot.paraF}}^{\rm m.t.m.}$ is invariant under the skew-symmetry and fusion
isomorphisms. The intertwining operator
$(Y_{{\rm tot}.\U(1)}^{\rm max})_{b,c}^{b+c}$ of the Gepner model should be chosen
from the 1-dimensional vector space so that it agrees with that of the
minimal tensor model on the states whose charges are shared by the two
theories, e.g., in $LL[(\vec{sh},\vec{\widetilde{sh}})_b]_{\rm m.t.m.} \cap
LL[(\vec{sh},\vec{\widetilde{sh}})_b]_{\rm Gepn.}$.

%%%%%%%%%%%%%%%%%%%%%%%%%%%%%%%%%%%%%%%%%%%%%%%%%%%%
% \subsection*{Acknowledgements}  % in BrE
 \subsection*{Acknowledgments}   % in AmE
%%%%%%%%%%%%%%%%%%%%%%%%%%%%%%%%%%%%%%%%%%%%%%%%%%%%%

The authors thank Masaki Okada for discussions, who participated in the early stage of this research. They also thank K. Hori, K. Kawabata and Y. Tachikawa for useful comments and discussions. 
This work is supported in part by WPI Initiative (KN and TW) and a Grant-in-Aid for Scientific Research (proj. \# 26K07082) (TW), MEXT, Japan.

%%%%%%%%%%%%%%%%%%%%%%%%%%%%%%%%%%%%%%%%%%%%%%%%%%%%%%%%%%%%%%%%%%%%%%%%%%%%%
% \begin{figure}[tbp]
% \begin{center}
%   \includegraphics[width=.8\linewidth]{hyperelliptic_2}  
% \caption{\label{fig:xxxx} 
% }
% \end{center}
% \end{figure}
%%%%%%%%%%%%%%%%%%%%%%%%%%%%%%%%%%%%%%%%%%%%%%%%%%%%%%%%%%%%%%%%%%%%%%%%%%%%%%

%
% \begin{thebibliography}{99}
%

% \end{thebibliography}

\bibliographystyle{alpha}
\bibliography{rationalCFT.bib}
\end{document}